\pdfoutput=1
\documentclass[Master, Submit, ngerman, UKenglish]{scrbook}

\usepackage[texlive=2016]{ubonn-thesis}
\usepackage{ubonn-biblatex}

\usepackage{thesis_defs}
\usepackage{physics}

 \hypersetup{colorlinks=false,
   linkbordercolor=blue,citebordercolor=magenta,urlbordercolor=darkgreen
 }

\graphicspath{%
  {figs/}%
  {figs/cover/}%
}

\ifthenelse{\equal{\ThesisVersion}{Draft}}{%
  \usepackage{background}
  \ifthenelse{\texlive < 2013}{%
    \SetBgContents{DRAFT}
    \SetBgColor{blue!30}
  }{%
    \backgroundsetup{contents=DRAFT, color=blue!30}
  }
}

\begin{document}

\ifthenelse{\equal{\ThesisVersion}{PILibrary}}{%
  \typeout{Document \jobname, Info: PI library version of thesis}
  \input{cover/\ThesisType_Cover}
}{}

\frontmatter

\ifthenelse{\equal{\ThesisType}{Unknown}}{%
  \typeout{Document \jobname, Error: Unknown thesis type - no title page printed}
}{%
  \ifthenelse{\equal{\ThesisType}{Bachelor}}{%
    \typeout{Document \jobname, Info: Bachelor thesis}
    \input{cover/\ThesisType_Title}
  }{%
    \ifthenelse{\equal{\ThesisVersion}{Final} \OR \equal{\ThesisVersion}{PILibrary}}{%
      \typeout{Document \jobname, Info: Final version of a \ThesisType  thesis}
      \input{cover/\ThesisType_Final_Title}
    }{
      \input{cover/\ThesisType_Submit_Title}
      \typeout{Document \jobname, Info: Draft/submission version of a \ThesisType  thesis}
    }
  }
}

\pagestyle{scrplain}


\tableofcontents

\mainmatter
\pagestyle{scrheadings}

\ifthenelse{\equal{\ThesisVersion}{Draft}}{%
  \ifthenelse{\texlive < 2013}{%
    \SetBgContents{}
  }{%
    \backgroundsetup{contents={}}
  }
}{}


\chapter{Introduction}
\label{sec:intro}
\textit{“Not only is the Universe stranger than we think, it is stranger than we can think.”} - Werner Heisenberg

\vspace{0.5cm}
The formation of the universe is explained by cosmological and particle physics models. The field of high energy physics tries to understand the nature and the formation of the universe. It was started in 1897 when the discovery of electron~\cite{electron} let humanity to step into the subatomic scale. Later on, the quantum mechanics and general relativity were established well enough to answer the open questions which pushed the boundaries of the scientific limits further. The two theories which can precisely describe the universe are the Standard Model of particle physics (SM) and the Lambda Cold Dark Matter ($\Lambda$CDM) model of cosmology. These theories are confirmed to high precision by the Large Hadron Collider (LHC) and the Planck Mission~\cite{planck} experiments. The LHC was built from 1998 to 2008~\cite{lhc} to collide high-energy protons to probe the behaviour of the particles.

A great milestone in the history of particle physics was reached when the Higgs boson was discovered by the ATLAS~\cite{higgsatlas} and CMS~\cite{higgscms} experiments at the LHC in 2012. It was the last piece of the Standard Model to be discovered. With this, the Standard Model is considered to be the most fundamental theory, which is accurate and precise in all of its predictions. However, the fact is that it is still not a complete theory because of the limitation of not able to explain some important phenomena. One of them is the instability of the mass of the Higgs boson. The radiative corrections to the mass of the Higgs boson conflict with the experimental value of the Higgs boson mass, which is $\SI{125.7}{\giga\electronvolt}$~\cite{higgsatlas}~\cite{higgscms}. These corrections quadratically diverge up to the cut-off scale of new physics. This is known as the Higgs mass hierarchy problem. This leads physicists to think beyond the scope of the Standard Model. There are various theories which are proposed to explain this phenomenon. Some of these theories predict the existence of an exotic particle called vector-like quark (VLQ) which couples with the Standard Model third generation quarks. VLQ can solve the Higgs mass hierarchy problem by introducing an additional loop correction which cancels out the divergence of these corrections. The search for VLQ is carried out in both ATLAS and CMS data collected so far from $pp$ collisions at the LHC. VLQ can either be produced singly or in pairs. The single VLQ production is a dominant process at high VLQ masses~\cite{wulzer}. In this thesis, the single production of two such types of VLQs, $T$ and $Y$ quark are presented in $T/Y\rightarrow Wb$ analysis.

The main topic discussed in this thesis is the background estimation in the all-hadronic channel of $T/Y\rightarrow Wb$ analysis by using a data-driven method. This thesis is organised as follows. A brief introduction of the theoretical concepts which are important in particle physics is given in Chapter \ref{sec:theory}. A description of the Standard Model is provided and familiarises the reader with the idea of vector-like quarks. In Chapter \ref{sec:lhcandatlas}, an overview of the LHC accelerator and a detailed description of the ATLAS detector are given. Jets and their reconstruction algorithm are discussed in Chapter \ref{sec:jetsandtaggers}. The two different types of jet algorithms are also discussed in detail which are used in this thesis. In Chapter \ref{sec:analysisstrategy}, the strategy used for this analysis is described, including the study of all the possible backgrounds, data preparation for the analysis and the selection of interesting events. The topics essential to physics analyses, such as a description of the data and Monte Carlo samples used and reconstruction of physics objects with the ATLAS detector are also covered. An introduction to a data-driven method called the ABCD method is given in Chapter \ref{sec:abcd}. The ABCD method is used to estimate the multijet background.  An overview of the method and its implementation on the data along with the further approaches to improve the estimate are shown in this chapter. In Chapter \ref{sec:uncertainty_result}, all the uncertainties which are taken into account while performing the multijet estimation are described, which include both statistical and systematic uncertainties. In Chapter \ref{sec:results}, the final result of the multijet estimate, including all the uncertainties is presented. A comparison between the performances of different $W$-taggers and the jet collections are also shown. Finally, in Chapter \ref{sec:conclusion}, a brief summary and the concluding remarks about the research that is documented in this thesis are provided.



\chapter{Theoretical concepts}
\label{sec:theory}

The purpose of this chapter is to provide a brief introduction to the underlying concepts and terminology used in particle physics. An overview of the Standard Model of particle physics is given, which describes the fundamental particles and the forces responsible for their interaction. Some drawbacks of the Standard Model are also discussed. Finally, a new exotic particle called vector-like quark is introduced.
\section{Basic concepts}%
\label{sec:theory:basicconcepts}

\subsection*{Natural units}%
\label{sec:theory:basicconcepts:naturalunits}\index{basicconcepts}
The S.I.\ units are comprised of [\si{\kilogram}, \si{\metre}, \si{\second}], which form a fundamental basis for the measurement of different macroscopic phenomena. However, it is not a common choice for the representation of the properties of sub-atomic particles. The system of units used in particle physics is known as \textit{natural units}. 

In natural units, \si{\planckbar} = \si{\clight} = 1 which means [$\si{\kilogram}$, $\si{\metre}$, $\si{\second}$] can be substituted by [$\si{\electronvolt}$, $\si{\electronvolt^{-1}}$, $\si{\electronvolt^{-1}}$], where $\si{\planckbar}$ is the reduced Planck's constant, $\si{\clight}$ is the speed of light in vacuum. For example, mass of a proton can be expressed as $\SI{1}{\giga\electronvolt}$.~\cite{thomson}

\subsection*{Cross-section}%
\label{sec:theory:basicconcepts:crossection}\index{crossection}
In particle physics, particle interactions such as scattering or decay processes are mainly concerned. Cross-section is a significant observable measured in the experiments. It is defined as the probability of occurrence of a given particle physics process. The cross-section of a given process is theoretically proportional to:

\begin{equation} \label{eqn:theory:basicconcepts:crossection}
	\sigma \propto \int {\lvert \mathcal{M} \rvert}^2 d\rho \,,
\end{equation}
where ${\lvert \mathcal{M} \rvert}^2$ is the square of the matrix element of the process. It is a measure of the transition amplitude from initial to the final state, and $\int d\rho$ is an integral over the phase space \cite{thesis:anji}. 

The cross-section has dimensions of area and is often expressed with \textit{barn} as its units where $\SI{1}{\barn} = \SI{1E-28}{\metre^2}$. The cross-sections for particle physics processes are typically at the higher energy scale, in the range of \textit{picobarn} (\si{\pico\barn}) to \textit{femtobarn} (\si{\femto\barn}), where \SI{1}{\pico\barn} = \SI{1E-12}{\barn} and \SI{1}{\femto\barn} = \SI{1E-15}{\barn} \cite{thomson}. 

\subsection*{Luminosity}%
\label{sec:theory:basicconcepts:Luminosity}\index{Luminosity}
The quantity that measures the ability of a particle accelerator to produce the required number of interactions is called the instantaneous luminosity $\mathcal{L}(t)$. It is defined as the proportionality factor between the number of events per second $\frac{dN}{dt}$ and the cross-section $\sigma$, given by Eqn.\ \ref{eqn:theory:basicconcepts:luminosity}

\begin{equation} \label{eqn:theory:basicconcepts:luminosity}
	\frac{dN}{dt} = \sigma \mathcal{L}(t) \,.
\end{equation} 
It has a dimension of events per time per area and is usually expressed in the units of $\si{\centi\metre^{-2}\sec^{-1}}$. $\mathcal{L}(t)$ also depends on the particle beam parameters and the target properties. The integral of the instantaneous luminosity with respect to time is called the integrated luminosity $\mathcal{L}_{\text int}$, as shown below:
\begin{equation} \label{eqn:theory:basicconcepts:integratedluminosity}
\mathcal{L}_{\text int} = \int \mathcal{L}(t) \, dt \,.
\end{equation}
$\mathcal{L}_{\text int}$ is often used for characterising the performance of a particle accelerator. In particular, collider experiments aim to maximise their integrated luminosities. Higher integrated luminosity means more events available for analysis.~\cite{thomson}

\subsection*{Decay width and Branching fraction}%
\label{sec:theory:basicconcepts:decaywidth}\index{decaywidth}
The \textbf{Decay width $\Gamma$} is a physical observable measured in decay processes. It is inversely proportional to the lifetime $\tau$ of the original particle which decays into daughter particles and is expressed in units of $\si{\electronvolt}$, given by Eqn.\ \ref{eqn:theory:basicconcepts:decaywidth}

\begin{equation} \label{eqn:theory:basicconcepts:decaywidth}
\Gamma = \frac{1}{\tau} \,.
\end{equation} 
Particles usually decay via different processes depending on the allowed interactions between mediator particles. If $\Gamma_{\text{i}}$ denotes the decay width of process i, then the total decay width $\Gamma_{\text{total}}$ can be written as: 

\begin{equation} \label{eqn:theory:basicconcepts:totaldecaywidth}
\Gamma_{\text{total}} = \sum_{i}^{}\Gamma_{\text{i}} \,.
\end{equation} 

\textbf{Branching ratio $\mathcal{B}_{\text{i}}$} of a given process can be defined as the ratio between the decay width $\Gamma_{\text{i}}$ and the total decay width. This can be expressed as:

\begin{equation} \label{eqn:theory:basicconcepts:branchingfraction}
\mathcal{B}_{\text{i}} = \frac{\Gamma_{\text{i}}}{\Gamma_{\text{total}}} \,.
\end{equation}

\subsection*{Center-of-mass energy}%
\label{sec:theory:basicconcepts:centerofmass}\index{centerofmass}
Center-of-mass energy $\sqrt{s}$ is the total amount of energy available to produce new particles in collisions. It is mathematically defined by Eqn.\ \ref{eqn:theory:basicconcepts:centerofmass}

\begin{equation}
\sqrt{s} = \sqrt{\left(\sum_{\text{i}}E_{\text{i}}\right)^{\text{2}} - \left(\sum_{i}\vec{p}_{\text{i}}\right)^{\text{2}}} \,,
\label{eqn:theory:basicconcepts:centerofmass}
\end{equation}
where $E_{\text{i}}$ and $\vec{p}_{\text{i}}$ denote the energy and the momentum vector of particle i respectively. In the case of two colliding beams with the same energy, and neglecting the masses, Eqn.\ \ref{eqn:theory:basicconcepts:centerofmass} reduces to $\sqrt{s} = 2E$, where $E$ is now the beam energy.~\cite{thesis:chris}  

\section{The Standard Model}%
\label{sec:theory:standardmodel}
The Standard Model (SM) of particle physics was developed in the late 1960s to combine all our current understandings of fundamental particles and forces. The SM is a quantum field theory which is described by Lagrangian dynamics. It postulates that all the matter in the universe is made up of fundamental elementary particles called fermions that interact via particle fields mediated by gauge bosons.~\cite{thomson} 

The SM is the theoretical framework used in particle physics to describe and predict the interactions of particles. Some of the predictions made by the SM have been verified to very high accuracy by the experiments. This section outlines the ingredient of the SM and their interactions and provides a brief overview of the Higgs mechanism responsible for the electroweak symmetry breaking of the SM.

\subsection{Particle content}%
\label{sec:theory:standardmodel:particle}
In the Standard Model, fundamental particles are described as the excitation of quantum fields. The fundamental particles are divided into two groups based on a quantum number which represents their intrinsic angular momentum, known as spin. 

The matter is composed of half-integer spin particles called fermions which are constrained by the Pauli exclusion principle. The interactions between the fermions are mediated by integer spin particles called bosons. The energy distribution of bosons is described by the Bose-Einstein statistics~\cite{thomson}. The bosons are responsible for the three interactions described within the SM: the strong interaction, the weak interaction and the electromagnetic interaction, which are described in the next sections. 

\begin{figure}[hbt!]
	\centering
	\includegraphics[width=0.85\linewidth]{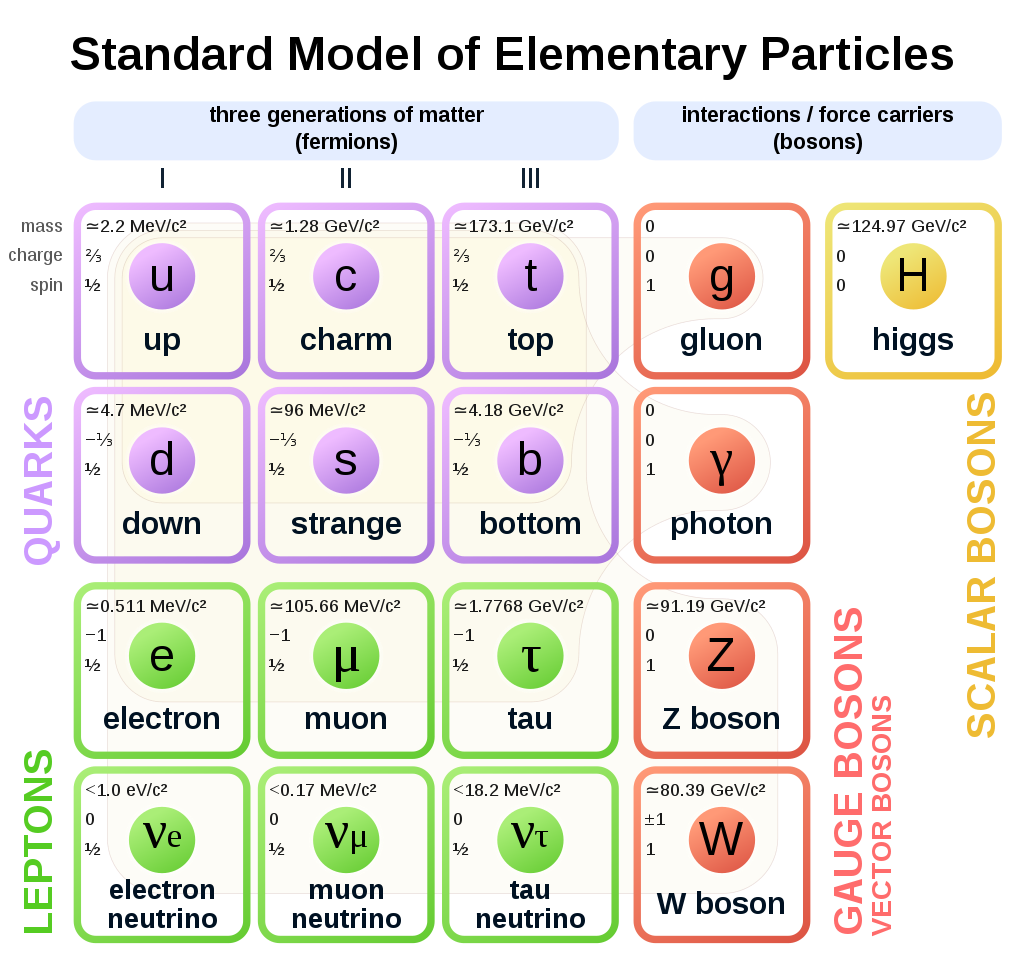}
	\caption{Overview of all the ingredients of the Standard Model which constitutes all the fermions and the gauge bosons. Elementary properties like mass, electric charge and spin of all the particles are also shown in the image.~\cite{standardmodelpicture}}
	\label{fig:theory:standardmodel}
\end{figure}

The energy distribution of fermions is described by the Fermi-Dirac statistics~\cite{thomson}. The fermions are further divided based on their sensitivity to the strong interaction. Fermions that are sensitive to the strong interaction are called quarks; otherwise, they fall into the category known as leptons. There are six types of quarks and leptons each, which are arranged in the form of generations in the SM. First-generation fermions include \Pup, \Pdown, \Pelectron and \Pnue. Second-generation fermions include \Pcharm, \Pstrange, \Pmuon and \Pnum. Third-generation fermions include \Ptop, \Pbottom, $\Ptau^{-}$ and \Pnut. The mass of the particles increases as one goes from first- to third-generation. Fig.\ \ref{fig:theory:standardmodel} shows all the fermions on the left side and the bosons on the right side. All six quarks are shown in purple and leptons in green. The SM also postulates a fundamental scalar boson responsible for the electroweak symmetry breaking within the SM, known as the Higgs boson as shown in yellow. Some elementary properties of these particles like their masses, electric charge and spin are also shown in the picture. The experimental evidence predicts the restriction on the number of generations of the lepton family to be three, but as far as concerned, there is no such restriction on the quark family.

The negative energy solutions of the Dirac equation predict the existence of the antiparticles~\cite{thomson}. So, all the 12 fermions have their antiparticles associated with them which exhibits precisely the same properties but carries an opposite charge of their corresponding particles.

\subsection{Gauge groups}%
\label{sec:theory:standardmodel:gauge}
The Standard Model is formed from the combination of three different local symmetry groups which can be written as~\cite{halzen}:
\begin{equation}
\text{SU(3)}_{\text{c}} \times \text{SU(2)}_{\text{L}} \times \text{U(1)}_{\text{Y}} \,,
\end{equation}
where SU(3)$_{\text{c}}$ symmetry group is generated by the colour charge associated with the strong interaction, described in section \ref{sec:theory:standardmodel:strong}. The SU(2)$_{\text{L}} \times$ U(1)$_{\text{Y}}$ symmetry groups represent the left-handed isospin and hypercharge symmetries within the SM, which are generated by the unified electroweak interaction, described in section \ref{sec:theory:standardmodel:weak}. 

These local symmetries are applied to the SM Lagrangian ($\mathcal{L}_{\text{SM}}$) which correspond to local gauge invariance. The SM Lagrangian can be written as~\cite{halzen}:

\begin{equation}
\mathcal{L}_{\text{SM}} = \mathcal{L}_{\text{Boson}} + \mathcal{L}_{\text{Fermion}} + \mathcal{L}_{\text{Yukawa}} + \mathcal{L}_{\text{Higgs}} \,,
\end{equation}
where $\mathcal{L}_{\text{Boson}}$ + $\mathcal{L}_{\text{Fermion}}$ represents the kinetic energies, electroweak interaction and self-interaction of fermions and bosons. $\mathcal{L}_{\text{Yukawa}}$ is a term from Yukawa coupling which gives mass to the particles. $\mathcal{L}_{\text{Higgs}}$ describes the spontaneous symmetry breaking by Higgs field.

\subsection{Particle interaction}%
\label{sec:theory:standardmodel:interaction}

\subsubsection{Electromagnetic interaction}%
\label{sec:theory:standardmodel:em}
The electromagnetic interaction (EM) is described by one of the Quantum Field Theory (QFT), known as Quantum electrodynamics (QED). It is based on the $\text{U(1)}_{\text{Y}}$ gauge group which represents the interaction between the charged fermions and a massless gauge boson known as a photon. Photon (\Pphoton) is a vector gauge boson which acts as a force carrier of the EM interaction. Only neutrinos (and their antiparticles) do not interact via the EM interaction because they are electrically neutral. $\text{U(1)}_{\text{Y}}$ is abelian in nature which means it is a commuting gauge group that does not allow any self-interaction terms for the photon field.

The QED Lagrangian for a charged fermion ($\psi$) of mass $m$ can be written as:
\begin{equation}
\mathcal{L}_{\text{QED}} = \bar{\psi} \, (i\textbf{D}_{\text{QED}} - m) \, \psi - \frac{1}{4}F_{\mu\nu}F^{\mu\nu} \,,
\end{equation}
where $F$ is the electric field tensor which can be expressed in terms of the photon vector field by $F^{\mu\nu} = \partial^{\mu}A^{\nu} - \partial^{\nu}A^{\mu}$ and $\textbf{D}_{\text{QED}}$ is a covariant derivative which also depends on the photon vector field $A$.~\cite{halzen}

\subsubsection{Strong interaction}%
\label{sec:theory:standardmodel:strong}
The strong interaction is described by a theory called the Quantum Chromodynamics (QCD). It explains the interaction between partons which include quarks and gluons. Gluons (\Pgluon) are vector gauge boson which mediates the strong interaction. Like electric charge in the QED, QCD also introduces a new type of property called colour charge. There are three different types of colour charges: red, blue and green. So, all the six types of quarks and their antiparticles can be represented as colour triplets. Particles carrying a colour charge can feel the effects of the strong interaction. The strong force is described by the $\text{SU(3)}_{\text{c}}$, which is a non-abelian gauge group that means it allows the self-interaction terms for the gluons. Due to the non-abelian nature of the force, it is more complex than the EM force, and give rise to eight different coloured gluons.~\cite{halzen}

The strong interaction is the strongest among all the three interactions described in the SM. One of the consequences of the QCD is the process of hadronisation which means that partons can only exist in bound colour neutral states called hadrons, such as the proton. The strength of the coupling constant of the strong interaction increases with the distance between the coloured particles due to the gluon self-interaction. If one tries to separate the two quarks within a hadron, a large amount of energy is required, and at some point it spontaneously produces a quark-antiquark pair, turning the initial hadron into a pair of hadrons instead of producing an isolated coloured quark. This phenomenon is known as \textit{colour confinement}. The process in which the hadrons are formed from the partons is called hadronisation. It cannot be described using standard perturbation theory since it takes place in the non-perturbative regime of the QCD where the strong coupling is larger than one. The strength of the coupling constant of the strong interaction is low at small distances, so the quarks can exist freely inside the hadrons. This feature of the QCD is known as \textit{asymptotic freedom}.~\cite{thomson}

There are two groups of hadrons: baryons and mesons. Baryons are made up of three quarks or antiquarks while mesons are made from a quark-antiquark pair. The effects of the hadronisation can be seen as clusters in the particle detectors, which are further reconstructed as jets. Jets are the important physics object for this thesis and are discussed in detail in Chapter \ref{sec:jetsandtaggers}.

The QCD Lagrangian for a coloured charged fermion ($\psi$) of mass $m$ can be written as:
\begin{equation}
\mathcal{L}_{\text{QCD}} = \bar{\psi}_{j} \, (i\gamma^{\mu}\textbf{D}_{\text{QCD,}\,\mu} - m_{j}) \, \psi_{j} - \frac{1}{4}G_{\mu\nu}^{\alpha}G^{\mu\nu}_{\alpha} \,,
\end{equation}
where $G_{\mu\nu}^{\alpha}$ is a gluon field tensor which includes the additional self-interaction term of the gluon vector fields and $\alpha$ is denoted as the eight coloured gluons. $\textbf{D}_{\text{QCD,}\mu}$ is a covariant derivative which can be expressed in terms of the gluon gauge field $A_{\mu}^{\alpha}$ and the strong coupling constant $\alpha_{\text{s}}$. Index $j$ is indicated as all the coloured flavours of the quarks.~\cite{halzen}

\subsubsection{Weak interaction and Electroweak unification}%
\label{sec:theory:standardmodel:weak}
The weak interaction is described by a theory called the Quantum Flavourdynamics (QFD). It explains the interaction between particles that is responsible for the radioactive decay of atoms. The Lagrangian of the weak interaction is constructed only from left-handed doublets in the $\text{SU(2)}_{\text{L}}$ gauge group since right-handed particles in the SM appeared as singlets and are not affected by the weak force. The weak force is better understood in terms of the Electroweak theory (EWT).~\cite{halzen}

In the Standard Model, the electromagnetic and weak interactions are unified at an energy of electroweak scale and described by the EWT proposed by Glashow, Salam and Weinberg. The electroweak interaction is described by $\text{SU(2)}_{\text{L}} \times \text{U(1)}_{\text{Y}}$ gauge group where L denotes the SU(2) doublet and singlet representation of left-handed particles and Y is denoted as the weak hypercharge, given as $\text{Y/2}=q-T_{3}$. Here, $T_{3}$ is the third component of the weak isospin and $q$ is electric charge.~\cite{halzen}

The electroweak Lagrangian for fermions can be written as:
\begin{equation}
\mathcal{L}_{\text{EW}} = \bar{\psi}_{f} \, (i\gamma^{\mu}\textbf{D}_{\text{EW,}\,\mu}) \, \psi_{f} \,,
\end{equation}
where the fermionic fields $\psi_{f}$ are organised in $\text{SU(2)}_{\text{L}}$ doublets and singlet, as shown below:

\begin{equation}
\psi_{\text{lepton}} = \begin{pmatrix} \Pnue \\ e \end{pmatrix}_{L}, \begin{pmatrix} \Pnum \\ \mu \end{pmatrix}_{L}, \begin{pmatrix} \Pnut \\ \tau \end{pmatrix}_{L}, e_{R}, \mu_{R}, \tau_{R} \,,
\end{equation}

\begin{equation}
\psi_{\text{quark}} = \begin{pmatrix} u \\ d \end{pmatrix}_{L}, \begin{pmatrix} c \\ s \end{pmatrix}_{L}, \begin{pmatrix} t \\ b \end{pmatrix}_{L}, u_{R}, d_{R}, c_{R}, s_{R}, t_{R}, b_{R} \,.
\end{equation}
The EWT is a parity-violating theory, so right-handed neutrinos are omitted in the SM. There are four vector fields which contribute to the electroweak interaction. A linear combination of $W^{1}_{\mu}$ and $W^{2}_{\mu}$ describes the two charged $W$ bosons, as shown below:
\begin{equation}
W^{\pm}_{\mu} = \frac{1}{\sqrt{2}}(W^{1}_{\mu} \mp iW^{2}_{\mu}) \,.
\end{equation}
And the linear combination of $W^{3}_{\mu}$ and $B_{\mu}$ represents the two neutral bosons, the $Z$ boson and the photon as shown below:
\begin{align}
A_{\mu} = \sin\theta_{W}W^{3}_{\mu} + \cos\theta_{W}B_{\mu} \,, \\
Z_{\mu} = \cos\theta_{W}W^{3}_{\mu} - \sin\theta_{W}B_{\mu} \,,
\end{align}
where $\theta_{W}$ is called the Weinberg angle and is given by $\tan\theta_{W} = \frac{g}{g'} \,,$ and $g$ and $g'$ are the gauge coupling constants.~\cite{halzen}

\subsection{Higgs mechanism}%
\label{sec:theory:standardmodel:higgs}
The addition of mass term to the Lagrangian shown before breaks the gauge symmetry of the model, so a different process is required for particles to become massive. The process which describes how particles acquire their mass, without breaking the gauge invariance of the SM, is called the Higgs mechanism. The Higgs mechanism introduces \textit{spontaneous symmetry breaking} of the $\text{SU(2)}_{\text{L}} \times \text{U(1)}_{\text{Y}}$ gauge group by adding a new complex scalar doublet field, called the Higgs field.~\cite{halzen} 

Let us consider a weak isospin doublet of two complex fields ($\Phi$):
\begin{equation}
\Phi = \begin{pmatrix} \phi^{+} \\ \phi^{o} \end{pmatrix} \,.
\end{equation}
Then the potential $\mathcal{U}(\Phi)$ given by the complex field can be written as:
\begin{equation}
\mathcal{U}(\Phi) = \mu^{2}(\Phi^{*}\Phi) + \lambda(\Phi^{*}\Phi)^{2} \,.
\end{equation}
Only solution with $\lambda>0$ and $\mu^{2}<0$ is considered so that the energy of the potential is bounded below and leads to a non-unique ground state energy. The minimum of this potential is not at the zero value of the field and is given by $v = \sqrt{\mu^{2}/\lambda}$. Now the Lagrangian can be written as~\cite{halzen}:
\begin{equation}
\mathcal{L}_{Higgs} = |\textbf{D}_{\mu}\Phi|^{2} - \mu^{2}(\Phi^{*}\Phi) - \lambda(\Phi^{*}\Phi)^{2} \,.
\end{equation}
By expanding the Lagrangian around the vacuum expectation value ($vev$), a mass term is introduced for the vector bosons. The masses of the bosons predicted by this mechanism are related in the following way:

\begin{align}
m_{\text{W}^{\pm}} = \frac{1}{2}vg \,, \\
m_{\text{Z}} = \frac{1}{2}v\sqrt{g^{2} + g'^{2}} \,, \\
\cos\theta_{W} = \frac{m_{\text{W}}}{m_{\text{Z}}} \,, \\
m_{\text{H}} = \sqrt{2}\mu \,.
\end{align}
The Higgs mechanism
proposes the existence of the Higgs boson ($H$) with $m_{\text{H}}=\SI{125}{\giga\electronvolt}$ which was confirmed in 2012 by the ATLAS and CMS collaborations at CERN.~\cite{higgsatlas}~\cite{higgscms} Fermions also acquire masses through the Higgs mechanism by interactions between the fermionic and Higgs scalar fields, called Yukawa couplings.~\cite{halzen}

\subsection{Feynman diagram}%
\label{sec:theory:standardmodel:feynmandiagram}\index{feynmandiagram}
A Feynman diagram is a graphical representation of the particle interaction. For calculating the probability of the particle interaction to happen, the scattering matrix has to be evaluated. By satisfying the conservation laws of energy and momentum, the scattering matrix ends up with the invariant Matrix Element |$\mathcal{M}$|. A Feynman diagram is a technique which is used to evaluate |$\mathcal{M}$|. It consists of different components which are associated with some mathematical expressions that are defined by Feynman rules.~\cite{thesis:rui}

A Feynman diagram consists of a point, called vertex, and lines attached to the vertex. A vertex represents the interaction of particles, which could either be the emission or absorption of a particle or the flavour change of a particle. There are three different types of lines: internal lines which connect two vertices, also known as a propagator, and the two types of external lines in which first the incoming lines that extend to a vertex representing an initial state and the second when the outgoing lines extend from a vertex representing the final state of the process. The external lines denote the real visible particles (or antiparticles), whereas internal lines denote virtual particles. The main difference between the two types of particles is that virtual particles do not obey the energy-momentum relation, whereas the real particles obey. 

Each vertex represents a point of interaction which could either be one of the three interactions. The strength of the interaction is given by the coupling constant $g$, which is shown for all the three interactions below:

\[\text{EM: }g = qe \,, \hspace{2cm} \text{Weak: }g = g_{\text{W}} \,, \hspace{2cm} \text{Strong: }g = \sqrt{\alpha_{\text{s}}} \,, \]
where $q$ is denoted as the electric charge, $g_{\text{W}}$ is represented as the coupling constant of the weak interaction and $\alpha_{\text{s}}$ is denoted as dimensionless strong interaction coupling constant. At each vertex, the conservation of energy, momentum, angular momentum, charge, lepton number and baryon number should be followed.

\subsection{Drawbacks of the Standard Model}%
\label{sec:theory:standardmodel:drawbacks}\index{drawbacks}
The Standard Model is the most rigorous theory of particle physics. It is incredibly precise and accurate in its predictions. It gives a reliable predictive model of the particles and their interactions. However, it cannot explain several phenomena. In order to solve this, several theories have been proposed. Unfortunately, there has been no experimental evidence of these theories yet. Some of the phenomena are discussed in detail below:

\begin{itemize}
\item \textbf{Hierarchy problem:} all the particles in the SM gain their masses by their interaction with the Higgs boson. However, the radiative corrections to the mass of the Higgs boson conflict with the experimental value of Higgs mass. These corrections diverge quadratically  up to the cut-off scale of new physics $\Lambda$, where the SM is no longer valid.

The radiative corrections appear in virtual loop diagrams of particles that couple to the Higgs field. This means that a considerable amount of \enquote{fine-tuning} is required to keep the Higgs mass at the electroweak scale ($\approx\SI{100}{\giga\electronvolt}$) if the SM is to remain valid up to high energies. The unnatural amount of fine-tuning to the Higgs mass is known as the \textit{hierarchy problem}. The radiative corrections to the Higgs mass can be expressed as:

\begin{equation}
	\Delta m_{H}^{2} = -\frac{\lambda^{2}_{f}}{8\pi^{2}} \left[\Lambda^{2} + ... \right] \,,
\end{equation}
where $\lambda_{f}$ is the Yukawa coupling of any fermion. The Hierarchy problem implies that there should be new physics at the \si{\tera\electronvolt} scale that eliminates the large loop contributions. Some of the possibilities for this to happen are supersymmetery (SUSY)~\cite{hierarchy2}, Gauge-Higgs unification~\cite{gaugehiggs}, composite Higgs~\cite{hierarchy2}, extra dimensions~\cite{extradim}, etc.

\item \textbf{Neutrino masses:} the Standard Model predicts that neutrinos should be massless like photons. However, experiments with solar, atmospheric, and accelerator neutrinos have provided compelling evidence that the three neutrinos oscillate between their flavours as they move. This is only possible if the neutrinos have some mass. These neutrino masses are quite small ($<\SI{0.1}{\electronvolt}$). Some experimental results have also suggested that there might be a fourth type of neutrino called the sterile neutrinos which are still yet to discover \cite{neutrinomass}.

\item \textbf{Matter-antimatter asymmetry:} whenever matter is created, usually its counterpart antimatter is also formed. It is believed that at the time of the Big Bang, matter and antimatter were produced in equal parts. But what we see today is that matter is dominant. This behaviour is unexplained by the SM. To achieve this asymmetry, theories should have to violate the CP (charge-parity), baryon number and lepton number conservation laws.

\item \textbf{Dark matter and dark energy:} In the universe, galaxies usually rotate with high speed, which means that the gravitational force generated by their matter should not be able to balance the centripetal force. But what we observe is that it holds them together, which means that there should be an extra force generated by some additional matter. This unknown matter is called dark matter.~\cite{darkmatter} It is believed that dark matter constitutes roughly 27\% of the universe, whereas dark energy represents 68\% of the universe, leaving about 5\% of the matter. However, the SM can only explain this tiny fraction of matter which we can see.

\item \textbf{Gravity:} the SM could not be able to explain the fourth type of fundamental force known as the gravitational force. It does not seem to have any impact on the subatomic interactions. But some theories state that there is an existence of a subatomic particle called graviton, which might transmit the gravitational force the same way as photons mediate the electromagnetic force.~\cite{drawbacks}

\item \textbf{Free parameters:} the SM has 19 free parameters whose values are experimentally determined, but there is no theoretical prediction of these values. This gives us a strong indication to extend our knowledge of the Standard Model to explain these results \cite{thesis:anji}.
\end{itemize}

\clearpage
\section{Vector-like quarks (VLQs)}%
\label{sec:theory:vectorlikequarks}

With the discovery of the Higgs boson, the Standard Model is the most comprehensive theory which can explain all the known elementary particles and their interactions. However, there are some shortcomings to this theory which are discussed in the last section. There are some theories which were formulated by considering the principles which are beyond the scope of the SM, called Beyond Standard Model (BSM) theories. These theories are based on several models which predict the existence of new particles that can cancel the divergence of the radiative corrections to the Higgs mass and solve the Higgs mass hierarchy problem. These new particles include the fourth-generation quarks, called vector-like quarks.

Vector-like quarks (VLQs) are hypothetical spin $\frac{1}{2}$ coloured particles. The chiral fourth-generation quarks (\Ptop' and \Pbottom'), which have identical features like the SM third-generation quarks (\Ptop and \Pbottom), are eliminated because their existence impact a significant effect on the production cross-section of the Higgs. But, VLQs on the otherhand show a much smaller impact on the production cross-section of the Higgs, which are even comparable to the uncertainty of the current measurements. Several models of new physics, such as Little Higgs models~\cite{little_higgs}, or those predicting the composite Higgs boson~\cite{composite_higgs}, also include VLQs. They do not receive their mass by the Higgs boson Yukawa coupling. Many models also assume their mixing with the SM third-generation quarks because of the large masses of \Ptop- and \Pbottom-quark.~\cite{vlqpaper}

SM quarks are chiral in nature because their left-handed and right-handed components transform differently under $\text{SU(2)}_{\text{L}} \times \text{U(1)}_{\text{Y}}$ gauge groups, whereas for VLQs, both the components transform similarly. The Lagrangian of the charged current can be written as:
\begin{equation}
	\mathcal{L}_{W} = \frac{g}{\sqrt{2}}(J^{\mu+}W_{\mu}^{+} + J^{\mu-}W_{\mu}^{-}) \,,
\end{equation}
where $J^{\mu\pm} = J^{\mu\pm}_{L} + J^{\mu\pm}_{R}$ and is known as the current density.

SM quarks only have contributions from the left-handed component, which is a mixture of both vector and axial-vector quantities, known as V-A interaction. Whereas in the case of VLQs, both left- and right-handed components have a contribution which together cancels out the axial-vector term and results in the vector quantity and therefore called vector-like quarks.~\cite{vlqinteraction}

\subsection{Multiplet representation}%
\label{sec:theory:multipletrepresentation}

There are four types of vector-like quarks which are listed below along with their electric charges:
\[T  = +\frac{2}{3} \,, B = -\frac{1}{3} \,, \]
\[X  = +\frac{5}{3} \,, Y = -\frac{4}{3} \,. \]
They are represented by different multiplet models, which depend on their weak isospin states. Table \ref{table:theory:multipletrepresentation} shows the different multiplet models and the VLQ candidates along with their weak isospin and the hypercharge. $T$ quark can belong to any multiplet, while $Y$ quark can only exist under doublet and triplet models.

\begin{table}[hbt!]
	\centering
	\begin{tabular}{c | c | c | c} 
		\toprule
		& Singlet & Doublet & Triplet \\
		\midrule
		$Q_{q}$ & 
		$T_{\text{$+\frac{2}{3}$}}$ \hspace{0.4cm} $B_{\text{$-\frac{1}{3}$}}$ & 
		$\begin{pmatrix} X_{\text{$+\frac{5}{3}$}} \\ T_{\text{$+\frac{2}{3}$}} \\ \end{pmatrix}$ \hspace{0.4cm}
		$\begin{pmatrix} T_{\text{$+\frac{2}{3}$}} \\ B_{\text{$-\frac{1}{3}$}} \\ \end{pmatrix}$ \hspace{0.4cm}  
		$\begin{pmatrix} B_{\text{$-\frac{1}{3}$}} \\ Y_{\text{$-\frac{4}{3}$}} \\ \end{pmatrix}$ & 
		$\begin{pmatrix} X_{\text{$+\frac{5}{3}$}} \\ T_{\text{$+\frac{2}{3}$}} \\ B_{\text{$-\frac{1}{3}$}} \\ \end{pmatrix}$ \hspace{0.4cm} 
		$\begin{pmatrix} T_{\text{$+\frac{2}{3}$}} \\ B_{\text{$-\frac{1}{3}$}} \\ Y_{\text{$-\frac{4}{3}$}} \\ \end{pmatrix}$ \\ 
		\midrule
		
		$T_{\text{3}}$ & 0 \hspace{0.6cm} 0 & $\frac{1}{2}$ \hspace{1.1cm} $\frac{1}{2}$ \hspace{1.1cm} $\frac{1}{2}$ & 1 \hspace{1.1cm} 1 \\ 

		$Y$ & $+\frac{4}{3}$ \hspace{0.4cm} $-\frac{2}{3}$ & $+\frac{7}{3}$ \hspace{0.9cm} $-\frac{1}{3}$ \hspace{0.9cm} $+\frac{5}{3}$ & $+\frac{4}{3}$ \hspace{0.9cm} $-\frac{2}{3}$ \\ 
		\bottomrule
	\end{tabular}
	\caption{Overview of the different multiplet models of VLQs along with their isospin $T_{\text{3}}$, hypercharge $Y$ and electric charge $q$.}
	\label{table:theory:multipletrepresentation}
\end{table}

This thesis focuses on the production of $T$ or $Y$ quark decaying into $Wb$, for which the interesting candidates are $T$ quark from ($T$) singlet, $Y$ quark from ($B, Y$) doublet and only $Y$ quark from ($T, B, Y$) triplet. $T$ quark from ($T, B, Y$) triplet does not couple to $Wb$. $T$ quark can decay into either $Wb$, $Zt$ or $Ht$ but at the high mass limit, the branching ratio converges to 2:1:1 ($Wb: Zt: Ht$). On the other hand, $Y$ quark can only decay into $Wb$ because of its charge and therefore, $\mathcal{B}(Y \rightarrow Wb) = 100\%$. That implies that $T$ quark can be produced by the fusion of either $Wb$, $Zt$ or $Ht$ but $Y$ quark can only be produced by the fusion of $W$ boson and $b$-quark.

\subsection{Production of VLQs}%
\label{sec:theory:production}

VLQs can either be can be produced via single and pair production in proton-proton ($pp$) collisions at the LHC. 

\begin{figure}[hbt!]
	\centering
	\includegraphics[width=\linewidth]{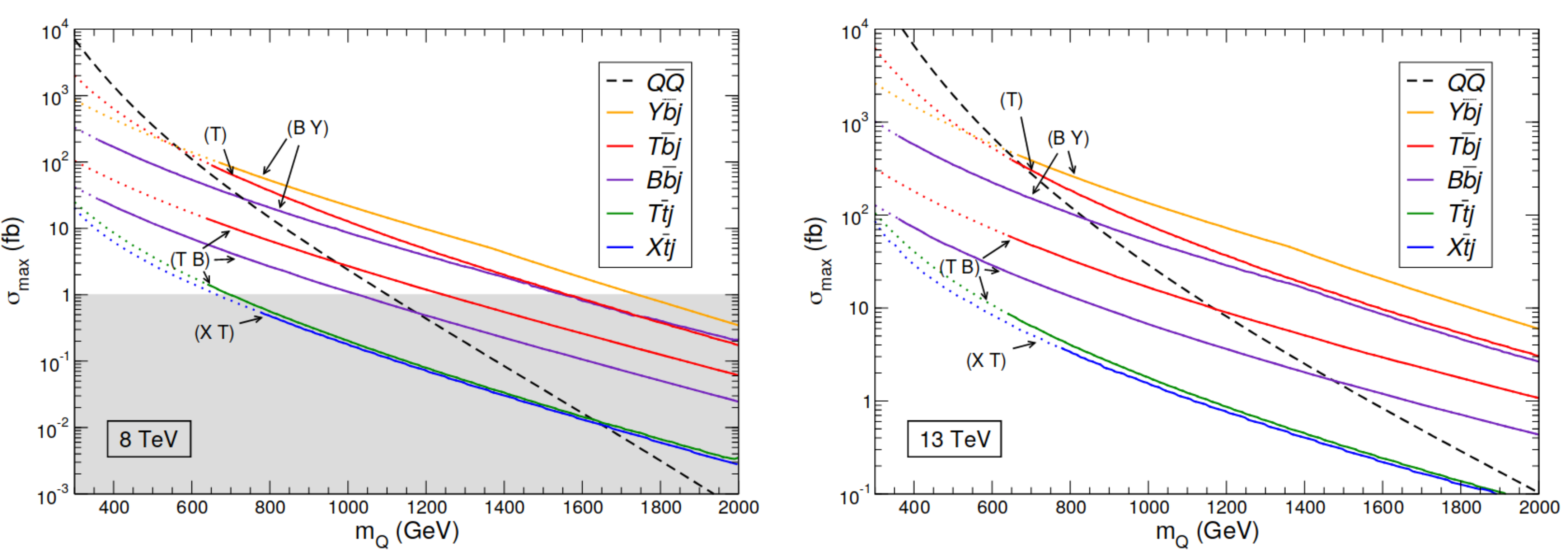}
	\caption{Production cross-section for single VLQ production and VLQ pair production as a function of VLQ mass at $\sqrt{s}=\SI{8}{\tera\electronvolt}$ and $\sqrt{s}=\SI{13}{\tera\electronvolt}$. The pair production process is denoted by dotted line whereas coloured lines show different single VLQ processes. Both the plots show that single production processes have larger production cross-section at high VLQ mass.~\cite{aguilar}}
	\label{fig:theory:production}
\end{figure}

\begin{itemize}
\item \textbf{Single VLQ production:} VLQs can be produced by the fusion of the SM gauge boson and third-generation quark, which is enabled by their strong coupling to the SM quarks. Therefore, searches for singly produced VLQs can be used to probe these couplings as a function of the VLQ mass. Single VLQ production is a dominant process at high VLQ masses in both $\sqrt{s}=\SI[per-mode=symbol]{8}{\tera\electronvolt}$ and $\sqrt{s}=\SI[per-mode=symbol]{13}{\tera\electronvolt}$, which can be seen from Fig.\ \ref{fig:theory:production}. The Feynman diagram for two possible single VLQ production processes is shown in Fig \ref{fig:theory:production:single}. In this thesis, the focus is the single production process of $T$ or $Y$ quark, as shown in Fig.\ \ref{fig:theory:production:single:ty}.

\begin{figure}[hbt!]
	\centering
	\begin{subfigure}{.4\textwidth}
		\centering
		\includegraphics[width=\linewidth,height=\textheight,keepaspectratio]{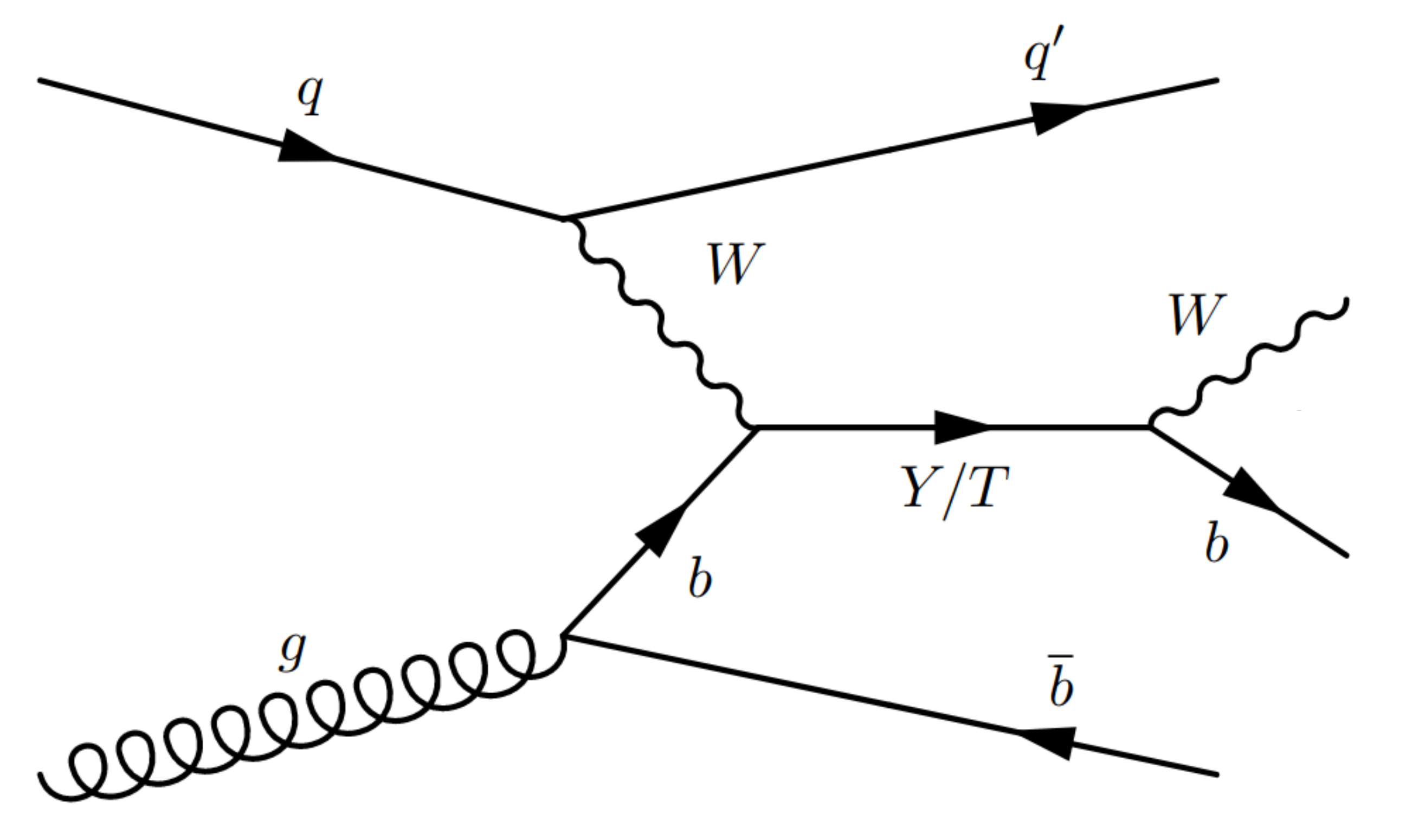}
		\caption{}
		\label{fig:theory:production:single:ty}
	\end{subfigure}\hspace{1cm}
	\begin{subfigure}{.35\textwidth}
		\centering
		\includegraphics[width=\linewidth,height=\textheight,keepaspectratio]{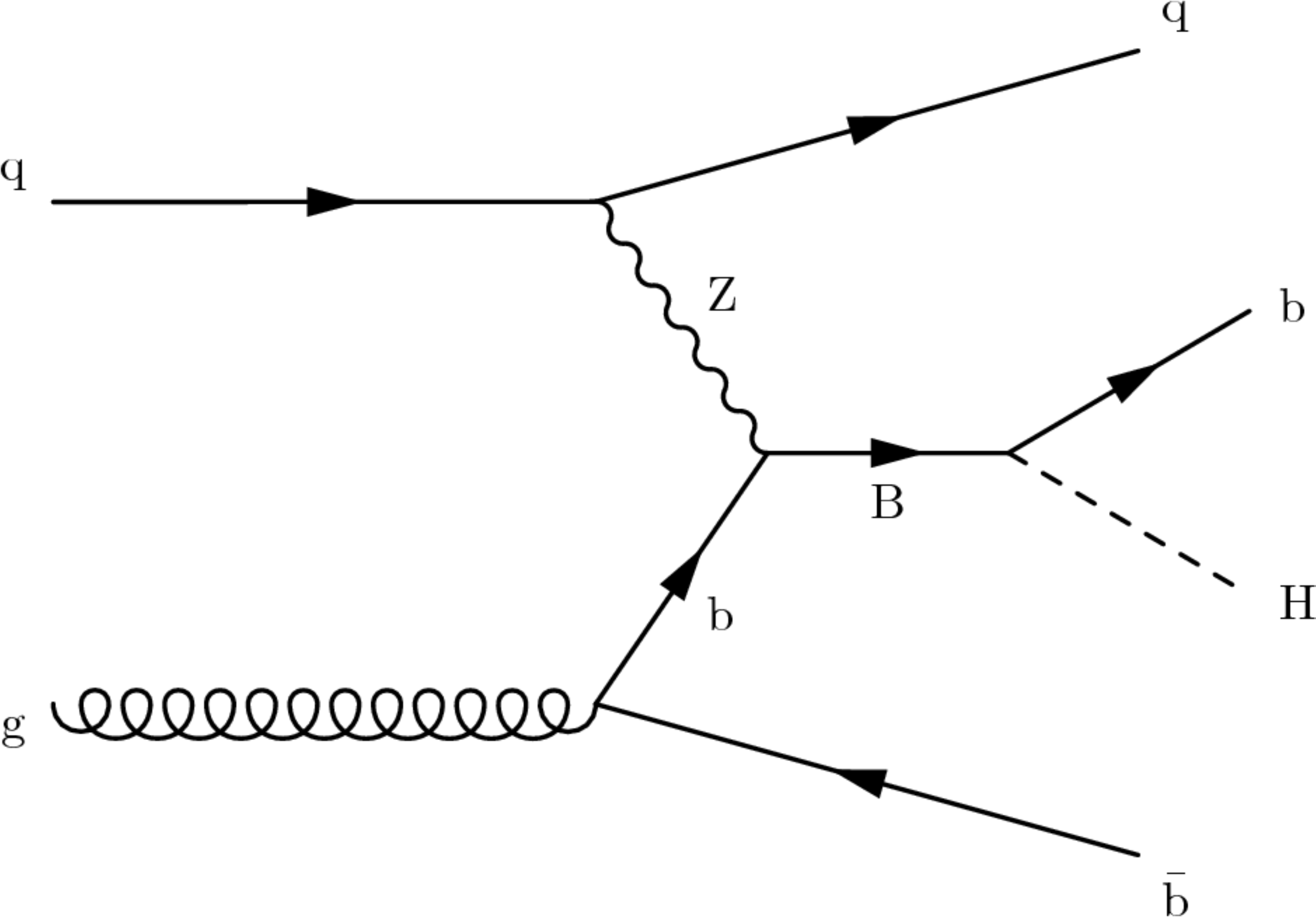}
		\caption{}
		\label{fig:theory:production:single:b}
	\end{subfigure}
	\caption{Feynman diagrams for single VLQ production, where (a) shows the production of $T/Y$ quark by the fusion of $W$ and $b$-quark and (b) shows the production of $B$ quark by the fusion of $Z$ and $b$-quark.~\cite{vlqpaper}}
	\label{fig:theory:production:single}
\end{figure}

\item \textbf{VLQ pair production:} VLQs can be produced in pair which allows to set a limit on VLQ masses. These masses are insensitive to the coupling because they are produced through the strong interaction. Pair production is a dominant process at lower VLQ masses in both  $\sqrt{s}=\SI[per-mode=symbol]{8}{\tera\electronvolt}$ and $\sqrt{s}=\SI[per-mode=symbol]{13}{\tera\electronvolt}$, which can be seen from Fig. \ref{fig:theory:production}. The Feynman diagram for a pair production process is shown in Fig.\ \ref{fig:theory:production:pair}.

\begin{figure}[hbt!]
	\centering
	\begin{subfigure}{.4\textwidth}
		\centering
		\includegraphics[width=\linewidth,height=\textheight,keepaspectratio]{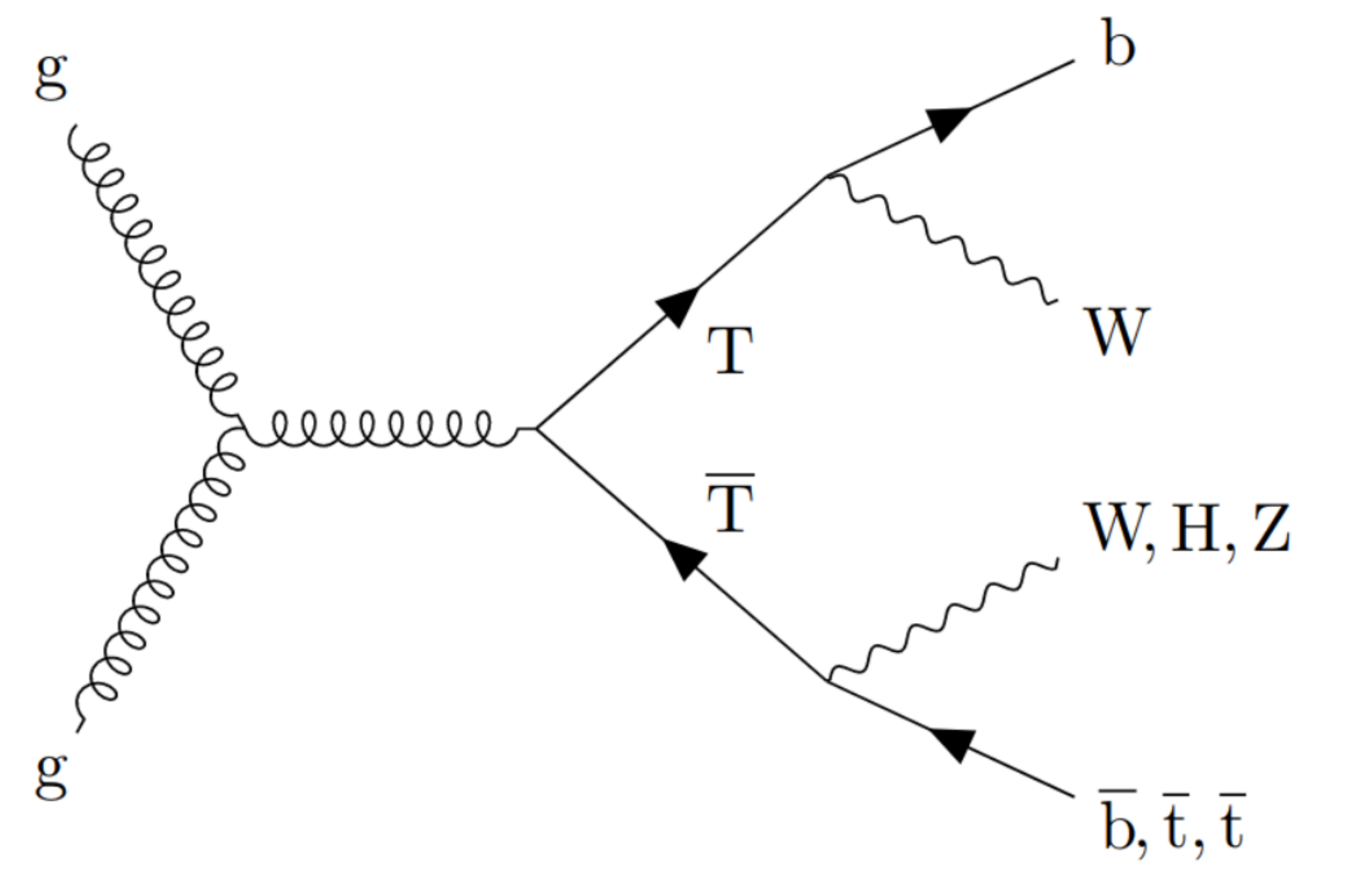}
		\caption{}
		\label{fig:theory:production:pair:t}
	\end{subfigure}
	\caption{Feynman diagram for VLQ pair production where $T\bar{T}$ is produced in pair, which can further decay into $Wb$, $Ht$ or $Zt$ pair.~\cite{pairproductiont}}
	\label{fig:theory:production:pair}
\end{figure}

\end{itemize}

\subsection{Models}%
\label{sec:theory:models}

Several theories have been proposed for describing the characteristics of VLQs. In this thesis, two different models have been presented that use different formulations of the Lagrangian. The Lagrangian depends on the different parameterisation that explains these new particles and their interactions. 

\begin{itemize}
\item \textbf{Renormalisable theory:} 
In this model, the Lagrangian is parameterised by a mixing angle $\theta_{\text{L,R}}$, which describes the mixing between the SM quarks and VLQs. This theory gives a complete information about the dependence of branching ratios on the multiplet dimensions. Hence, a given value of either left- or right-handed mixing angle $\theta_{\text{L}}$ or $\theta_{\text{R}}$ entirely determines all the branching ratios of VLQ at any mass point.~\cite{aguilar}

\item \textbf{General theory:} According to this theory, the Lagrangian is parameterised by a non-renormalisable coupling term $c_{\text{L,R}}^{\text{Wb}}$. This theory can predict the production cross-section of VLQs at next-to-leading-order terms (NLO).~\cite{wulzer}
\end{itemize}

By comparing the respective Lagrangians of the renormalisable model and general model, a relation is established between the coupling terms $c_{\text{L,R}}^{\text{Wb}}$ and the mixing angles $\theta_{\text{L,R}}$ within a given multiplet and is shown by the equations below:

\begin{equation}
c_{\text{L,R}}^{\text{Wb}} = \sqrt{2}\sin\theta_{\text{L,R}} \,,
\label{eqn:theory:models:singlet}
\end{equation}

\begin{equation}
c_{\text{L}}^{\text{Wb}} = 2\sin\theta_{\text{L}} \,,
\label{eqn:theory:models:triplet}
\end{equation}
where Eqn.\ \ref{eqn:theory:models:singlet} is true for $T$ singlet and ($B,Y$) doublet model, and Eqn.\ \ref{eqn:theory:models:triplet} is valid for ($T,B,Y$) triplet model. These relations are only valid within the renormalisable formulation, and if one considers only the interaction between $T/Y$, \PW and \Pbottom.~\cite{vlqpaper}



\chapter{LHC and the ATLAS detector}
\label{sec:lhcandatlas}
The purpose of this chapter is to describe the machinery behind the production of the data. A brief overview of the Large Hadron Collider and the ATLAS detector is given. An introduction to the coordinate system, which is used by the ATLAS detector and an explanation of all the sub-detector components of the detector is given in the last section of the chapter.

\section{The Large Hadron Collider (LHC)}%
\label{sec:lhcandatlas:lhc}
The Large Hadron Collider (LHC)~\cite{lhc} is a proton-proton collider situated at the Swiss-French border. The collider is built in a \SI{26.7}{\kilo\metre} circular tunnel which is approximately \SI{100}{\metre} underground. It is designed to collide protons at $\sqrt{s}$=\SI{13}{\tera\electronvolt}, which makes it the highest energy particle collider ever built. The high centre-of-mass energy is essential for the possible production of new exotic particles. The collider is designed to have high luminosity of particle collisions, $\SI{1e34}{\centi\metre^{-2}\sec^{-1}}$. The high luminosity means that rare events (with low cross-section) will occur at relatively high rates. The LHC was operational for two significant periods of time, called Run 1 and Run 2 respectively. Run 1 corresponds to the data collected from 2009 to 2013 at $\sqrt{s}=\SI{7}{\tera\electronvolt}$ and $\sqrt{s}=\SI{8}{\tera\electronvolt}$. Then it was shut down for two years. After that, the data collected from 2015 to 2018 at $\sqrt{s}=\SI{13}{\tera\electronvolt}$ corresponds to Run 2 phase. In this thesis, the data collected during Run 2 is used for the analysis.

\begin{figure}[hbt!]
	\centering
	\includegraphics[width=0.9\linewidth]{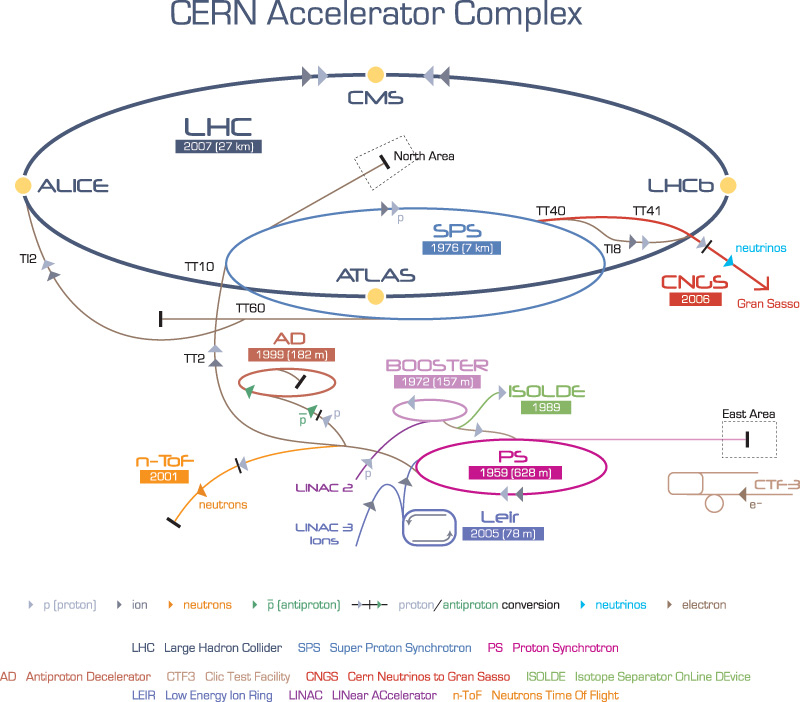}
	\caption{A schematic showing the layout of the accelerator complex, which includes pre-accelerator and the LHC ring. The four interaction points are shown in yellow dots around the LHC ring where the four main experiments are built.~\cite{lhc_picture}}
	\label{fig:lhcandatlas:lhc}
\end{figure}

To achieve such high energy, several accelerating machines are required. An overview of the main components of the accelerator complex is shown in Fig.\ \ref{fig:lhcandatlas:lhc}. Protons are first injected to the pre-accelerator, which includes linear accelerator \enquote{LINAC 2} where protons are accelerated to an energy of \SI{1.4}{\giga\electronvolt} and then injected to the Booster. The Booster also groups the protons in bunches of around \num{1e11} protons before injecting them into the Proton Synchrotron (PS). The PS accelerates the proton bunches up to an energy of \SI{25}{\giga\electronvolt} before injecting them into the Super Proton Synchroton (SPS). The SPS accelerates the protons up to an energy of \SI{450}{\giga\electronvolt}. After the proton bunches are injected into the LHC, where they are accelerated to a maximum energy of \SI{6.5}{\tera\electronvolt} per beam. Before reaching this energy, protons are injected into two separate beam pipes. Protons in one of these pipes circulate clock-wise, while protons in the other beam pipe travel in the opposite direction. The LHC uses approximately \num{1300} dipole magnets to circulate the proton bunches around the ring about $\approx\num{11000}$ times a second, which generates a magnetic field of \SI{8.3}{\tesla}. The LHC also uses over \num{300} quadrupole magnets to keep the beams focused. The proton bunches become uncollimated over time. Therefore, another \num{5000} correcting magnets are used to keep the protons in bunches, and also to make orbital corrections as the protons circulate the LHC. With this setup, energies reached at the LHC during Run 2 resulted in center-of-mass energy of \SI{13}{\tera\electronvolt}, with an instantaneous luminosity of $\SI{1e34}{\centi\metre^{-2}\sec^{-1}}$. During Run 2, a total integrated luminosity of $\SI{156}{\femto\barn^{-1}}$ was delivered by the LHC, but $\SI{139}{\femto\barn^{-1}}$ was considered good enough for physics analyses.~\cite{lhc}~\cite{thesis:fletcher}

The proton beams are collided at different interaction points along the LHC ring, where the particle detectors are built to detect particles created during a collision. The four main experiments are shown in Fig.\ \ref{fig:lhcandatlas:lhc}. The two largest particle detectors, A Toroidal LHC ApparatuS (ATLAS)~\cite{atlas} and Compact Muon Solenoid (CMS)~\cite{cms} are general-purpose particle detectors, which have a wide-ranging physics searches which include both SM and BSM searches. Large Hadron Collider beauty (LHCb)\cite{lhcb} focuses on $b$-physics, which is a physics of rare decays of $b$-mesons and precise measurements of CP-violation in the SM. The fourth particle physics experiment, A Large Ion Collider Experiment (ALICE)~\cite{alice} is a heavy-ion experiment that is specifically designed to look at lead-lead ion and proton-lead ion collisions. It focuses on quark-gluon plasma, a primordial state of matter, which is thought to have existed in the early universe.

\section{The ATLAS detector}%
\begin{figure}[hbt!]
	\centering
	\includegraphics[width=0.9\linewidth]{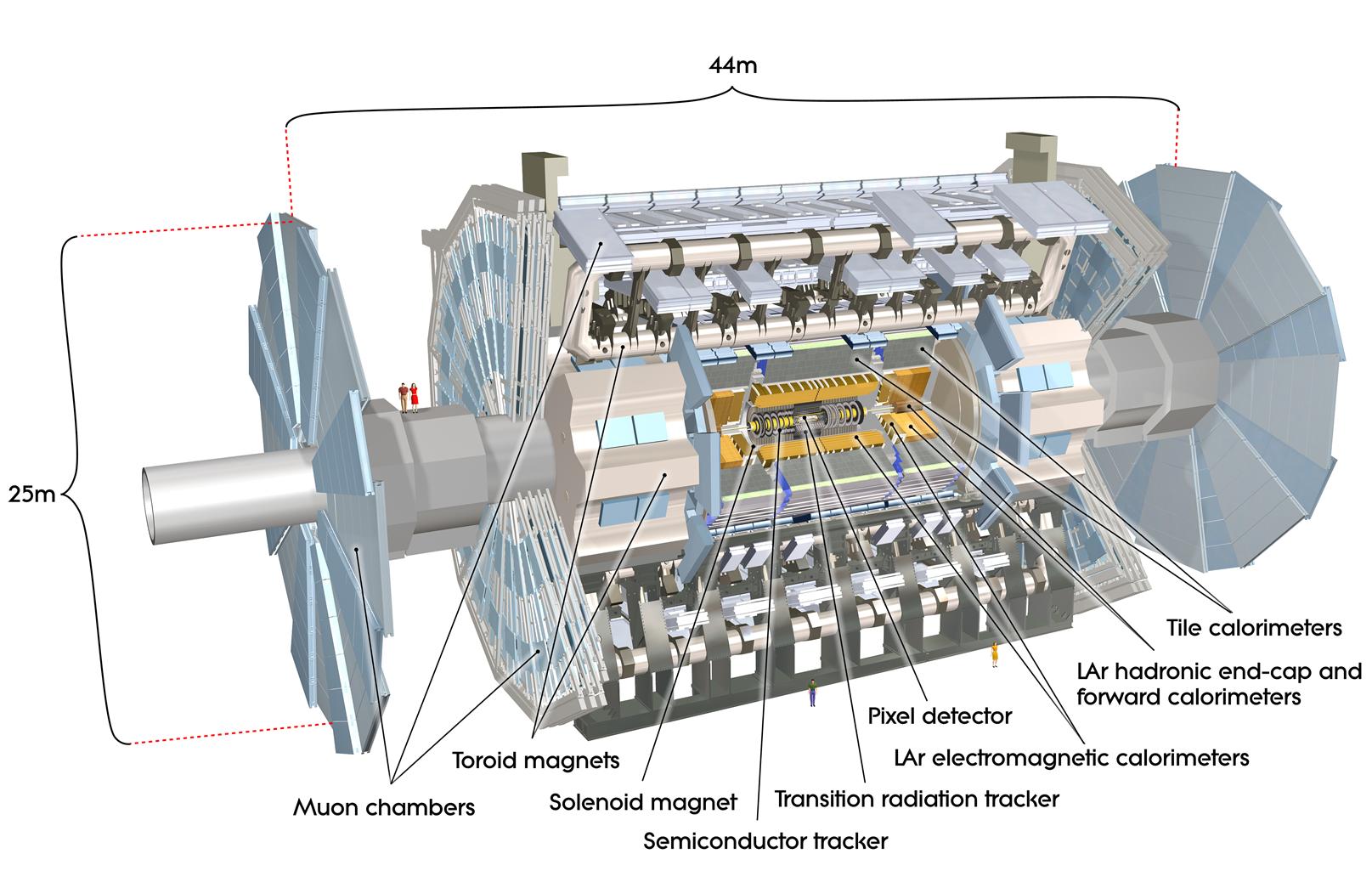}
	\caption{An overview of the ATLAS detector showing all the components of the detector.~\cite{detector}}
	\label{fig:lhcandatlas:atlas}
\end{figure}

\label{sec:lhcandatlas:atlas}
The ATLAS detector (A Toroidal LHC ApparatuS)~\cite{atlas} is a multi-purpose particle detector which has dimensions of a total length of \SI{46}{\meter}, a height of \SI{25}{\meter} and a total weight of \SI{7000}{\tonne}. The ATLAS detector comprises of several layers of sub-detectors, where each is serving a specific function to detect one or more properties of the particles passing through them. These sub-detectors can be ordered into four categories: the inner detector (ID), the calorimeter system, the muon spectrometer and the magnet system. An overview of the detector is shown in Fig.\ \ref{fig:lhcandatlas:atlas}.~\cite{detector}

\subsection{Coordinate system and kinematic observables}%
\begin{figure}[hbt!]
	\centering
	\includegraphics[width=0.7\linewidth]{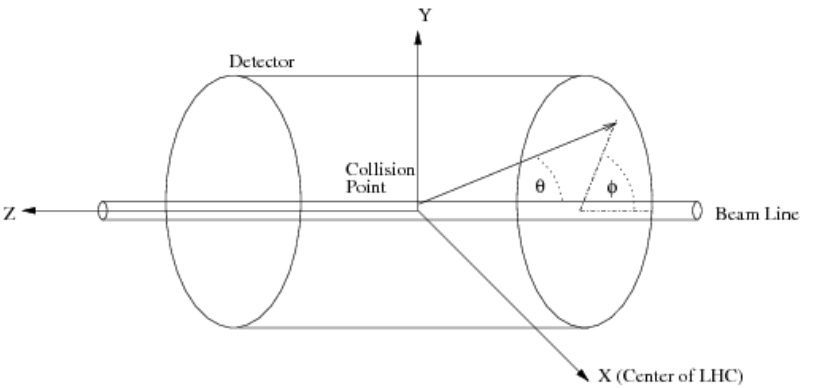}
	\caption{A schematic showing the coordinate system of the ATLAS detector.~\cite{atlas_coordinate}}
	\label{fig:lhcandatlas:atlas:observables}
\end{figure}

\label{sec:lhcandatlas:atlas:observables}
The ATLAS detector is based on a right-handed cartesian coordinate system with the x-axis pointing into the centre of the LHC collider ring, the y-axis denoting the vertical direction towards the surface and the z-axis along the LHC beam pipe.~\cite{detector} The ATLAS detector obeys a concentric cylindrical symmetry. Therefore, the spherical polar angles of $\theta$ and $\phi$ are used. The azimuthal angle ($\phi$) is measured around the beam axis in the z-axis direction, and the polar $\theta$ angle is measured as the angle from the beam axis. A schematic of the ATLAS coordinate system is shown in Fig.\ \ref{fig:lhcandatlas:atlas:observables}.
 
In the $pp$ collision at the LHC, partons are colliding with each other, where each of these partons carries a part of the proton momentum. Since the initial momentum is zero in the transverse plane, so it is convenient to define a quantity called transverse momentum:
\begin{equation}
	p_{\text{T}} = \sqrt{p_{\text{x}}^{\text{2}} + p_{\text{y}}^{\text{2}}} \,,
\end{equation}
where $p_{\text{x}}$ and $p_{\text{y}}$ are denoted as the momentum components in x and y directions, respectively. According to the conservation of momentum, the total momentum in the transverse plane should result in zero. Therefore, if the sum of all transverse momenta does not yield zero, the missing transverse momentum can be calulated as:
\begin{equation}
	p_{\text{T}}^{\text{miss}} = -\sum_{i}^{}p_{\text{T}i} \,,
\end{equation}
where $i$ is denoted as all the detected particles. It is used to find neutrinos since they pass through the detector undetected.

The centre-of-mass frame of the collision is usually Lorentz boosted in the z-axis, but the physics analyses need quantities that are invariant under a Lorentz boost. One such quantity which is invariant under a boost along the z-axis and gives information about the angular distribution of the particles, is rapidity $y$:
\begin{equation}
	y = \frac{1}{2}\ln(\frac{E+p_{\text{z}}}{E-p_{\text{z}}}) \,,	
\end{equation}
where $E$ is denoted is the energy of the particle and $p_{\text{z}}$ is represented as the momentum of the particle along the z-axis.

Pseudorapidity ($\eta$) is widely used instead of rapidity since the momentum of the partons along the z-axis is unknown, and it is defined as:
\begin{equation}
	\eta = -\ln{\tan(\frac{\theta}{2})} \,.
\end{equation}

It is also crucial to define a quantity that measures the distance between physics objects for the jet formation and to resolve the overlapped physics objects in the detector. The Lorentz invariant geometric variable $\Delta R$ is used for this, which is defined as:
\begin{equation}
	\Delta R = \sqrt{\Delta \eta^{\text{2}} + \Delta \phi^{\text{2}}} \,.
\end{equation}

\subsection{Detector overview}%
\label{sec:lhcandatlas:atlas:components}

\subsubsection{Inner detector}%
\begin{figure}[hbt!]
	\centering
	\includegraphics[width=0.7\linewidth]{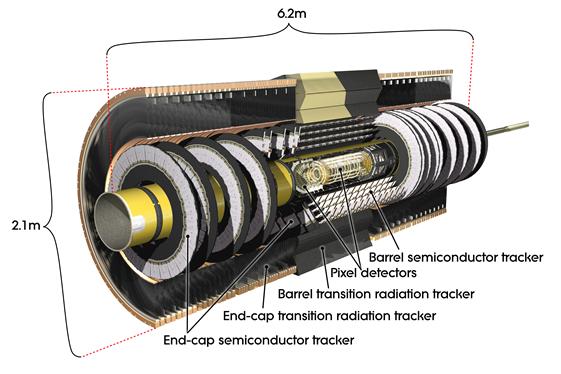}
	\caption{A schematic showing the inner detector of the ATLAS detector.~\cite{atlas}}
	\label{fig:lhcandatlas:atlas:inner}
\end{figure}

\label{sec:lhcandatlas:atlas:inner}
The Inner Detector (ID) is a cylindrical tracking detector which is placed close to the interaction point to reconstruct the primary vertex and measure the momentum of the tracks. The ID consists of the insertable B-layer (IBL), pixel detector, the semiconductor tracker (SCT) and the transition radiation tracker (TRT). The pixel detector is composed of the silicon pixel modules and the SCT is composed of the silicon microstrip sensors. They both are precision silicon tracking detectors and cover a pseudorapidity range of $|\eta|<2.5$. The TRT is a drift chamber made up of gas-filled straw tubes that includes a pseudorapidity range of $|\eta|<2.0$. The entire inner detector is kept in a magnetic field of \SI{2}{\tesla}, generated by the ATLAS solenoid magnet. The magnetic field curves the trajectory of charged particles when they pass through the inner detector, which is used to measure the particle momentum from the radius of curvature of the resultant track. A schematic of the ATLAS inner detector is shown in Fig.\ \ref{fig:lhcandatlas:atlas:inner}.~\cite{atlas}

\begin{itemize}
	\item  \textbf{IBL:} the IBL is an insertable B-layer, placed at the innermost layer of the pixel detector. It was inserted for the LHC Run 2 operation. This layer is composed of silicon pixel modules arranged on carbon fibre staves at a radius of \SI{33}{\milli\meter} surrounding the beam pipe. The IBL provides an additional measurement point closer to the interaction point, and hence improving the impact parameter reconstruction and vertexing.~\cite{ibl}
	
	\item \textbf{Pixel detector:} the pixel detector is a silicon detector, placed at the centre of the ID and has the highest granularity of all the detectors in the ID. The detector is built from 60 million silicon pixels, which have a pixel size of $50\times400$ $\si{\micro\metre^{\text{2}}}$. The high granularity of the pixel detector is used to reconstruct vertices from particle collisions with high resolution. The pixel detector is one of the essential components used to identify jets that have originated from $b$-quarks by precisely reconstructing the primary and secondary vertices.~\cite{atlas}
	
	\item \textbf{SCT:} the SCT is a silicon detector which lies outside the pixel detector around $\SI{300}{\milli\meter}$ away from the beam pipe. It consists of 6 million silicon wafer strip each $\SI{80}{\micro\meter}$ wide, which is designed to track charged particles. It has a track resolution of $\approx\SI{200}{\micro\meter}$, which is not as good as the pixel detector. The SCT has four different layers of modules that are kept around the barrel of the inner detector, and two endcap sections which have nine disks of modules. With the endcap sections, the SCT can provide charged particle tracking up to a pseudorapidity of $|\eta|=2.5$. The SCT contributes to the measurement of 8 different track parameters, including momentum, impact parameter and vertex position.~\cite{atlas}
	
	\item \textbf{TRT:} the TRT is a straw tube detector which covers the largest part of the ID. It provides further tracking information about charged particles. The TRT is made up of straw-tubes which has a diameter of $\SI{4}{\milli\meter}$ which can provide up to 32 hits per track, thus making a considerable contribution to the momentum measurement of the track. The TRT also has some particle identification capabilities through the amount of transition radiation (TR) particle deposits in the detector. TR is the radiation emitted by a relativistic charged particle when it passes through an inhomogeneous medium.~\cite{atlas}
\end{itemize}

\subsubsection{Calorimeters}%
\label{sec:lhcandatlas:atlas:calorimeters}
Calorimeter is used to measure the energy of the particle and identification of the tracks. The ATLAS detector uses two different calorimeter systems, i.e.\ electromagnetic and hadronic calorimeter. These calorimeters measure the deposition of energy from particles. They are sampling calorimeters built from alternating layers of an active material, where the energy measurements are recorded, and an absorbing region, which is used to contain the particles within the calorimeter. The ATLAS calorimeters provide full angular coverage for particles up to $|\eta|<4.9$, but precise energy measurements only for particles with pseudorapidity of $|\eta|<2.5$.~\cite{atlas}

\begin{itemize}
	\item \textbf{Electromagnetic calorimeter:} the electromagnetic calorimeter (ECAL) is used to measure the energy of particles interacting predominantly via electromagnetic interaction with the calorimeter material. These particles include either charged particles that are typically electrons which interact via bremsstrahlung or neutral particles like photons which interacts mainly via electron-positron production. ECAL uses Liquid Argon (LAr) as its active medium, and lead as its absorbing medium. It is divided into three separate regions: the barrel region (covering $|\eta|<1.5$) and two end-cap regions (covering $1.4<|\eta|<3.2$). Since the inner detector only extends to $|\eta|<2.5$, this restricts the precision of measurements in the electromagnetic calorimeter outside this $\eta$ region. This led to the design of the barrel region of the electromagnetic calorimeter having the highest level of granularity.~\cite{atlas} Energy from particles is deposited into four different layers of the electromagnetic calorimeter. The first layer is pre-sampler layer which is placed in front of the electromagnetic calorimeter ($|\eta|<1.8$), and is used to correct for energy loss caused by the ID and solenoid. Then comes, the first sampling layer which has the highest level of granularity, and is used to provide the very high $\eta-\phi$ resolution required to differentiate between photon and \Pgpz. The second sampling layer is the primary energy deposition layer in the electromagnetic calorimeter. The $\eta-\phi$ resolution of the second layer is around a tenth of the first sampling layer, as the very high $\eta-\phi$ resolution is no longer required. At last, the third sampling layer in which cells are coarser than the other layers of the calorimeter because only very high energy electrons and photons would able to make it to this layer of the electromagnetic calorimeter.~\cite{atlas}
	
	\item \textbf{Hadronic calorimeter:} the hadronic calorimeter (HCAL) is used to measure the energy of particles which interacts predominantly via strong interaction with the calorimeter material, for example, hadrons like protons and neutrons pass through the HCAL, they interact with the nucleus of the material of the HCAL and produce a hadronic shower. HCAL measures the energy deposited by the hadronic shower on the material of the HCAL. It is comprised of five different sections: the barrel tile section ($|\eta|<1.0$), two tile end-cap sections, called the extended barrel ($1.0<|\eta|<1.7$) and two hadronic end-caps ($1.5<|\eta|<3.2$). In the tile regions, thick scintillating plastic panels of \SI{3}{\milli\meter} are used as the active material, and iron plates are used as the absorbing material. These tiles are arranged in a staggered fashion to avoid any gaps within sections of the calorimeter. The scintillation light caused by the tiles is collected by wavelength shifting fibres, which transport the photons to the photomultiplier tubes. The hadronic end-cap calorimeter (HEC) is made from LAr placed between copper absorption layers. LAr acts as an active medium and copper acts as a passive medium. HEC is composed of two disks on either side of the detector. Each disk has 32 modules divided into four different sampling layers.~\cite{atlas}
	
	\item \textbf{Forward calorimeter:} the forward calorimeter (FCAL) is located in the forward region of the ATLAS detector ($3.1<|\eta|<4.9$). It uses LAr as its active material and consists of three parts with different absorption materials. The first part uses copper as a passive material to measure the forward electromagnetic particles. The two other parts are made from Tungsten, which has a very high radiation length and best suited for measuring the forward hadronic particles.~\cite{atlas}
\end{itemize}
 
\subsubsection{Muon spectrometer}%
\label{sec:lhcandatlas:atlas:muonspectrometer}
Muons can pass through the calorimeters without losing much of their energy, so it is difficult to track them. The ATLAS muon spectrometer (MS) is designed to track and measure the \pt of muons originating from the interaction point. An intense magnetic field is required for the muon chambers to be effective over a wide range of \pt and $\eta$. This magnetic field is produced by the toroidal magnetic system which covers the entire muon system.

The muon system consists of two different types of muon chambers: muon drift tubes (MDT) and the cathode-strip chambers (CSC). The muon drift chamber is located in the barrel region of the detector, which provides precise measurements of the location of ionisation tracks from muons passing through the chamber. The most crucial track position is in the primary bending direction of the toroidal magnetic field, as this direction is used to calculate the curvature of the muon track, and hence the \pt of the muon. The cathode-strip chambers are located in the forward region of the ATLAS detector and are designed to handle a high occupancy of low \pt forward muons. The CSCs are multiwire proportional chambers, where the wires are oriented in the radial direction pointing to the centre of the muon CSC wheel. Resistive plate chambers (RPCs) are used to trigger on muons in the barrel section, and thin gap chambers (TGCs) are used to trigger on muons in the end-cap sections.~\cite{atlas}

\subsubsection{Magnets}%
\label{sec:lhcandatlas:atlas:magnets}
The ATLAS detector uses two different magnet systems to measure the momentum of the charged particles accurately. The first magnet system is a solenoid magnet which provides an axial magnetic field of \SI{2}{\tesla}. It covers the whole inner detector. The second magnetic system is toroidal, which produces magnetic fields in the muon system. It is divided into two parts: barrel region which is located around the central calorimeter and the end-caps which are situated at the two ends of the detector. The magnetic field strength in the barrel region is \SI{0.5}{\tesla} which increases to \SI{1}{\tesla} in the end cap regions of the muon system.~\cite{atlas}

\subsubsection{Trigger system}%
\label{sec:lhcandatlas:atlas:trigger}
Due to the high luminosity of the LHC in Run 2, a large storage system is required to store the information about every event. There are a lot of events which are not interesting for the physics analysis for various reasons. So, the ATLAS detector needs a trigger system to only select and store those events which are interesting for the physics analysis. The ATLAS trigger system is divided into two parts: Level-1 trigger and the High-Level trigger (HLT). Level-1 is entirely a hardware-based trigger and is constructed from specialised electronics. It uses the raw information from the muon systems and calorimeters to search for high energy physics objects, such as jets, electrons, photons, muons and decayed taus. The HLT is an event filter which includes processing of the data using the full detector information. Events that pass the Level-1 trigger are reduced to the event rate of \SI{100}{\kilo\hertz} from \SI{40}{\mega\hertz}. Then, they are passed to the HLT, which further reduces the event rate to \SI{1}{\kilo\hertz}.~\cite{atlas}



\chapter{Jets and tagging}
\label{sec:jetsandtaggers}
The aim of this chapter is to provide a brief introduction about the jet formation and their reconstruction process using a jet reconstruction algorithm. Two different types of jet collections which are used in this thesis are discussed in detail. After the reconstruction of jets, they are identified by using the jet tagging algorithm. Two different types of jet tagging algorithms are described in the last section of this chapter. 
\section{Jets}
\label{sec:jetsandtaggers:jets}
Jets are extended particle showers originating from the hadronisation of a parton. Quarks and gluons are constrained to the confinement. Therefore, they hadronise into hadrons. These hadrons can be grouped into jets by a jet reconstruction algorithm. The jet reconstruction algorithm uses the energy deposited by the particle shower of charged and neutral particles in the calorimeter along with the information of tracks and vertices, to reconstruct a jet.

\subsection{Jet reconstruction algorithm}
\label{sec:jetsandtaggers:jets:algorithm}
The jet algorithm is used to reconstruct a four-vector from the hadronisation products of a parton that corresponds to the underlying parton four-vector. The jets that are considered in this thesis are reconstructed using the anti-$k_{\text{T}}$ jet algorithm.~\cite{antikt} 

The anti-$k_{\text{T}}$ algorithm is a sequential jet algorithm which takes four-vectors of individual particles as input and clusters them together until a certain threshold requirement is fulfilled. First, the distance between each object is calculated. This distance does not only depend on the geometry but also on the kinematic properties. The distance between two objects, $i$ and $j$ is given by the following expression:
\begin{equation}
	d_{ij} = \text{min}(p_{\text{T}i}^{-2} , p_{\text{T}j}^{-2})\frac{\Delta R_{ij}^{2}}{R} \,,
\end{equation}
where $p_{\text{T}}$ is the transverse momentum of each object and $\Delta R_{ij}$ is the angular distance calculated by $\Delta R_{ij} = \sqrt{\Delta \eta^{2}+\Delta\phi^{2}}$. R is called the \textit{radius parameter}, which defines a threshold value on a radius of the cone in which the objects should be considered.~\cite{thesis:tanja} For small-$R$ jets, it is set to be $R=0.4$, and for large-$R$ jets, it is set to be $R=1.0$. In this thesis, both small and large-$R$ jets have been used.

\begin{figure}[hbt!]
	\centering
	\includegraphics[width=0.6\linewidth]{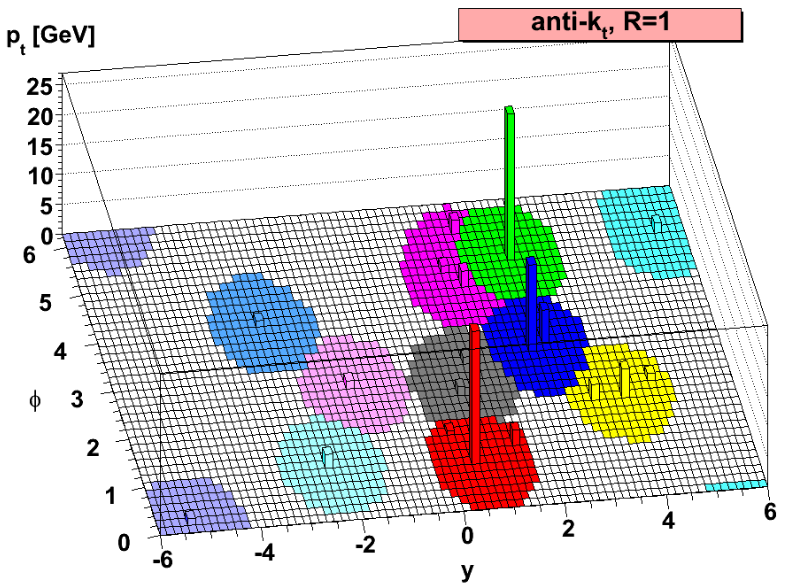}
	\caption{A schematic showing the simulated event with several particles. Jets (in colour) are reconstructed using the anti-$k_{\text{T}}$ algorithm.~\cite{antikt}}
	\label{fig:jetsandtaggers:jets:algorithm}
\end{figure}

Then, the distance between the object and the beam is calculated by $d_{iB} = p_{\text{T}i}^{-2}$. If this distance is greater than $d_{ij}$, then the objects $i$ and $j$ are combined to form a jet. This process is performed till $d_{iB}<d_{ij}$, then the formed jet is considered as the final jet, and it is then removed from the list. This procedure is repeated until every object is taken out and all final jets have been obtained. An illustration of the behaviour of the anti-$k_{\text{T}}$ algorithm in $\eta-\phi$ plane is shown in Fig.\ \ref{fig:jetsandtaggers:jets:algorithm} in an event with few particles.~\cite{thesis:tanja}

Jet energy has to be corrected before using the jets for event reconstruction because the detector response and the jet reconstruction algorithm are not perfect. The correction is performed such that the energy of the reconstructed jet corresponds to the energy of a jet reconstructed from the true stable particles in the detector. This procedure is called jet energy calibration.~\cite{jet_calibration}

Jets can be contaminated with partons from pile-up or underlying events, which can lead to a distortion of the jet four-momentum vector, which can cause not a good momentum resolution. This is usually common in large-$R$ jets because the probability to include hadrons that are not associated with the hard part of the interaction increases for a larger jet area. There are several methods to remove the contamination from the jets after reconstruction. One of the methods that are used in the ATLAS experiment is jet trimming.~\cite{jet_trimming} The jet trimming method uses the fact that the contribution from the pile-up or underlying event is usually softer than the contribution from the hard scattering.~\cite{thesis:ruth} 

Since a jet algorithm takes four-momenta as input, one can specify which information to use for the construction of the four-momenta. The ATLAS experiment uses calibrated calorimeter clusters to determine the four-momenta. Jets which are reconstructed based on calorimeter information are called calorimeter jets. Another method is to use the information of tracks along with the calorimeter information to determine the four-momenta for jet reconstruction. Jet reconstructed by using this method, is called particle flow jets. They both are discussed in detail in the next sections. The standard jet collection with a radius parameter of $R=0.4$ used by the ATLAS collaboration is a calorimeter jet.~\cite{thesis:ruth}

\subsection{Topo-cluster jets}
\label{sec:jetsandtaggers:jets:topo}
A calorimeter is divided into small units called cells. The signal produced in these cells corresponds to the input to a clustering algorithm to form a cluster. The topo-cluster formation algorithm starts from a seed cell, whose signal-to-noise (S/N) ratio is above a threshold value. Here, noise refers to an average expected noise in that cell. Another threshold value is set in order to include the neighbouring cells to the seed to form a protocluster. This process takes place iteratively where the S/N of all the neighbouring cells of the protocluster is evaluated. If it passes the threshold, it is merged to the protocluster. Finally, all calorimeter cells neighbouring the formed protocluster are added, which forms the topo-cluster. The topo-cluster algorithm efficiently suppresses the calorimeter noise. All the threshold are optimised for the clustering algorithm.~\cite{thesis:tanja}

\begin{figure}[hbt!]
	\centering
	\includegraphics[width=0.6\linewidth]{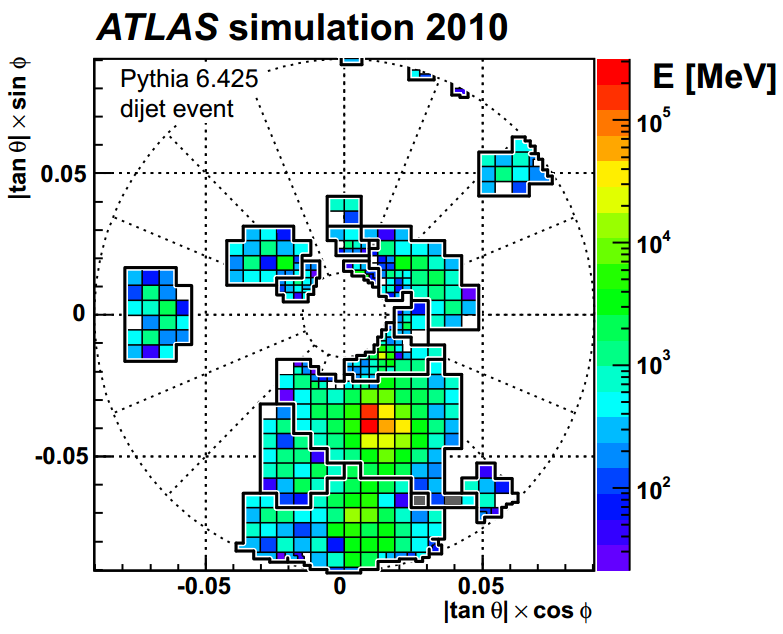}
	\caption{A schematic showing an example of the formation of topo-cluster using a clustering algorithm.~\cite{topocluster}}
	\label{fig:jetsandtaggers:jets:topo}
\end{figure}

The topo-cluster algorithm also includes a splitting step to optimise the separation of showers from other close-by particles. In a topo-cluster, all cells are searched for a local maxima (maximum energy) with a threshold of \SI{500}{\mega\electronvolt}. The local maxima are then used as seeds for a new iteration of topological clustering, which splits the original cluster into more topo-clusters. The jet finding algorithm considers only topo-clusters with positive energy. Fig.\ \ref{fig:jetsandtaggers:jets:topo} shows an example of the formation of topological clusters where each small unit represents a calorimeter cell, and its colour represents the amount of energy deposited in it. These cells are grouped into topo-clusters shown by black lines that represent the borders of the topo-clusters.~\cite{thesis:tanja} 

After the clusters are formed by using the topo-cluster algorithm, anti-$k_{\text{T}}$ algorithm is used to reconstruct the topo-cluster jets. Two different jet calibrations are used in this thesis.  Small-$R$ jets are calibrated using the electromagnetic (EM) scheme, whereas large-$R$ jets are calibrated using the local cluster weighting (LCW) scheme. The main difference between the LCW and EM calibration schemes is that the LCW calibration classifies clusters as either being electromagnetic or hadronic in origin, whereas the EM calibration does not. Then, based on this classification, the LCW applies an energy correction that is dependent on the type of particle which created the calorimeter deposit.

\subsection{Particle flow jets}
\label{sec:jetsandtaggers:jets:pflow}
Particle flow (PFlow) jet reconstruction is introduced to take full advantage of all the particle sub-detectors to improve the energy resolution of reconstructed physics objects. The primary motivation for using particle flow is that at low energy, the tracking detectors provide a better \pt resolution for charged particles than the calorimeters. The \pt resolution ($\sigma_{\pt}$) of a charged particle in the ATLAS tracking detectors can be expressed as:~\cite{atlas}

\begin{equation}
	\frac{\sigma_{\pt}}{\pt} = 0.05\%\pt \oplus 1\% \,,
\end{equation}
where \pt is in units of \si{\giga\electronvolt}.

\begin{figure}[hbt!]
	\centering
	\includegraphics[width=0.6\linewidth]{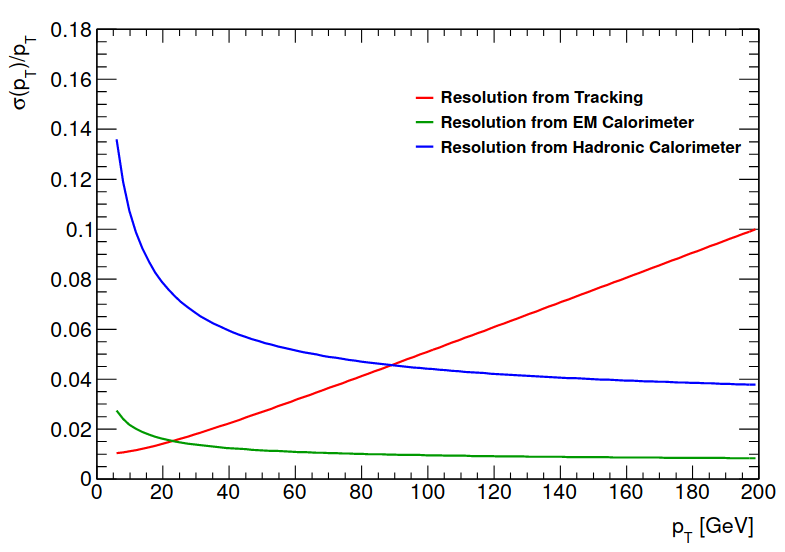}
	\caption{A plot showing the \pt resolution response as a function of \pt for the charged particles. It shows the response for EM and hadronic calorimeter along with the tracking detector.~\cite{atlas}}
	\label{fig:jetsandtaggers:jets:pflow1}
\end{figure}

Fig.\ \ref{fig:jetsandtaggers:jets:pflow1} shows the \pt resolution as a function of \pt for charged particles in the EM and hadronic calorimeters compared to the tracking \pt resolution. It can be seen in this figure that at low \pt, the \pt resolution of particles is better in the tracker than the calorimeters. Therefore, if the \pt measurement from the tracker is used instead of the calorimeter deposits for low-\pt charged particles, there should be an improved \pt resolution for those charged particles.

At the LHC, particle flow has been used by the CMS experiment. In many searches in the ATLAS experiment, particle flow jets are used especially in SUSY and BSM searches, where it is potential to improve the jet energy resolution.

The particle flow algorithm is divided into four different stages, which are described in detail below:

\begin{itemize}
	\item \textbf{Track-cluster matching:} the tracks reconstructed from the hits in the ATLAS inner detector are extrapolated to the calorimeter. This provides the impact parameter coordinates of the extrapolated tracks to different layers of the calorimeters. These extrapolated impact parameter coordinates are then used to find the topo-cluster (described in previous section) that are closest to the extrapolated track. If no track is matched to the topo-cluster, it remained unmodified, and the measurements from the calorimeters are used. The cluster made from the neutral particle is treated as a neutral cluster by the particle flow algorithm. However, if tracks are matched to the cluster, then the cluster continued to the second step, which is charged shower subtraction process.~\cite{pflow}
	
	\item \textbf{Charged shower subtraction:} it is the removal of energy from calorimeter cell deposits from the clusters with associated tracks to avoid any double-counting of energy already deposited by the charged particles in the calorimeters. The amount of energy removed is determined by measuring the calorimeter response ($E/p$). It is defined as the ratio of the energy of a cluster in the calorimeter to the momentum of an associated track. $E/p$ varies depending on the region of the calorimeter and the energy of the incident particle. The energy is subtracted from calorimeter cells in the topo-cluster until the total amount of energy subtracted from the cluster is consistent with the fraction from the $E/p$ distribution.~\cite{pflow} 
	
	\item \textbf{Cluster annihilation:} it includes the removal of calorimeter clusters with associated tracks. If the remaining energy after subtraction is zero, the cluster is removed.
	
	\item \textbf{Neutral particle calibration:} the particle flow algorithm is performed before the calibration is performed on topo-clusters. So the energy deposits in the calorimeter are at the EM scale. After the charged shower subtraction process, only neutral deposits remained in the calorimeter. Therefore, these clusters should be fully calibrated to the appropriate energy scale. Otherwise, jets formed with the remaining topo-clusters would not be properly calibrated.
	
\end{itemize}

\section{Tagging}
\label{sec:jetsandtaggers:taggers}
After the jets are reconstructed, tagging is performed on the jets. Tagging is an identification of the jets originating from the decay of a particle. In this, working point (WP) describes the efficiency of tagging, for example, if tagging is performed at 70\% WP that means 70\% of all the jets which are originating from the decay of that particle are getting tagged by the jet tagging algorithm. This section summarises the two different types of tagging used in this thesis which include different jet tagging algorithm or taggers.

\subsection{$W$-tagging}
\label{sec:jetsandtaggers:taggers:w}
$W$-tagging is an identification of a jet originating from the decay of $W$ boson. It is implemented on large-$R$ jets because all-hadronic decay products of $W$ boson are boosted and can be reconstructed within a single large-$R$ jet. It is performed by a $W$-tagging algorithm which uses the jet substructure information to perform $W$-tagging. The two $W$-tagging algorithms which are based on the jet substructure variables and are used to perform $W$-tagging in this thesis are discussed in detail below:~\cite{wtagger}

\subsubsection{Two-variable tagger}
\label{sec:jetsandtaggers:taggers:twovariable}
The two-variable tagger is used for $W$ tagging of anti-$k_{\text{T}}$ large-$R$ jets of $R=1.0$.~\cite{wtagger} It is locally calibrated for two working points 50\% and 80\% of flat signal efficiency by using the topo-clusters trimmed jets as inputs. The tagger uses two substructure variables to tag $W$ jets: the combined jet mass $m^{\text{comb}}$ and $D_{2}$ variable. $D_{2}$ is a ratio of energy-correlation functions, approximately given by:~\cite{d2}

\begin{equation}
	D_{2} \approx \frac{p_{2}}{p^{2}} \text{ max}(\theta^{2},\theta_{2}^{2}) \,,
\end{equation}
where $p$ is denoted as the momentum of emission with which the jet mass is dominated and $p_{2}$ is denoted as the sum of the momentum of all the other emissions, and $\theta$ and $\theta_{2}$ are the emission angles of the two momenta respectively.

The tagger is optimised with respect to signal jets that are matched to a truth $W$ boson and that have a truth groomed jet mass of $\SI{50}{\giga\electronvolt} < m^{\text{comb}} < \SI{100}{\giga\electronvolt}$ for $W$-tagging. The tagger is optimised using the flat \pt signal sample of $W'\rightarrow WZ\rightarrow qqqq$ against QCD multijet events in MC16d.~\cite{wtagger}

\begin{figure}[hbt!]
	\centering
	\includegraphics[width=0.5\linewidth]{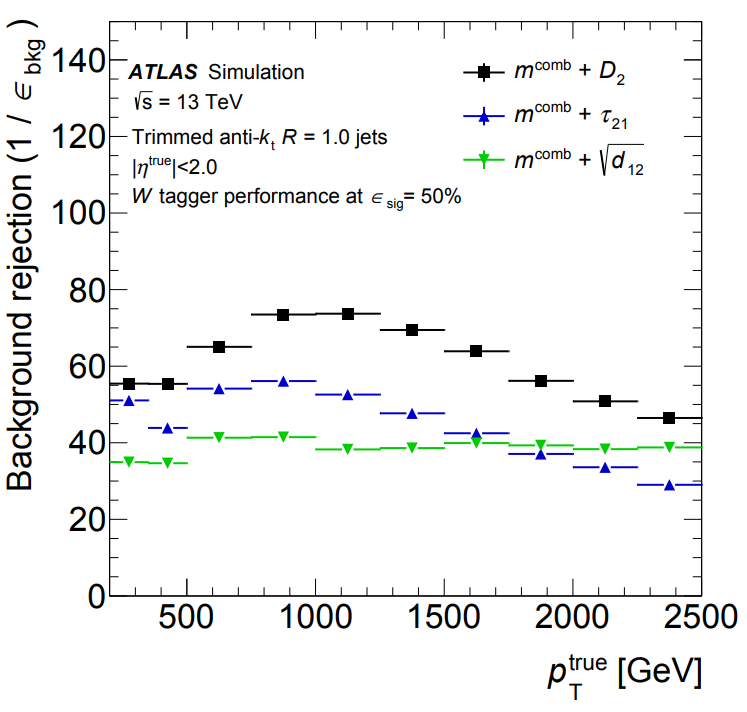}
	\caption{A plot shows the background rejection of $W$-tagging as a function of \pt at a fixed 50\% signal efficiency working point. It shows the response for three different combinations of jet substructure variables.~\cite{wtagging}}
	\label{fig:jetsandtaggers:taggers:twovariable}
\end{figure}

Fig.\ \ref{fig:jetsandtaggers:taggers:twovariable} shows the resulting background rejections as a function of jet \pt for a selection of the most powerful two-variable combinations. Based on this plot, in the case of $W$-tagging, the combination of $m^{\text{comb}}$ and $D_{2}$ is most powerful in the kinematic range of interest and is taken as the baseline pairing for $W$-tagging. However, at higher jet \pt, where the power of $D_{2}$ decreases, it retains constant discrimination power. $W$-tagging for topo-cluster jets calibrated to LCW scheme are only supported up to \SI{2.5}{\tera\electronvolt} due to the degradation of the combined jet mass definition.

\subsubsection{Three-variable tagger}
\label{sec:jetsandtaggers:taggers:threevariable}
The three-variable tagger is also used for $W$-tagging of large-$R$ jets which is locally calibrated on two working points 50\% and 80\% of flat signal efficiency. The tagger uses three substructure variables to perform $W$-tagging: the combined jet mass $m^{\text{comb}}$, $D_{2}$ variable and Ntrk. Ntrk is a number of ghost tracks associated with the ungroomed jet of $\pt>\SI{0.5}{\giga\electronvolt}$. The tagger is optimised using the flat \pt signal sample of $W'\rightarrow WZ\rightarrow qqqq$ against QCD multijet events in MC16d.~\cite{wtagger}

\begin{figure}[hbt!]
	\centering
	\begin{subfigure}{.45\textwidth}
		\centering
		\includegraphics[width=\linewidth,height=\textheight,keepaspectratio]{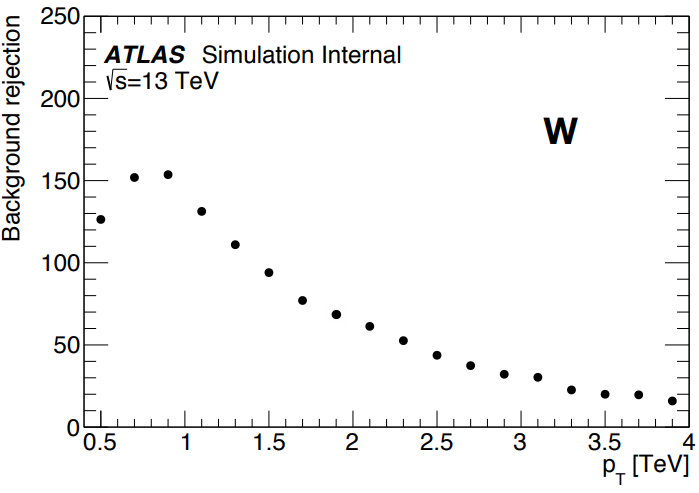}
		\caption{}
		\label{fig:jetsandtaggers:taggers:threevariable:50}
	\end{subfigure}\hspace{0.3cm}
	\begin{subfigure}{.45\textwidth}
		\centering
		\includegraphics[width=\linewidth,height=\textheight,keepaspectratio]{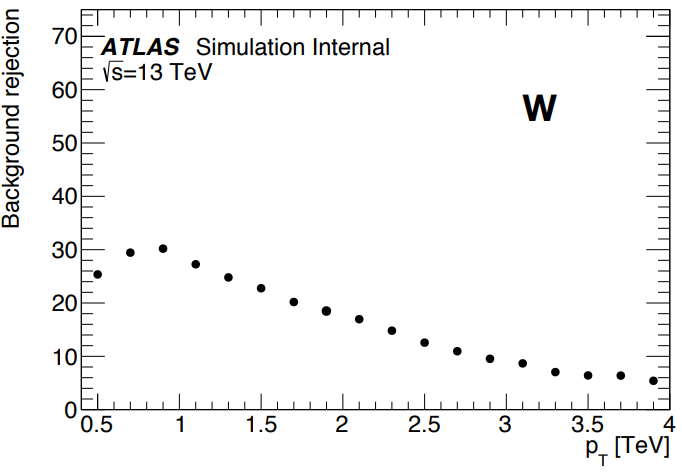}
		\caption{}
		\label{fig:jetsandtaggers:taggers:threevariable:80}
	\end{subfigure}
	\caption{The plots are showing the background rejection of three-variable tagger at a fixed (a) 50\% and (b) 80\% signal efficiency working point as a function of jet \pt.~\cite{wtagger}}
	\label{fig:jetsandtaggers:taggers:threevariable}
\end{figure}

Fig.\ \ref{fig:jetsandtaggers:taggers:threevariable} shows background rejection of the three-variable tagger as a function of jet \pt. It shows the response at two signal efficiency working points, i.e.\ 50\% and 80\%. One can compare the background rejection response of three-variable tagger with the two-variable tagger shown in Fig.\ \ref{fig:jetsandtaggers:taggers:twovariable} and notice that three-variable tagger has better rejection power at 50\% working point. These taggers show the best results at around \SI{1}{\tera\electronvolt} and then rejection power gradually decreases at higher \pt. At 80\% working point, the rejection power decreases, as expected, since it has lower selection criteria as compared to 50\% working point.

\begin{figure}[hbt!]
	\centering
	\includegraphics[width=0.5\linewidth]{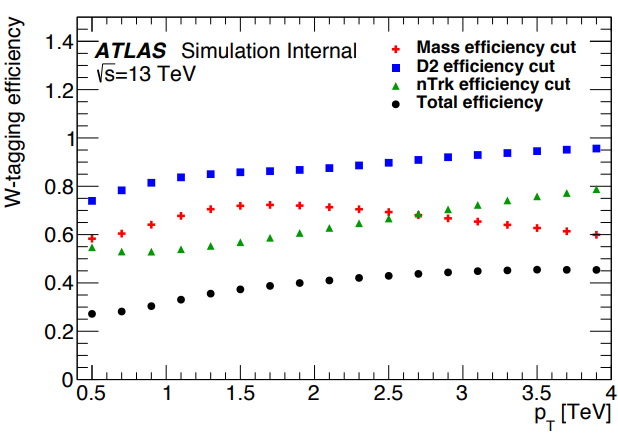}
	\caption{A plot is showing the $W$-tagging efficiency of three-variable tagger as a function of jet \pt. It also shows an individual response of all the three jet substructure variables.~\cite{wtagger}}
	\label{fig:jetsandtaggers:taggers:threevariable:signal}
\end{figure}

Fig.\ \ref{fig:jetsandtaggers:taggers:threevariable:signal} shows the $W$-tagging efficiency of the three-variable tagger as a function of \pt along with the individual responses of all the three jet substructure variables. One can see that the mass efficiency cut started dropping after \SI{1.5}{\tera\electronvolt} whereas the other efficiencies increase at high \pt. The efficiency of $D_{2}$ variable has a higher $W$-tagging efficiency among the other two substructure variable. That is why it is the main input of these two $W$-tagging algorithms. 

\subsection{$b$-tagging}%
\label{sec:analysisstrategy:physicsobjets:bjets}
$b$-tagging is an identification of jets originating from the decay of $b$-quark. It is one of the important selection criteria for this analysis. Jets originating from bottom quarks (called $b$-jets) are identified by reconstructing secondary and tertiary vertices from the tracks associated with the jets. The idea behind this is when a $b$-quark is produced, it hadronises and produce $B$-hadrons. They have an interesting property of a long lifetime. That means they can travel a measurable distance before they further decay. This distance is in order of few millimetres. So, if there is a secondary vertex (where $B$-hadron decays) which is displaced from the primary vertex (point of collision) in a jet, the jet can be tagged as $b$-jet. Moreover, due to the large mass of $B$-hadron, the decay products of $B$-hadron may have large \pt and large opening angle.~\cite{thomson}

The $b$-tagging algorithm used in this analysis is a multivariate algorithm called MV2c10, where training is performed on signal $b$-jets against a mixture background of about $90\%$ light-flavoured jets and $10\%$ $c$-jets. There are several working points which are provided for $b$-tagging. They are defined by a single cut on the MV2c10 output so that a particular $b$-jet efficiency is satisfied.~\cite{thesis:rui} Table \ref{table:analysisstrategy:physicsobjets:bjets} summarises the efficiencies for different working points. It also shows the cut value on the output for each of these working points. The rejection of $c$-jets and light-flavoured jets are also shown in the table. Rejection is calculated such that one out of this number of jets is misidentified as $b$-jet. 

\begin{table}[hbt!]
	\centering
	\begin{tabular}{c | c | c | c} 
		\toprule
		Working point & Cut value on output & $c$-jet rejection & Light-jet rejection \\
		\midrule
		60\% & 0.9349 & \num{34} & \num{1538} \\
		70\% & 0.8244 & \num{12} & \num{381} \\
		77\% & 0.6459 & \num{6} & \num{134} \\
		85\% & 0.1758 & \num{3.1} & \num{33} \\
		\bottomrule
	\end{tabular}
	\caption{Efficiency and cut value on output for each of the working points of the MV2c10 algorithm. Rejection of $c$-jets and light-flavoured jets are also shown. The values are estimated using a simulated $t\bar{t}$ sample.~\cite{thesis:rui}}
	\label{table:analysisstrategy:physicsobjets:bjets}
\end{table}

In this analysis, $70\%$ working point is chosen to balance $b$-tagging efficiency and rejection power. It requires the MV2c10 score to be higher than \num{0.8244}, which corresponds to a rejection factor of $c$-jets and light-flavoured jets to be 12 and 381 respectively. The $b$-tagging is performed on small-$R$ jets and the small-$R$ jets satisfying these criteria are referred to as $b$-jets.


\chapter{Analysis strategy}
\label{sec:analysisstrategy}
The purpose of this chapter is to provide a brief overview of the $T/Y\rightarrow Wb$ process. The contribution from the Standard Model backgrounds is discussed in detail in section \ref{sec:analysisstrategy:backgrounds}. The generators used for the production of MC samples are also outlined. Moreover, the physics objects are described in addition to the event selection used for this analysis. Finally, a motivation for the need of a data-driven method is given.

\section{Overview of $T/Y\rightarrow Wb$ }%
\label{sec:analysisstrategy:t/y}
The analysis concentrates on a search for single production of vector-like quarks $Q$ by the fusion of $W$ boson and $b$-quark in $pp$ collisions ($pp\rightarrow Qqb$) with a subsequent decay of $Q \rightarrow Wb$. Here $Q$ can either be $T$ quark, $Y$ quark or their antiquarks ($\bar{Q}=\bar{T}\text{ or }\bar{Y}$). Since it is an inclusive analysis so the search includes both the final states $W^{\text{+}}b$ and $W^{\text{-}}\bar{b}$ coming from the decay of $Q$ and $\bar{Q}$ respectively.

There are two decay modes of $W$ boson. It can either decay leptonically or hadronically depending on the branching fraction of a decay mode. The analysis targets two channels depending on the decay of $W$ boson. Both the channels are described in detail in the section below:

\subsection{Leptonic channel}%
\begin{figure}[hbt!]
	\centering
	\includegraphics[width=0.6\linewidth]{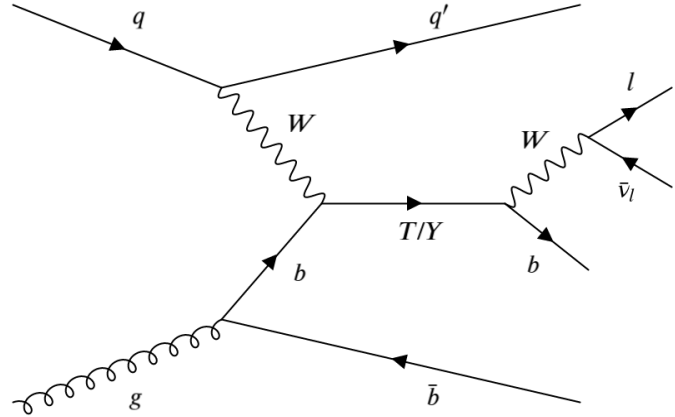}
	\caption{A leading-order Feynman diagram for single production of $T/Y$ quark in $pp \rightarrow Q\rightarrow Wb$ process where $W$ boson further decays leptonically.}
	\label{fig:analysisstrategy:leptonicchannel}
\end{figure}

\label{sec:analysisstrategy:leptonicchannel}
This section describes the leptonic channel of $T/Y\rightarrow Wb$ process. In this, $W$ boson decays leptonically, giving a lepton + jets signature which is characterised by the presence of exactly one electron (or muon), two (or more jets) and missing transverse momentum from the neutrino.~\cite{vlqpaper}. A leading-order (LO) Feynman diagram for the leptonic channel is shown in Fig.\ \ref{fig:analysisstrategy:leptonicchannel}. 

The main background processes from a single-lepton signature arise from ($t\bar{t}$), single top and $W$ boson production in association with jets ($W$+jets), with smaller contributions from $Z$ boson production in association with jets ($Z$+jets) and diboson ($WW$, $WZ$, $ZZ$) production. Multijet events can also contribute to the background if a jet is misidentified as an electron or non-prompt electron (or muon).~\cite{vlqpaper} Processes like $t\bar{t}W$, $t\bar{t}Z$ and $t\bar{t}H$ are also considered which can also produce a very small effect to the background.

\begin{figure}[hbt!]
	\centering
	\includegraphics[width=0.55\linewidth]{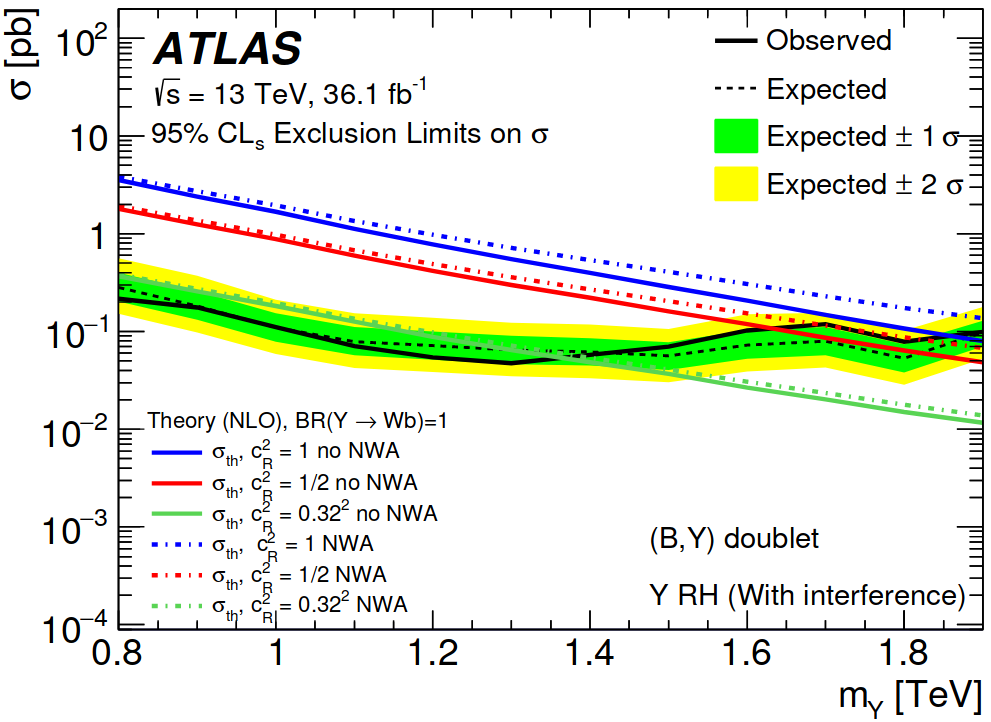}
	\caption{The expected and observed 95\% CL limits for right-handed $Y$ quark from ($B,Y$) doublet as a function of VLQ mass are shown for cross-section times branching ratio $\mathcal{B}(Y\rightarrow Wb)$.~\cite{vlqpaper}}
	\label{fig:analysisstrategy:leptonicchannel:limit}
\end{figure}

The analysis for leptonic channel with the ATLAS 2015-16 data corresponding to integrated luminosity of $\SI{36.1}{\femto\barn^{\text{-1}}}$ results in setting the mass-coupling limits.~\cite{vlqpaper} Fig.\ \ref{fig:analysisstrategy:leptonicchannel:limit} shows 95\% confidence level (CL) limits on cross-section times branching ratio of the right-handed $Y$ quark for a ($B,Y$) doublet model as a function of VLQ mass. The branching ratio $\mathcal{B}(Y\rightarrow Wb)$ is set to one for theoretical predictions. The area above the solid line gives the excluded region. This result is from the previous analysis and now a new analysis with the ATLAS 2015-18 data corresponding to an integrated luminosity of $\SI{139.1}{\femto\barn^{\text{-1}}}$ with better and improved methods is also being done.

\subsection{Hadronic channel}%
\begin{figure}[hbt!]
	\centering
	\includegraphics[width=0.8\linewidth]{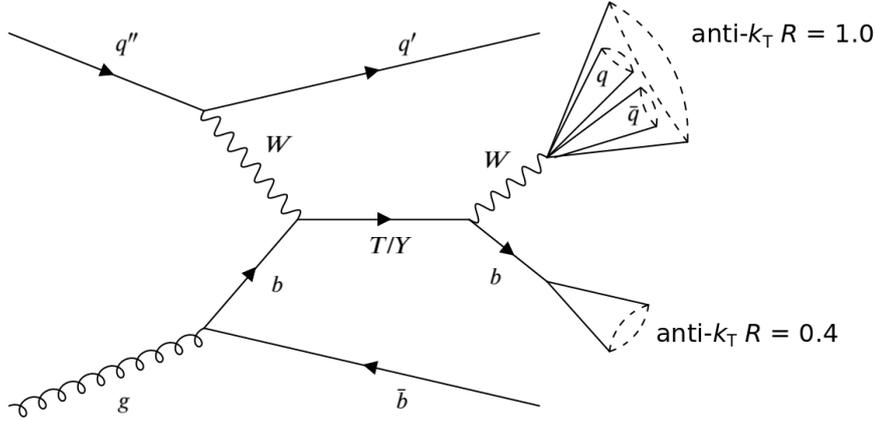}
	\caption{A leading-order Feynman diagram for single production of $T/Y$ quark in $pp \rightarrow Q\rightarrow Wb$ process where $W$ boson further decays hadronically.}
	\label{fig:analysisstrategy:hadronicchannel}
\end{figure}

\label{sec:analysisstrategy:hadronicchannel}
This channel is optimised to a search for $T/Y$ quark which decays into a high $p_{\text{T}}$ $W$ boson and $b$-quark, where $W$ boson decays hadronically. Both $W$ boson and $b$-quark are approximately back-to-back in the transverse plane since both originate from the decay of heavy particle. Fig.\ \ref{fig:analysisstrategy:hadronicchannel} shows a leading-order Feynman diagram of the hadronic decay of $W$ boson in $T/Y \rightarrow Wb$ process. There is an additional $b$-quark which is produced from the gluon splitting can be observed in either the forward or central region. This $b$-quark usually produces a jet of low energy which falls outside the acceptance region of the detector. There is also a light-flavour quark $q'$ in the process which often produces a jet in the forward region of the detector.

All the main backgrounds with an all-jets signature arising from different SM processes are described in section \ref{sec:analysisstrategy:backgrounds}.

\subsection{Motivation for hadronic channel}%
\label{sec:analysisstrategy:motivation}
This thesis is focused on the hadronic channel of $T/Y\rightarrow Wb$. So, it is important to mention why it is interesting to study this decay mode. Some of the important points are described below:
\begin{itemize}
	\item $W$ boson is a vector boson, which has a very short lifetime. This means that within the ATLAS detector, it is never observed directly, only the decay products can be measured. $W$ boson decays into two fermions, i.e.\ either into a lepton and anti-neutrino ($\Plepton\APnulepton$) or a quark anti-quark ($q\bar{q}$) pair, where \Plepton includes all the three lepton flavours (\Pelectron, \Pmuon and \Ptauon). The branching fraction of $\mathcal{B}(W\rightarrow q\bar{q})$ is higher than $\mathcal{B}(W\rightarrow \Plepton\APnulepton)$, which makes the hadronic decay of $W$ boson an interesting channel to probe the kinematics of an analysis. 
	
	\item The decay of a heavy particle $T/Y$ produces high \pt $W$ boson and $b$-quark. So, the hadronic decay products of $W$ boson are boosted, which allow us to use large-$R$ jets to reconstruct $W$ boson. It also allows to use the taggers developed for such decays. Two of these taggers which are used in this analysis are already discussed in section \ref{sec:jetsandtaggers:taggers}. 
	
	\item Moreover, the leptonic analysis with 2015-18 ATLAS data for setting mass-coupling limits is being done. So, an addition of hadronic analysis is used to increase the sensitivity.~\cite{leptonic:twiki}
\end{itemize}

\section{Backgrounds}%
\label{sec:analysisstrategy:backgrounds}
This section covers all the possible backgrounds which are significant for the hadronic channel of $T/Y\rightarrow Wb$ that are studied for this analysis. The final state consists of $W$ boson with $b$-quarks, which resembles with the final state of many SM processes, for example, single top, $t\bar{t}$, $W$ boson and $Z$ boson accompanied with jets and multijet, which is a dominant background. The samples which are used to predict these backgrounds are produced from the Monte Carlo (MC) simulation, which is described in detail in section \ref{sec:analysisstrategy:mc:background}. The contribution from diboson processes ($WW$, $WZ$, $ZZ$) is negligible compared to the other backgrounds. Also, the MC samples for the hadronic decay of dibsoson process are not yet modelled. Therefore, they are not mentioned explicitly. 
\subsubsection{Single top}%
\begin{figure}[hbt!]
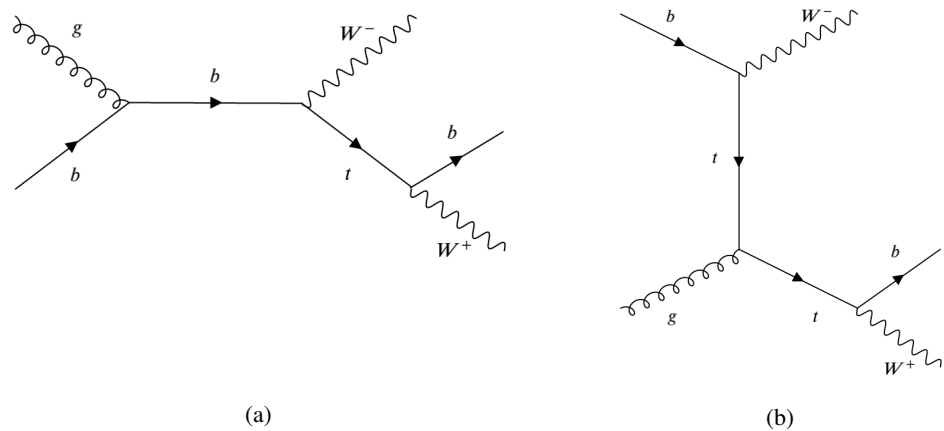

	\centering
	\begin{subfigure}{.45\textwidth}
		\centering
		\includegraphics[width=\linewidth,height=\textheight,keepaspectratio]{singletop.png}\vspace{1.5cm}
		\caption{}
		\label{fig:analysisstrategy:singletop:s}
	\end{subfigure}\hspace{1cm}
	\begin{subfigure}{.3\textwidth}
		\centering
		\includegraphics[width=\linewidth,height=\textheight,keepaspectratio]{singletop1.png}
		\caption{}
		\label{fig:analysisstrategy:singletop:t}
	\end{subfigure}
	\caption{A leading-order Feynman diagram of (a) $s$-channel like and (b) $t$-channel like $tW$ production in a five-flavour scheme.}
	\label{fig:analysisstrategy:singletop}
\end{figure}

\label{sec:analysisstrategy:singletop}
Single top processes are the events in which top-quark is produced by the electroweak interaction. There are three main channels for the production of single top i.e.\ $s$-channel, $t$-channel and $tW$-channel. At the LHC, $t$-channel is a dominant process followed by $tW$-channel and then $s$-channel. In $tW$ production, the most distinct feature is a fermionic propagator. The $tW$ channel can be drawn as $s$-channel-like or $t$-channel-like. Fig.\ \ref{fig:analysisstrategy:singletop:s} shows a leading-order Feynman diagram of $s$-channel like $tW$ production, in which a bottom-quark first interacts with a gluon to become a virtual bottom-quark which then decays into $W$ boson and top-quark. Fig.\ \ref{fig:analysisstrategy:singletop:t} shows a leading-order Feynman diagram of $t$-channel like $tW$ production, in which a bottom-quark first emits a $W$ boson and turns to a virtual top-quark, which then interacts with a gluon to produce a top-quark. Both these diagrams require a bottom-quark inside a proton that means five flavours of parton should be treated as active inside a proton. It is known as a five-flavour scheme. The top-quark in both diagrams can further decay into $W$ boson and accompanied $b$-quark.~\cite{thesis:rui}  

\subsubsection{\ensuremath{t\bar{t}}}%
\begin{figure}[hbt!]
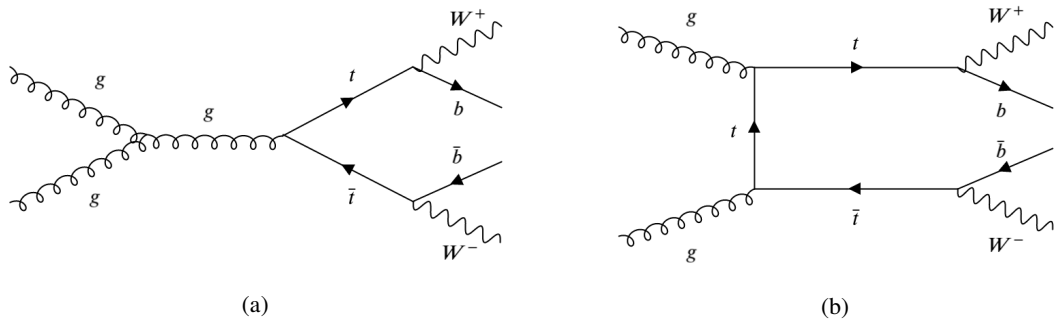

	\centering
	\begin{subfigure}{.45\textwidth}
		\centering
		\includegraphics[width=\linewidth,height=\textheight,keepaspectratio]{ttbar.png}
		\caption{}
		\label{fig:analysisstrategy:ttbar1}
	\end{subfigure}\hspace{1cm}
	\begin{subfigure}{.4\textwidth}
		\centering
		\includegraphics[width=\linewidth,height=\textheight,keepaspectratio]{ttbar1.png}
		\caption{}
		\label{fig:analysisstrategy:ttbar2}
	\end{subfigure}
	\caption{Some leading-order Feynman diagrams for $t\bar{t}$ production by $gg$ fusion.}
	\label{fig:analysisstrategy:ttbar}
\end{figure}

\label{sec:analysisstrategy:ttbar}
Top-quark pair production ($t\bar{t}$) can be produced either by gluon-gluon fusion ($gg$) or by annihilation of quark anti-quark pair ($q\bar{q}$). At the LHC, $gg$ fusion becomes a dominant process which contributes $\sim90\%$ of $t\bar{t}$ production. Some leading-order Feynman diagrams of $t\bar{t}$ production by $gg$ fusion are shown in Fig.\ \ref{fig:analysisstrategy:ttbar}. The final states $t$ and $\bar{t}$ can further decay into $\PWplus b$ and $\PWminus\bar{b}$ respectively and hence considered as one of the background processes for this analysis. The contribution from $t\bar{t}$ is higher than single top due to its larger production cross-section.~\cite{thesis:rui}

\subsubsection{$W$+jets}%
\begin{figure}[hbt!]
	\centering
	\includegraphics[width=0.35\linewidth]{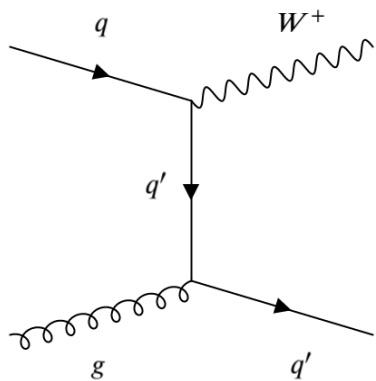}
	\caption{A leading-order Feynman diagram for the production of $W$ boson accompanied by jets.}
	\label{fig:analysisstrategy:wjets}
\end{figure}

\label{sec:analysisstrategy:w+jets}
Another potentially significant background for the hadronic analysis is the production of $W$ boson accompanied by jets. An example of a corresponding Feynman diagram is shown in Fig. \ref{fig:analysisstrategy:wjets} where a quark first emits a $W$ boson, changes its flavour and turns into a virtual quark which further interacts with a gluon to give a quark which can be $b$-quark. So, this process also imitates the $T/Y\rightarrow Wb$ final state similar to single top production. $W$+jets process has significantly higher production-cross-section than single top and $t\bar{t}$.

\subsubsection{$Z$+jets}%
\begin{figure}[hbt!]
	\centering
	\includegraphics[width=0.33\linewidth]{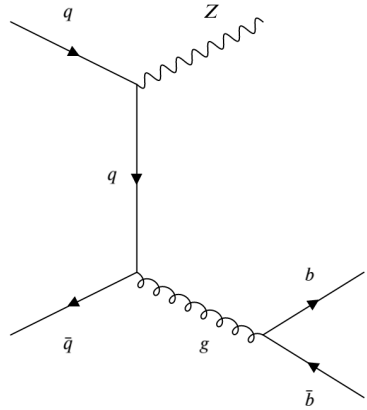}
	\caption{A leading-order Feynman diagram for the production of $Z$ boson accompanied by jets.}
	\label{fig:analysisstrategy:zjets}
\end{figure}

\label{sec:analysisstrategy:z+jets}
This process includes the production of $Z$ boson accompanied by jets, where jets can be of any flavour. The branching ratio of $Z$ boson decaying into hadrons $\mathcal{B}(Z\rightarrow\text{ hadrons})$ $\sim70\%$. So, the contribution from $Z$+jets is significant for the background when $Z$ boson decays into $q\bar{q}$ with the accompanied jets to be $b$-jets. An example of a leading-order Feynman diagram for $Z$+jets process is shown in Fig.\ \ref{fig:analysisstrategy:zjets}. In this diagram, $Z$ boson is emitted from a quark which further annihilates with anti-quark to produce a gluon, which decays into $b\bar{b}$ pair.

\subsubsection{Multijet}%
\label{sec:analysisstrategy:multijets}
The hadronic channel is not a clean channel like leptonic channel because of the contribution from the QCD processes which accounts for a significant background. These QCD processes are very challenging because their theoretical prediction is so difficult to describe and also there are several experimental challenges associated with them. A detailed discussion on QCD has been outlined in section \ref{sec:theory:standardmodel:strong}.

\begin{figure}[hbt!]
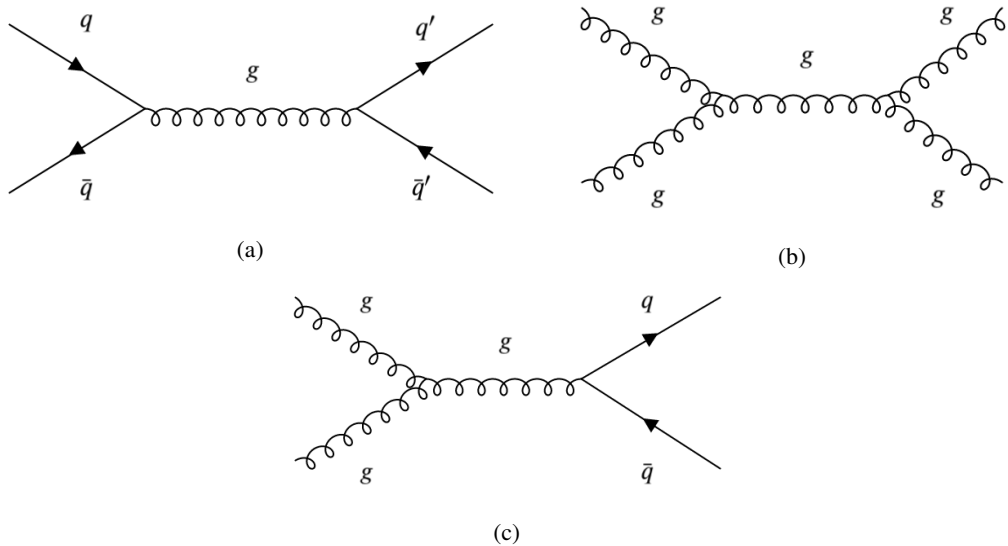

	\centering
	\begin{subfigure}{.45\textwidth}
		\centering
		\includegraphics[width=\linewidth,height=\textheight,keepaspectratio]{multijets1.png}
		\caption{}
		\label{fig:analysisstrategy:multijets1}
	\end{subfigure}\hspace{0.5cm}
	\begin{subfigure}{.4\textwidth}
		\centering
		\includegraphics[width=\linewidth,height=\textheight,keepaspectratio]{multijets2.png}
		\caption{}
		\label{fig:analysisstrategy:multijets2}
	\end{subfigure}\hspace{0.5cm}
	\begin{subfigure}{.4\textwidth}
	\centering
	\includegraphics[width=\linewidth,height=\textheight,keepaspectratio]{multijets3.png}
	\caption{}
	\label{fig:analysisstrategy:multijets3}
	\end{subfigure}
\caption{Some examples of a leading-order Feynman diagram for multijet production from gluon splitting.}
\label{fig:analysisstrategy:multijets}
\end{figure}

In the $pp$ collision at $\sqrt{s}=\SI{13}{\tera\electronvolt}$, there is a lot of hadronic activity going on which constitute a large number of jets. Multijet include jets of any flavour. There are several ways for the production of multijet at the LHC. Some of them are depicted in Fig.\ \ref{fig:analysisstrategy:multijets}. These diagrams show the production of $q\bar{q}$ pair and $gg$ pair from gluon splitting. Multijet production has a higher cross-section in order of magnitude as compared to the other backgrounds. That is why multijet is the dominant background in the all-hadronic channel of this analysis. They have a problem with modelling because of the QCD processes.

\section{Datasets}%
\label{sec:analysisstrategy:datasets}

\subsection{Data samples}%
\label{sec:analysisstrategy:datasets:data}
In this thesis, full ATLAS Run 2 data have been used that means the data were collected from 2015 to 2018 with the ATLAS detector at $\sqrt{s}$ = 13 \si{\tera\electronvolt}. The selected data were recorded during stable LHC conditions and with the full functionality of the ATLAS detector. The data also fulfils the condition of \enquote{Good Run Lists (GRL)}. The GRL selects data where all of the detector systems were in good quality. Good quality means that all of the detectors were performing sufficiently well for the data to be used in physics analyses. The total amount of data satisfying these conditions correspond to a total integrated luminosity of $\SI{139}{\femto\barn^{\text{-1}}}$. The integrated luminosities of data obtained in different years are summed in Table \ref{table:analysisstrategy:datasets:data}.
\begin{table}[hbt!]
	\centering
	\begin{tabular}{c | c} 
		\toprule
		Data taken in & $\mathcal{L}_{\text{int}}$ (\si{\femto\barn^{\text{-1}}}) \\
		\midrule
		2015-16 & 36.2 \\
		2017 & 44.3 \\ 
		2018 & 58.5 \\ 
		\bottomrule
	\end{tabular}
	\caption{Integrated luminosities of the data for all the years of LHC Run 2 operation.~\cite{intnote:zboson}}
	\label{table:analysisstrategy:datasets:data}
\end{table}

\subsection{Monte Carlo (MC) samples}%
\label{sec:analysisstrategy:mc}
In order to study physics, a well-simulated sample is needed, which can predict the contribution from each background processes as well as the expected signal process. For this, the Monte-Carlo simulation is used, which is based on the stochastic methods to simulate various processes. They are used to predict event rates, simulate possible backgrounds, study detector effects, etc. The predictions from the MC samples can be then compared to the results achieved from the data.

The generation of MC samples is divided into three steps. First, the underlying events are generated using the help of a general-purpose event generator, e.g.\ \textsc{Sherpa}~\cite{sherpa}. After this, the hadronisation process of these events is simulated by the help of parton showering software. At last, these events are passed to the detector simulation software called \textsc{Geant4}~\cite{geant}, to simulate the detector effects as if it is passed to the real ATLAS detector to detect the particles. The last step includes either a full simulation or through a faster simulation making use of parameterised showers in the calorimeters. 

The effects of both in-time and out-of-time pile-up\footnote{In-time pile-up means additional $pp$ collisions in the same bunch crossings, and out-of-time pile-up means additional $pp$ collisions in the nearby bunch crossings.} have been modelled in the MC samples by reweighting. The reweighting is done to match the pile-up conditions observed in data taken in the respective year. The MC samples are modelled by following the common Physics Modelling Group (PMG) recommendations. 

MC samples simulating Run 2 conditions are divided into three campaigns, labelled MC16a, MC16d and MC16e corresponding to the pile-up conditions of the data taken in 2015-16, 2017 and 2018 respectively. A general overview of the simulated samples used for the expected signal events and background events is given in the sections below:

\subsubsection{Signal}%
\label{sec:analysisstrategy:mc:signal}
The simulated events for signal processes are generated at leading-order in a four-flavour scheme with the \textsc{Madgraph5}~\cite{madgraph} generator, interfaced to \textsc{Pythia8}~\cite{pythia} for parton showering and hadronisation. Since the mass of $T/Y$ quark is not known, a search scan has to be performed for all the possible mass points. So, the signal samples are produced for masses ranging from $\SI{1.1}{\tera\electronvolt}$ to $\SI{2.3}{\tera\electronvolt}$ in steps of $\SI{200}{\giga\electronvolt}$ with $k_\text{T} = 1.0$. The coupling parameter $k_{\text{T}}$ in the model described in Ref.\ \cite{wulzer}, used for the signal production is related to the coupling parameters $c_{\text{L,R}}^{\text{Wb}}$ in section \ref{sec:theory:models} by: 
\begin{equation*}
k_{\text{T}}f(m) = \frac{c_{\text{L,R}}^{\text{Wb}}}{\sqrt{2}}, \hspace{1cm} \text{where } f(m)= \sqrt{\frac{1}{1+\mathcal{O}(m_{\text{VLQ}}^{\text{-4}})}} \,.
\end{equation*}
At good approximation, the above equation can be written as:

\begin{equation}
	k_{\text{T}} = \frac{c_{\text{L,R}}^{\text{Wb}}}{\sqrt{2}} \,.
	\label{key}
\end{equation}
These samples are processed through a full detector simulation using \textsc{Geant4}~\cite{geant}. The signal samples are generated in such a way that the samples for other coupling strengths can be achieved by reweighting these samples. Since the kinematic distributions of the decay products for $T$ quark and $Y$ quark in $Wb$ decay channel are same, only $Y$ quark signal samples are generated and they are used to derive the results for the $T$ quark signals. Other possible decay modes of $T$ quark ($T\rightarrow Zt$, $T\rightarrow Ht$) have negligible contribution in this search.~\cite{vlqpaper} 

A summary of all the signal sample for VLQ mass and their production cross-section in $pp$ collision is shown in Table \ref{table:analysisstrategy:mc:signal}.

\begin{table}[hbt!]
	\centering
	\begin{tabular}{c | c} 
		\toprule
		VLQ mass (\si{\tera\electronvolt}) & $\sigma$ (\si{\pico\barn}) \\
		\midrule
		1.1 & \SI{0.478}{} \\
		1.3 & \SI{0.231}{} \\
		1.5 & \SI{0.114}{} \\ 
		1.7 & \SI{0.057}{} \\ 
		1.9 & \SI{0.029}{} \\ 
		2.1 & \SI{0.015}{} \\ 
		2.3 & \SI{0.008}{} \\ 
		\bottomrule
	\end{tabular}
	\caption{Overview of all the signal samples for different VLQ masses along with their production cross-section in the $pp$ collision.~\cite{wulzer}}
	\label{table:analysisstrategy:mc:signal}
\end{table}

\subsubsection{Background}%
\label{sec:analysisstrategy:mc:background}
 All the possible background processes described in section \ref{sec:analysisstrategy:backgrounds} are listed in Table \ref{table:analysisstrategy:mc:background} along with the softwares used for their production.
 
 \begin{table}[hbt!]
 	\centering
 	\begin{tabular}{c | c | c} 
 		\toprule
 		Background & MC generator & Parton shower \\ [0.5ex]
 		\midrule
 		Single top & \textsc{Powheg-box} & \textsc{Pythia 8} \\
 		$t\bar{t}$ & \textsc{Powheg-box} & \textsc{Pythia 8} \\
 		$W$+jets & \textsc{Sherpa} & \textsc{Sherpa} \\ 
 		$Z$+jets & \textsc{Sherpa} & \textsc{Sherpa} \\ 
 		Multijets & \textsc{Pythia 8} & \textsc{Pythia 8} \\ 
 		\bottomrule
 	\end{tabular}
 	\caption{Overview of all the background samples used in this analysis along with their exact mode of MC production.}
 	\label{table:analysisstrategy:mc:background}
 \end{table}

\subsubsection{MC weights}%
\label{sec:analysisstrategy:mc:weights}
The simulated MC samples may not represent the actual conditions required for the analysis. Moreover, the selection efficiencies of the algorithm or method may not be efficient for different physics objects. In order to compensate these effects, MC simulated events have to be multiplied with a factor known as $weight$. There are several weights for the correction of various processes which are being calculated from different methods. The following weights are assigned to the MC events for this analysis:
\begin{itemize}
	\item $w_{\text{lumi}}:$ the MC generation needs a high computing power and the events generated are also in the order of millions. Due to the limiting availability of large computing requirement, MC samples are usually generated for a certain number of events depending on their cross-sections and then normalised to the required luminosity. This normalisation is done by multiplying it with a factor called $w_{\text{lumi}}$ which can be calculated by: \[w_{\text{lumi}} = \frac{\sigma\mathcal{L}_{\text{int}}}{N} \,,\]
	where $\sigma$ is cross-section of the process, $\mathcal{L}_{\text{int}}$ being the integrated luminosity to which the events have to be normalised and $N$ is defined as the number of events in the original MC sample. The value for $\mathcal{L}_{\text{int}}$ depends on the campaign of each event.
	
	\item $w_{\text{MC}}:$ in MC production, every event describes a possible physics process. The contribution from that process can be decided by its $w_{\text{MC}}$. So, the total number of events in MC sample is a sum of all the weighted events, where weighted events are the product of event and their corresponding $w_{\text{MC}}$. 
	
	\item $w_{\text{pileup}}:$ the run conditions in the production of MC samples are not same as for the recorded data. For example, <$\mu$>, which is defined as the average number of collisions in single bunch crossing, is different in MC simulation compared to the true data taking run conditions. Therefore, the MC events are reweighted to match the <$\mu$> in data.
	
	\item $w_{\text{trigger}}:$ accounts for differences in data and MC efficiencies due to trigger selections which are described in section \ref{sec:analysisstrategy:eventselection:preselection}.
	
	\item $w_{\text{jvt}}:$ it is associated with the reconstruction of jet-vertex in high pile-up environment.
	
	\item $w_{\text{b-tagging}}:$ it accounts for the $b$-tagging efficiency of a tagger described in section \ref{sec:analysisstrategy:physicsobjets:bjets}.
\end{itemize}

In this analysis, all the weights are close to one except $w_{\text{lumi}}$ and $w_{\text{MC}}$. The total weight for each event can be calculated by the product of all the individual weights, as shown in Eqn.\ \ref{eqn:analysisstrategy:mc:weights}.
\begin{equation}
w_{\text{event}} = w_{\text{MC}} \times w_{\text{pileup}} \times w_{\text{trigger}} \times w_{\text{jvt}} \times w_{\text{b-tagging}} \,.
\label{eqn:analysisstrategy:mc:weights}
\end{equation}

\section{Physics objects}%
\label{sec:analysisstrategy:physicsobjets}
The particles interact with a detector in different ways. Each particle has a specific signature. The offline reconstruction algorithms make use of all sub-detector information to reconstruct and identify the physics objects like jets, electrons, etc. This analysis relies on the reconstruction and identification of jets. For jet reconstruction, a radius parameter $R$ is used to define a small-$R$ jet and large-$R$ jet. In order to identify small-$R$ jet which originates from $b$-quark, the $b$-tagging algorithm is used. The objects are reconstructed by following the common Jet and Etmiss working group~\cite{wtagger} object definitions and calibrations and the samples are produced from the \textsc{Exot7} derivation.

\subsection*{Large-$R$ jets}%
\label{sec:analysisstrategy:physicsobjets:largerjets}
The jets considered for large-$R$ jets have $p_{\text{T}} > \SI{25}{\giga\electronvolt}$ and $|\eta| < 2.5$. They are reconstructed from topological calorimeter clusters using the anti-$k_{\text{T}}$ algorithm with $R=1.0$. The hadronic decay products of boosted $W$ boson can be captured inside a single large-$R$ jet. The large-$R$ jets are trimmed to minimise the impact from energy depositions from pile-up interactions which are not associated with the original shower. The trimming algorithm reconstructs $k_{\text{T}}$ jets with a radius parameter of $0.2$ inside the original large-$R$ jet and removes topo-clusters associated with the $k_{\text{T}}$ jet from the original large-$R$ jet if they contribute less than $5\%$ to the large-$R$ jet $p_{\text{T}}$. The new collection of topo-clusters is then used to calculate the jet kinematics and shower properties.~\cite{thesis:rui} In this thesis, the three-variable tagger is used to perform $W$-tagging on large-$R$ jets, which has been discussed in section \ref{sec:jetsandtaggers:taggers:threevariable}. The two-variable tagger described in section \ref{sec:jetsandtaggers:taggers:twovariable} is also used for $W$-tagging and then a performance comparison is evaluated between the two taggers.

\subsection*{Small-$R$ jets}%
\label{sec:analysisstrategy:physicsobjets:smallrjets}
The jets considered for small-$R$ jets have $p_{\text{T}} > \SI{25}{\giga\electronvolt}$ and $|\eta| < 4.5$. They are reconstructed by using the particle-flow algorithm as described in section \ref{sec:jetsandtaggers:jets:pflow} with radius parameter $R=0.4$. After the jets are reconstructed, a dedicated calibration is applied, which includes jet area pile-up suppression and scale factors based on the MC simulations. It brings the measured jet $p_{\text{T}}$ to the particle level.~\cite{thesis:rui} In this thesis, jets reconstructed from the topo-cluster algorithm as described in section \ref{sec:jetsandtaggers:jets:topo}, are also used for small-$R$ jets and performance comparison is evaluated between the two types of jet collections.

\section{Event selection}%
\label{sec:analysisstrategy:eventselection}
This section describes the event selection applied to the data and MC. Event selection is applied to filter the interesting events, at the same time, minimising the contribution from background events. It is performed in two stages:
\begin{enumerate}
	\item Preselection
	\item Definition of the signal (SR), validation (VR) and control region (CR)
\end{enumerate}

Preselection is a set of loose cuts on different parameters to filter out majority of the background, which are described in section \ref{sec:analysisstrategy:eventselection:preselection}. After analysing the kinematics of the physical observables at preselection, a set of more dense cuts are applied to define three different regions called the signal region which is enriched with signal-like events, validation region which is close to the signal region but not exactly the signal region and it is used to validate the method before applying it to the signal region. And, last but not least control region which is entirely dominated by background events. These regions are described in detail in the next chapter.

\subsection{Preselection}%
\label{sec:analysisstrategy:eventselection:preselection}
\begin{itemize}
	\item \textbf{Trigger:} in this analysis, the large-$R$ jets are passed through the high-level trigger (HLT). In this, an unprescaled single jet trigger HLT\_j360 is used in the 2015 data-taking period. It is the lowest unprescaled trigger. For 2016 data-taking period, HLTj\_420 is used and HLT\_j460 is applied for 2017-18 data taking period. The trigger HLTj\_460 reaches the 99.9\% efficiency plateau at around \SI{500}{\giga\electronvolt}.~\cite{intnote:wprime}
	
	\item \textbf{$W$-tagged jet:} an event is selected if it contains exactly one $W$-tagged jet of $p_{\text{T}}$ $\geq$ \SI{500}{\giga\electronvolt}. $W$-tagging is performed on large-$R$ jets by the three-variable tagger at 50\% working point, described in \ref{sec:jetsandtaggers:taggers:threevariable}. The secondary condition is that the leading large-$R$ jet should be $W$-tagged, that means $p_{\text{T}}$ of the $W$-tagged large-$R$ jet should be highest among all the other large-$R$ jets in that event.
	
	\item \textbf{$b$-tagged jet:} an event is selected if it contains atleast one $b$-tagged jet of $p_{\text{T}}$ $\geq$ \SI{200}{\giga\electronvolt}. $b$-tagging is performed on small-$R$ jets by the MV2c10 tagger at 70\% working point, described in section \ref{sec:analysisstrategy:physicsobjets:bjets}. So, in an event there can be more than one $b$-tagged jets but the leading $b$-tagged jet should have $p_{\text{T}}$ $\geq$ \SI{200}{\giga\electronvolt}.
	
	\item \textbf{Lepton veto:} an event is removed if it contains an electron or muon which fulfills the following requirements:
	\begin{itemize}
		\item Electron: $p_{\text{T}}$ $\geq$ \SI{25}{\giga\electronvolt}, $\abs{\eta}$ < 2.47, fulfilling tight likelihood identification criteria, with \enquote{FCTight} isolation requirements.~\cite{leptons}
		\item Muon: $p_{\text{T}}$ $\geq$ \SI{25}{\giga\electronvolt}, $\abs{\eta}$ < 2.47, fulfilling medium likelihood identification criteria, with \enquote{FCTight} isolation requirements.~\cite{leptons}
	\end{itemize}
	
	\item \textbf{Forward jet:} an event is selected if it contains atleast one forward jet. Forward jet is defined as a small-$R$ jet of $p_{\text{T}}$ $\geq$ \SI{25}{\giga\electronvolt}, $\abs{\eta}$ > 2.5 and it is not $b$-tagged at 70\% working point.
\end{itemize}

\begin{figure}[hbt!]
	\centering
	\begin{subfigure}{.45\textwidth}
		\centering
		\includegraphics[width=\linewidth,height=\textheight,keepaspectratio]{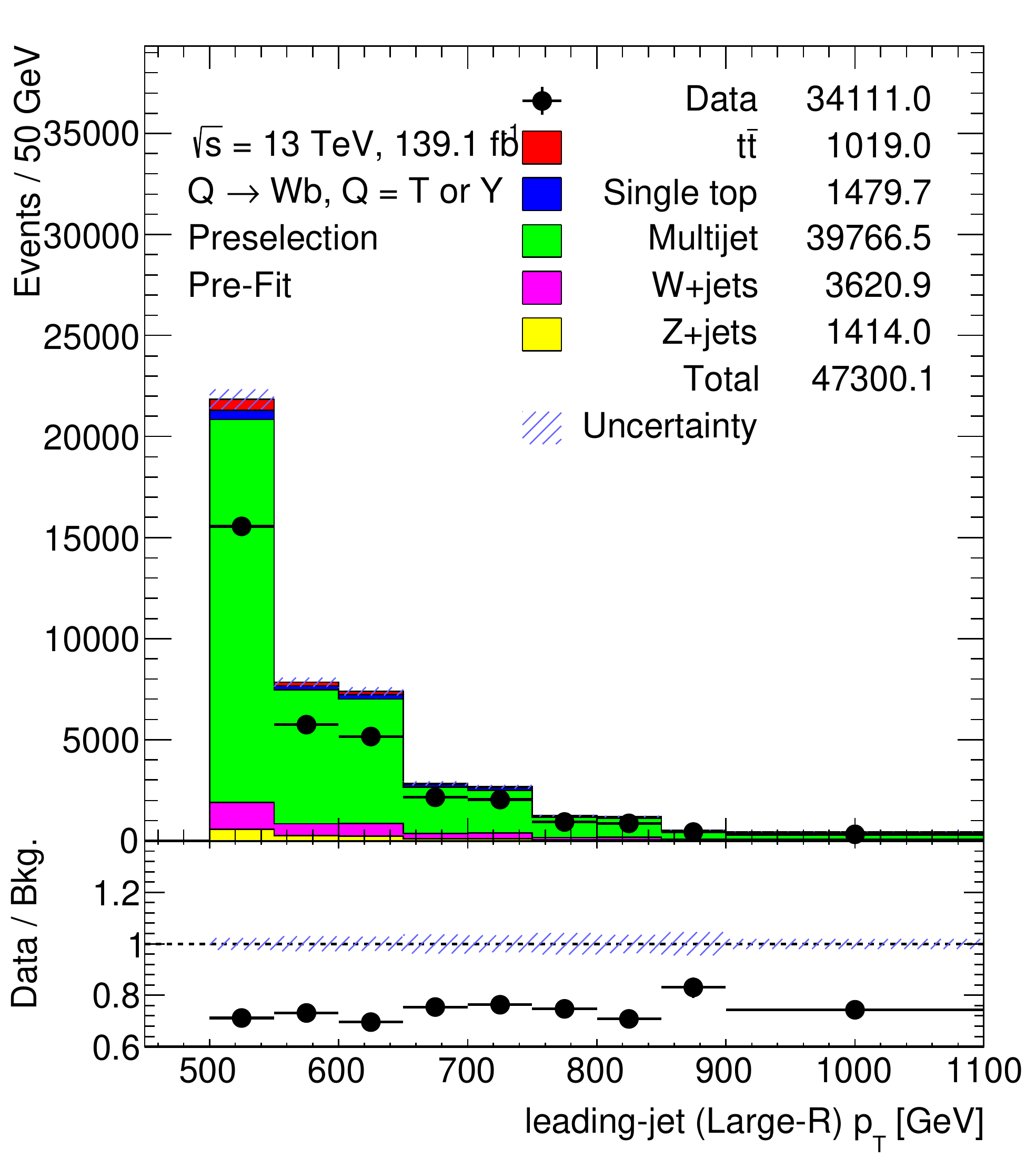}
		\caption{}
		\label{fig:analysisstrategy:eventselection:preselection:ljet}
	\end{subfigure}
	\begin{subfigure}{0.45\textwidth}
		\centering
		\includegraphics[width=\linewidth,height=\textheight,keepaspectratio]{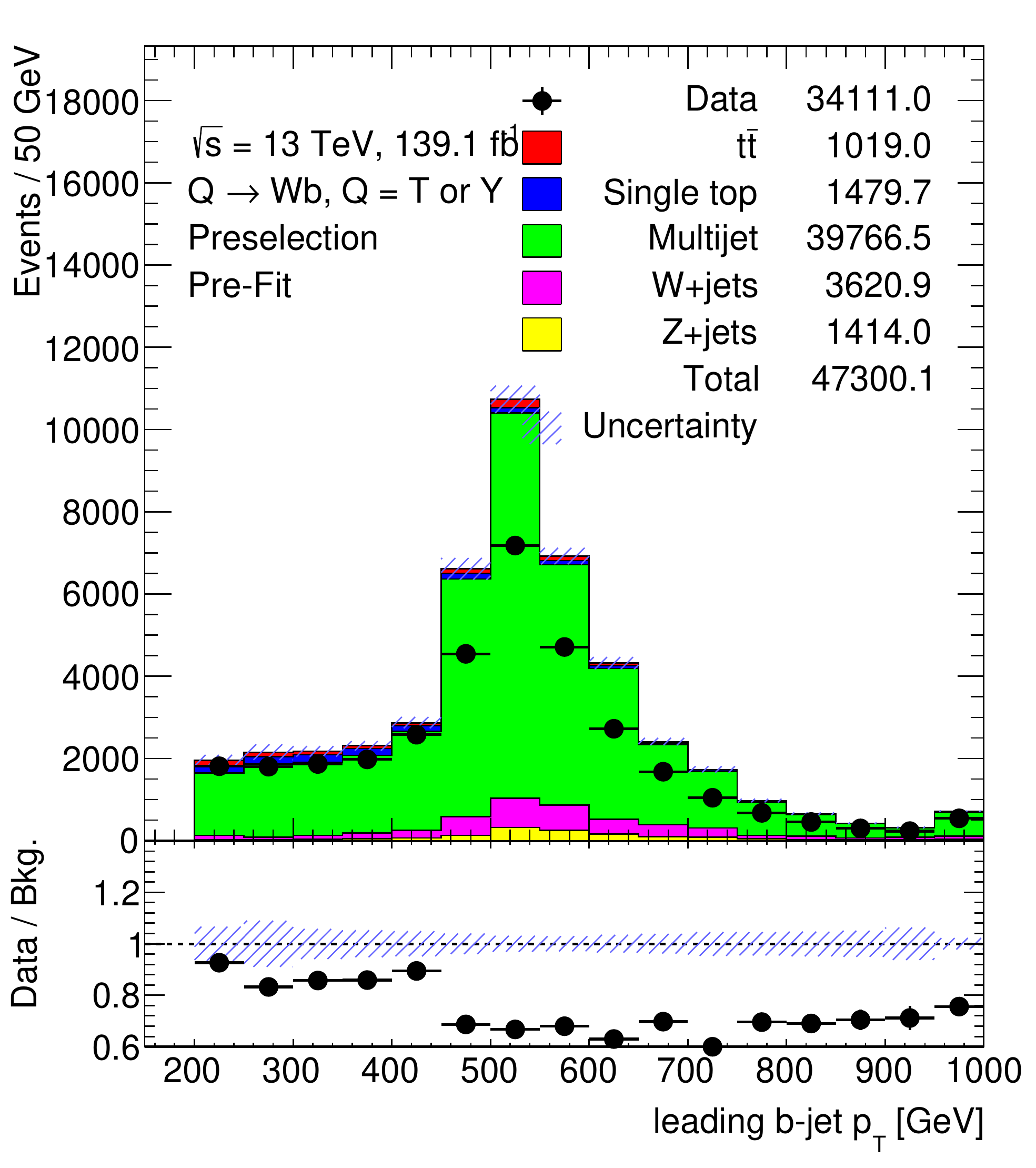}
		\caption{}
		\label{fig:analysisstrategy:eventselection:preselection:bjet}
	\end{subfigure}
	\begin{subfigure}{.45\textwidth}
		\centering
		\includegraphics[width=\linewidth,height=\textheight,keepaspectratio]{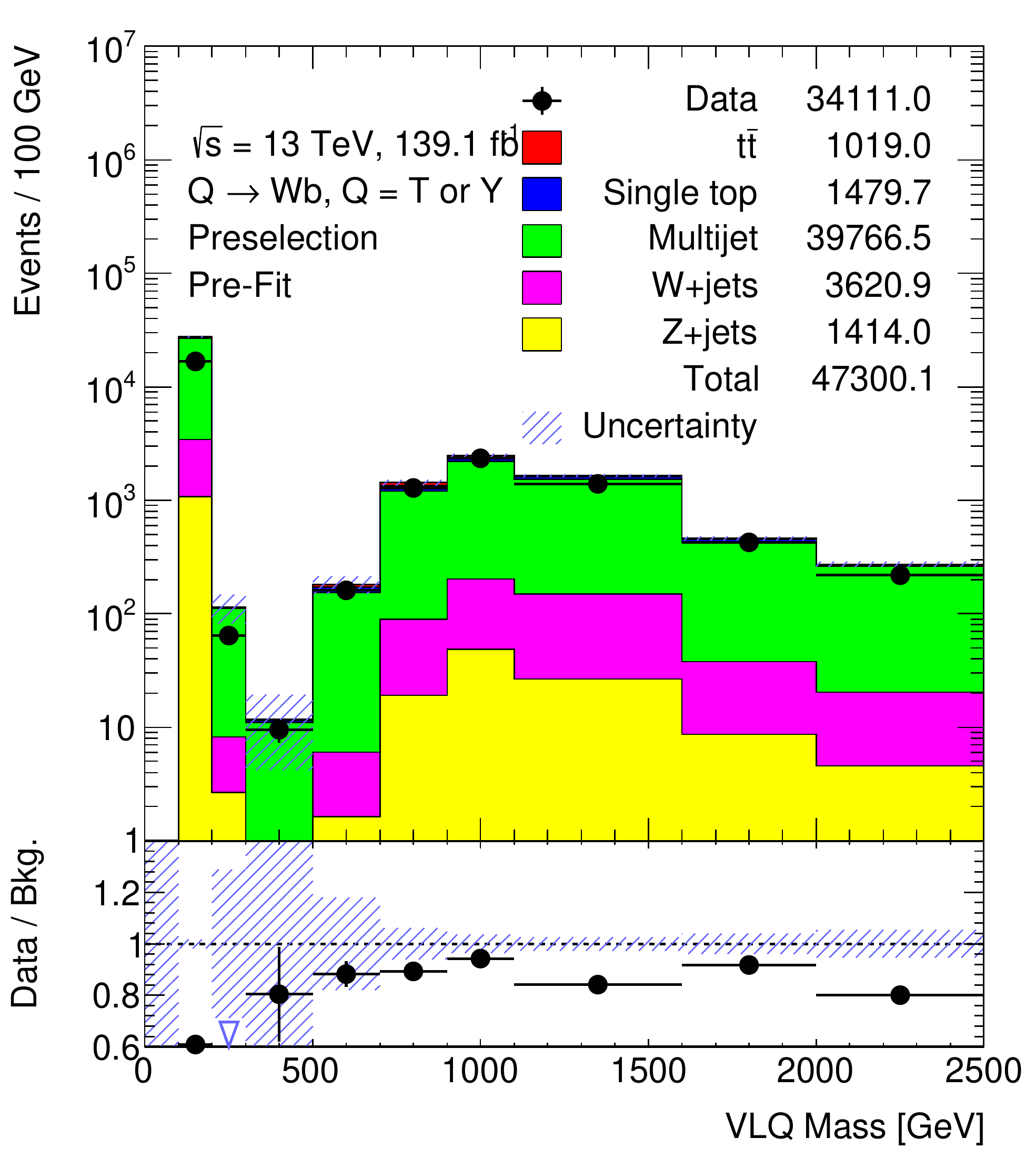}
		\caption{}
		\label{fig:analysisstrategy:eventselection:preselection:vlqm}
	\end{subfigure}
	\begin{subfigure}{.45\textwidth}
		\centering
		\includegraphics[width=\linewidth,height=\textheight,keepaspectratio]{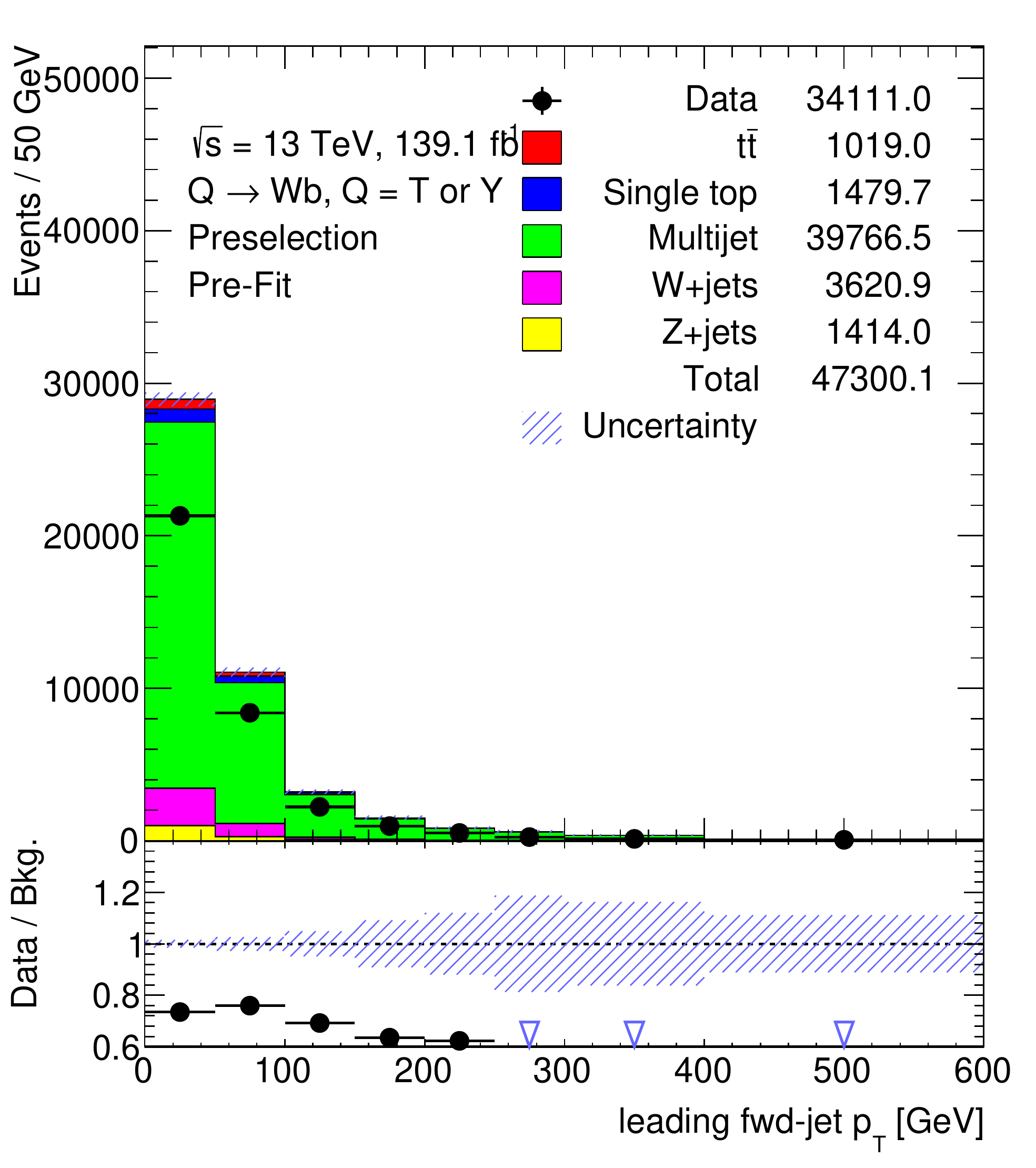}
		\caption{}
		\label{fig:analysisstrategy:eventselection:preselection:forwardjet}
	\end{subfigure}
	\caption{A data/MC comparison plots for different kinematic and reconstructed variables at the preselection level. All the backgrounds are produced from the MC simulation. It shows the distributions of (a) \pt of the $W$-tagged jet, (b) \pt of the leading $b$-tagged jet, (c) VLQ mass and (d) \pt of the leading forward jet.}
	\label{fig:analysisstrategy:eventselection:preselection}
\end{figure}

After having a well-defined preselection region, both the data and MC samples were processed. Some kinematic distributions of the physics objects are shown in Fig.\ \ref{fig:analysisstrategy:eventselection:preselection}. It includes $p_{\text{T}}$ distribution of $W$-tagged large-$R$ jet, leading $b$-tagged jet, VLQ mass distribution (which is reconstructed from the kinematics of the $W$-tagged jet and the leading $b$-tagged jet) and the leading forward jet distribution. These distributions show the contribution from each background as a stacked plot along with the contribution from the data. All the backgrounds shown in the plots are produced from the MC simulation. A ratio plot between the data and MC is also shown to evaluate the difference between the data events and predicted MC events. The overflow bin has been properly taken care of in these distributions by adding the content of it to the last bin. It can be observed from the distributions that the contribution from multijet is the main background, whereas the contribution from other backgrounds is very small. As described in \ref{sec:analysisstrategy:multijets}, it is because the cross-section of the multijet production is larger in order of magnitude compared to the cross-section of the other backgrounds. Also, since it is an all-hadronic analysis, so the contribution from the multijet background is dominant. Note that the VLQ mass in Fig.\ \ref{fig:analysisstrategy:eventselection:preselection:vlqm} is plotted on a log scale.

Since $W$ boson and $b$-quark are produced from the decay of heavy particle $T/Y$ quark, they are expected to be back-to-back. That is why a peak at around \SI{500}{\giga\electronvolt} can be observed in the \pt distribution of the leading $b$-tagged jet in Fig.\ \ref{fig:analysisstrategy:eventselection:preselection:bjet}. One can also expect $b$-jets with lower \pt i.e.\ \pt $\leq$ \SI{450}{\giga\electronvolt} because of the angular distribution of $W$ boson and $b$-quark. 

At the preselection, $W$-tagged jet and $b$-tagged jet have \pt $\geq$ \SI{500}{\giga\electronvolt} and \pt $\geq$ \SI{200}{\giga\electronvolt} respectively. They are chosen in such a way to reconstruct a VLQ mass of $m_{\text{VLQ}}$ $\geq$ \SI{700}{\giga\electronvolt} because the search at lower masses has been excluded by previous searches. So, this analysis targets at higher VLQ masses which can be seen in Fig.\ \ref{fig:analysisstrategy:eventselection:preselection:vlqm}. A peak at \SI{100}{\giga\electronvolt} < $m_{\text{VLQ}}$ < \SI{200}{\giga\electronvolt} is due to the events in which the leading $b$-jet is inside the $W$-tagged jet and the VLQ mass is reconstructed from them. Fig.\ \ref{fig:analysisstrategy:eventselection:preselection:forwardjet} shows \pt distribution of the leading forward jet, which is a light-flavoured jet. It is generally of low energy and lies in the forward region of the detector.

\section{Need for data-driven background estimation}%
\label{sec:analysisstrategy:datadriven}
The hadronic analysis is not always a clean channel because multijet is the dominant background. Multijet background comes predominantly from the QCD processes, in which a jet is identified as $W$-jet or $b$-jet. It is very difficult to model the multijet processes using the MC simulation because of the QCD effects, so it is usually mismodelled. Therefore, a better approach to estimate the multijet is needed in the hadronic analysis.

Moreover, a clear mismmodelling can be observed in the multijet MC at the preselection, as shown in Fig.\ \ref{fig:analysisstrategy:eventselection:preselection}. So, a better estimate of the multijet background is needed. In order to fix this, one can still think of scaling the multijet MC, but the ratio plots in these distributions show that just scaling the multijet MC to the data would not work. So, a dynamic approach has to be taken to predict the multijet background. This leads to perform a data-driven method for the estimation of the multijet background, which is discussed in detail in the next chapter.


\chapter{ABCD data-driven method}
\label{sec:abcd}
The goal of this chapter is to introduce one of the data-driven methods called the ABCD method, which is used to estimate multijet in this analysis. An overview of how a correction is applied to the results from the ABCD method, which leads to an introduction of the correction factor is given. Four different approaches of calculation of the correction factor are discussed in the last section of this chapter.

\section{Data-driven estimation of background}
\label{sec:abcd:data-driven}
Sometimes it is difficult to simulate a background process very precisely using the MC production because of several reasons. For example, limited understanding of the QCD process, inability to simulate hadronisation with precision, need large MC samples to simulate all the possible cases, not sufficient computing power, not an accurate understanding of all details in the detector and also when uncertainties from the simulation are significant or not known. So, one needs an alternative way of producing these background processes. The alternative approach is the extraction of background from the experimental data itself in a combination of the MC simulation. This technique is known as a data-driven method. Some standard data-driven methods are matrix method, ABCD method and the fake factor method. In this thesis, the ABCD method is used to estimate the multijet background, which is discussed in detail in the next section.

\section{The ABCD method}
\label{sec:abcd:method}
The ABCD method is generally used to estimate the backgrounds which are difficult to understand from the MC simulation. The idea behind this method is first to choose two uncorrelated variables, which means two mutually independent variables that can characterise the data in order to estimate the background. After that, four mutually exclusive regions are defined by making selections on these two variables. These regions are defined in such a way that one of the regions should be enriched in signal events where the contribution from the background events should be low. The other three regions should be signal-deficient regions where the background events should be in excess.~\cite{thesis:arshia}

\begin{figure}[hbt!]
	\centering
	\includegraphics[width=0.5\linewidth]{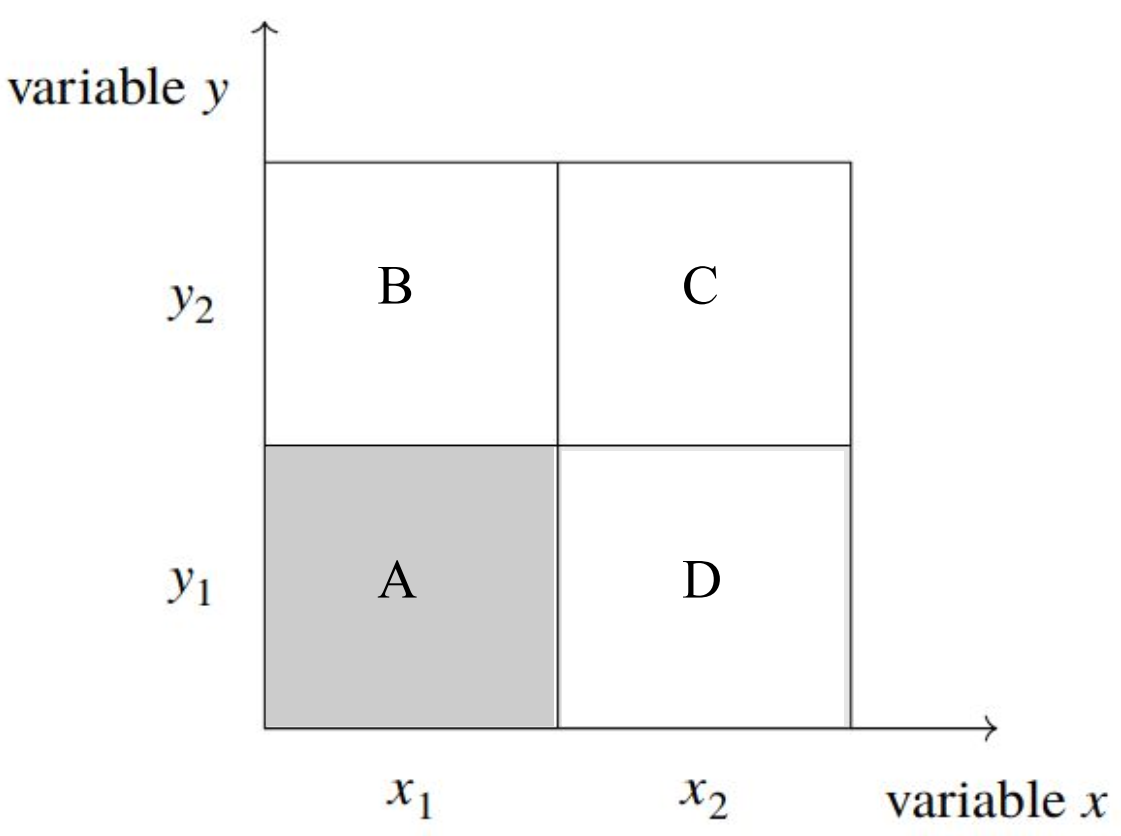}
	\caption{A pictorial representation of the ABCD method where four regions A, B, C and D are defined by the selections made on two orthogonal variables x and y. Region A (in grey) is a signal-enriched region and the other three regions (in white) are signal-deficient regions.}
	\label{fig:abcd:method}
\end{figure}

A schematic for visualisation is given in Fig.\ \ref{fig:abcd:method}, which shows two-dimensional phase space spanned by two uncorrelated variables x and y. Region A, B, C and D are defined in a manner such that region A is a signal-enriched region (where the background estimate has to be performed) and region B, C and D are signal-deficient regions. Then, the background estimate in region A can be estimated by extrapolating the background events from the signal-deficient regions by Eqn.\ \ref{eqn:abcd:method}:

\begin{equation}
	N_{\text{A}} = N_{\text{B}} \times \frac{N_{\text{D}}}{N_{\text{C}}} \,,
	\label{eqn:abcd:method}
\end{equation}
where $N_{\text{j}}$ denotes the number of background events in region j $\in$ \{A,B,C,D\}.

\subsubsection{Assumptions:}
There are some assumptions to the ABCD method, which need to be fulfilled before implementing the method to the data. The important assumptions are discussed in detail below:

\begin{enumerate}
	\item The two variables x and y should be chosen in such a way that they are perfectly uncorrelated.
	
	\item The foundation of this method relies on the definition of the regions. The signal-deficient regions should not contain any signal events. The signal contamination can be examined by calculating the signal-background ratio in all the regions. It is calculated by~\cite{signalbackgroundratio}: 
	\begin{equation}
		\text{Signal-background ratio = } \sqrt{2\left[(N_{\text{j}}^{\text{S}}+N_{\text{j}}^{\text{B}})\ln(1+\frac{N_{\text{j}}^{\text{S}}}{N_{\text{j}}^{\text{B}}})-N_{\text{j}}^{\text{S}}\right]} \approx \frac{N_{\text{j}}^{\text{S}}} {\sqrt{N_{\text{j}}^{\text{B}}}} \,,
		\label{eqn:abcd:sbratio}
	\end{equation}
	 where $N_{\text{j}}^{\text{S}}$ and $N_{\text{j}}^{\text{B}}$ denote the number of signal and background MC events respectively in region j $\in$ \{A,B,C,D\}.
	
	\item Since it is a data-driven method, so there should be enough data events in the signal-deficient regions to extrapolate the background behaviour precisely to region A. It is also useful to propagate the statistical uncertainty on the background estimate in region A.
	
\end{enumerate}

\section{Implementation of the ABCD method}
\label{sec:abcd:implementation}

\subsection{Choice of variables}
\label{sec:abcd:implementation:variables}
In this thesis, the ABCD method is used to estimate the multijet background. So, it is important to choose variables that ensure abundance of the multijet background in signal-deficient regions. The variables $W$-tagging efficiency and the $b$-jet multiplicity are chosen because according to one of the assumptions, the two variables should be perfectly uncorrelated. In $W$-tagging, it is explicitly made sure that no $b$-tagging information is used. Thus, choosing them as two uncorrelated variables for the ABCD method. 

The working point efficiency of $W$-tagging is chosen as one of the variables for the ABCD method. The two selected working points (WP) are $80\%$ referred as $loose$, and $50\%$ referred as $tight$. Tight is a subset of loose because it has a stronger criteria and therefore, less number of events have large-$R$ jets which pass tight $W$-tagging criteria. The three regions which are defined by making the selections based on these two working points are tight, loose not tight, and not loose. These regions are described in detail below:
\begin{itemize}
	\item Tight: it contains events which have large-$R$ jets that are $W$-tagged at 50\% WP.
	\item Loose not tight: it contains events which have large-$R$ jets that are $W$-tagged at 80\% WP but not tagged at 50\%.
	\item Not loose: it contains all the events which have large-$R$ jets that are not $W$-tagged at 80\% WP. 
\end{itemize}
There are two taggers for large-$R$ jets which are used in this thesis, i.e.\ two-variable tagger and three-variable tagger. In this chapter, large-$R$ jets which are tagged with three-variable tagger are used for the estimation of multijet background using the ABCD method. The estimation with two-variable tagger is also performed and used to compare the performance of the two taggers, which are described in section \ref{sec:results:taggers}.

The presence of $b$-jets can further characterise the signal and background events. So the multiplicity of $b$-tagged jets is chosen as the second variable. The two regions which are defined based on this are $0b$ and $\geq1b$, which are described in detail below:
\begin{itemize}
\item $0b$: it contains events which do not have any small-$R$ jets that are $b$-tagged.
\item $\geq1b$: it contains events which have at least one small-$R$ jet that is $b$-tagged.
\end{itemize}
There are two jet collections for small-$R$ jets which are used in this thesis, i.e.\ EMTopo and PFlow jets. In this chapter, PFlow jet collection is used for the estimation of multijet background using the ABCD method. The estimation with EMTopo jet collection is also performed and used to compare the performance of the two jet collections, which are described in section \ref{sec:results:jetcollections}.

\subsection{Definition of regions}
\label{sec:abcd:implementation:regions}
There are six regions which are defined by making selection cuts on these two variables. A pictorial representation is shown in Fig.\ \ref{fig:abcd:method:regions}. It shows $b$-jet multiplicity on the x-axis and $W$-tagging WP on the y-axis. The regions include signal region, validation region and control region, which are described in detail below:

\begin{figure}[hbt!]
	\centering
	\includegraphics[width=0.8\linewidth]{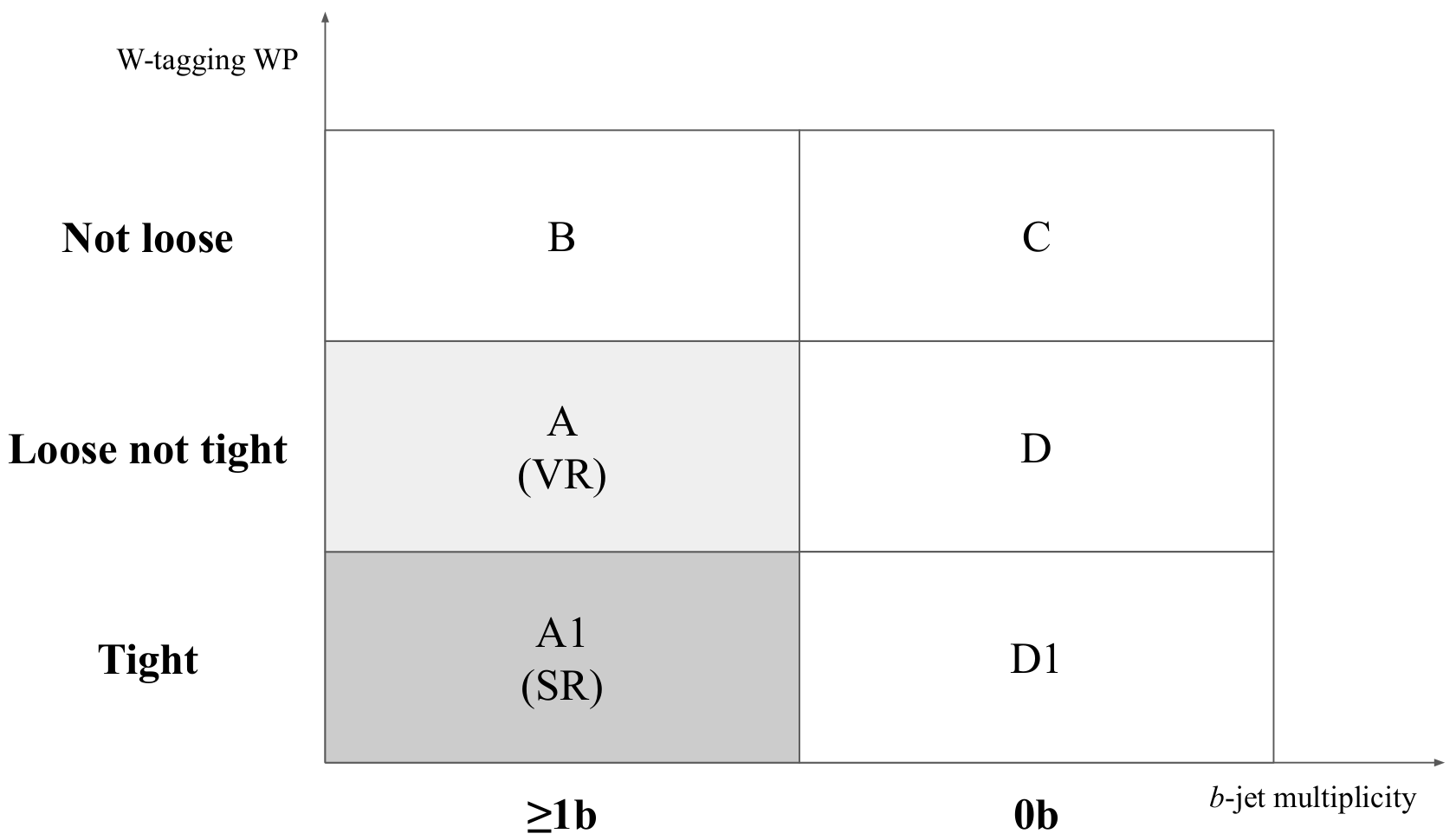}
	\caption{A representation of all the ABCD regions defined on two uncorrelated variables. Region A1 is signal region, region A is validation region and the other four regions B, C, D, D1 are control regions.}
	\label{fig:abcd:method:regions}
\end{figure}

\subsubsection{Signal region}
\label{sec:abcd:implementation:regions:sr}
Signal region (SR) is defined in such a way that it should be enriched with signal events. Region A1 is represented as SR, which should contain those events which pass the following criteria. The leading large-$R$ jet should be $W$-tagged at tight selection. Along with that, the events should have at least one $b$-tagged jet, and the leading $b$-tagged jet (denoted by $b_{\text{1}}$) should be a central jet which means it should lie within the central region of the detector which can be made sure by $|\eta_{b_{1}}| < 2.5$ cut. Also, the leading $b$-tagged jet should not be a part of $W$-tagged large-$R$ jet and should be well separated which is ensured by implementing $\Delta R(W,b_{\text{1}}) > 1.0$ cut. It was observed that after applying this cut, many events were filtered out because there were so many $b$-tagged jets inside the $W$-tagged large-$R$ jet. In this analysis, data is blinded in SR, which means all the calculations in this region can only be performed by using the MC samples.

\subsubsection{Validation region}
\label{sec:abcd:implementation:regions:vr}
Validation region (VR) is defined in such a way that kinematically it is close to the SR but not exactly the SR. So, region A is represented as VR to check and validate the method. VR includes selection cuts in which method can be applied both to the data and MC without exploiting the fact that it is biased to the SR. The selection cuts include leading large-$R$ jet to be $W$-tagged at loose not tight selection with a requirement of at least one $b$-tagged jet, where the leading $b$-tagged jet lies in a central region of the detector. The $\Delta R(W,b_{\text{1}}) > 1.0$ is also implemented to remove the events where the $W$-tagged jet and $b$-jets are not well separated. Multijet estimation is performed in VR using the ABCD method, and the data/bkg.\ comparison plots are shown to understand how well the estimation agrees with the data.

\subsubsection{Control region}
\label{sec:abcd:implementation:regions:cr}
Control region (CR) is defined in such a way that it should be enriched with background events. Region B, C, D, and D1 are referred to as CR. The selection cuts for all the four CRs are described in detail below:

\begin{itemize}
	\item \textbf{Region B:} it consists of the events which should at least have a large-$R$ jet, and the leading large-$R$ jet should not be $W$-tagged at the loose selection. Events should also have at least one $b$-tagged jet, where the leading $b$-tagged jet lies in a central region of the detector. The leading large-$R$ jet and the leading $b$-tagged jet should be well separated by imposing the condition $\Delta R(J_{\text{1}},b_{\text{1}}) > 1.0$, where $J_{\text{1}}$ is defined as the leading large-$R$ jet.
	
	\item \textbf{Regions C, D and D1:} all the three regions contain events in which none of the small-$R$ jets is $b$-tagged and the leading small-$R$ jet (denoted by $j_{\text{1}}$) should be a central jet. The selection cut which distinguishes the events in these three regions is the requirement of leading large-$R$ jet not to be $W$-tagged in the case of region C, and $W$-tagged at tight and loose not tight selection in the case of region D and D1 respectively. Furthermore, the leading large-$R$ jet (whether $W$-tagged or not) and the leading small-$R$ jet should be well separated by imposing a condition on $\Delta R(W/J_{\text{1}},j_{\text{1}}) > 1.0$.
\end{itemize}

\subsection{Closure tests}
\label{sec:abcd:implementation:closuretests}
Before proceeding with the estimation of the multijet background using the ABCD method, two preliminary closure tests are performed to check whether the defined regions are consistent with the assumptions. These are described in detail below: 

\subsubsection{Test I}
\label{sec:abcd:implementation:closuretests:testi}
The ABCD method is based on the assumption that the regions should be defined in such a way that the CRs should not have any signal events. In order to look at this, the MC signal samples are chosen, and the event yields are evaluated in all the six regions. This test is performed on the samples with different coupling values keeping the VLQ mass fixed, and also on the samples with different VLQ masses, keeping the coupling values constant. This is to make sure that the test is not biased to any mass and coupling values of the signal sample. So, four different MC signal samples are considered, which are shown in Table \ref{table:abcd:implementation:closuretests:testi:samples}.
\begin{table}[hbt!]
	\centering
	\begin{tabular}{c|c|c} 
		\toprule
		\textbf{Sample} & \textbf{VLQ mass} & \textbf{$k_\text{T}$} \\
		\midrule
		I & \SI{1.1}{\tera\electronvolt} & 0.1 \\
		II & \SI{1.5}{\tera\electronvolt} & 0.1 \\
		III & \SI{1.1}{\tera\electronvolt} & 0.3 \\
		IV & \SI{1.5}{\tera\electronvolt} & 0.3 \\
		\bottomrule
	\end{tabular}
	\caption{Four different MC signal samples along with their mass and coupling values which are used in closure test I. }
	\label{table:abcd:implementation:closuretests:testi:samples}
\end{table}

\begin{figure}[hbt!]
	\centering
	\graphicspath{{figs/chapter5/closuretest/}}
	\begin{subfigure}{.48\textwidth}
		\centering
		\includegraphics[width=\linewidth,height=\textheight,keepaspectratio]{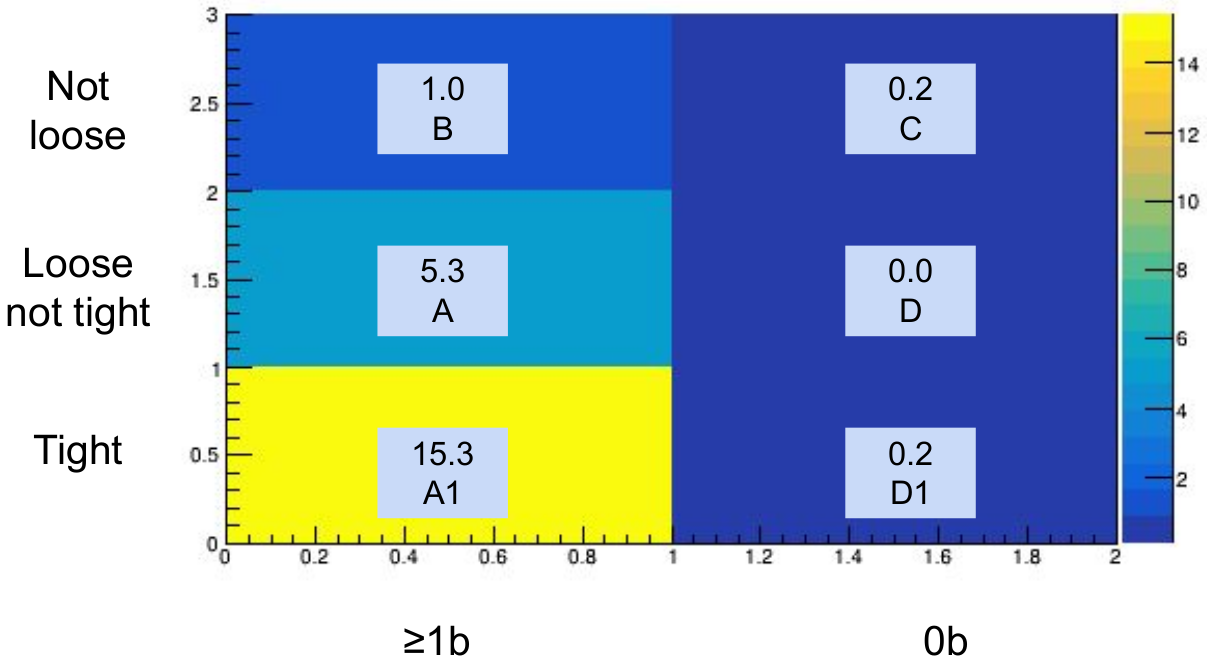}
		\caption{}
		\label{fig:abcd:implementation:closuretests:testi:samplei}
	\end{subfigure}\hspace{0.3cm}
	\begin{subfigure}{.48\textwidth}
		\centering
		\includegraphics[width=\linewidth,height=\textheight,keepaspectratio]{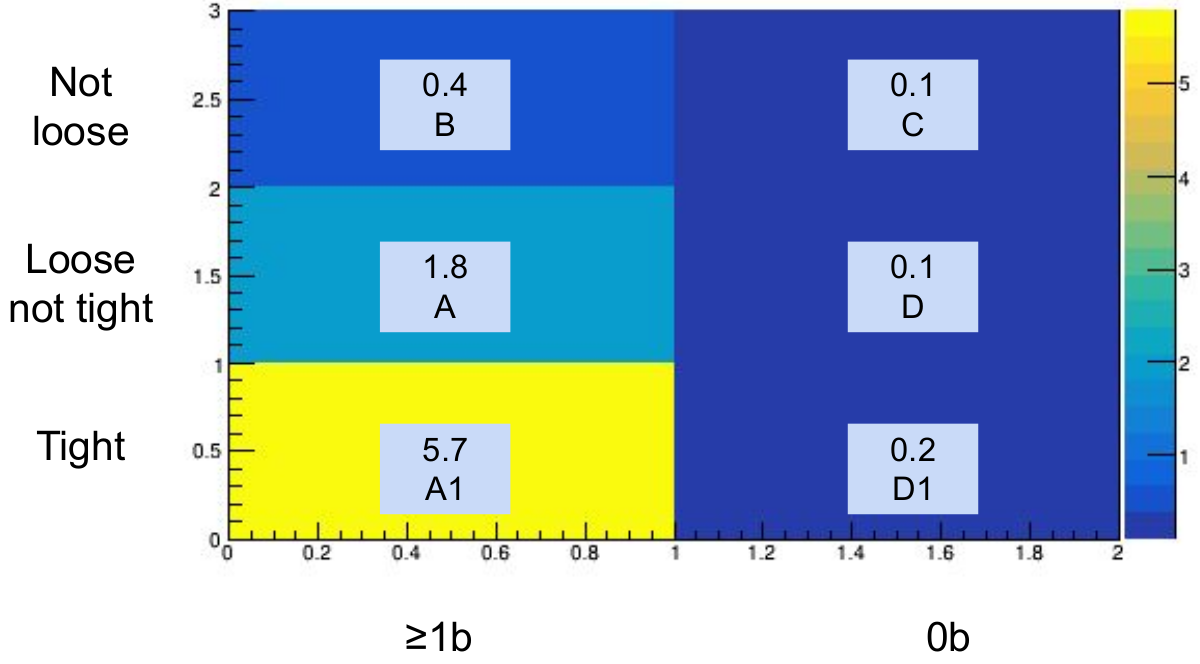}
		\caption{}
		\label{fig:abcd:implementation:closuretests:testi:sampleii}
	\end{subfigure}
	\begin{subfigure}{.48\textwidth}
		\centering
		\includegraphics[width=\linewidth,height=\textheight,keepaspectratio]{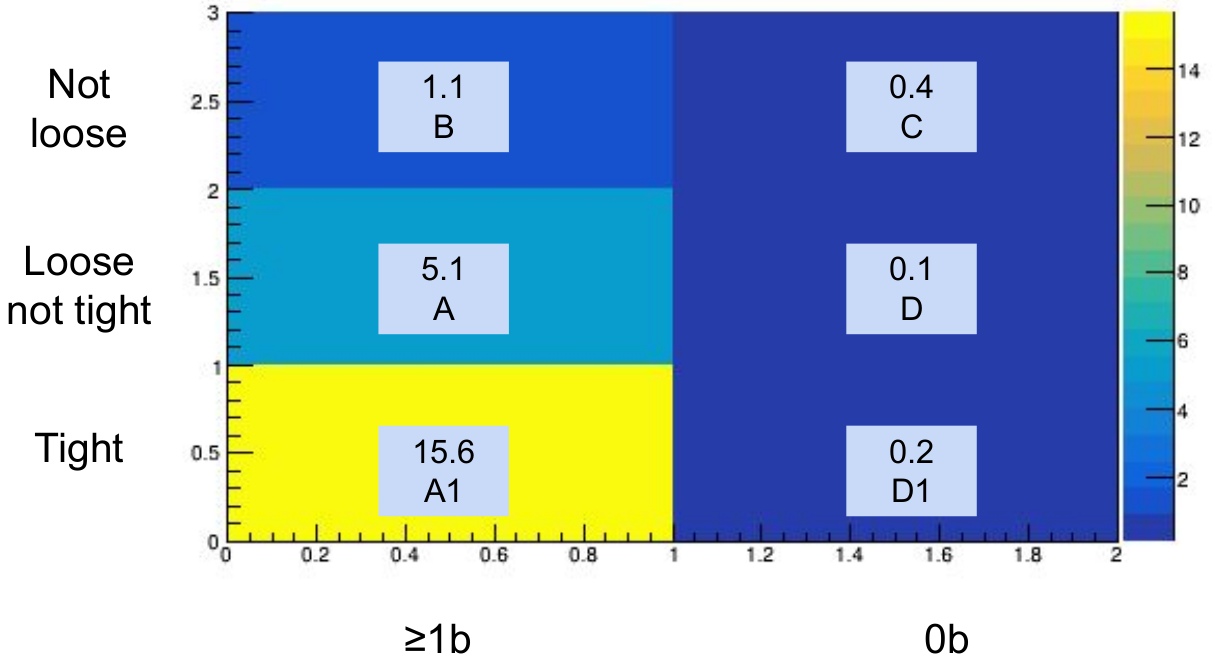}
		\caption{}
		\label{fig:abcd:implementation:closuretests:testi:sampleiii}
	\end{subfigure}\hspace{0.3cm}
	\begin{subfigure}{.48\textwidth}
		\centering
		\includegraphics[width=\linewidth,height=\textheight,keepaspectratio]{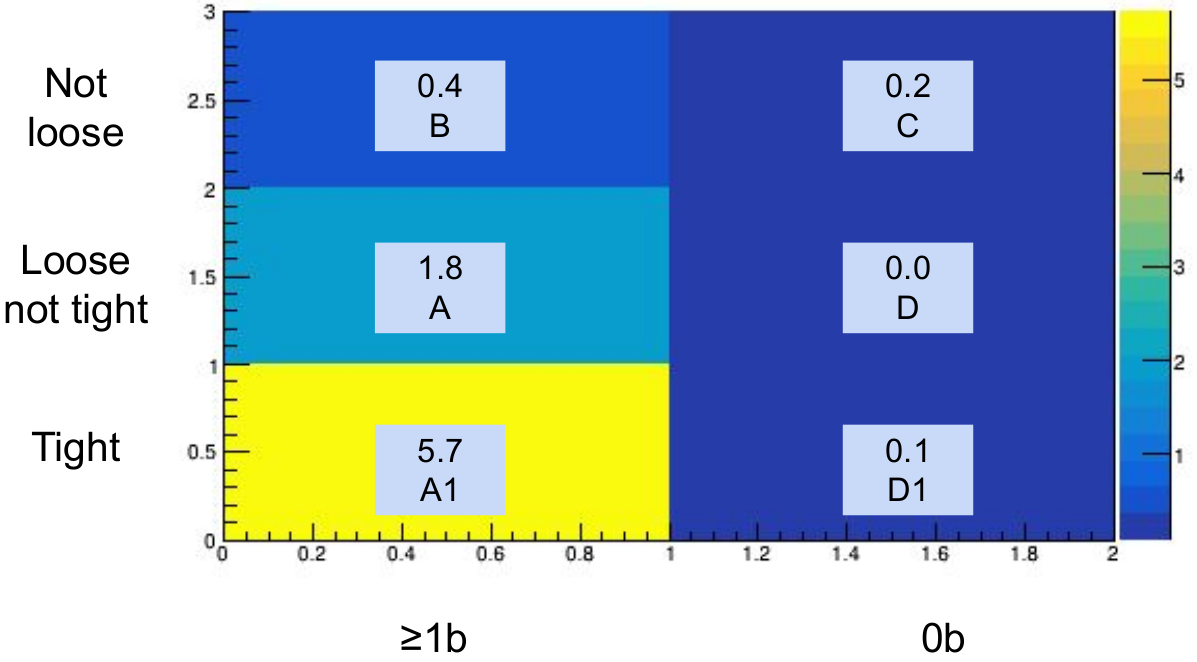}
		\caption{}
		\label{fig:abcd:implementation:closuretests:testi:sampleiv}
	\end{subfigure}
	\caption{Signal-background ratio in all the six ABCD regions for (a) Sample I, (b) Sample II, (c) Sample III and (d) Sample IV.}
	\label{fig:abcd:implementation:closuretests:testi}
\end{figure}

Table \ref{table:abcd:implementation:closuretests:testi} shows the event yields of the MC samples of all the backgrounds and the four MC simulated signal samples in all the six ABCD regions. Note that the multijet MC samples are the ones which are mismodelled. The event yield from the multijet MC samples is used to calculate the signal-background ratio, which is defined by Eqn.\ \ref{eqn:abcd:sbratio}, where $N^{\text{B}}_{\text{j}}$ is total background events from MC samples which are shown in Table \ref{table:abcd:implementation:closuretests:testi}. Fig.\ \ref{fig:abcd:implementation:closuretests:testi} shows the signal-background ratio for all the four signal samples. It can be observed in these plots that the choice of region A1 and A are in such a way that they are enriched in signal events, and the signal contamination in other regions (CRs) are negligible. So, the choice of the selection cuts to define these regions fulfill one of the main assumptions of the ABCD method.

\begin{sidewaystable}[!htbp]
		\centering
		\begin{tabular}{ c | c | c | c | c | c | c } 
			\toprule
			MC sample & SR & VR & \multicolumn{4}{c}{CR} \\ \cline{2-7}
			& Region A1 & Region A & Region B & Region C & Region D & Region D1 \\
			\midrule
			Single top & $\num{1130}\pm\num{18}$ & $\num{714}\pm\num{14}$ & $\num{861}\pm\num{15}$ & $\num{165}\pm\num{7}$ & $\num{9}\pm\num{1}$ & $\num{14}\pm\num{2}$ \\
			$t\bar{t}$ & $\num{568}\pm\num{14}$ & $\num{964}\pm\num{19}$ & $\num{5614}\pm\num{45}$ & $\num{894}\pm\num{18}$ & $\num{25}\pm\num{3}$ & $\num{15}\pm\num{2}$ \\
			$W$+jets & $\num{1257}\pm\num{80}$ & $\num{1022}\pm\num{82}$ & $\num{3937}\pm\num{178}$ & $\num{7627}\pm\num{234}$ & $\num{1540}\pm\num{94}$ & $\num{1991}\pm\num{109}$ \\
			$Z$+jets & $\num{317}\pm\num{27}$ & $\num{617}\pm\num{42}$ & $\num{4264}\pm\num{125}$ & $\num{2648}\pm\num{88}$ & $\num{744}\pm\num{43}$ & $\num{496}\pm\num{34}$ \\
			Multijet & $\num{15726}\pm\num{343}$ & $\num{75607}\pm\num{738}$ & $\num{395532}\pm\num{1667}$ & $\num{795961}\pm\num{2140}$ & $\num{119964}\pm\num{717}$ & $\num{23842}\pm\num{344}$ \\
			\midrule
			\textbf{Total bkg.} & $\num{19000}\pm\num{354}$ & $\num{78927}\pm\num{744}$ & $\num{410210}\pm\num{1681}$ & $\num{807297}\pm\num{2154}$ & $\num{122284}\pm\num{724}$ & $\num{26360}\pm\num{362}$ \\
			\midrule
			Signal I & $\num{2121}\pm\num{107}$ & $\num{1512}\pm\num{127}$ & $\num{653}\pm\num{54}$ & $\num{208}\pm\num{39}$ & $\num{32}\pm\num{8}$ & $\num{42}\pm\num{9}$ \\
			Signal II & $\num{794}\pm\num{29}$ & $\num{518}\pm\num{24}$ & $\num{299}\pm\num{17}$ & $\num{92}\pm\num{9}$ & $\num{35}\pm\num{6}$ & $\num{39}\pm\num{5}$ \\
			Signal III & $\num{2153}\pm\num{60}$ & $\num{1432}\pm\num{54}$ & $\num{723}\pm\num{31}$ & $\num{197}\pm\num{17}$ & $\num{45}\pm\num{7}$ & $\num{68}\pm\num{9}$ \\
			Signal IV & $\num{788}\pm\num{17}$ & $\num{516}\pm\num{20}$ & $\num{297}\pm\num{10}$ & $\num{99}\pm\num{6}$ & $\num{29}\pm\num{3}$ & $\num{43}\pm\num{3}$ \\
			\bottomrule
		\end{tabular}
		\caption{Event yields of all the backgrounds and signals from the MC simulation in all the six ABCD regions. The errors shown here are all statistical uncertainty.}
		\label{table:abcd:implementation:closuretests:testi}
		\vspace{1cm}
		\centering
		\begin{tabular}{ c | c | c | c | c | c | c } 
			\toprule
			Multijet & SR & VR & \multicolumn{4}{c}{CR} \\ \cline{2-7}
			& Region A1 & Region A & Region B & Region C & Region D & Region D1 \\
			\midrule
			MC yields & $\num{15726}\pm\num{343}$ & $\num{75607}\pm\num{738}$ & $\num{395532}\pm\num{1667}$ & $\num{795961}\pm\num{2140}$ & $\num{119964}\pm\num{717}$ & $\num{23842}\pm\num{344}$ \\
			Expected yields & $\num{11847}\pm\num{267}$ & $\num{59612}\pm\num{558}$ & - & - & - & - \\
			\midrule
			\textbf{\% diff.\ to MC} & 24\% & 21\% & - & - & - & - \\
			\bottomrule
		\end{tabular}
		\caption{Event yields of multijet background from the MC sample in all the ABCD regions and the expected yield calculated from Eqn.\ \ref{eqn:abcd:implementation:closuretests:testii} in SR A1 and VR A.}
		\label{table:abcd:implementation:closuretests:testii}
\end{sidewaystable}

\subsubsection{Test II}
\label{sec:abcd:implementation:closuretests:testii}
The ABCD method needs to be validated first on the multijet MC before applying it to the data. To accomplish this, the method is applied to the multijet MC sample. In VR A and SR A1, the ABCD method is used to obtain an expression for the expected number of multijet events, given by Eqn.\ \ref{eqn:abcd:implementation:closuretests:testii}. 
\begin{equation}
	N_{\text{A,A1}}^{\text{Exp.}} = N_{\text{B}}^{\text{MC}} \times \frac{N_{\text{D,D1}}^{\text{MC}}}{N_{\text{C}}^{\text{MC}}} \,,
	\label{eqn:abcd:implementation:closuretests:testii}
\end{equation}
where $N_{\text{A,A1}}^{\text{Exp.}}$ is denoted as the expected number of multijet background events in VR and SR, and $N_{\text{j}}^{\text{MC}}$ is the number of multijet events from MC in region $j\in\{B, C, D, D1\}$. 

The expected number of events can be further compared with the actual number of events from the MC multijet samples in VR A and SR A1. Table \ref{table:abcd:implementation:closuretests:testii} shows event yields of the multijet background from MC samples in all the six regions along with the expected yields from the ABCD method in VR A and SR A1.

One can observe that the difference between the actual yield of the multijet MC and the expected yield from the ABCD method in VR A and SR A1 is almost $25\%$. This shows that one of the assumptions which state that the two variables should be perfectly uncorrelated does not hold completely with the choice of the variables and a correction might be required.
\section{Preliminary multijet estimate}
\label{sec:abcd:estimate}
As the ABCD method is validated in the aforementioned regions, now it is implemented on the data to estimate the multijet in VR A. Before using the ABCD equation on data, a slight modification is made to Eqn.\ \ref{eqn:abcd:method}. Instead of directly using the event yields from the data for CRs $N_{\text{B}}$, $N_{\text{C}}$ and $N_{\text{D}}$, the contribution from all the other backgrounds in the respective regions should be subtracted from them because the ABCD equation only holds for multijet background. This leads to Eqn.\ \ref{eqn:abcd:estimate}, which is used to estimate multijet background events in VR A, denoted by $N_{\text{A}}^{\text{Est.\ multijet}}$.

\begin{equation}
	N_{\text{A}}^{\text{Est.\ multijet}}[i] = (N_{\text{B}}^{\text{Data}}[i]-N_{\text{B}}^{\text{Other bkg}}[i]) \times \frac{(N_{\text{D}}^{\text{Data}}[i]-N_{\text{D}}^{\text{Other bkg}}[i])}{(N_{\text{C}}^{\text{Data}}[i]-N_{\text{C}}^{\text{Other bkg}}[i])} \,,
\label{eqn:abcd:estimate}
\end{equation}
where $N_{\text{j}}^{\text{Other bkg}}[i]$ is event yield from all the other backgrounds which include single top, $t\bar{t}$, $W$+jets and $Z$+jets in region j and $N_{\text{j}}^{\text{Data}}[i]$ is event yield from the data in region j where j $\in$ \{B,C,D\}. Here [$i$] denotes that this calculation is performed bin-by-bin\footnote{Bin-by-bin means when a calculation is performed for each bin separately by taking the number of events in that bin within a distribution.} (where $i=$ bin) to reconstruct a distribution for the estimated multijet in VR A, and it is implemented separately for each distribution. So, the event yields for the estimated multijet would be different for each distribution.

\begin{figure}[hbt!]
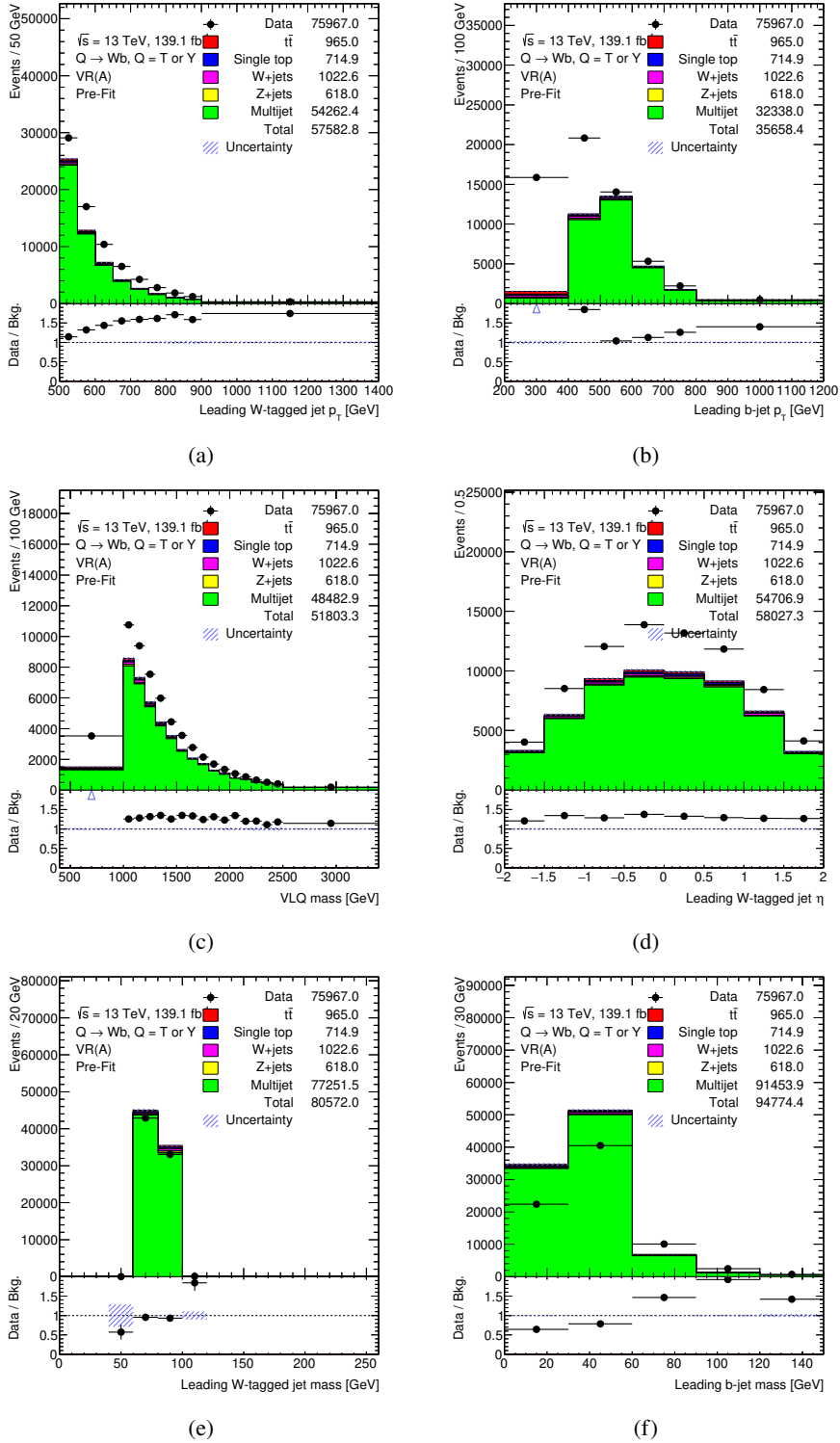

	\centering
	\graphicspath{{figs/chapter5/nocorr/}}
	\begin{subfigure}{.35\textwidth}
		\centering
		\includegraphics[width=\linewidth,height=\textheight,keepaspectratio]{VR_B_ljet_pt.pdf}
		\caption{}
		\label{fig:abcd:estimate:ljet_pt}
	\end{subfigure}\hspace{0.6cm}
	\begin{subfigure}{.35\textwidth}
		\centering
		\includegraphics[width=\linewidth,height=\textheight,keepaspectratio]{VR_B_jet_pt.pdf}
		\caption{}
		\label{fig:abcd:estimate:jet_pt}
	\end{subfigure}
	\begin{subfigure}{.35\textwidth}
		\centering
		\includegraphics[width=\linewidth,height=\textheight,keepaspectratio]{VR_B_VLQM.pdf}
		\caption{}
		\label{fig:abcd:estimate:VLQM}
	\end{subfigure}\hspace{0.6cm}
	\begin{subfigure}{.35\textwidth}
		\centering
		\includegraphics[width=\linewidth,height=\textheight,keepaspectratio]{VR_B_ljet_eta.pdf}
		\caption{}
		\label{fig:abcd:estimate:ljet_eta}
	\end{subfigure}
	\begin{subfigure}{.35\textwidth}
		\centering
		\includegraphics[width=\linewidth,height=\textheight,keepaspectratio]{VR_B_ljet_m.pdf}
		\caption{}
		\label{fig:abcd:estimate:ljet_m}
	\end{subfigure}\hspace{0.6cm}
	\begin{subfigure}{.35\textwidth}
		\centering
		\includegraphics[width=\linewidth,height=\textheight,keepaspectratio]{VR_B_jet_m.pdf}
		\caption{}
		\label{fig:abcd:estimate:jet_m}
	\end{subfigure}
	\caption{A data/bkg.\ comparison of kinematic and reconstructed variables in VR A where multijet background (in green) is estimated from the ABCD method and other backgrounds are from the MC simulation. The variables include (a) $p_{\text{T}}$ of $W$-tagged large-$R$ jet, (b) $p_{\text{T}}$ of leading $b$-tagged small-$R$ jet, (c) VLQ mass reconstructed from the kinematics of $W$-tagged large-$R$ jet and leading $b$-tagged small-$R$ jet, (d) $\eta$ distribution of $W$-tagged large-$R$ jet, (e) mass of $W$-tagged large-$R$ jet, and (f) mass of leading $b$-tagged small-$R$ jet.}
	\label{fig:abcd:estimate}
\end{figure}

The multijet estimation is performed in six kinematic variables, which are shown in Fig.\ \ref{fig:abcd:estimate}. These plots show the data events (in dots), estimated multijet (in green) and other MC background samples. The last bin of each plot includes the events from the overflow bin. A data/bkg.\ ratio plot is also shown, which includes the ratio of data and the stacked plot. The stacked plot consists of the estimated multijet and the other MC background events. Since the multijet is a dominant background, the contribution from all the other backgrounds is very small, which can be seen from the plots. Fig.\ \ref{fig:abcd:estimate:ljet_pt} shows the $p_{\text{T}}$ distribution of $W$-tagged large-$R$ jet where a general disagreement between data and estimated multijet can be seen. Fig.\ \ref{fig:abcd:estimate:jet_pt} shows the $p_{\text{T}}$ distribution of leading $b$-tagged jet which peaks at around \SI{500}{\giga\electronvolt} to balance the momentum from $W$-tagged large-$R$ jet. The data/bkg.\ disagreement can be seen at low $p_{\text{T}}$. This is because the bin content of the first few bins does not have enough events in the CR distributions to properly estimate the multijet in those bins. The CR distributions are shown in Appendix \ref{sec:app}. A similar kind of behaviour can also be seen in the first bin of Fig.\ \ref{fig:abcd:estimate:VLQM}, which shows the VLQ mass distribution. Fig.\ \ref{fig:abcd:estimate:ljet_eta} shows the $\eta$ distribution of $W$-tagged large-$R$ jet, which is, in general, a more uniform distribution. The data/bkg.\ disagreement is also flat in $\eta$. Fig.\ \ref{fig:abcd:estimate:ljet_m} shows the mass distribution of $W$-tagged large-$R$ jet, which peaks at the mass of $W$ boson and shows quite a good data/bkg.\ agreement. Fig.\ \ref{fig:abcd:estimate:jet_m} shows the mass distribution of leading $b$-tagged jet, where multijet is overestimated at low mass because, in CR from where the multijet is extrapolated, the leading small-$R$ jet is not $b$-tagged. Also, the data/bkg.\ disagreement at high mass is basically coming from low statistics.

\subsection{Need for correction}
\label{sec:abcd:estimate:need}
From the multijet estimate using the ABCD method shown in Fig.\ \ref{fig:abcd:estimate}, a general disagreement in data/bkg.\ can be seen. This is because one of the assumptions is not fulfulled. The choice of the variables in this thesis is in such a way to make sure there should be a minimal correlation which can be neglected. However, from the disagreement, it can be seen that there is a small amount of correlation in the two variables which needs to be taken into account. 

Moreover, the regions defined on the multiplicity of $b$-jets have different kinematics of the leading small-$R$ jet. This leading small-$R$ jet is further used to reconstruct the VLQ mass. The difference can be seen quite well from the distributions in the CRs shown in Appendix \ref{sec:app}. In that case, the ABCD relation given by Eqn.\ \ref{eqn:abcd:method} does not hold anymore and should be re-written as:

\begin{equation}
\frac{N_{\text{A}}}{N_{\text{B}}} \propto \frac{N_{\text{D}}}{N_{\text{C}}} \,,
\label{eqn:abcd:estimate:need}
\end{equation}
where $N_{\text{j}}$ is denoted as the number of background events in region j $\in$ \{A,B,C,D\}.

Also, the nature of this correction should be taken into account. Fig.\ \ref{fig:abcd:estimate:ljet_eta} shows that the correction should be implemented in such a way that a scale factor should be introduced. Then the distribution can be multiplied by this scale factor. In contrast, other distributions which are shown in Fig.\ \ref{fig:abcd:estimate} (a), (b), (c) and (f) say that this correction should be implemented for each bin separately in order to fix the shape of the distribution.

\section{Correction factor}
\label{sec:abcd:correctionfactor}
From the previous section, it has been seen that there is a need to introduce a correction factor to the ABCD equation. So, Eqn.\ \ref{eqn:abcd:estimate} can be re-written as :

\begin{equation}
N_{\text{A}}^{\text{Est.\ multijet}}[i] = \R \times (N_{\text{B}}^{\text{Data}}[i]-N_{\text{B}}^{\text{Other bkg}}[i]) \times \frac{(N_{\text{D}}^{\text{Data}}[i]-N_{\text{D}}^{\text{Other bkg}}[i])}{(N_{\text{C}}^{\text{Data}}[i]-N_{\text{C}}^{\text{Other bkg}}[i])} \,,
\label{eqn:abcd:correctionfactor}
\end{equation}
where \R is a correction factor which is calculated from multijet MC samples. It can be calculated in two different ways depending upon the nature of the correction.

\begin{itemize}
	\item \textbf{Normalisation method:} \R is calculated to compensate for the difference in the normalisation of the estimated multijet and the data. It is calculated by the expression given below:
	
	\begin{equation}
		\R = \frac{\N{A}{multijet MC}}{\N{B}{multijet MC}} \times \frac{\N{C}{multijet MC}}{\N{D}{multijet MC}} \,,
	\end{equation}
	where \N{j}{multijet MC} is the number of multijet events from MC sample in region $j\in\{A, B, C, D\}$, which can be calculated by taking the integral of the distribution. So, in the end, \R is a number which can be used in Eqn.\ \ref{eqn:abcd:correctionfactor} to scale the estimated multijet.
	
	\item \textbf{Shape method:} \R is calculated to fix the shape of the distributions by calculating it bin-by-bin. A corresponding expression can be written as:
	
	\begin{equation}
	\R[i] = \frac{\N{A}{multijet MC}[i]}{\N{B}{multijet MC}[i]} \times \frac{\N{C}{multijet MC}[i]}{\N{D}{multijet MC}[i]} \,,
	\end{equation}
	where $[i]$ shows that the calculation is performed for each bin separately within a distribution ($i=$ bin). So, it leads to a separate \R$[i]$ distribution for each kinematic variables, which can further be used to correct the estimated multijet.
\end{itemize}

The difference between the two methods is that \R from the normalisation method would be a number which is independent of distribution and can be used to scale the estimated multijet of all the kinematic distributions whereas \R$[i]$ from the shape method is a distribution dependent correction factor and would be different for every kinematic distribution.

The correction factor is calculated from both the methods. A value for the correction factor comes out to be \R = 1.26 in the case of normalisation method, which is used to scale the estimated multijet from the ABCD method in all the six distributions shown in Fig.\ \ref{fig:abcd:estimate}. Similarly, \R$[i]$ from shape method is also evaluated, and six different distributions are obtained, which are further multiplied with their respective six distributions to get the estimated multijet. 

\begin{figure}[hbt!]
	\centering
	\begin{subfigure}{.35\textwidth}
		\centering
		\includegraphics[width=\linewidth,height=\textheight,keepaspectratio]{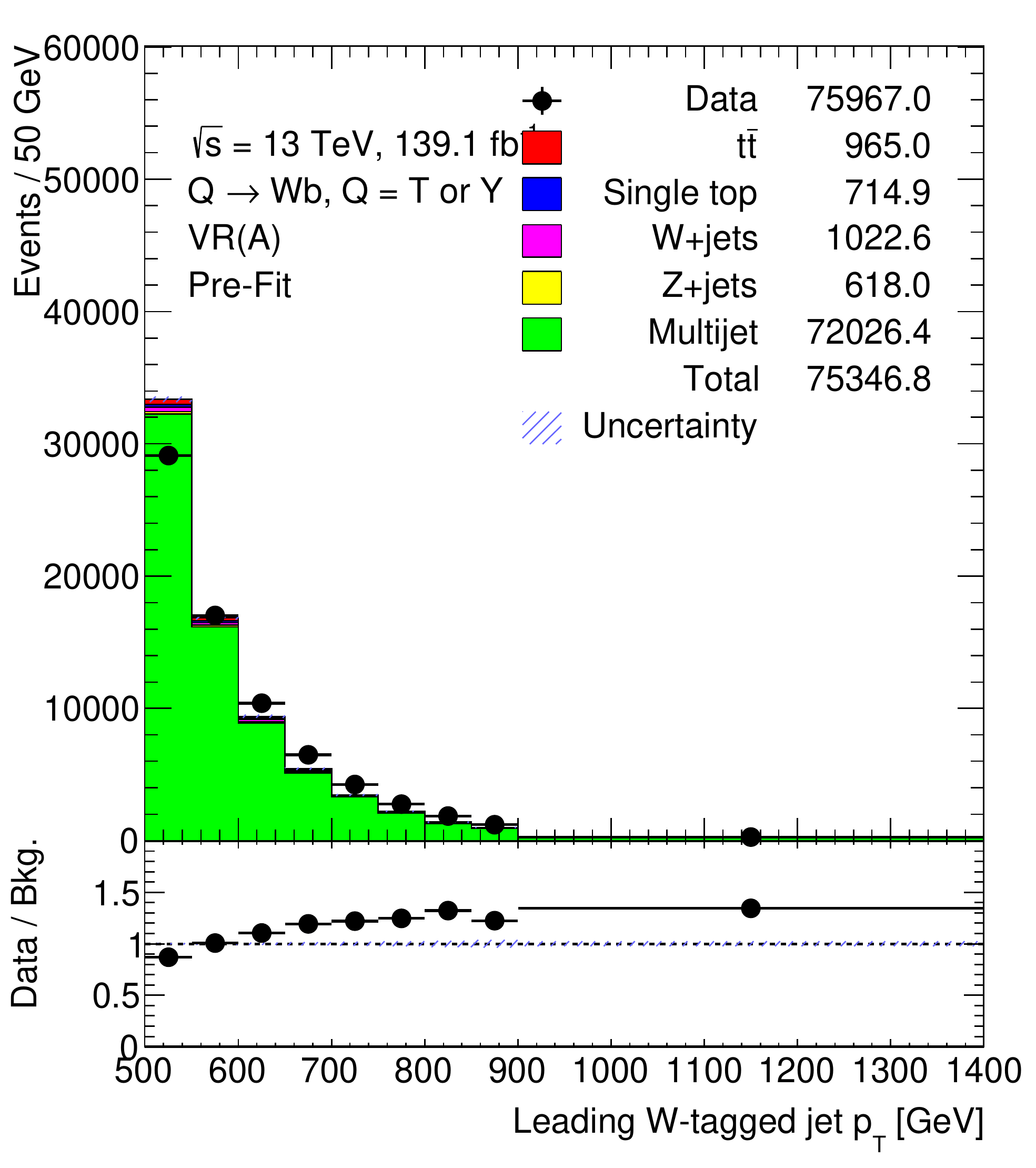}
		\caption{}
		\label{fig:abcd:correctionfactor:integral:ljet_pt}
	\end{subfigure}\hspace{0.6cm}
	\begin{subfigure}{.35\textwidth}
		\centering
		\includegraphics[width=\linewidth,height=\textheight,keepaspectratio]{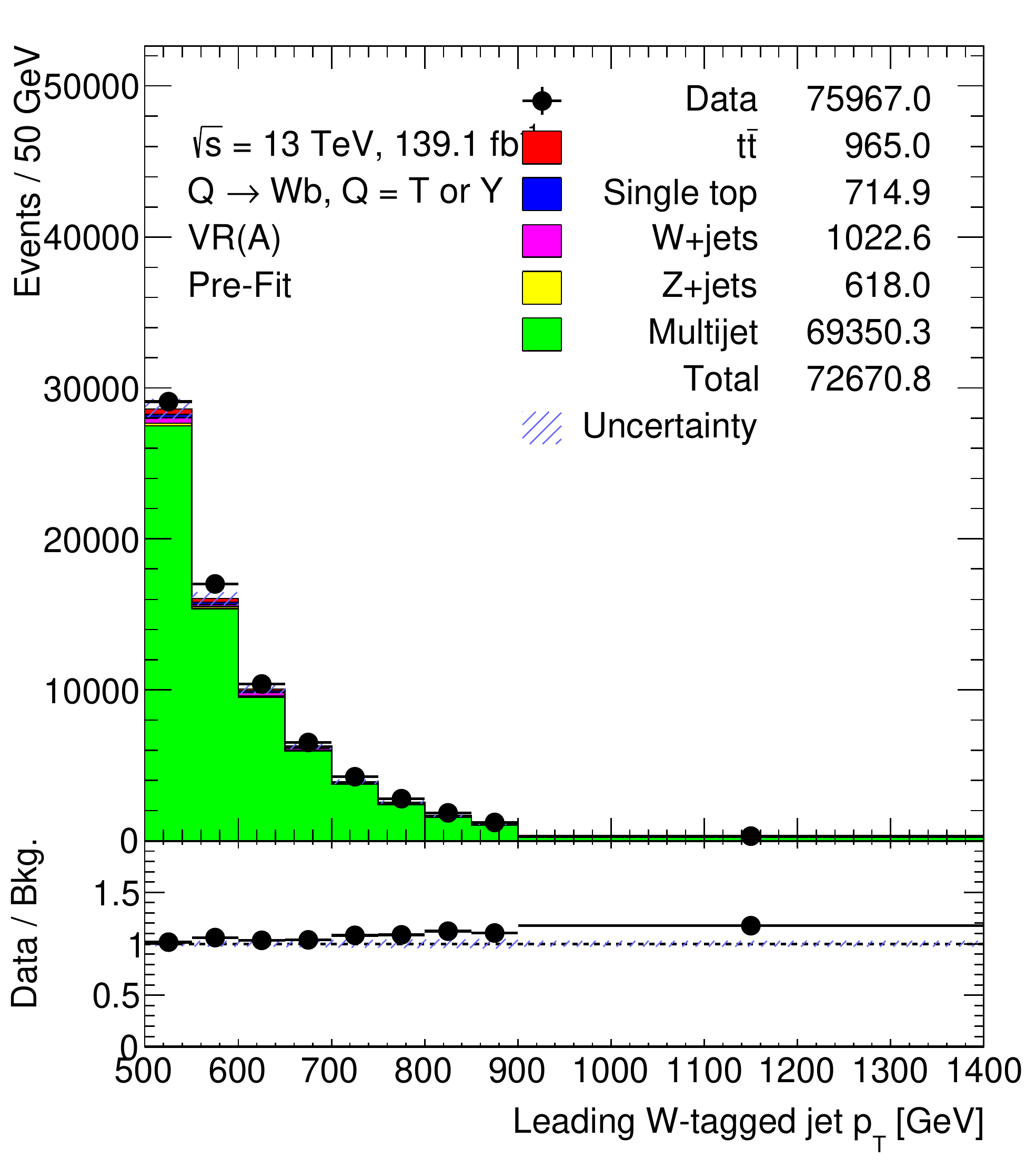}
		\caption{}
		\label{fig:abcd:correctionfactor:bin:ljet_pt}
	\end{subfigure}
	\caption{Comparison between the estimated multijet when \R is calculated by (a) normalisation method and (b) shape method in the \pt distribution of $W$-tagged large-$R$ jet.}
	\label{fig:abcd:correctionfactor:ljet_pt}
\end{figure}

\begin{figure}[hbt!]
	\centering
	\begin{subfigure}{.35\textwidth}
		\centering
		\includegraphics[width=\linewidth,height=\textheight,keepaspectratio]{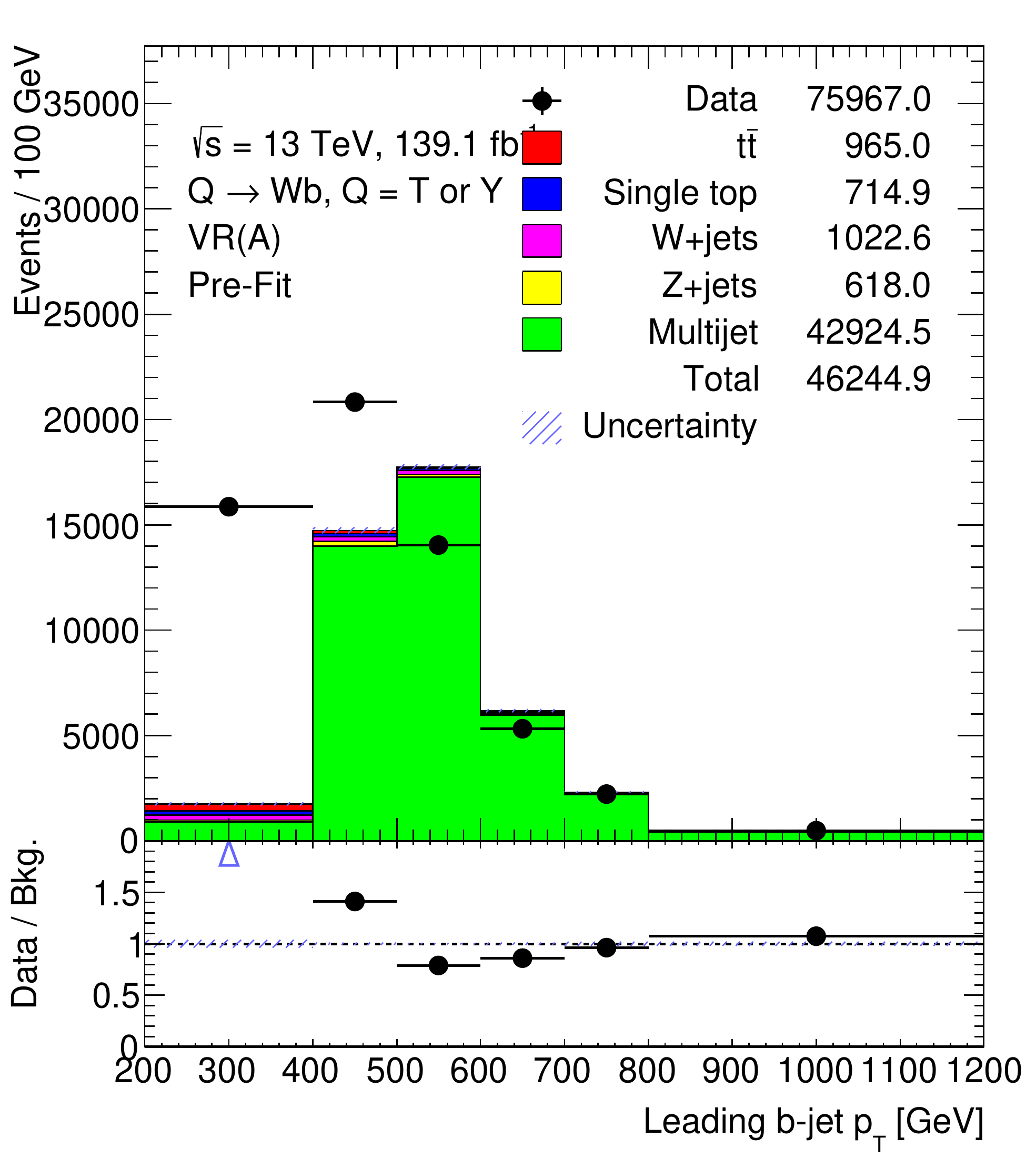}
		\caption{}
		\label{fig:abcd:correctionfactor:integral:jet_pt}
	\end{subfigure}\hspace{0.6cm}
	\begin{subfigure}{.35\textwidth}
		\centering
		\includegraphics[width=\linewidth,height=\textheight,keepaspectratio]{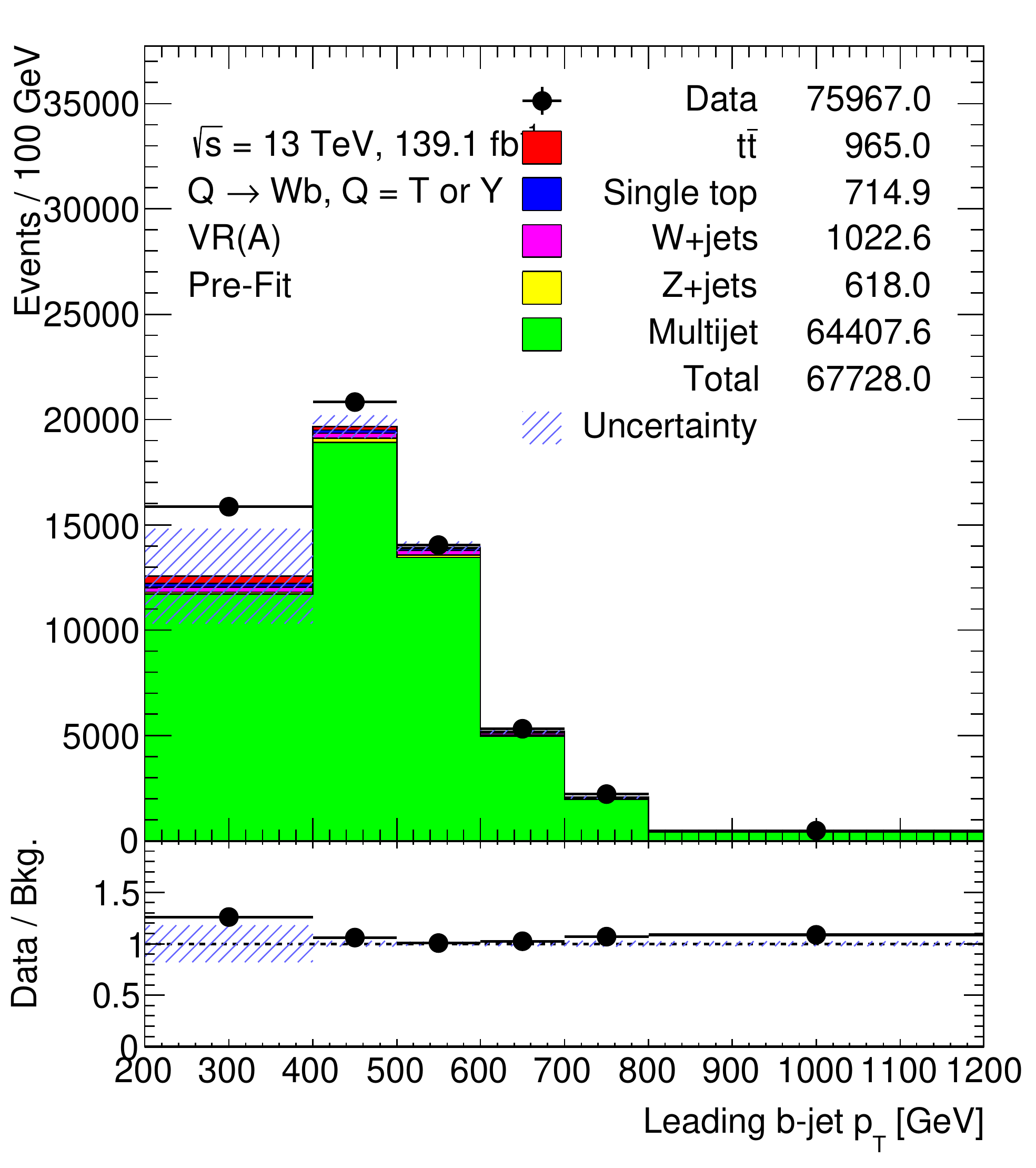}
		\caption{}
		\label{fig:abcd:correctionfactor:bin:jet_pt}
	\end{subfigure}
	\caption{Comparison between the estimated multijet when \R is calculated by (a) normalisation method and (b) shape method in the \pt distribution of leading $b$-tagged jet.}
	\label{fig:abcd:correctionfactor:jet_pt}
\end{figure}

\begin{figure}[hbt!]
	\centering
	\begin{subfigure}{.35\textwidth}
		\centering
		\includegraphics[width=\linewidth,height=\textheight,keepaspectratio]{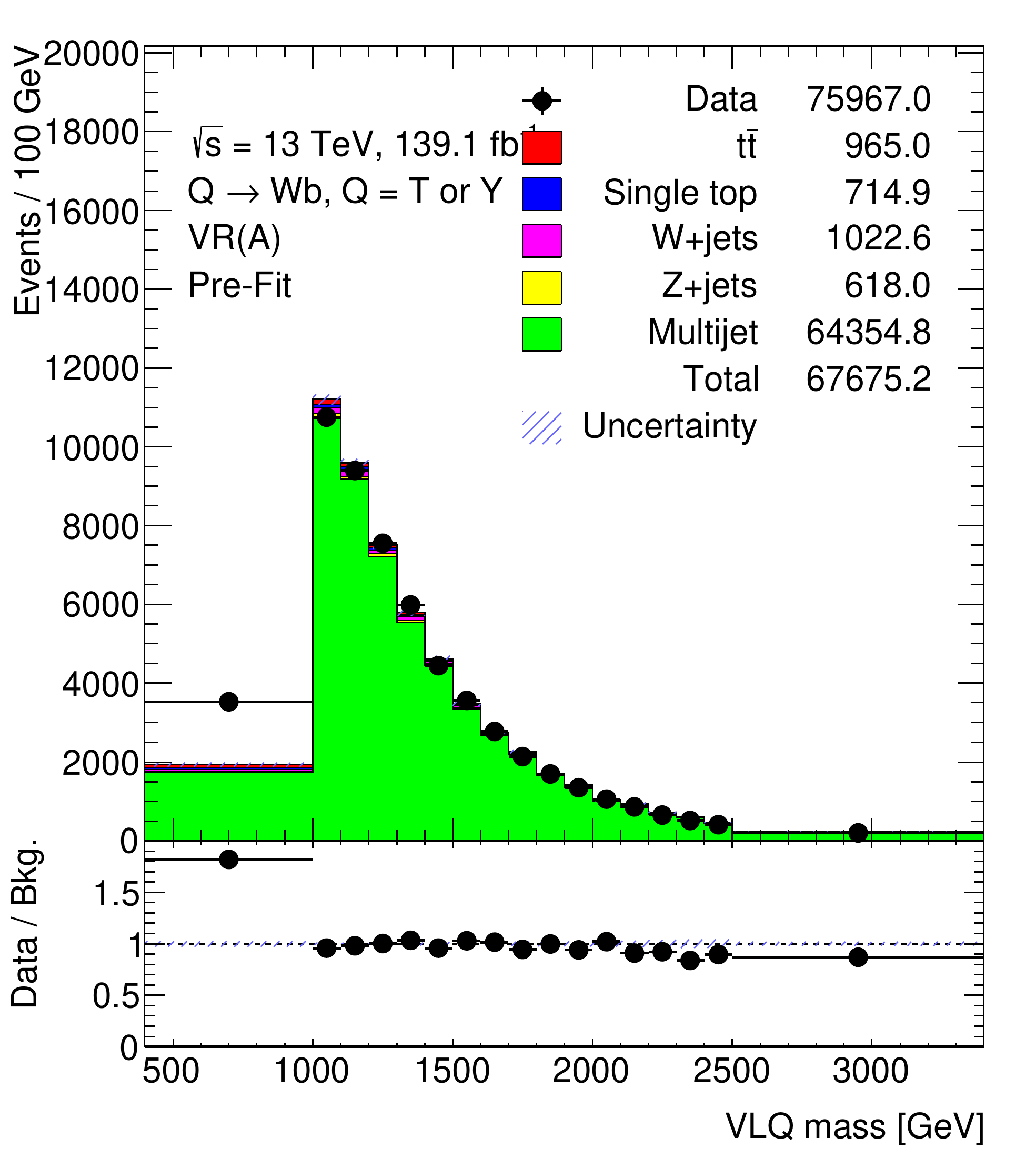}
		\caption{}
		\label{fig:abcd:correctionfactor:integral:VLQM}
	\end{subfigure}\hspace{0.6cm}
	\begin{subfigure}{.35\textwidth}
		\centering
		\includegraphics[width=\linewidth,height=\textheight,keepaspectratio]{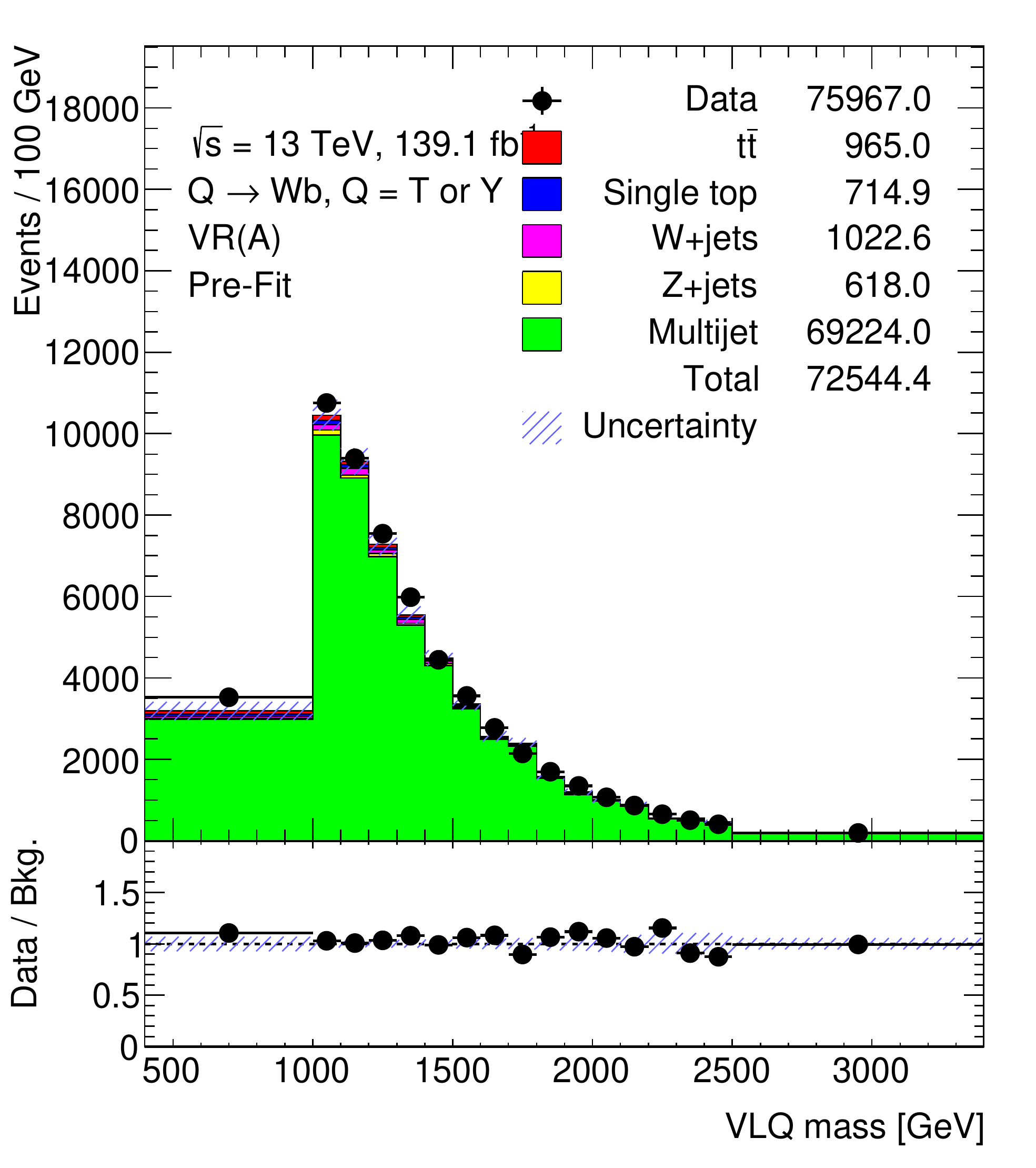}
		\caption{}
		\label{fig:abcd:correctionfactor:bin:VLQM}
	\end{subfigure}
	\caption{Comparison between the estimated multijet when \R is calculated by (a) normalisation method and (b) shape method in the VLQ mass distribution.}
	\label{fig:abcd:correctionfactor:VLQM}
\end{figure}

\begin{figure}[hbt!]
	\centering
	\begin{subfigure}{.35\textwidth}
		\centering
		\includegraphics[width=\linewidth,height=\textheight,keepaspectratio]{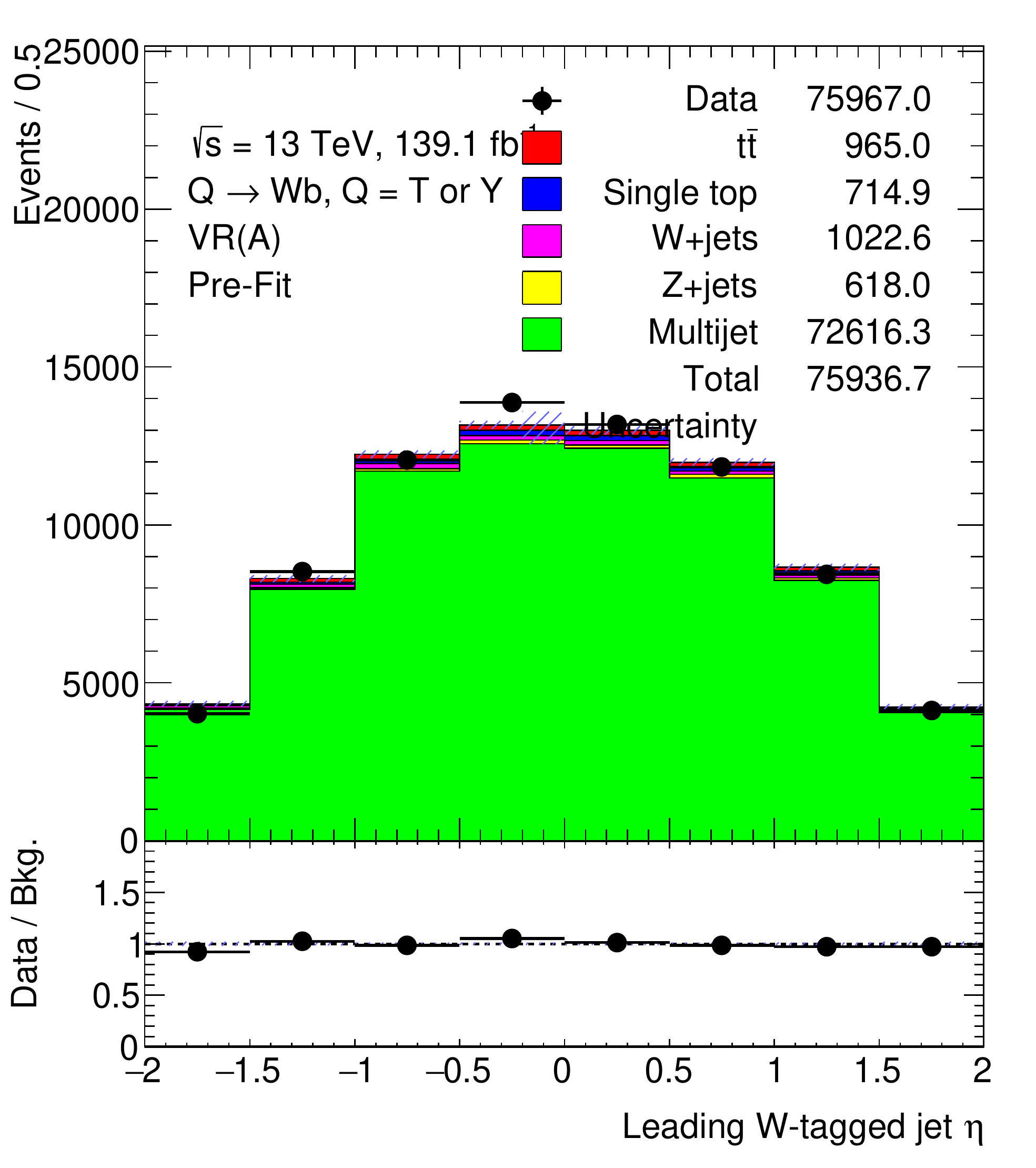}
		\caption{}
		\label{fig:abcd:correctionfactor:integral:ljet_eta}
	\end{subfigure}\hspace{0.6cm}
	\begin{subfigure}{.35\textwidth}
		\centering
		\includegraphics[width=\linewidth,height=\textheight,keepaspectratio]{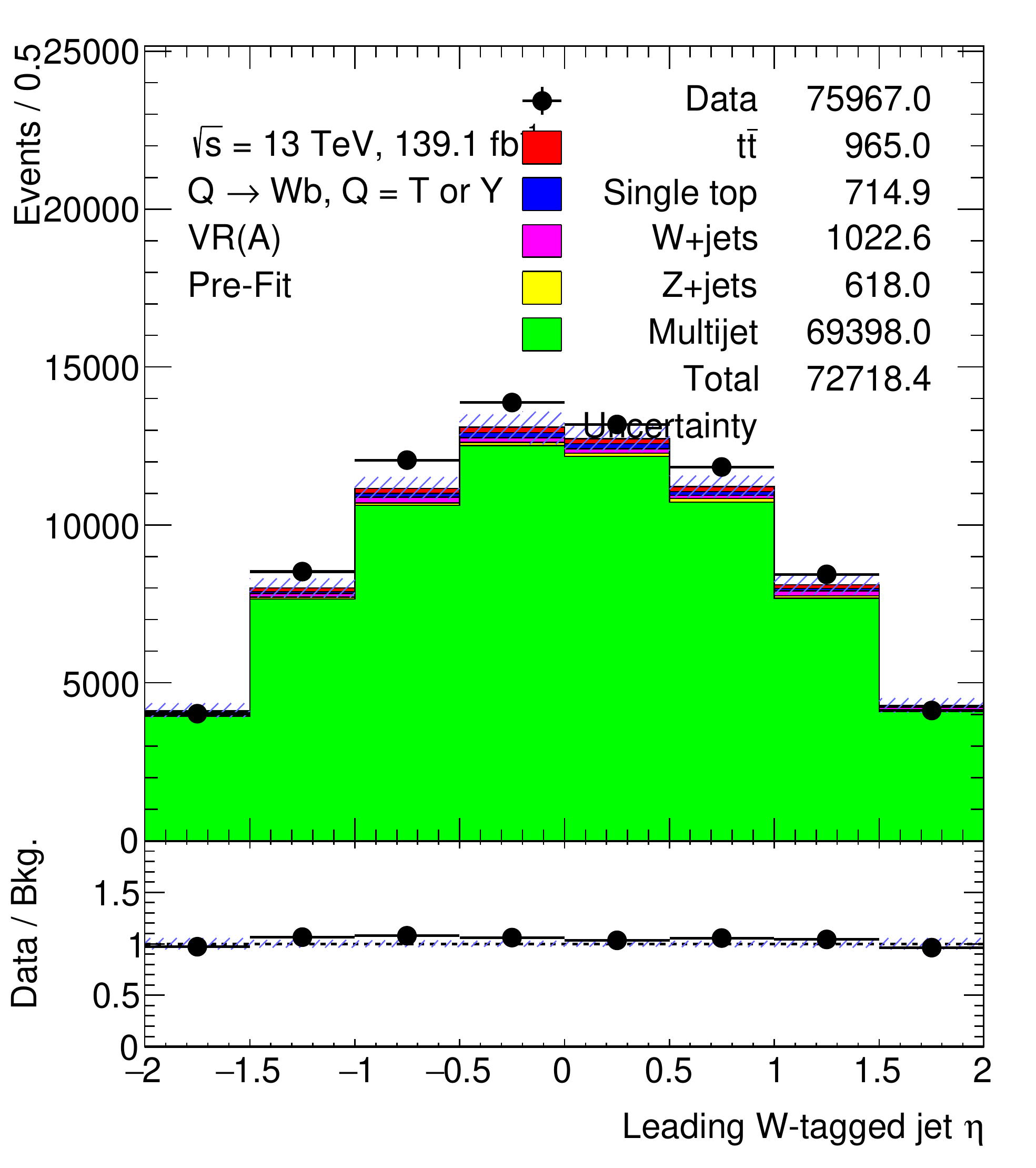}
		\caption{}
		\label{fig:abcd:correctionfactor:bin:ljet_eta}
	\end{subfigure}
	\caption{Comparison between the estimated multijet when \R is calculated by (a) normalisation method and (b) shape method in the $\eta$ distribution of $W$-tagged large-$R$ jet.}
	\label{fig:abcd:correctionfactor:ljet_eta}
\end{figure}

Fig.\ \ref{fig:abcd:correctionfactor:ljet_pt} shows the \pt distribution of $W$-tagged large-$R$ jet when \R is calculated from both the methods. The estimated multijet agrees well with the data when \R is calculated using the shape method as shown in Fig.\ \ref{fig:abcd:correctionfactor:bin:ljet_pt}. A significant difference can be seen in the \pt distribution of leading $b$-tagged jet in Fig.\ \ref{fig:abcd:correctionfactor:jet_pt} where \R has compensated well at low \pt when it is calculated by the shape method. A similar kind of effect can also be seen at the first bin of VLQ mass distribution in Fig.\ \ref{fig:abcd:correctionfactor:VLQM}. On the other hand, $\eta$ distribution of $W$-tagged large-$R$ in Fig.\ \ref{fig:abcd:correctionfactor:ljet_eta} shows that there is not much difference between the two methods. This is because the $\eta$ distribution is a uniform distribution and scaling the distribution or applying correction bin-by-bin does create a significant difference. However, looking at the event yields of multijet in the two distributions, it can be inferred that calculation from normalisation method shows better results than the shape method. The priority is bin-by-bin agreement of the data and estimated multijet in all the distributions. The error shown in the plots is all statistical uncertainty which is discussed in detail in the next chapter.

\clearpage
\section{Further improvement}
\label{sec:abcd:furtherimprovement}
After applying the \R, the multijet estimate shows a good agreement with the data. However, there is still a room for improvement. At preselection, it has been observed that the data/MC mismodelling is flat in \pt and VLQ mass, as shown in Fig.\ \ref{fig:analysisstrategy:eventselection:preselection}. Also, in the hadronic channel, one cannot trust the multijet MC  because of the mismodelling. But the correction factor in the previous section was calculated from these mismodelled multijet MC. So, in order to make the estimate purely data-driven process, a likelihood fit is performed to fit multijet MC to the data. So, in this section, the process of fitting the multijet MC to data is discussed along with how it has improved the estimation.
\subsection{Fitting}
\label{sec:abcd:furtherimprovement:fitting}
The aim is to determine the correction factor from the data. The task can be encoded into a statistical model where multijet behaves according to the observed data which can be represented by a parameter. A maximum likelihood fit~\cite{likelihood} of the model to data is used to obtain the parameter estimate. Then, the multijet MC is scaled with this parameter estimate. The whole process is implemented by the maximum likelihood method.

\paragraph{Maximum likelihood method} 
Let us consider the Poisson distribution that describes the observed event yields in an experiment.~\cite{thesis:rui} 
\begin{equation}
	\mathcal{P}(n|\lambda) = \lambda^{n} \frac{e^{-\lambda}}{n!} \,,
\end{equation}
where $n$ is denoted as the observed event yields and $\lambda$ is represented as the expected event yields.

Let us suppose an observable $x$ follows a distribution $f(x)$, the likelihood of the parameter $\lambda$ can be expressed as:
\begin{equation}
\mathcal{L}(\lambda) = \mathcal{P}(n|\lambda) \prod_{\text{event}}^{n} f(x) \,.
\end{equation}
The function $f(x)$ is usually described by histograms, where the observable $x$ is binned to discrete values. The event yields in each bin are Poisson distributed, making the likelihood a product of multiple Poisson distributions. It is also known as the \textit{binned likelihood} and can be written as:

\begin{equation}
\mathcal{L}(\lambda_{\text{b}}) = \prod_{\text{b$\in$bins}}^{} \mathcal{P}(n_{\text{b}}|\lambda_{\text{b}}) \,.
\end{equation}

Likelihood is multiplicative, so when it is performed for more than one region together, such as various control regions together, then it can be written as:

\begin{equation}
\mathcal{L}(\lambda_{\text{b,r}}) = \prod_{\text{r$\in$regions}}^{}\prod_{\text{b$\in$bins}}^{} \mathcal{P}(n_{\text{b,r}}|\lambda_{\text{b,r}}) \,,
\end{equation}
where $n_{\text{b,r}}$ and $\lambda_{\text{b,r}}$ are denoted as the number of observed events and the true events in a given bin of a given region respectively.

In most cases, log-likelihood functions are used because it converts the product of likelihoods into a simple summation which can be written as:

\begin{equation}
\text{ln }\mathcal{L}(\lambda_{\text{b,r}}) = \sum_{\text{r$\in$regions}}^{}\sum_{\text{b$\in$bins}}^{} \text{ln }\mathcal{P}(n_{\text{b,r}}|\lambda_{\text{b,r}}) \,.
\label{eqn:abcd:furtherimprovement:fitting}
\end{equation}
Now, the expression can be maximised to give an estimate of $\lambda_{\text{b,r}}$ denoted as $\hat{\lambda}_{\text{b,r}}$.

\begin{equation}
	\frac{\partial\text{ln } \mathcal{L}(\lambda_{\text{b,r}})}{\partial \lambda_{\text{b,r}}} \bigg|_{\lambda_{\text{b,r}}=\hat{\lambda}_{\text{b,r}}} = 0 \,.
\end{equation}

Sometimes it is challenging to perform this calculation analytically. In such cases, a set of computer programs which are based on numerical methods are generally preferred. In this thesis, a package called \textsc{TrexFitter}, which is based on \textsc{RooStats}~\cite{roostats} is used to perform the likelihood fit.
\subsection{Implementation on the multijet MC}
\label{sec:abcd:furtherimprovement:scaledcorr}
The maximumm likelihood method is performed to fit multijet MC to the data. It can only be applied in VR A since the data is blinded in SR A1. So, region A, B, C and D are taken into account. The fit is performed in region A and B together and region C and D together instead of fitting it individually because the regions should be correlated, and the multijet MC should be scaled by the same amount in the two regions where their normalisation is completely correlated.

In Eqn.\ \ref{eqn:abcd:furtherimprovement:fitting}, the expected event yield is predicted by the MC samples which contain both signal and background events. So, $\lambda_{\text{b,r}}$ can be expressed as:
\begin{equation}
	\lambda_{\text{b,r}} = \mu\lambda_{\text{b,r}}^{\text{sig. MC}} + (\lambda_{\text{b,r}}^{\text{other bkg. MC}} + \lambda_{\text{b,r}}^{\text{multijet MC}} ) \,,
\end{equation}
where $\mu$ is signal strength. The fit is performed as a background only fit, i.e.\ $\mu = 0$, while keeping the normalisation of the multijet as a free parameter. The normalisation of the other backgrounds is kept constant.

So, the log-likelihood function for region A and B together can be written as: 
\begin{equation}
\text{ln }\mathcal{L}(\lambda_{\text{b,r}}) = \sum_{\text{r$\in$\{A,B\}}}^{}\sum_{\text{b$\in$bins}}^{} \text{ln }\mathcal{P}(n_{\text{b,r}}|\lambda_{\text{b,r}}) \,.
\end{equation}
Similarly, the log-likelihood function for region C and D together can be written as: 
\begin{equation}
\text{ln }\mathcal{L}(\lambda_{\text{b,r}}) = \sum_{\text{r$\in$\{C,D\}}}^{}\sum_{\text{b$\in$bins}}^{} \text{ln }\mathcal{P}(n_{\text{b,r}}|\lambda_{\text{b,r}}) \,.
\end{equation}
These likelihood functions are maximised to get an estimate of $\hat{\lambda}_{\text{b,r}}$. It is performed for all the six variables separately to get six distributions where the multijet MC are fitted to the data, so we call it scaled multijet MC. The distribution for some kinematic variables for both multijet MC and scaled multijet MC in all the four regions A, B, C and D are shown in Appendix \ref{sec:app2}.

\subsection{Calculation of \R from the scaled multijet MC}
\label{sec:abcd:furtherimprovement:rcorr}
Now, \R is calculated from the scaled multijet MC (as discussed in the previous section) again by using two methods. It is calculated for each kinematic distribution separately.

\begin{itemize}
	\item \textbf{Normalisation method:} it is calculated by the expression given below:
	
	\begin{equation}
	\R = \frac{\N{A}{scaled multijet MC}}{\N{B}{scaled multijet MC}} \times \frac{\N{C}{scaled multijet MC}}{\N{D}{scaled multijet MC}} \,,
	\end{equation}
	where \N{j}{scaled multijet MC} is the number of events from the scaled multijet MC in region $j\in\{A, B, C, D\}$, which can be calculated by taking the integral of the distribution. So, in the end, \R is a number which can be used in Eqn.\ \ref{eqn:abcd:correctionfactor} to scale the estimated multijet. Note that now this number is not same for all the kinematic distributions because the multijet MC in each kinematic distribution is scaled differently in the scaled multijet MC distributions. Table \ref{table:abcd:furtherimprovement:rcorr} shows the value of \R for all the six distributions.
	
	\begin{table}[hbt!]
		\centering
		\begin{tabular}{c|c|c|c|c|c|c} 
			\toprule
			Kinematic  & \multicolumn{3}{c}{$W$-tagged jet} \vline & \multicolumn{2}{c}{leading $b$-tagged jet} \vline & VLQ mass\\ \cline{2-6}
			distribution & \pt & mass & $\eta$ & \pt & mass & \\ 
			\midrule
			\R & \num{1.31} & \num{1.32} & \num{1.33} & \num{1.34} & \num{1.33} & \num{1.33} \\
			\bottomrule
		\end{tabular}
		\caption{Value of \R for all the six kinematic distributions calculated from the scaled multijet MC by using normalisation method.}
		\label{table:abcd:furtherimprovement:rcorr}
	\end{table}
	
	\item \textbf{Shape method:} a corresponding expression can also be written as:
	\begin{equation}
	\R[i] = \frac{\N{A}{scaled multijet MC}[i]}{\N{B}{scaled multijet MC}[i]} \times \frac{\N{C}{scaled multijet MC}[i]}{\N{D}{scaled multijet MC}[i]} \,,
	\end{equation}
	where $[i]$ shows that the calculation is performed for each bin separately within a distribution ($i=$ bin). So, it leads to a separate \R$[i]$ distribution for each kinematic distribution, which can further be used to correct the estimated multijet in Eqn.\ \ref{eqn:abcd:correctionfactor}.
\end{itemize}

Summing up, in this thesis, \R is calculated from four different methods:
\begin{enumerate}
	\item From multijet MC by using normalisation method. (described in section \ref{sec:abcd:correctionfactor})
	\item From multijet MC by using shape method. (described in section \ref{sec:abcd:correctionfactor})
	\item From scaled multijet MC by using normalisation method.
	\item From scaled multijet MC by using shape method.
\end{enumerate}

\begin{figure}[hbt!]
	\centering
	\begin{subfigure}{.35\textwidth}
		\centering
		\includegraphics[width=\linewidth,height=\textheight,keepaspectratio]{figs/chapter5/prefitintegral/VR_B_VLQM.pdf}
		\caption{}
		\label{fig:abcd:furtherimprovement:scaledcorr:prefit:integral:VLQM}
	\end{subfigure}\hspace{0.6cm}
	\begin{subfigure}{.35\textwidth}
		\centering
		\includegraphics[width=\linewidth,height=\textheight,keepaspectratio]{figs/chapter5/prefitbinbybin/VR_B_VLQM.pdf}
		\caption{}
		\label{fig:abcd:furtherimprovement:scaledcorr:prefit:bin:VLQM}
	\end{subfigure}
	\begin{subfigure}{.35\textwidth}
		\centering
		\includegraphics[width=\linewidth,height=\textheight,keepaspectratio]{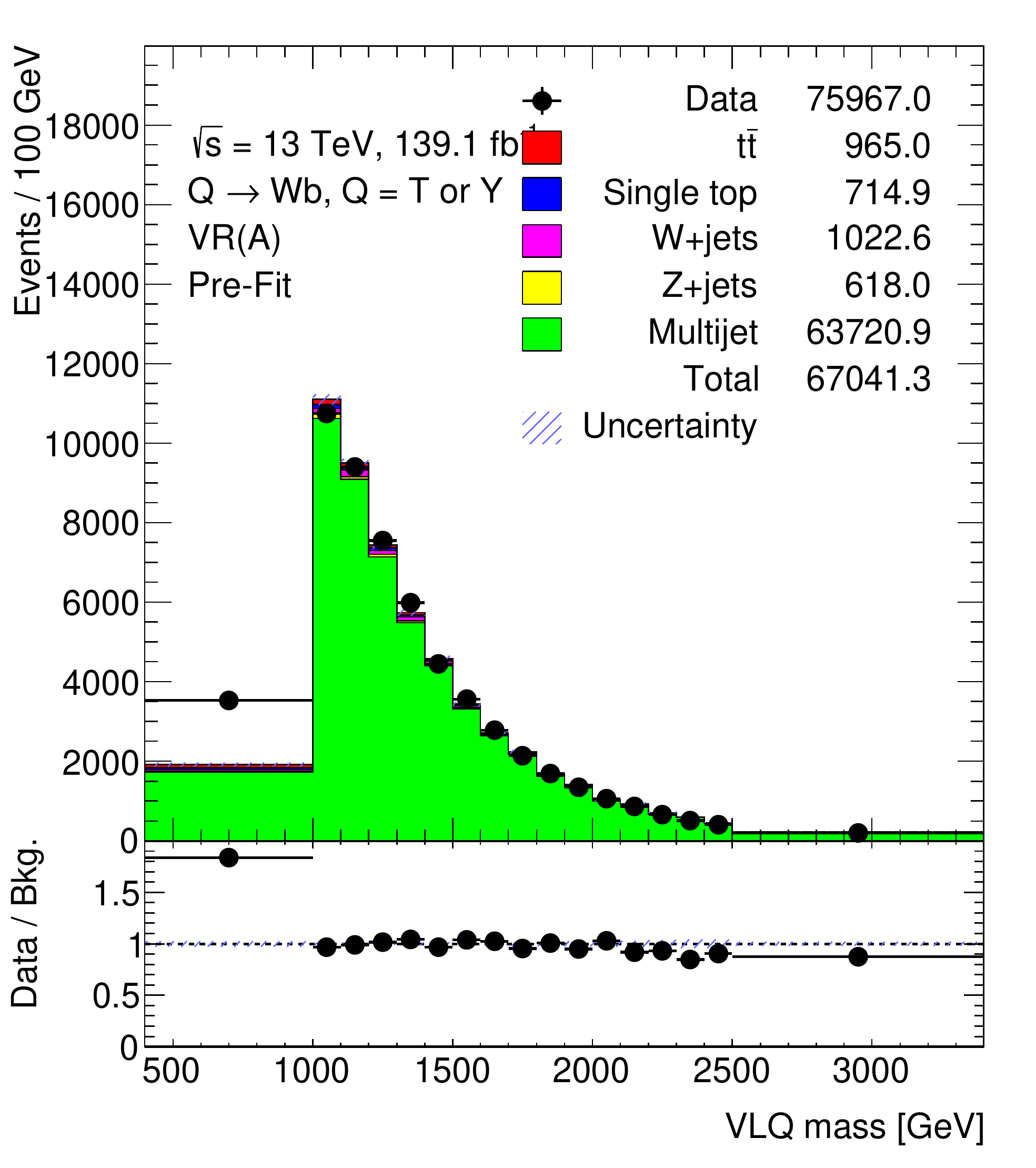}
		\caption{}
		\label{fig:abcd:furtherimprovement:scaledcorr:postfit:integral:VLQM}
	\end{subfigure}\hspace{0.6cm}
	\begin{subfigure}{.35\textwidth}
		\centering
		\includegraphics[width=\linewidth,height=\textheight,keepaspectratio]{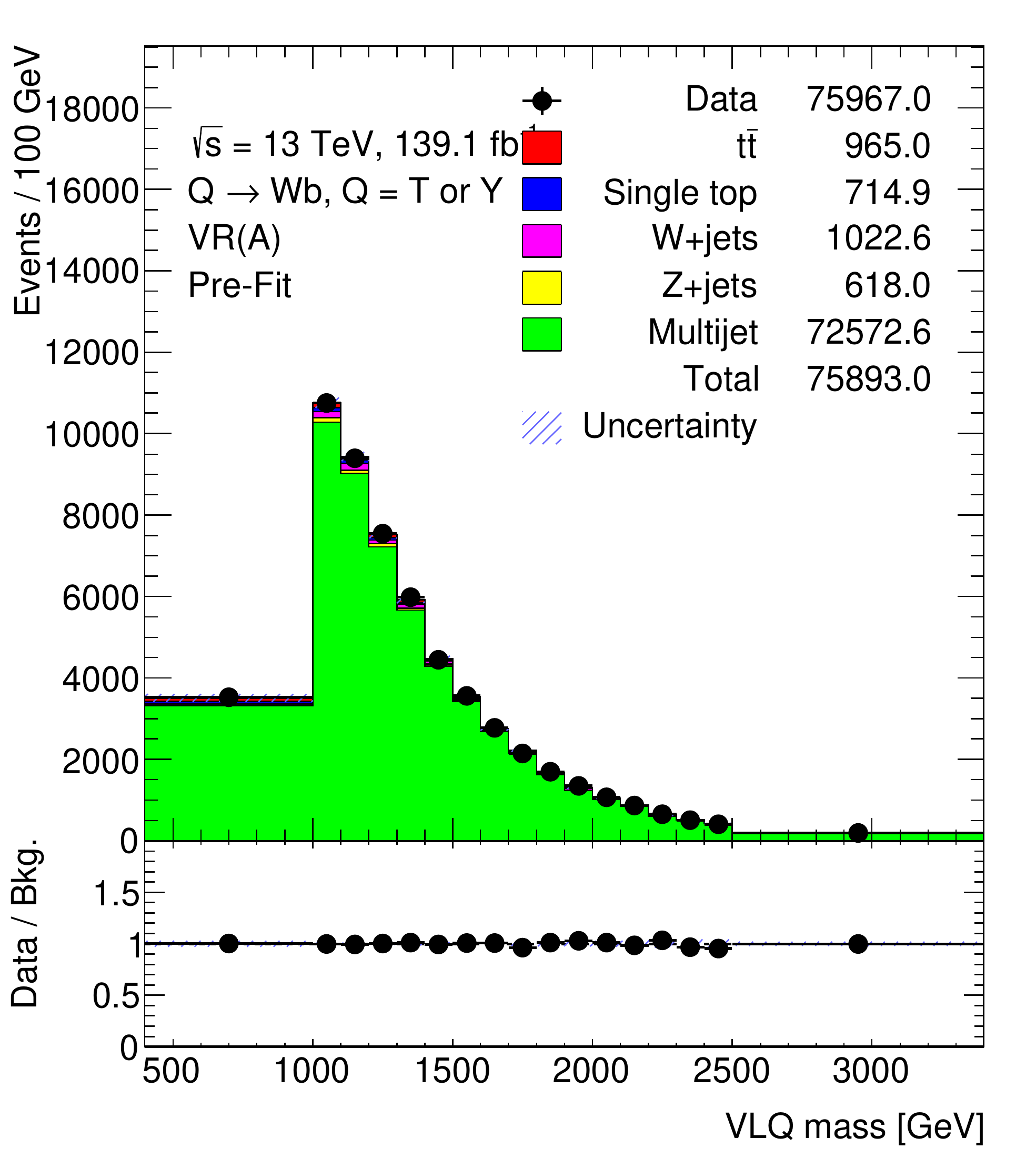}
		\caption{}
		\label{fig:abcd:furtherimprovement:scaledcorr:postfit:bin:VLQM}
	\end{subfigure}
	\caption{VLQ mass distribution where multijet background is estimated when \R is calculated from the multijet MC by using (a) normalisation method and (b) shape method, and from the scaled multijet MC by (c) normalisation method and (d) shape method.}
	\label{fig:abcd:furtherimprovement:scaledcorr:VLQM}
\end{figure}

A data/bkg.\ comparison plot is shown in Fig.\ \ref{fig:abcd:furtherimprovement:scaledcorr:VLQM}, where the estimated multijet is shown in VLQ mass distribution when \R is calculated from all the four methods. It can be observed that \R calculated from normalisation method does not produce any significant difference when it is calculated from either multijet MC or scaled multijet MC. However, better results in the estimate can be seen from the shape method, especially in the event yields of the estimated multijet. A detailed event yields of the estimated multijet when \R is calculated from all the four methods are shown in Table \ref{table:abcd:furtherimprovement:scaledcorr}. 

\begin{figure}[hbt!]
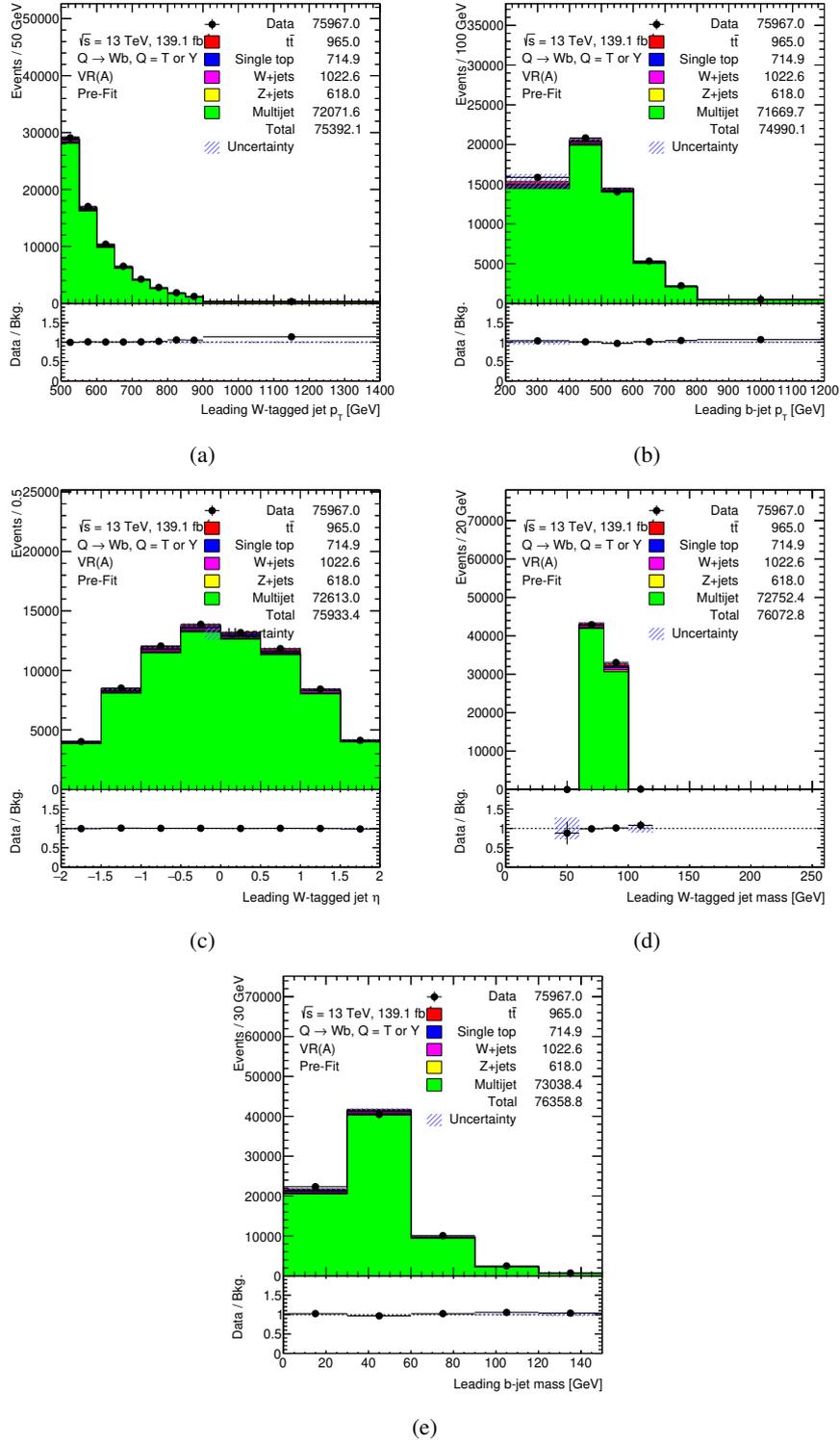

	\centering
	\graphicspath{{figs/chapter5/postfitbinbybin/}}
	\begin{subfigure}{.35\textwidth}
		\centering
		\includegraphics[width=\linewidth,height=\textheight,keepaspectratio]{VR_B_ljet_pt.pdf}
		\caption{}
		\label{fig:abcd:furtherimprovement:scaledcorr:ljet_pt}
	\end{subfigure}\hspace{0.6cm}
	\begin{subfigure}{.35\textwidth}
		\centering
		\includegraphics[width=\linewidth,height=\textheight,keepaspectratio]{VR_B_jet_pt.pdf}
		\caption{}
		\label{fig:abcd:furtherimprovement:scaledcorr:jet_pt}
	\end{subfigure}
	\begin{subfigure}{.35\textwidth}
		\centering
		\includegraphics[width=\linewidth,height=\textheight,keepaspectratio]{VR_B_ljet_eta.pdf}
		\caption{}
		\label{fig:abcd:furtherimprovement:scaledcorr:ljet_eta}
	\end{subfigure}\hspace{0.6cm}
	\begin{subfigure}{.35\textwidth}
		\centering
		\includegraphics[width=\linewidth,height=\textheight,keepaspectratio]{VR_B_ljet_m.pdf}
		\caption{}
		\label{fig:abcd:furtherimprovement:scaledcorr:ljet_m}
	\end{subfigure}
	\begin{subfigure}{.35\textwidth}
		\centering
		\includegraphics[width=\linewidth,height=\textheight,keepaspectratio]{VR_B_jet_m.pdf}
		\caption{}
		\label{fig:abcd:furtherimprovement:scaledcorr:jet_m}
	\end{subfigure}
	\caption{A data/bkg.\ comparison of kinematic and reconstructed variables in VR A where multijet background is estimated from the ABCD method in which \R is calculated by using shape method from the scaled multijet MC. The variables include (a) $p_{\text{T}}$ of $W$-tagged large-$R$ jet, (b) $p_{\text{T}}$ of leading $b$-tagged small-$R$ jet, (c) $\eta$ distribution of $W$-tagged large-$R$ jet, (d) mass of $W$-tagged large-$R$ jet, and (e) mass of leading $b$-tagged small-$R$ jet. Errors shown in the plots are all statistical uncertainty.}
	\label{fig:abcd:furtherimprovement:scaledcorr}
\end{figure}

So, in general Fig.\ \ref{fig:abcd:furtherimprovement:scaledcorr:postfit:bin:VLQM} shows the best estimate of multijet background where \R is calculated from the scaled multijet MC by using shape method. The other distributions are also shown in Fig.\ \ref{fig:abcd:furtherimprovement:scaledcorr} to confirm that the method works efficiently in all the other distributions as well. So, this multijet estimate is regarded as the final estimate.

\begin{table}[hbt!]
	\centering
	\begin{tabular}{c|c|c|c|c} 
		\toprule
		Sample & \multicolumn{2}{c}{Multijet MC} \vline & \multicolumn{2}{c}{Scaled multijet MC} \\ \cline{2-5} 
		 & Normalisation & Shape & Normalisation & Shape \\ 
		\midrule
		Data & 75967 & 75967 & 75967 & 75967 \\
		Other bkg. & $\num{3320}\pm\num{95}$ & $\num{3320}\pm\num{95}$ & $\num{3320}\pm\num{95}$ & $\num{3320}\pm\num{95}$ \\
		Est. multijet & $\num{72616}\pm\num{304}$ & $\num{69398}\pm\num{914}$ & $\num{71901}\pm\num{301}$ & $\num{72613}\pm\num{305}$ \\
		\midrule
		Total bkg. & $\num{75936}\pm\num{319}$ & $\num{72718}\pm\num{919}$ & $\num{75221}\pm\num{316}$ & $\num{75933}\pm\num{319}$ \\
		\bottomrule
	\end{tabular}
	\caption{Event yields of the estimated multijet background in VR A from the ABCD method when \R is calculated from all the four methods. The errors shown here are from statistical uncertainty.}
	\label{table:abcd:furtherimprovement:scaledcorr}
\end{table}

\chapter{Uncertainties}
\label{sec:uncertainty_result}
The purpose of this chapter is to discuss all the uncertainties taken into account for the multijet estimation. It includes both the statistical and systematic uncertainties. A detailed procedure for the calculation of different systematic uncertainties and their individual contribution to the estimated multijet is also described in detail. 
\section{Statistical uncertainties}
\label{sec:uncertainties:statistical}
Statistical uncertainties can reliably be estimated by repeating measurements. If the sample follows the Poisson distribution, then the statistical uncertainty reduces to $1/\sqrt{N}$ where $N$ is the sample size. When a histogram is filled with events, the statistical uncertainty associated with each bin of a histogram is given by $\sqrt{\text{bin content}}$. When the arithmetic operations are performed between the histograms, the statistical uncertainties propagate and follow the Gaussian error propagation. 

In this thesis, the statistical uncertainties are arising from two different sources: first from the ABCD method where the arithmetic operations (like subtraction, multiplication and division) are performed between the histograms according to the ABCD equation shown in \ref{eqn:abcd:correctionfactor}. In this, the statistical uncertainty in each bin of the distributions propagates by Gaussian error propagation and contributes to the estimated multijet distribution.

The other statistical uncertainty arises from the calculation of \R. In this, the likelihood fit is performed to fit the multijet MC to the data, and the output of the fit, which is the scaled multijet MC is used to calculate \R. The statistical uncertainty in the distribution of the scaled multijet MC contributes less because the fit constrains the uncertainty associated with each bin. However, when the arithmetic operations are performed on the scaled multijet MC for the calculation of \R, the uncertainties in each bin of the distribution follow the Gaussian error propagation and contribute further to the statistical uncertainty. A detailed contribution of the total statistical uncertainty to the multijet estimate in all the distributions is given in Table \ref{table:uncertainties:systematics:detail}.
\section{Systematic uncertainties}
\label{sec:uncertainties:systematics}
Systematic uncertainties are the ones which are not directly due to the statistics of the sample. They can arise from any source, for example, limited knowledge of backgrounds or the badly modelled backgrounds. Sometimes, they can also come from the imperfect detector resolutions or some other external factors affecting the detector acceptance. In general, they are difficult to determine because they cannot be calculated solely from the sampling fluctuations. So, different approaches have to be performed in order to estimate various systematic uncertainties. They are broadly divided into three categories:

First is theoretical uncertainties which arise from the theoretical prediction of different observables. The choice of the MC generator and their parameters are empirical for modelling the signal and background samples. So, different choices lead to different results which can affect the measurement. It includes uncertainties on the calculation of the matrix element, uncertainty on the parton shower and the hadronisation process, correction of different radiation levels (like initial- and final-state radiation), uncertainty due to the choice of the parton distribution function etc. A general procedure to estimate the uncertainties which depend on the choice of a generator is to replace a generator by another one or vary the values of the parameters, and simulate full samples again and then evaluate their contribution to the uncertainty.

Second is detector and algorithm performances uncertainties which arise from the imperfect detector and reconstruction algorithms, which are usually studied by the various performance groups in ATLAS. These include uncertainties associated with different efficiencies, trigger and scale factors. It also consists of inaccuracy in energy calibration and the measurement of jet energy scale and jet energy resolution. Uncertainties in the reconstruction and tagging algorithms also contribute a lot, especially when $b$-tagging is used in the analysis. The procedure of estimating these uncertainties might differ depending on the analysis.

The third is luminosity and cross-section uncertainties. One should assume uncertainty on luminosity since the luminosity is also not always constant throughout the data taking period, and there might be some fluctuations which need to be taken into account. The various modelling groups in ATLAS usually give the uncertainty on the theoretical values of the cross-section of different processes which needs to be taken into account in the analysis.

The contribution from these uncertainties leads to one of the two possible types of effects on the distributions. The first one is normalisation and shape effect that means some uncertainties (when taken into account) leads to a difference in the event yields as well as fluctuations in the shape of the distribution. The second one is shape only effect where one can only see the difference in the shape of the distribution without any change in the total number of events. Although it does not affect the measurement (or the multijet estimate in this case), it is important to consider this type of uncertainty when comparing an estimate with the data at the distribution level.

Due to limited time and availability of the MC samples, the study of all the systematic uncertainties is not feasible, where some of them are even beyond the scope of this thesis.  So, the systematic uncertainties which are studied in this thesis are described in detail below:

\subsection{$b$-tagging uncertainty}
\label{sec:uncertainties:systematics:btagging}
In this thesis, the estimation of multijet relies on the choice of two uncorrelated variables where one of the variables is $b$-jet multiplicity. The tagging of jets which are actually originating from the decay of $b$-quark is purely objective because sometimes the $b$-tagging algorithm mistagged $c$-jets and other light flavoured jets as $b$-jets.  So, in that case, the uncertainty associated with the performance of the $b$-tagging algorithm has to be taken into account.

The $b$-tagging uncertainties are estimated by varying five parameters in the MC simulation, which contributes the most. These parameters affect the $b$-tagging scale factor, which changes the $b$-tagging weight $w_{\text{b-tagging}}$ in Eqn.\ \ref{eqn:analysisstrategy:mc:weights}. And these weights are then applied to the events of all MC simulated backgrounds to calculate the total event yield.

The $b$-tagging uncertainty has been implemented by changing all the five parameters one by one in the MC simulation of all the backgrounds. For each of the parameters, first the MC is simulated by taking up variation and then by taking the down variation into account. So that means, in total, ten different MC samples for all the backgrounds are produced (five for both up and down variations). Then, the ABCD method is performed by using each of these MC background samples one by one, and the multijet background is estimated in the validation region. The multijet estimate varies in all these ten cases because the MC backgrounds have different $b$-tagging weights in all these cases (because of the change in the parameter). The difference in the normalisation of each of these ten multijet estimates with the final multijet estimate is taken as the $b$-tagging uncertainty. Since both the up and down variations of five parameters are taken into account, so the absolute mean of them is taken into account. 

The five parameters which are varied to calculate the $b$-tagging uncertainty are described in detail below along with their individual contribution towards the $b$-tagging uncertainty:

\begin{itemize}
	\item \textit{eigenvars\_b:} the scale factors are applied in order to correct for mismodelling in the output of the $b$-tagging algorithm. These scale factors are derived separately for $b$-jets, $c$-jets and light-flavour jets. Using an eigenvector decomposition~\cite{btagging:uncertainties}, these uncertainties are decomposed into different eigenvectors of $b$-jets, $c$-jets and light-flavoured jets. \textit{eigenvars\_b} is defined as the eigenvectors for $b$-jet efficiency. Fig.\ \ref{fig:uncertainties:systematics:btagging:b} shows the contribution of uncertainty from the eigenvector of $b$-jet efficiency in \pt distribution of the leading $b$-jet and VLQ mass. It can be observed that this uncertainty does not contribute much to the normalisation difference but contributes towards the shape difference.
	\begin{figure}[hbt!]
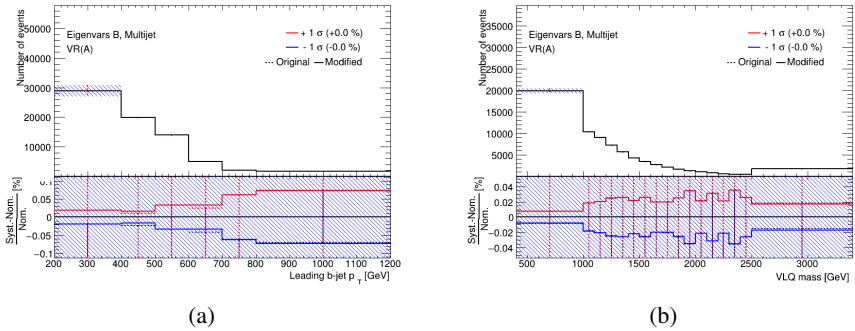

		\centering
		\graphicspath{{figs/chapter6/Systematics/EigenvarsB/}}
		\begin{subfigure}{.35\textwidth}
			\centering
			\includegraphics[width=\linewidth,height=\textheight,keepaspectratio]{VR_B_jet_pt_Multijets.png}
			\caption{}
			\label{fig:uncertainties:systematics:btagging:b:jetpt}
		\end{subfigure}\hspace{0.6cm}
		\begin{subfigure}{.35\textwidth}
			\centering
			\includegraphics[width=\linewidth,height=\textheight,keepaspectratio]{VR_B_VLQM_Multijets.png}
			\caption{}
			\label{fig:uncertainties:systematics:btagging:b:vlqm}
		\end{subfigure}
		\caption{Comparison between the Systematics-Nominal and Nominal, where Systematics-Nominal denotes the estimated multijet when both up (in red) and down (in blue) variations in \textit{eigenvars\_b} are taken into account, and Nominal denotes the final estimated multijet (in black). The difference between the two is considered as the systematic uncertainty from the eigenvectors of the $b$-jet efficiency. It is shown for (a) leading $b$-tagged jet \pt and (b) VLQ mass distribution.}
		\label{fig:uncertainties:systematics:btagging:b}
	\end{figure}

	\item \textit{eigenvars\_c:} it is defined as the eigenvectors for $c$-jets which describes the scale factors of $c$-jets mistagged as $b$-jets. Fig.\ \ref{fig:uncertainties:systematics:btagging:c} shows the contribution of uncertainty from the eigenvector of $c$-jet efficiency in \pt distribution of the leading $b$-jet and VLQ mass. One can see that \pt distribution shows 0.1\% difference in the normalisation of the distribution, whereas VLQ mass distribution only contributes to the difference in the shape of the distribution.
	\begin{figure}[hbt!]
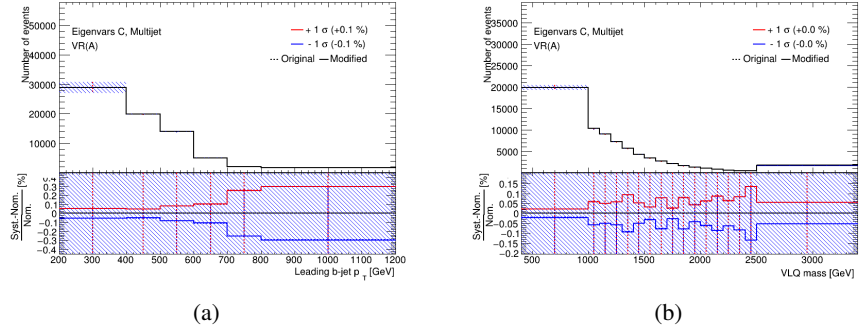

		\centering
		\graphicspath{{figs/chapter6/Systematics/EigenvarsC/}}
		\begin{subfigure}{.35\textwidth}
			\centering
			\includegraphics[width=\linewidth,height=\textheight,keepaspectratio]{VR_B_jet_pt_Multijets.png}
			\caption{}
			\label{fig:uncertainties:systematics:btagging:c:jetpt}
		\end{subfigure}\hspace{0.6cm}
		\begin{subfigure}{.35\textwidth}
			\centering
			\includegraphics[width=\linewidth,height=\textheight,keepaspectratio]{VR_B_VLQM_Multijets.png}
			\caption{}
			\label{fig:uncertainties:systematics:btagging:c:vlqm}
		\end{subfigure}
		\caption{Comparison between the Systematics-Nominal and Nominal, where Systematics-Nominal denotes the estimated multijet when both up (in red) and down (in blue) variations in \textit{eigenvars\_c} are taken into account, and Nominal denotes the final estimated multijet (in black). The difference between the two is considered as the systematic uncertainty from the eigenvectors of $c$-jets. It is shown for (a) leading $b$-tagged jet \pt and (b) VLQ mass distribution.}
		\label{fig:uncertainties:systematics:btagging:c}
	\end{figure}

	\item \textit{eigenvars\_light:} it is defined as the eigenvectors for light-flavoured jets which describes the scale factors of light-flavoured jets mistagged as $b$-jets. Fig.\ \ref{fig:uncertainties:systematics:btagging:light} shows the contribution of uncertainty from the eigenvector of light-flavoured jet in \pt distribution of the leading $b$-jet and VLQ mass. It contributes to the difference in the normalisation of all the kinematic distributions except the $\eta$ distribution.
	\begin{figure}[hbt!]
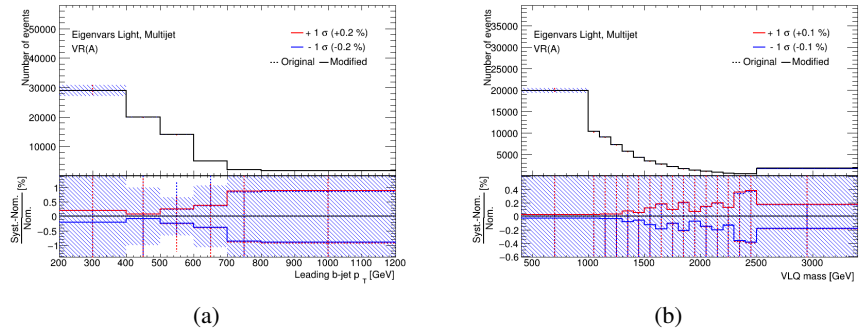

		\centering
		\graphicspath{{figs/chapter6/Systematics/EigenvarsLight/}}
		\begin{subfigure}{.35\textwidth}
			\centering
			\includegraphics[width=\linewidth,height=\textheight,keepaspectratio]{VR_B_jet_pt_Multijets.png}
			\caption{}
			\label{fig:uncertainties:systematics:btagging:light:jetpt}
		\end{subfigure}\hspace{0.6cm}
		\begin{subfigure}{.35\textwidth}
			\centering
			\includegraphics[width=\linewidth,height=\textheight,keepaspectratio]{VR_B_VLQM_Multijets.png}
			\caption{}
			\label{fig:uncertainties:systematics:btagging:light:vlqm}
		\end{subfigure}
		\caption{Comparison between the Systematics-Nominal and Nominal, where Systematics-Nominal denotes the estimated multijet when both up (in red) and down (in blue) variations in \textit{eigenvars\_light} are taken into account, and Nominal denotes the final estimated multijet (in black). The difference between the two is considered as the systematic uncertainty from the eigenvectors of light-flavoured jets. It is shown for (a) leading $b$-tagged jet \pt and (b) VLQ mass distribution.}
		\label{fig:uncertainties:systematics:btagging:light}
	\end{figure}

	\item \textit{extrapolation}: the precise measurement of $b$-tagging efficiency for jets is only implemented up to a certain jet \pt. The evaluation of the uncertainty in $b$-tagging efficiency beyond this \pt range can be calculated by \textit{extrapolation}, which describes the extrapolation uncertainties of \pt.~\cite{btagging:uncertainties} Fig.\ \ref{fig:uncertainties:systematics:btagging:extra} shows the contribution of the extrapolation uncertainty in \pt distribution of the leading $b$-jet and VLQ mass.
	\begin{figure}[hbt!]
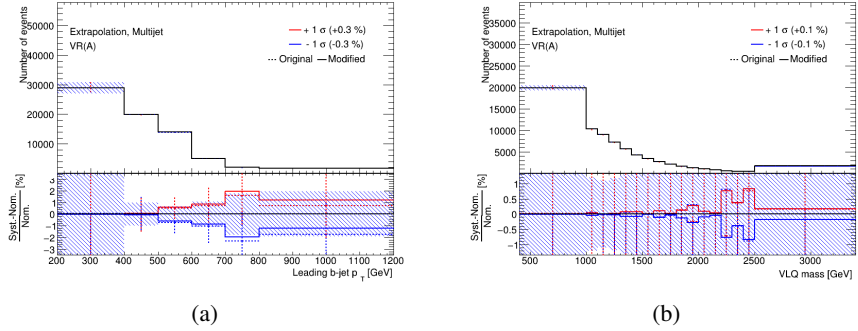

		\centering
		\graphicspath{{figs/chapter6/Systematics/Extrapolation/}}
		\begin{subfigure}{.35\textwidth}
			\centering
			\includegraphics[width=\linewidth,height=\textheight,keepaspectratio]{VR_B_jet_pt_Multijets.png}
			\caption{}
			\label{fig:uncertainties:systematics:btagging:extra:jetpt}
		\end{subfigure}\hspace{0.6cm}
		\begin{subfigure}{.35\textwidth}
			\centering
			\includegraphics[width=\linewidth,height=\textheight,keepaspectratio]{VR_B_VLQM_Multijets.png}
			\caption{}
			\label{fig:uncertainties:systematics:btagging:extra:vlqm}
		\end{subfigure}
		\caption{Comparison between the Systematics-Nominal and Nominal, where Systematics-Nominal denotes the estimated multijet when both up (in red) and down (in blue) variations in \textit{extrapolation} are taken into account, and Nominal denotes the final estimated multijet (in black). The difference between the two is considered as the systematic uncertainty from the extrapolation of \pt. It is shown for (a) leading $b$-tagged jet \pt and (b) VLQ mass distribution.}
		\label{fig:uncertainties:systematics:btagging:extra}
	\end{figure}

	\item \textit{extrapolation\_c:} it describes the extrapolation uncertainty for $c$-jets mistagged as $b$-jets. Fig.\ \ref{fig:uncertainties:systematics:btagging:extra:c} shows the contribution of the extrapolation uncertainty from the mistagging of $c$-jets in \pt distribution of the leading $b$-jet and VLQ mass. These plots show that \textit{extrapolation\_c} uncertainty does not contribute to the difference in normalisation of the distributions.
	\begin{figure}[hbt!]
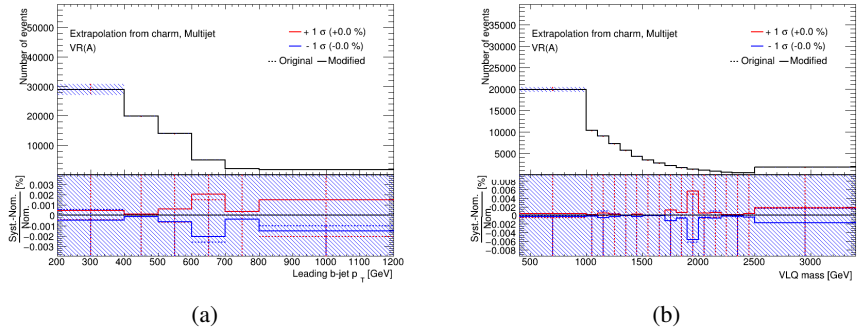

		\centering
		\graphicspath{{figs/chapter6/Systematics/Extrapolationfromcharm/}}
		\begin{subfigure}{.35\textwidth}
			\centering
			\includegraphics[width=\linewidth,height=\textheight,keepaspectratio]{VR_B_jet_pt_Multijets.png}
			\caption{}
			\label{fig:uncertainties:systematics:btagging:extra:c:jetpt}
		\end{subfigure}\hspace{0.6cm}
		\begin{subfigure}{.35\textwidth}
			\centering
			\includegraphics[width=\linewidth,height=\textheight,keepaspectratio]{VR_B_VLQM_Multijets.png}
			\caption{}
			\label{fig:uncertainties:systematics:btagging:extra:c:vlqm}
		\end{subfigure}
		\caption{Comparison between the Systematics-Nominal and Nominal, where Systematics-Nominal denotes the estimated multijet when both up (in red) and down (in blue) variations in \textit{extrapolation\_c} are taken into account, and Nominal denotes the final estimated multijet (in black). The difference between the two is considered as the systematic uncertainty from the extrapolation of \pt when $c$-jets are mistagged. It is shown for (a) leading $b$-tagged jet \pt and (b) VLQ mass distribution.}
		\label{fig:uncertainties:systematics:btagging:extra:c}
	\end{figure}
\end{itemize}

It was expected that the $b$-tagging uncertainties should be contributing the most because it is one of the uncorrelated variables which is used to define the ABCD regions. But however, the $b$-tagging uncertainties are only associated with the MC simulated backgrounds. The MC simulated backgrounds include multijet MC and the other MC backgrounds. The multijet MC is used for the calculation of \R, in which the likelihood fit is performed to fit the multijet MC to the data. The output of this fit, which is the scaled multijet MC, is used for the calculation of \R. In this, the likelihood fit constrains the $b$-tagging uncertainty in the scaled multijet MC. So, the contribution from the $b$-tagging uncertainty in the \R is suppressed, and the only contribution from the $b$-tagging uncertainty comes from the ABCD method, where the other MC backgrounds are used to subtract their contributions from the data according to Eqn.\ \ref{eqn:abcd:correctionfactor}. 

Also, $b$-tagging uncertainties contribute differently to all the six distributions because the likelihood fit and the calculation of the \R is performed separately for all the six distributions, but roughly it contributes to <1\% in the event yield of the multijet estimate in the validation region. Detailed values for all the distributions are given in Table \ref{table:uncertainties:systematics:detail}.

\begin{table}[hbt!]
	\centering
	\begin{tabular}{c|c|c|c|c|c|c} 
		\toprule
		Multijet & \multicolumn{3}{c}{$W$-tagged jet} \vline & \multicolumn{2}{c}{leading $b$-tagged jet} \vline & VLQ mass\\ \cline{2-6}
		Region A & \pt & mass & $\eta$ & \pt & mass & \\ 
		\midrule
		Est.\ Events & \num{72072} & \num{72752} & \num{72613} & \num{71670} & \num{73038} & \num{72573} \\
		Stat.\ uncertainty & $\pm$0.4\% & $\pm$0.5\% & $\pm$0.4\% & $\pm$2.6\% & $\pm$0.4\% & $\pm$0.8\% \\
		\midrule
		eigenvars\_b & $\pm$0.0\% & $\pm$0.0\% & $\pm$0.0\% & $\pm$0.0\% & $\pm$0.0\% & $\pm$0.0\% \\
		eigenvars\_c & $\pm$0.1\% & $\pm$0.0\% & $\pm$0.0\% & $\pm$0.1\% & $\pm$0.1\% & $\pm$0.0\% \\
		eigenvars\_light & $\pm$0.1\% & $\pm$0.1\% & $\pm$0.0\% & $\pm$0.2\% & $\pm$0.4\% & $\pm$0.1\% \\
		extrapolation & $\pm$0.1\% & $\pm$2.1\% & $\pm$0.1\% & $\pm$0.3\% & $\pm$1.2\% & $\pm$0.1\% \\
		extrapolation\_c & $\pm$0.0\% & $\pm$0.0\% & $\pm$0.0\% & $\pm$0.0\% & $\pm$0.0\% & $\pm$0.0\% \\
		\midrule
		Multijet MC & $\pm$3.8\% & $\pm$6.2\% & $\pm$4.4\% & $\pm$10.1\% & $\pm$4.6\% & $\pm$4.6\% \\
		Closure & $\pm$0.0\% & $\pm$0.0\% & $\pm$0.0\% & $\pm$0.0\% & $\pm$0.0\% & $\pm$0.0\% \\
		Cross-section & $\pm$0.2\% & $\pm$0.2\% & $\pm$0.2\% & $\pm$0.2\% & $\pm$0.2\% & $\pm$0.2\% \\
		\bottomrule
	\end{tabular}
	\caption{Overview of statistical and systematic uncertainties on the final estimated multijet when \R is calculated from the scaled multijet MC for all the six distributions.}
	\label{table:uncertainties:systematics:detail}
\end{table}

\subsection{Multijet MC uncertainty}
\label{sec:uncertainties:systematics:multijet}
\begin{figure}[hbt!]
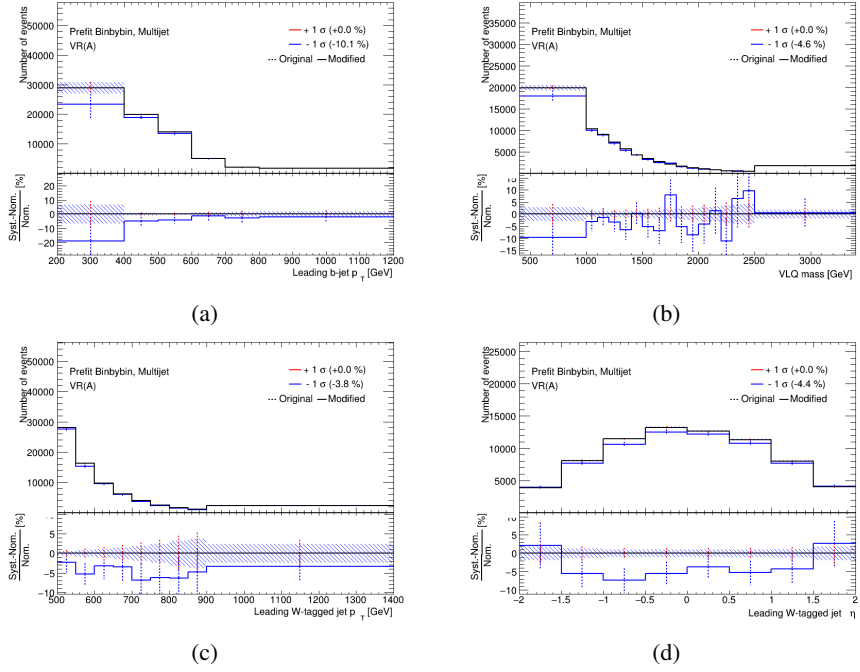

	\centering
	\graphicspath{{figs/chapter6/Systematics/PrefitBinbybin/}}
	\begin{subfigure}{.35\textwidth}
		\centering
		\includegraphics[width=\linewidth,height=\textheight,keepaspectratio]{VR_B_jet_pt_Multijets.png}
		\caption{}
		\label{fig:uncertainties:systematics:multijet:jetpt}
	\end{subfigure}\hspace{0.6cm}
	\begin{subfigure}{.35\textwidth}
		\centering
		\includegraphics[width=\linewidth,height=\textheight,keepaspectratio]{VR_B_VLQM_Multijets.png}
		\caption{}
		\label{fig:uncertainties:systematics:multijet:vlqm}
	\end{subfigure}
	\begin{subfigure}{.35\textwidth}
		\centering
		\includegraphics[width=\linewidth,height=\textheight,keepaspectratio]{VR_B_ljet_pt_Multijets.png}
		\caption{}
		\label{fig:uncertainties:systematics:multijet:ljetpt}
	\end{subfigure}\hspace{0.6cm}
	\begin{subfigure}{.35\textwidth}
		\centering
		\includegraphics[width=\linewidth,height=\textheight,keepaspectratio]{VR_B_ljet_eta_Multijets.png}
		\caption{}
		\label{fig:uncertainties:systematics:multijet:ljeteta}
	\end{subfigure}
	\caption{Comparison between the Systematics-Nominal and Nominal, where Systematics-Nominal (in blue) denotes the estimated multijet when \R is calculated from the multijet MC (which are mismodelled), and Nominal denotes the final estimated multijet (in black) when \R is calculated from the scaled multijet MC. The difference between the two is considered as the multijet MC uncertainty. It is shown for (a) leading $b$-tagged jet \pt, (b) VLQ mass, (c) \pt and (d) $\eta$ distribution of leading $W$-tagged jet.}
	\label{fig:uncertainties:systematics:multijet}
\end{figure}

As discussed before in section \ref{sec:abcd:correctionfactor}, the multijet estimate from the ABCD method where \R is calculated from the multijet MC samples (which are mismodelled) by using the bin-by-bin method can also be a result because it is also estimated from the same ABCD method. So, the difference between this estimate and the final estimate, which uses the scaled multijet MC for the calculation of \R is taken as multijet MC uncertainty. This uncertainty contributes a lot to the normalisation of the distribution. So, it is regarded as normalisation and shape uncertainty. The symmetry of multijet MC uncertainty is computed as one-sided since it only contributes towards the down variation and the up variation is added as the mirrored version of it. Fig.\ \ref{fig:uncertainties:systematics:multijet} shows the effect of this uncertainty in four different distributions. One can see that it adds differently to all the distributions in the validation region. Detailed values for all the distributions are given in Table \ref{table:uncertainties:systematics:detail}. The multijet MC uncertainty contributes most to the \pt distribution of the leading $b$-tagged jet where the first bin corresponds to almost 20\% uncertainty.

\subsection{Closure uncertainty}
\label{sec:uncertainties:systematics:closure}
In the closure uncertainty, the ABCD method is performed on the multijet MC instead of the data to estimate the multijet background in the validation region by using Eqn.\ \ref{eqn:uncertainties:systematics:closure}.
\begin{equation}
N_{\text{A}}^{\text{Est.\ multijet}}[i] = N_{\text{B}}^{\text{multijet MC}}[i] \times \frac{N_{\text{D}}^{\text{multijet MC}}[i]}{N_{\text{C}}^{\text{multijet MC}}[i]} \,,
\label{eqn:uncertainties:systematics:closure}
\end{equation}
where $N_{\text{j}}^{\text{Multijet MC}}[i]$ is event yield from the multijet MC in region j where j $\in$ \{B,C,D\}. Here [$i$] denotes that this calculation is performed bin-by-bin (where $i=$ bin).
 
Since multijet MC (which are mismodelled) are used for the multijet estimate, it is called MC-driven ABCD estimate instead of data-driven ABCD estimate. Here, data-driven ABCD estimate refers to the final multijet estimate when the \R is calculated from the scaled multijet MC by using the shape method. Note that there is no correction factor introduced in Eqn.\ \ref{eqn:uncertainties:systematics:closure} because previously in the data-driven ABCD estimate, the \R was calculated from the scaled multijet MC which was produced by performing a likelihood fit on the multijet MC. This cannot be performed in MC-driven ABCD estimate since the ABCD method itself is applied on the multijet MC. So, the normalisation of the MC-driven ABCD estimate would be less as compared to the normalisation of the data-driven ABCD estimate. Then, the normalisation of the MC-driven ABCD estimate is scaled to the normalisation of the data-driven ABCD estimate so that there would not be any difference in the normalisation of the two multijet estimates, but there would only be a shape variation. The difference in the shape of the distributions of the MC-driven ABCD estimate and the data-driven ABCD estimate is considered as the closure uncertainty. It is purely a shape uncertainty.

\begin{figure}[hbt!]
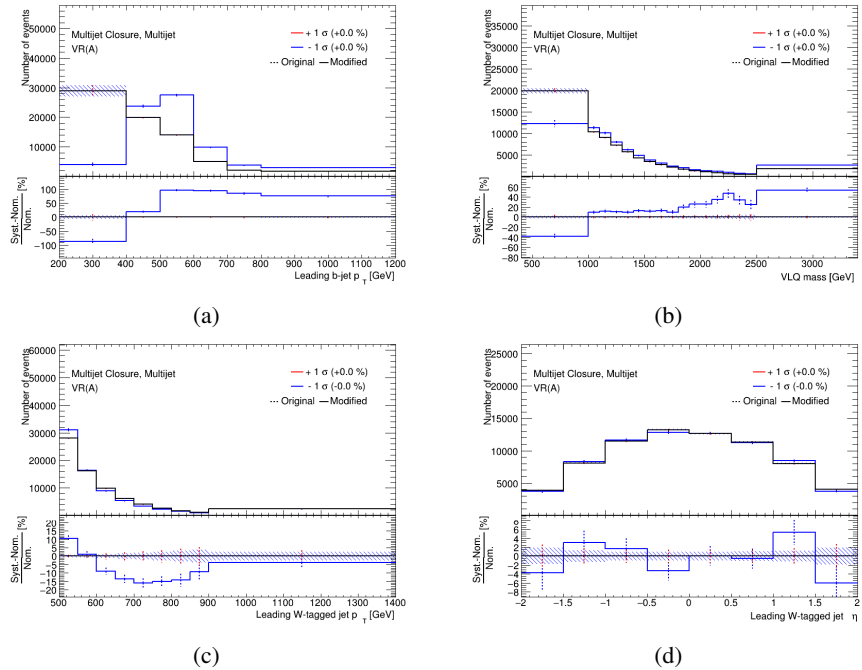

	\centering
	\graphicspath{{figs/chapter6/Systematics/DijetsClosure/}}
	\begin{subfigure}{.35\textwidth}
		\centering
		\includegraphics[width=\linewidth,height=\textheight,keepaspectratio]{VR_B_jet_pt_Multijets.png}
		\caption{}
		\label{fig:uncertainties:systematics:closure:jetpt}
	\end{subfigure}\hspace{0.6cm}
	\begin{subfigure}{.35\textwidth}
		\centering
		\includegraphics[width=\linewidth,height=\textheight,keepaspectratio]{VR_B_VLQM_Multijets.png}
		\caption{}
		\label{fig:uncertainties:systematics:closure:vlqm}
	\end{subfigure}
	\begin{subfigure}{.35\textwidth}
		\centering
		\includegraphics[width=\linewidth,height=\textheight,keepaspectratio]{VR_B_ljet_pt_Multijets.png}
		\caption{}
		\label{fig:uncertainties:systematics:closure:ljetpt}
	\end{subfigure}\hspace{0.6cm}
	\begin{subfigure}{.35\textwidth}
		\centering
		\includegraphics[width=\linewidth,height=\textheight,keepaspectratio]{VR_B_ljet_eta_Multijets.png}
		\caption{}
		\label{fig:uncertainties:systematics:closure:ljeteta}
	\end{subfigure}
	\caption{Comparison between the Systematics-Nominal and Nominal, where Systematics-Nominal (in blue) denotes the multijet from the MC-driven ABCD estimate, and Nominal denotes the final estimated multijet (in black) which is a data-driven ABCD estimate. The difference between the two is considered as the closure uncertainty. It is shown for (a) leading $b$-tagged jet \pt, (b) VLQ mass, (c) \pt and (d) $\eta$ distribution of leading $W$-tagged jet.}
	\label{fig:uncertainties:systematics:closure}
\end{figure}

Fig.\ \ref{fig:uncertainties:systematics:closure} shows the effect of the closure uncertainty in four different distributions. The symmetry of the closure uncertainty is computed as one-sided since in some bins it contributes towards up variation, and in some bins, it contributes towards down variation as it can be noticed from the plots. One can observe a considerable shape variation in the \pt distribution of the leading $b$-jet, especially in the first bin where the difference is almost 100\%. This also shows how the \R corrects these bins where the difference is so significant.

\subsection{Cross-section uncertainty}
\label{sec:uncertainties:systematics:crossection}
The uncertainty in the values of a cross-section of different background processes is discussed in this section. The scale uncertainties on all background processes are given in Table \ref{table:uncertainties:systematics:crossection} by Physics Modelling Group (PMG)~\cite{crossection:top}~\cite{crossection:w}. 

\begin{table}[hbt!]
	\centering
	\begin{tabular}{c|c} 
		\toprule
		Background & Scale uncertainty \\ 
		\midrule
		Single top & $\pm$2.5\% \\
		$t\bar{t}$ & +2.4\% -3.5\% \\
		$W$+jets & $\pm$6\% \\
		$Z$+jets & $\pm$6\% \\
		\bottomrule
	\end{tabular}
	\caption{Scale uncertainty values for all the background processes.}
	\label{table:uncertainties:systematics:crossection}
\end{table}

\begin{figure}[hbt!]
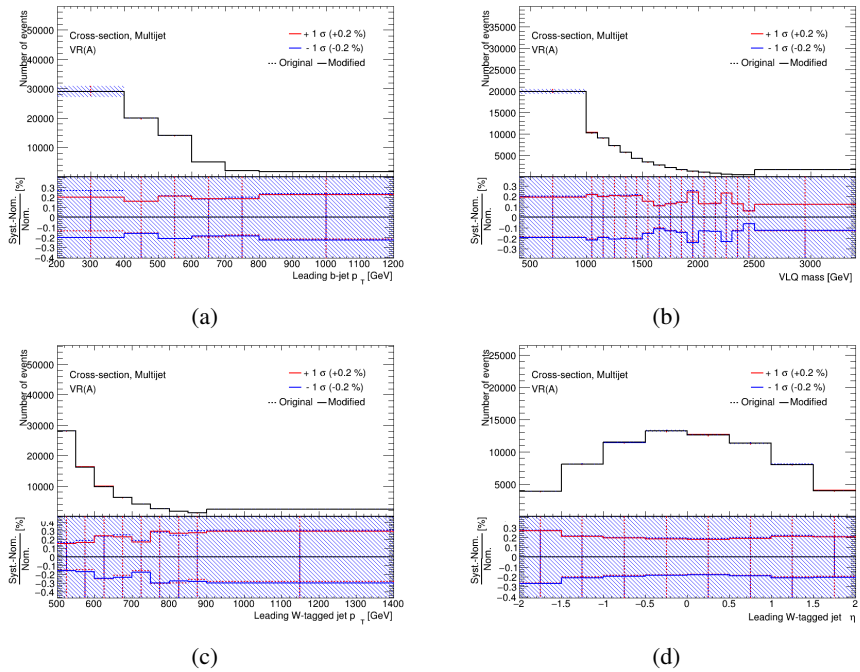

	\centering
	\graphicspath{{figs/chapter6/Systematics/Crossection/}}
	\begin{subfigure}{.35\textwidth}
		\centering
		\includegraphics[width=\linewidth,height=\textheight,keepaspectratio]{VR_B_jet_pt_Multijets.png}
		\caption{}
		\label{fig:uncertainties:systematics:crossection:jetpt}
	\end{subfigure}\hspace{0.6cm}
	\begin{subfigure}{.35\textwidth}
		\centering
		\includegraphics[width=\linewidth,height=\textheight,keepaspectratio]{VR_B_VLQM_Multijets.png}
		\caption{}
		\label{fig:uncertainties:systematics:crossection:vlqm}
	\end{subfigure}
	\begin{subfigure}{.35\textwidth}
		\centering
		\includegraphics[width=\linewidth,height=\textheight,keepaspectratio]{VR_B_ljet_pt_Multijets.png}
		\caption{}
		\label{fig:uncertainties:systematics:crossection:ljetpt}
	\end{subfigure}\hspace{0.6cm}
	\begin{subfigure}{.35\textwidth}
		\centering
		\includegraphics[width=\linewidth,height=\textheight,keepaspectratio]{VR_B_ljet_eta_Multijets.png}
		\caption{}
		\label{fig:uncertainties:systematics:crossection:ljeteta}
	\end{subfigure}
	\caption{Comparison between the Systematics-Nominal and Nominal, where Systematics-Nominal denotes the estimated multijet when both up (in red) and down (in blue) variations in cross-section uncertainties of all the other backgrounds are taken into account, and Nominal denotes the final estimated multijet (in black). The difference between the two is considered as the multijet MC  uncertainty. It is shown for (a) leading $b$-tagged jet \pt, (b) VLQ mass, (c) \pt and (d) $\eta$ distribution of leading $W$-tagged jet.}
	\label{fig:uncertainties:systematics:crossection}
\end{figure}

Both up and down variations are taken into account by adding or subtracting the scale uncertainty on the cross-section of these backgrounds. Then, the ABCD method is performed in both the cases and the absolute mean of it is considered as the cross-section uncertainty. The scale uncertainty on the multijet MC is not taken into account since the ABCD estimate is only affected by the scale uncertainty on the cross-section of other backgrounds. Fig.\ \ref{fig:uncertainties:systematics:crossection} shows the impact of this uncertainty in four different distribution. It contributes to 0.2\% of an event yield of the multijet estimate in the validation region. Detailed values for all the distributions are given in Table \ref{table:uncertainties:systematics:detail}.

\chapter{Results}
\label{sec:results}
The aim of this chapter is to present the final results of the multijet estimate using the ABCD method, including all the uncertainties discussed in the previous chapter. A performance comparison with the use of different taggers on the multijet estimation is shown in section \ref{sec:results:taggers}. A similar kind of performance comparison is also studied by using different jet collections, which is described in section \ref{sec:results:jetcollections}.
 
\section{Estimated multijet including uncertainties}
\label{sec:results:estimation}
\begin{figure}[hbt!]
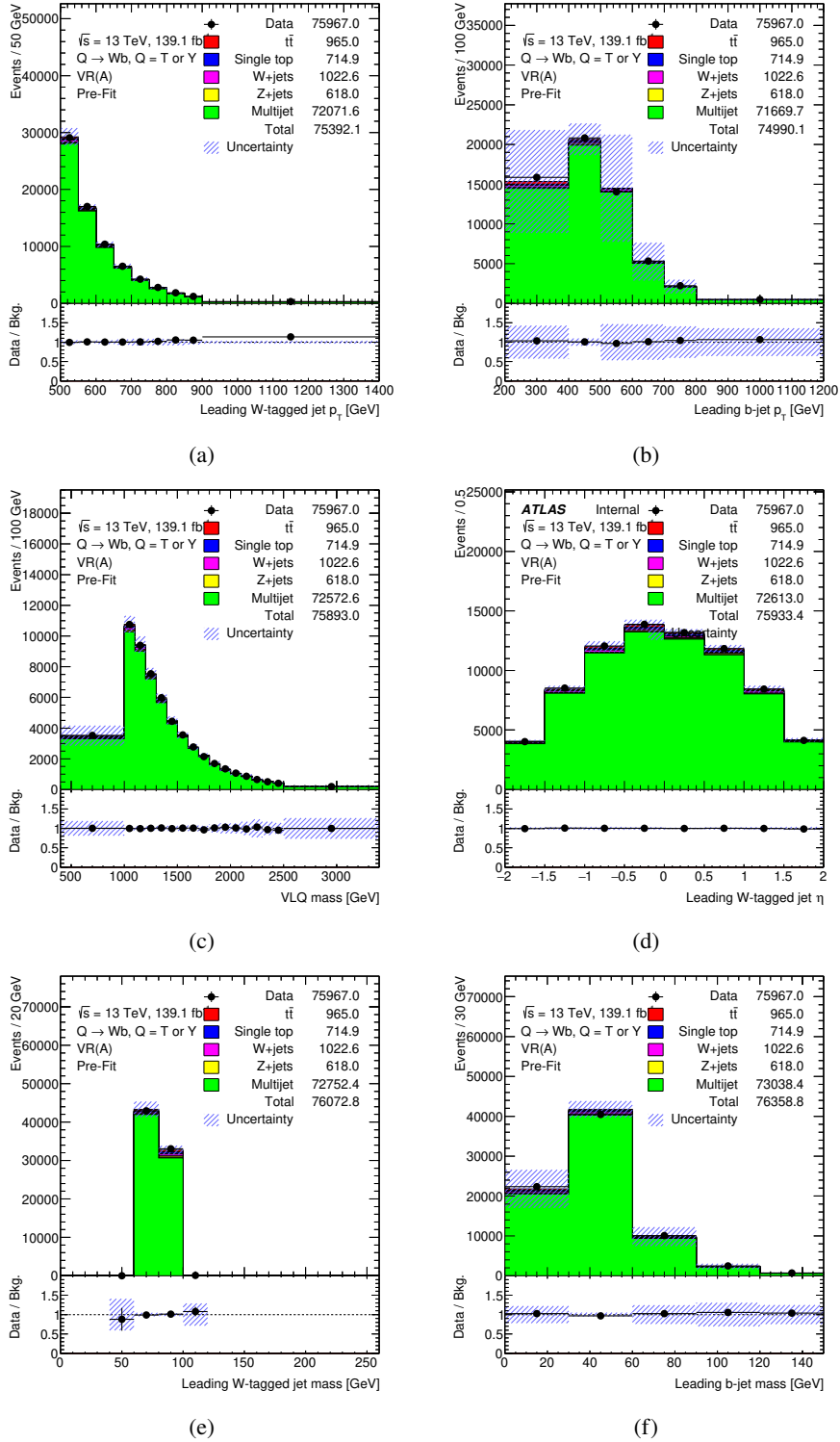

	\centering
	\graphicspath{{figs/chapter6/final/}}
	\begin{subfigure}{.35\textwidth}
		\centering
		\includegraphics[width=\linewidth,height=\textheight,keepaspectratio]{VR_B_ljet_pt.pdf}
		\caption{}
		\label{fig:results:estimation:ljet_pt}
	\end{subfigure}\hspace{0.6cm}
	\begin{subfigure}{.35\textwidth}
		\centering
		\includegraphics[width=\linewidth,height=\textheight,keepaspectratio]{VR_B_jet_pt.pdf}
		\caption{}
		\label{fig:results:estimation:jet_pt}
	\end{subfigure}
	\begin{subfigure}{.35\textwidth}
		\centering
		\includegraphics[width=\linewidth,height=\textheight,keepaspectratio]{VR_B_VLQM.pdf}
		\caption{}
		\label{fig:results:estimation:VLQM}
	\end{subfigure}\hspace{0.6cm}
	\begin{subfigure}{.35\textwidth}
		\centering
		\includegraphics[width=\linewidth,height=\textheight,keepaspectratio]{VR_B_ljet_eta.pdf}
		\caption{}
		\label{fig:results:estimation:ljet_eta}
	\end{subfigure}
	\begin{subfigure}{.35\textwidth}
		\centering
		\includegraphics[width=\linewidth,height=\textheight,keepaspectratio]{VR_B_ljet_m.pdf}
		\caption{}
		\label{fig:results:estimation:ljet_m}
	\end{subfigure}\hspace{0.6cm}
	\begin{subfigure}{.35\textwidth}
		\centering
		\includegraphics[width=\linewidth,height=\textheight,keepaspectratio]{VR_B_jet_m.pdf}
		\caption{}
		\label{fig:results:estimation:jet_m}
	\end{subfigure}
	\caption{Estimated multijet background from the ABCD method including all the uncertainties in the validation region. The distributions include (a) $p_{\text{T}}$ of $W$-tagged large-$R$ jet, (b) $p_{\text{T}}$ of leading $b$-tagged small-$R$ jet, (c) VLQ mass reconstructed from the kinematics of $W$-tagged large-$R$ jet and leading $b$-tagged small-$R$ jet, (d) $\eta$ distribution of $W$-tagged large-$R$ jet, (e) mass of $W$-tagged large-$R$ jet, and (f) mass of leading $b$-tagged small-$R$ jet.}
	\label{fig:results:estimation}
\end{figure}

The multijet estimate in the validation region using the ABCD method, including all the discussed uncertainties is shown in Fig.\ \ref{fig:results:estimation}. It shows the estimated multijet in all the six kinematic distributions where each distribution is having different event yields and different uncertainties. The difference between the estimated multijet (and other backgrounds) and the data is within the described uncertainty. In Fig.\ \ref{fig:results:estimation:jet_pt}, the significant amount of uncertainty can be seen in \pt distribution of the leading $b$-jet because of the shape uncertainty, which is primarily from the closure uncertainty as discussed in section \ref{sec:uncertainties:systematics:closure}. A similar kind of behaviour can also be seen in the mass distributions. Table \ref{table:results:estimation} shows the event yield of the estimated multijet along with their uncertainties for all the distributions. It can be observed that the multijet (from the ABCD method) plus the contribution from other backgrounds (from the MC simulation) agree well with the data within the described uncertainties.

\begin{table}[hbt!]
	\centering
	\begin{tabular}{c|c|c} 
		\toprule
		Distribution & Event yield & Plots\\ 
		\midrule
		$W$-tagged jet \pt & $\num{72071} \pm \num{288} \text{ (stat.)} \pm \num{2745} \text{ (sys.)}$ & Fig.\ \ref{fig:results:estimation:ljet_pt} \\
		$W$-tagged jet mass & $\num{72752} \pm \num{363} \text{ (stat.)} \pm \num{4765} \text{ (sys.)}$ & Fig.\ \ref{fig:results:estimation:ljet_m} \\
		$W$-tagged jet $\eta$ & $\num{72613} \pm \num{290} \text{ (stat.)} \pm \num{3199} \text{ (sys.)}$ & Fig.\ \ref{fig:results:estimation:ljet_eta} \\
		leading $b$-tagged jet \pt & $\num{71669} \pm \num{1863} \text{ (stat.)} \pm \num{7245} \text{ (sys.)}$ & Fig.\ \ref{fig:results:estimation:jet_pt} \\
		leading $b$-tagged jet mass & $\num{73038} \pm \num{292} \text{ (stat.)} \pm \num{3488} \text{ (sys.)}$ & Fig.\ \ref{fig:results:estimation:jet_m} \\
		VLQ mass & $\num{72572} \pm \num{580} \text{ (stat.)} \pm \num{3343} \text{ (sys.)}$ & Fig.\ \ref{fig:results:estimation:VLQM} \\
		\midrule
		Other bkg.\ & $\num{3320} \pm \num{95} \text{ (stat.)}$ & \\ 
		\midrule
		\textbf{Data} & \textbf{\num{75967}} & \\
		\bottomrule
	\end{tabular}
	\caption{Result showing the estimated multijet with all the uncertainties in the validation region for all the kinematic distributions.}
	\label{table:results:estimation}
\end{table}

\section{Performance comparison of two different $W$-taggers}
\label{sec:results:taggers}

A performance comparison is evaluated between the choice of the $W$-tagger. As described in section \ref{sec:jetsandtaggers:taggers:twovariable}, the large-$R$ jets also are tagged with two-variable tagger. Then, the entire ABCD method is performed again by using these $W$-tagged jets instead of previously using the three-variable tagger for $W$-tagged jets. While performing the ABCD method, all the other variables and parameters are kept similar to the one in the case of the three-variable tagger. \R is calculated from scaled multijet MC (as described in section \ref{sec:abcd:furtherimprovement:rcorr}) by using the shape method for both the cases. All the uncertainties are calculated in exactly a similar way, how it is calculated in the case of the three-variable tagger. 

\begin{figure}[hbt!]
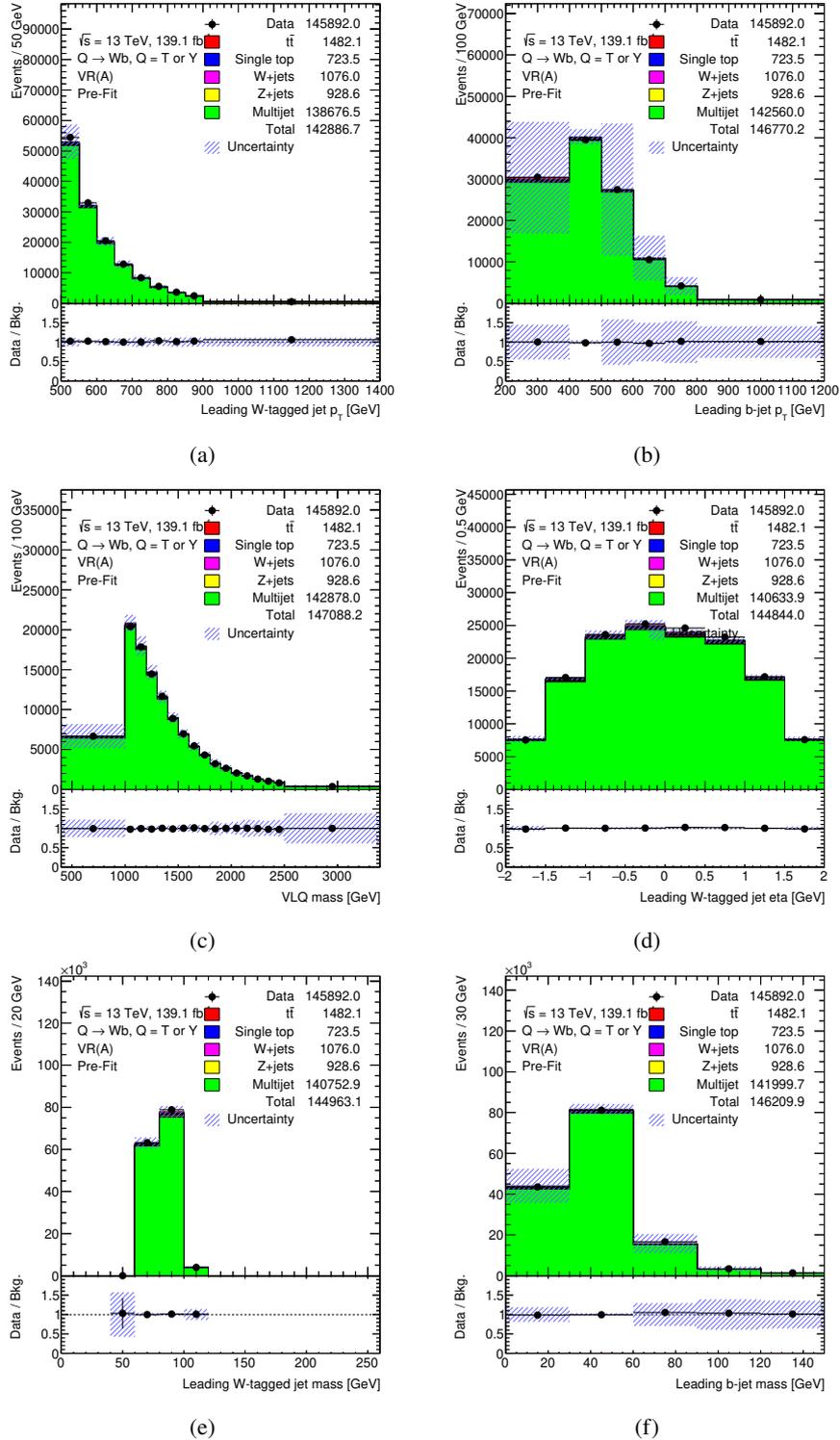

	\centering
	\graphicspath{{figs/chapter6/twovariable/}}
	\begin{subfigure}{.35\textwidth}
		\centering
		\includegraphics[width=\linewidth,height=\textheight,keepaspectratio]{VR_B_ljet_pt.pdf}
		\caption{}
		\label{fig:results:taggers:ljet_pt}
	\end{subfigure}\hspace{0.6cm}
	\begin{subfigure}{.35\textwidth}
		\centering
		\includegraphics[width=\linewidth,height=\textheight,keepaspectratio]{VR_B_jet_pt.pdf}
		\caption{}
		\label{fig:results:taggers:jet_pt}
	\end{subfigure}
	\begin{subfigure}{.35\textwidth}
		\centering
		\includegraphics[width=\linewidth,height=\textheight,keepaspectratio]{VR_B_VLQM.pdf}
		\caption{}
		\label{fig:results:taggers:VLQM}
	\end{subfigure}\hspace{0.6cm}
	\begin{subfigure}{.35\textwidth}
		\centering
		\includegraphics[width=\linewidth,height=\textheight,keepaspectratio]{VR_B_ljet_eta.pdf}
		\caption{}
		\label{fig:results:taggers:ljet_eta}
	\end{subfigure}
	\begin{subfigure}{.35\textwidth}
		\centering
		\includegraphics[width=\linewidth,height=\textheight,keepaspectratio]{VR_B_ljet_m.pdf}
		\caption{}
		\label{fig:results:taggers:ljet_m}
	\end{subfigure}\hspace{0.6cm}
	\begin{subfigure}{.35\textwidth}
		\centering
		\includegraphics[width=\linewidth,height=\textheight,keepaspectratio]{VR_B_jet_m.pdf}
		\caption{}
		\label{fig:results:taggers:jet_m}
	\end{subfigure}
	\caption{Results when $W$-tagging is performed by using the two-variable tagger in the estimate of multijet background from the ABCD method including all the uncertainties in the validation region. The distributions include (a) $p_{\text{T}}$ of $W$-tagged large-$R$ jet, (b) $p_{\text{T}}$ of leading $b$-tagged small-$R$ jet, (c) VLQ mass reconstructed from the kinematics of $W$-tagged large-$R$ jet and leading $b$-tagged small-$R$ jet, (d) $\eta$ distribution of $W$-tagged large-$R$ jet, (e) mass of $W$-tagged large-$R$ jet, and (f) mass of leading $b$-tagged small-$R$ jet.}
	\label{fig:results:taggers}
\end{figure}

Fig.\ \ref{fig:results:taggers} shows all the six the kinematic distributions in the validation region when $W$-tagging is performed by the two-variable tagger. A data/bkg.\ comparison is shown in these distributions where multijet is estimated by the ABCD method. The uncertainties shown in the plots include both statistical and systematics uncertainties which are calculated in a similar way for the two-variable tagger case. One can see more number of events (almost double) in data and estimated multijet event yield, which means that two-variable tagger is tagging the leading large-$R$ jet in more events than the three-variable tagger at the same working point. Also, a similar trend in the uncertainties can be observed in the distributions which has been observed with the three-variable tagger.

Table \ref{table:results:taggers1} shows the event yield of the estimated multijet along with their uncertainties for the $\eta$ distribution of $W$-tagged jet. It is shown for both the multijet estimate when the $W$-tagging is performed by two-variable tagger and three-variable tagger. It can be observed that the multijet (from the ABCD method) plus the contribution from other backgrounds (from the MC simulation) agree well with the data within the described uncertainties for the results from two-variable tagger as well.

\begin{table}[hbt!]
	\centering
	\begin{tabular}{c | c | c } 
		\toprule
		 & Two-variable tagger & Three-variable tagger \\
		\midrule
		Est.\ multijet & $\num{140634} \pm \num{506} \text{ (stat.)} \pm \num{5288} \text{ (sys.)}$ & $\num{72613} \pm \num{290} \text{ (stat.)} \pm \num{3199} \text{ (sys.)}$ \\
		Other bkg.\ & $\num{4210} \pm \num{101} \text{ (stat.)}$ & $\num{3320} \pm \num{95} \text{ (stat.)}$ \\
		\midrule
		\textbf{Data} & \textbf{\num{145892}} & \textbf{\num{75967}} \\
		\bottomrule
	\end{tabular}
	\caption{Result showing the estimated multijet with all the uncertainties in the validation region for the $\eta$ distribution of $W$-tagged jet when two-variable and three-variable tagger are used for $W$-tagging.}
	\label{table:results:taggers1}
\end{table}

Table \ref{table:results:taggers} shows the event yield of the data in different ABCD regions when $W$-tagging is performed both by two-variable tagger and three-variable tagger. Since the ABCD method is a data-driven method and one of the assumptions of this method as described in section \ref{sec:abcd:method}, is that there should be enough number of data events in the control region to extrapolate the background behaviour precisely in the distributions of the validation region. That means, more statistics in the data in region B, C and D can extrapolate the multijet background precisely. In the table, it can be seen clearly that when the $W$-tagging is performed by the three-variable tagger, there are more data events in the control regions (except region D). That is why three-variable tagger is used for $W$-tagging in the final estimate of the multijet background. Moreover, the three-variable tagger is developed to be sensitive to the QCD rejection, which has higher QCD rejection than the two-variable tagger.

\begin{table}[hbt!]
	\centering
	\begin{tabular}{c | c | c | c | c | c | c} 
		\toprule
		Event yield in data & SR & VR & \multicolumn{4}{c}{CR} \\ \cline{2-7}
		& Region A1 & Region A & Region B & Region C & Region D & Region D1 \\
		\midrule
		Two-variable tagger & - & \num{145892} & \num{267559} & \num{478436} & \num{196851} & \num{11625} \\
		Three-variable tagger & - & \num{75967} & \num{365089} & \num{605052} & \num{95173} & \num{18719} \\
		\bottomrule
	\end{tabular}
	\caption{Result showing the event yields of the data in all the ABCD regions when $W$-tagging is performed by both two-variable and three-variable tagger.}
	\label{table:results:taggers}
\end{table}

\section{Performance comparison of two different jet collections}
\label{sec:results:jetcollections}
A performance comparison is also evaluated between the choice of the jet collection for small-$R$ jets. As described in section \ref{sec:jetsandtaggers:jets:topo}, the topo-cluster jet collection is also used for small-$R$ jets which are calibrated to electromagnetic scale and the $b$-tagging is performed on these small-$R$ jets. Another difference is that there are different versions of \textsc{AnalysisTop}~\cite{analysistop} used for EMTopo and particle flow jets, e.g.\ \textsc{AnalysisTop 21.2.84}~\cite{analysistop} is used for EMTopo jet collection and \textsc{AnalysisTop 21.2.92}~\cite{analysistop} is used for particle flow jet collections. 

The entire ABCD method is performed again by using these $b$-tagged jets (where the small-$R$ jets are from EMTopo jet collections) instead of previously using the $b$-tagged jets from particle flow jet collection. While performing the ABCD method, all the other variables and parameters are kept similar to the one in the particle flow case. \R is calculated from scaled multijet MC (as described in section \ref{sec:abcd:furtherimprovement:rcorr}) by using the shape method for both the cases. All the uncertainties are calculated in exactly a similar way, how it is calculated in the case of the particle flow jet collection for small-$R$ jets.  

\begin{figure}[hbt!]
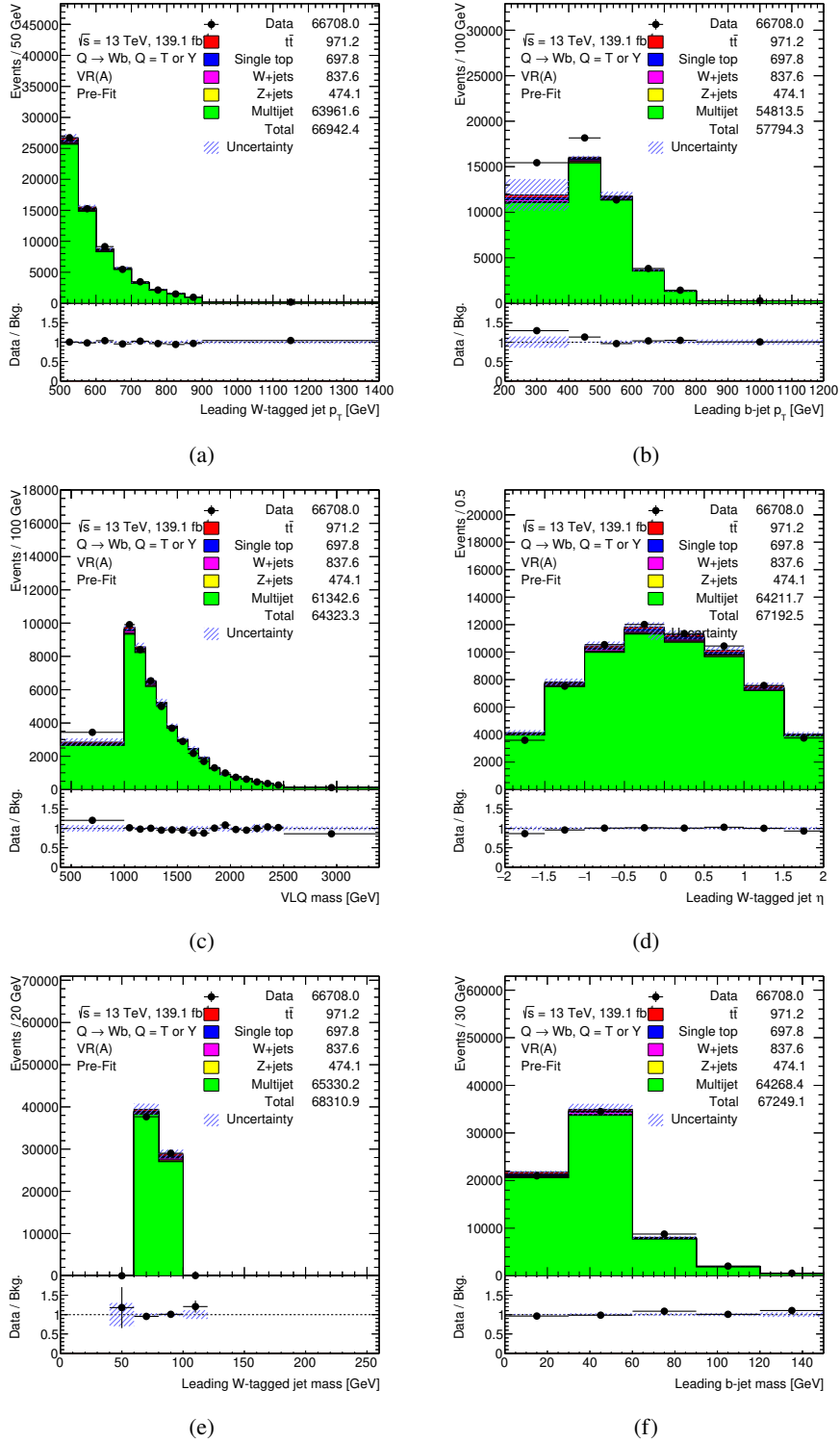

	\centering
	\graphicspath{{figs/chapter6/topo/}}
	\begin{subfigure}{.35\textwidth}
		\centering
		\includegraphics[width=\linewidth,height=\textheight,keepaspectratio]{VR_B_ljet_pt.pdf}
		\caption{}
		\label{fig:results:jetcollections:ljet_pt}
	\end{subfigure}\hspace{0.6cm}
	\begin{subfigure}{.35\textwidth}
		\centering
		\includegraphics[width=\linewidth,height=\textheight,keepaspectratio]{VR_B_jet_pt.pdf}
		\caption{}
		\label{fig:results:jetcollections:jet_pt}
	\end{subfigure}
	\begin{subfigure}{.35\textwidth}
		\centering
		\includegraphics[width=\linewidth,height=\textheight,keepaspectratio]{VR_B_VLQM.pdf}
		\caption{}
		\label{fig:results:jetcollections:VLQM}
	\end{subfigure}\hspace{0.6cm}
	\begin{subfigure}{.35\textwidth}
		\centering
		\includegraphics[width=\linewidth,height=\textheight,keepaspectratio]{VR_B_ljet_eta.pdf}
		\caption{}
		\label{fig:results:jetcollections:ljet_eta}
	\end{subfigure}
	\begin{subfigure}{.35\textwidth}
		\centering
		\includegraphics[width=\linewidth,height=\textheight,keepaspectratio]{VR_B_ljet_m.pdf}
		\caption{}
		\label{fig:results:jetcollections:ljet_m}
	\end{subfigure}\hspace{0.6cm}
	\begin{subfigure}{.35\textwidth}
		\centering
		\includegraphics[width=\linewidth,height=\textheight,keepaspectratio]{VR_B_jet_m.pdf}
		\caption{}
		\label{fig:results:jetcollections:jet_m}
	\end{subfigure}
	\caption{Results when EMTopo jet collection is used for for small-$R$ jets in the estimate of multijet background from the ABCD method including all the uncertainties in the validation region. The distributions include (a) $p_{\text{T}}$ of $W$-tagged large-$R$ jet, (b) $p_{\text{T}}$ of leading $b$-tagged small-$R$ jet, (c) VLQ mass reconstructed from the kinematics of $W$-tagged large-$R$ jet and leading $b$-tagged small-$R$ jet, (d) $\eta$ distribution of $W$-tagged large-$R$ jet, (e) mass of $W$-tagged large-$R$ jet, and (f) mass of leading $b$-tagged small-$R$ jet.}
	\label{fig:results:jetcollections}
\end{figure}

Fig.\ \ref{fig:results:jetcollections} shows the kinematic distributions in the validation region when the EMTopo jet collection is used for small-$R$ jets. A data/bkg.\ comparison is shown in these distributions where multijet is estimated by the ABCD method. One can see less number of events in data and estimated multijet event yield. The uncertainties shown in the plots include both statistical and systematic uncertainties which are already discussed for the particle flow case.

Table \ref{table:results:jetcollections1} shows the event yield of the estimated multijet along with their uncertainties for the $\eta$ distribution of $W$-tagged jet. It is shown for both the multijet estimate when the small-$R$ jets are from EMTopo jets collection and particle flow jet collection. Here, $W$-tagging is performed by three-variable tagger in both the cases. It can be observed that the multijet (from the ABCD method) plus the contribution from other backgrounds (from the MC simulation) agree well with the data within the described uncertainties for the results from EMTopo jet collection as well. But the particle flow jet collection gives better statistics and shows better agreement with the data. 

\begin{table}[hbt!]
	\centering
	\begin{tabular}{c | c | c } 
		\toprule
		& EMTopo jets & Particle flow jets \\
		\midrule
		Est.\ multijet & $\num{64212} \pm \num{229} \text{ (stat.)} \pm \num{3981} \text{ (sys.)}$ & $\num{72613} \pm \num{290} \text{ (stat.)} \pm \num{3199} \text{ (sys.)}$ \\
		Other bkg.\ & $\num{2981} \pm \num{89} \text{ (stat.)}$ & $\num{3320} \pm \num{95} \text{ (stat.)}$ \\
		\midrule
		\textbf{Data} & \textbf{\num{66708}} & \textbf{\num{75967}} \\
		\bottomrule
	\end{tabular}
	\caption{Result showing the estimated multijet with all the uncertainties in the validation region for the $\eta$ distribution of $W$-tagged jet when EMTopo and particle flow jet collections are used for small-$R$ jets.}
	\label{table:results:jetcollections1}
\end{table}

Table \ref{table:results:jetcollections} shows the event yield of the data in different ABCD regions when both EMTopo and particle flow jet collections are used one-by-one for small-$R$ jets.
From the event yield in the data, one can see that there is more number of data events in the control regions when EMTopo jet collection is used for small-$R$ jets than the particle flow jet collection. 

\begin{table}[hbt!]
	\centering
	\begin{tabular}{c | c | c | c | c | c | c} 
		\toprule
		Event yield in data & SR & VR & \multicolumn{4}{c}{CR} \\ \cline{2-7}
		& Region A1 & Region A & Region B & Region C & Region D & Region D1 \\
		\midrule
		EMTopo jets & - & \num{66708} & \num{1590130} & \num{4313450} & \num{100573} & \num{19911} \\
		Particle flow jets & - & \num{75967} & \num{365089} & \num{605052} & \num{95173} & \num{18719} \\
		\bottomrule
	\end{tabular}
	\caption{Result showing the event yields of the data in all the ABCD regions when EMTopo and particle flow jet collections are used for small-$R$ jets.}
	\label{table:results:jetcollections}
\end{table}

\chapter{Conclusion}
\label{sec:conclusion}
In this thesis, the estimation of multijet background is presented in the all-hadronic decay channel of $T/Y\rightarrow Wb$ analysis. It is performed by studying the data of $pp$ collisions at $\sqrt{s}=\SI{13}{\tera\electronvolt}$ which is taken by using the ATLAS detector in 2015-18, corresponding to an integrated luminosity of $\SI{139}{\femto\barn^{-1}}$. 

Vector-like quarks are coloured spin $\frac{1}{2}$ hypothetical particles, which are predicted by various BSM theories to solve the phenomena which could not be explained by the SM. One of the phenomena is the Higgs mass hierarchy problem, which can be solved by introducing VLQs. They can be produced either singly or in pair, but the single VLQ production is a dominant production mode at high VLQ masses, which brings us to study the single production of two types of VLQs, $T$ and $Y$ quark. Since VLQs are heavy particles, $T/Y$ quark decays into $W$ boson and $b$-quark, where $W$ boson can decay either leptonically or hadronically. In this thesis, the hadronic decay channel has been studied. 

Since an all-hadronic channel is being probed, the contribution from multijet is a dominant background because of the QCD processes. At the preselection (in Fig.\ \ref{fig:analysisstrategy:eventselection:preselection}), a significant amount of mismodelling in the multijet MC was observed, which cannot be fixed by scaling the multijet MC. Therefore, a data-driven method has to be performed to get a better estimate of multijet background. For this, one of the data-driven methods called the ABCD method has been used.

The ABCD method should be performed in the signal region, but since we are blinded to the data in the signal region, the method is performed in the validation region to validate the method. It was observed from the multijet estimate from the ABCD method in the validation region that there was a need for applying correction to the multijet estimate because one of the main assumptions of the ABCD method was not fulfiled by choice of the uncorrelated variables. So, a correction factor has been introduced, which was calculated from the multijet MC. It was calculated by using two different methods called normalisation and shape method. After applying the correction factor, the estimated multijet from the ABCD method shows a better agreement with the data.

The second part of this thesis is to improve the multijet estimate by improvising the ABCD method because the correction factor was calculated from the multijet MC, which are mismodelled. In order to fix this, a likelihood fit is performed to fit multijet MC to the data while keeping the other backgrounds constant. Then this scaled multijet MC has been used to calculate the correction factor by using both the methods. So, in total, four different ways of calculation of the correction factor have been shown in this thesis. An improvement has been seen in the estimated multijet after applying the correction factor, which is calculated from the scaled multijet MC by using shape method. So, this multijet estimate is regarded as a final estimate.

Furthermore, statistical uncertainties have been studied, which come from the ABCD method itself as well as the calculation of the correction factor. Along with that, systematic uncertainties have been implemented, which include the uncertainties from $b$-tagging. Some systematic uncertainties have also been assigned such as uncertainty on the cross-section of the other backgrounds, closure uncertainty, etc.

The final event yield of the estimated multijet from the ABCD method including all the uncertainties in the validation region comes out to be: 

\begin{table}[hbt!]
	\centering
	\begin{tabular}{c|c} 
		\toprule
		\textbf{Multijet ABCD estimate} & \textbf{$\num{72600} \pm \num{300} \text{ (stat.)} \pm \num{3200} \text{ (sys.)}$} \\
		Other backgrounds & $\num{3320} \pm \num{95} \text{ (stat.)}$ \\
		\midrule
		Data & $\num{75967}$ \\
		\bottomrule
	\end{tabular}
	\label{table:conclusion}
\end{table}

The estimated multijet is consistent with the data within the uncertainties. The event yield stated here is from the $\eta$ distribution of $W$-tagged jet since it is a more consistent distribution, but the difference in the event yields of the estimated multijet across the distributions is just less than 2\%.

So, the results observed in the validation region shows that the ABCD method works well, and the method has been validated in the validation region. Now it can be applied in the signal region to estimate the multijet background. And then the estimate in the signal region can further be used for setting mass-coupling limits.



\ifthenelse{\texlive > 2010} {%
	\printbibliography[heading=bibintoc]
}{%
	{\raggedright
		\bibliographystyle{refs/atlasBibStyleWithTitle.bst}
		\bibliography{mythesis.bib} 
	}
}
\appendix
\chapter{Control region distributions}
\label{sec:app}
In this appendix, the distributions for all the kinematic and reconstructed variables from all the four control regions B, C, D and D1 are provided. 

In the control region, the other backgrounds are from the MC simulation and the multijet background is calculated from the equation below:

\begin{equation}
	N_{\text{j}}^{\text{Multijet}}[i] = N_{\text{j}}^{\text{Data}}[i] - N_{\text{j}}^{\text{Other bkg.}}[i] \,,
	\label{eqn:app}
\end{equation}
where $\text{j}\in\{\text{B, C, D, D1}\}$ and $i=$ bin.

\begin{figure}[hbt!]
	\centering
	\graphicspath{{figs/appendix/CRB/}}
	\begin{subfigure}{.35\textwidth}
		\centering
		\includegraphics[width=\linewidth,height=\textheight,keepaspectratio]{CR_B_ljet_pt.pdf}
		\caption{}
		\label{fig:app:cr_b:ljet_pt}
	\end{subfigure}\hspace{0.6cm}
	\begin{subfigure}{.35\textwidth}
		\centering
		\includegraphics[width=\linewidth,height=\textheight,keepaspectratio]{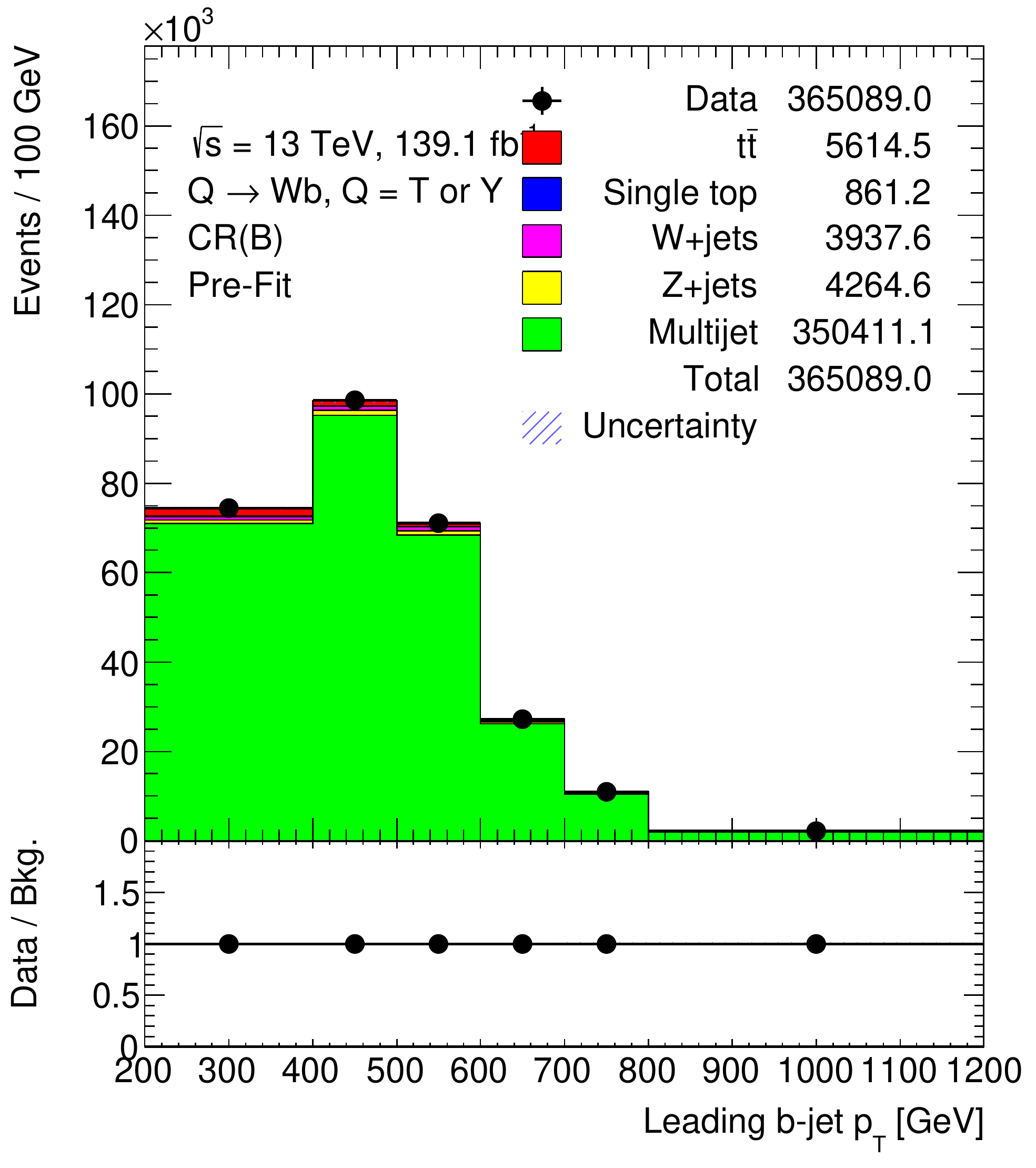}
		\caption{}
		\label{fig:app:cr_b:jet_pt}
	\end{subfigure}
	\begin{subfigure}{.35\textwidth}
		\centering
		\includegraphics[width=\linewidth,height=\textheight,keepaspectratio]{CR_B_VLQM.pdf}
		\caption{}
		\label{fig:app:cr_b:VLQM}
	\end{subfigure}\hspace{0.6cm}
	\begin{subfigure}{.35\textwidth}
		\centering
		\includegraphics[width=\linewidth,height=\textheight,keepaspectratio]{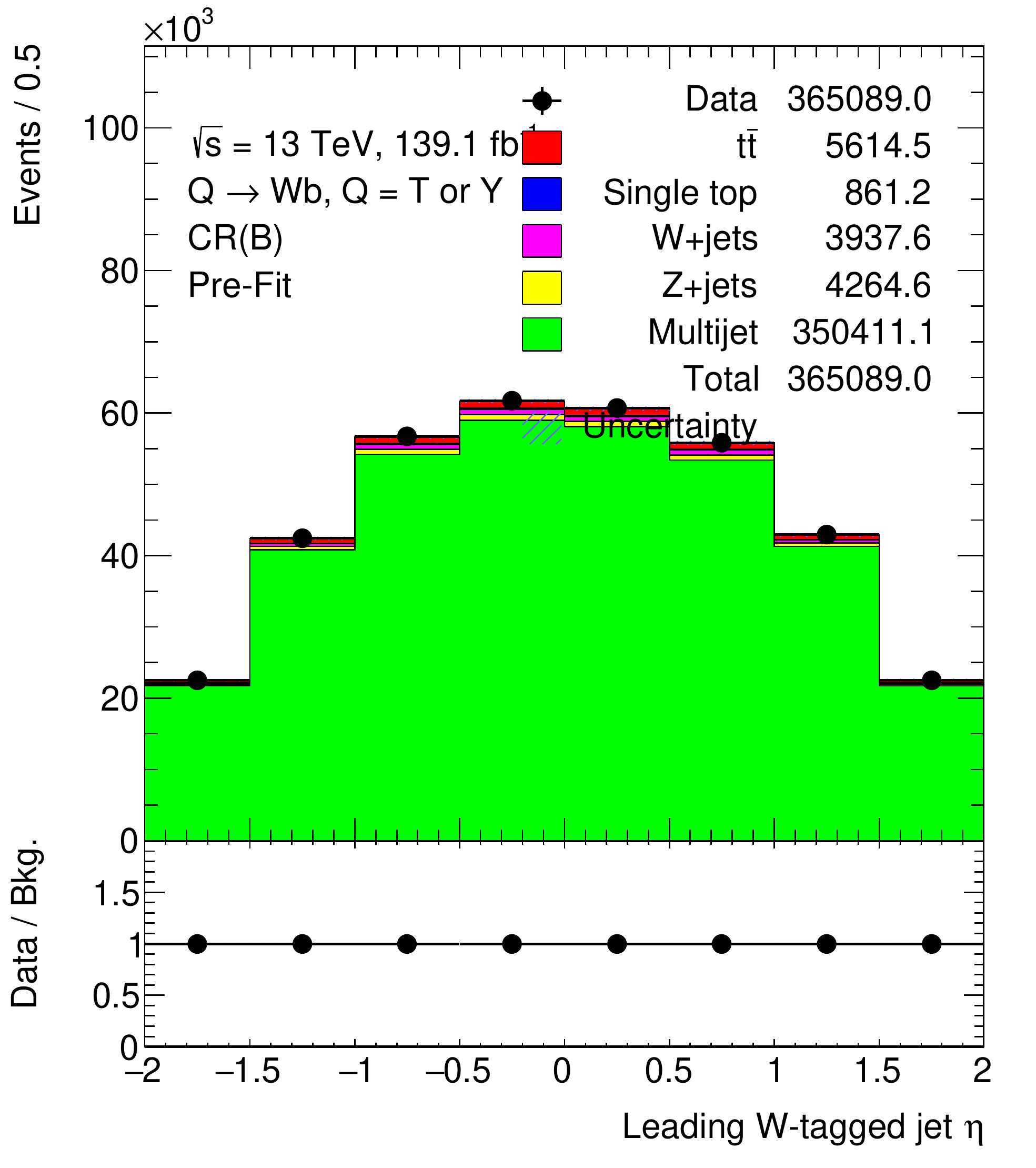}
		\caption{}
		\label{fig:app:cr_b:ljet_eta}
	\end{subfigure}
	\begin{subfigure}{.35\textwidth}
		\centering
		\includegraphics[width=\linewidth,height=\textheight,keepaspectratio]{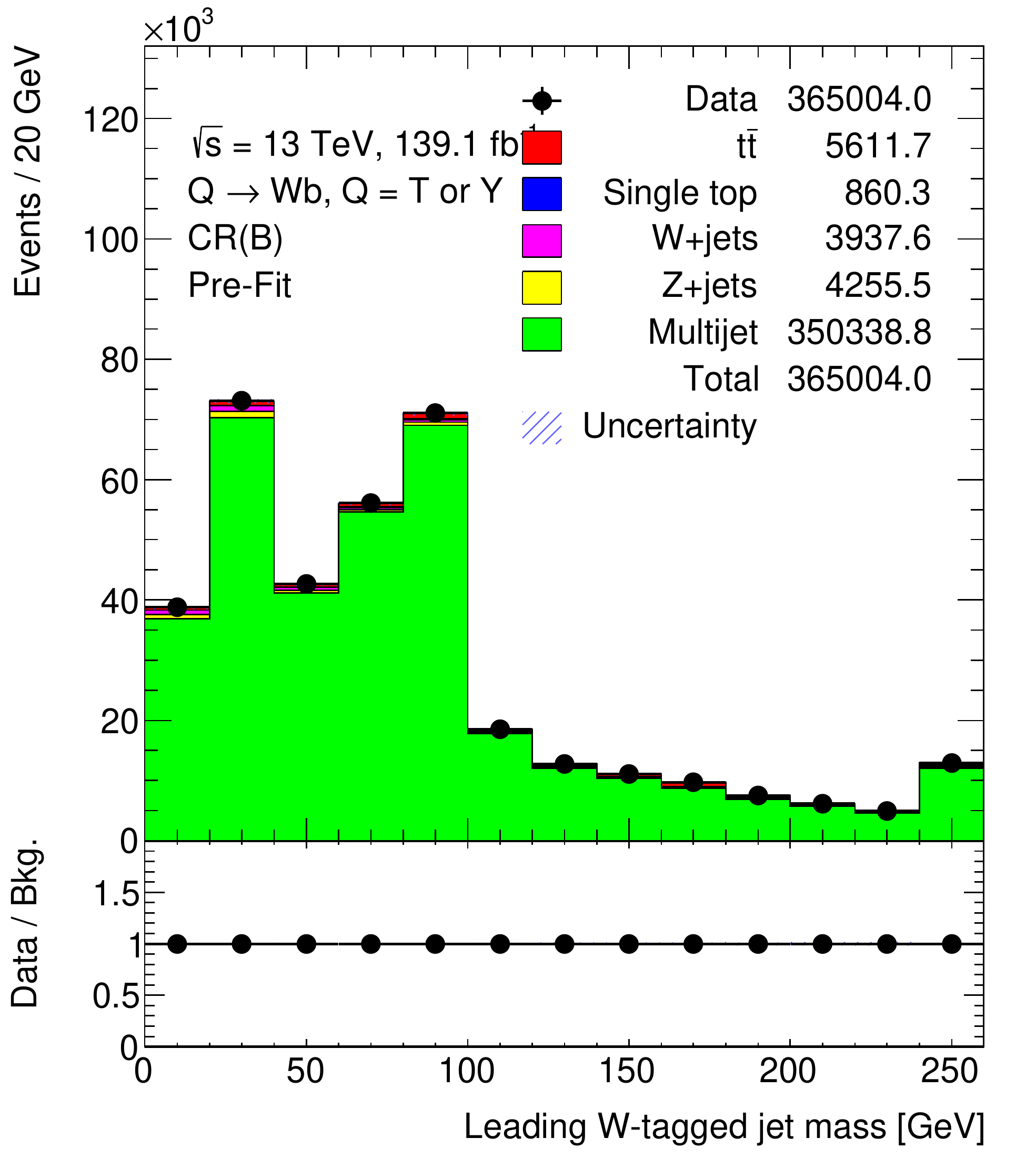}
		\caption{}
		\label{fig:app:cr_b:ljet_m}
	\end{subfigure}\hspace{0.6cm}
	\begin{subfigure}{.35\textwidth}
		\centering
		\includegraphics[width=\linewidth,height=\textheight,keepaspectratio]{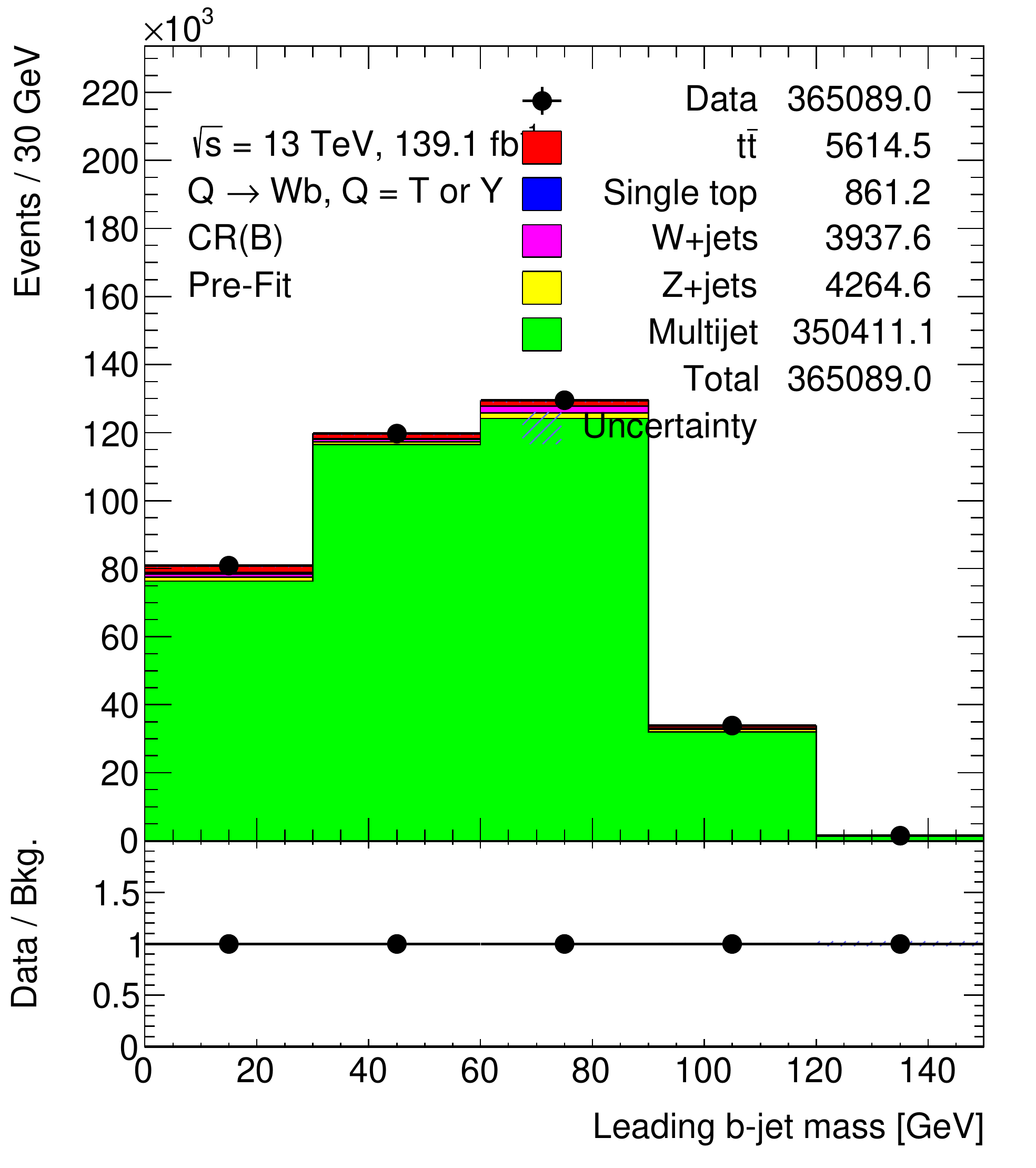}
		\caption{}
		\label{fig:app:cr_b:jet_m}
	\end{subfigure}
	\caption{A data/bkg.\ comparison of kinematic and reconstructed variables in CR B where the multijet background (in green) is calculated by Eqn.\ \ref{eqn:app} and the other backgrounds are from the MC simulation. The variables include (a) $p_{\text{T}}$ of $W$-tagged large-$R$ jet, (b) $p_{\text{T}}$ of leading $b$-tagged small-$R$ jet, (c) VLQ mass reconstructed from the kinematics of $W$-tagged large-$R$ jet and leading $b$-tagged small-$R$ jet, (d) $\eta$ distribution of $W$-tagged large-$R$ jet, (e) mass of $W$-tagged large-$R$ jet, and (f) mass of leading $b$-tagged small-$R$ jet.}
	\label{fig:app:cr_b}
\end{figure}

\begin{figure}[hbt!]
	\centering
	\graphicspath{{figs/appendix/CRC/}}
	\begin{subfigure}{.35\textwidth}
		\centering
		\includegraphics[width=\linewidth,height=\textheight,keepaspectratio]{CR_C_ljet_pt.pdf}
		\caption{}
		\label{fig:app:cr_c:ljet_pt}
	\end{subfigure}\hspace{0.6cm}
	\begin{subfigure}{.35\textwidth}
		\centering
		\includegraphics[width=\linewidth,height=\textheight,keepaspectratio]{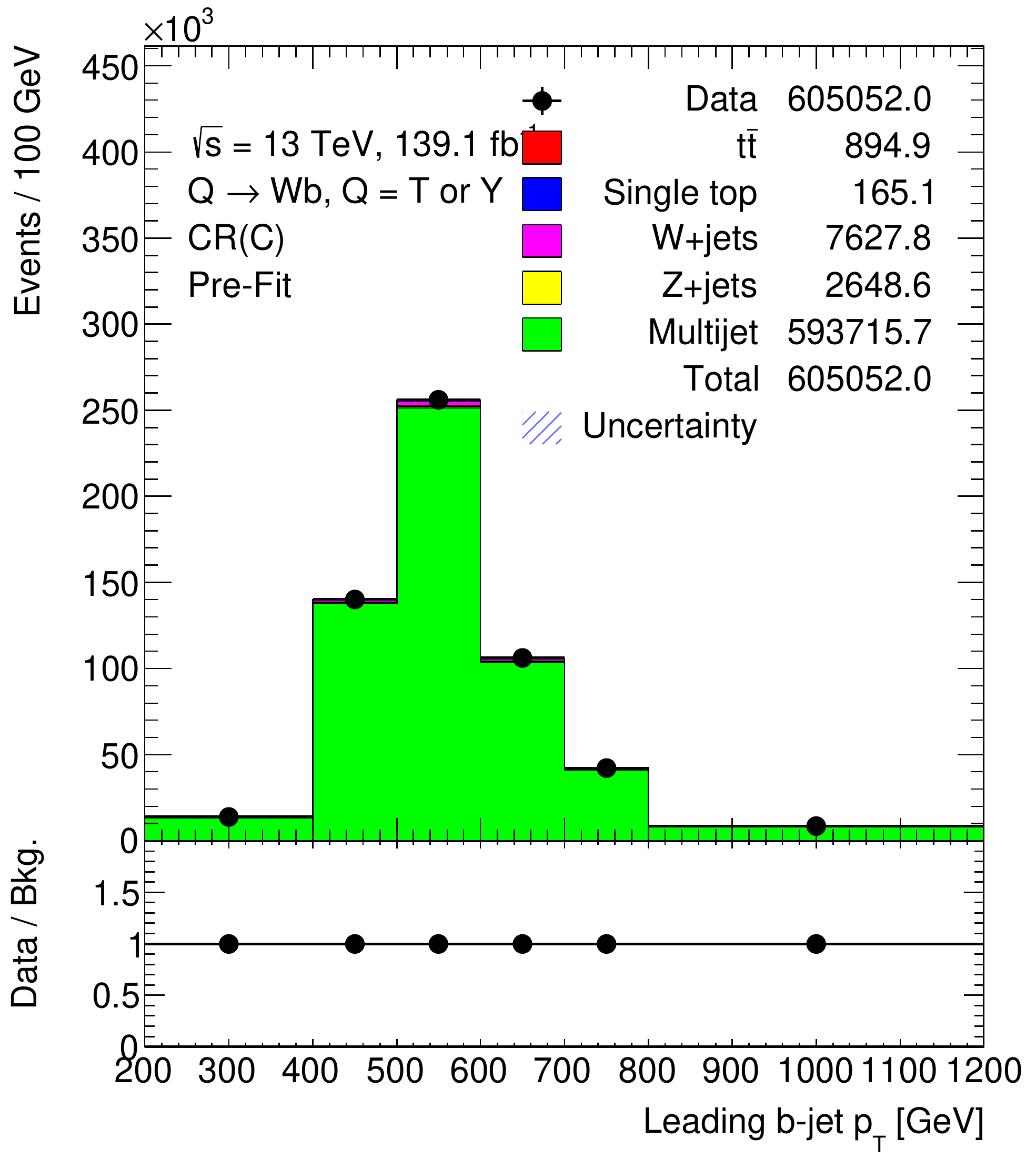}
		\caption{}
		\label{fig:app:cr_c:jet_pt}
	\end{subfigure}
	\begin{subfigure}{.35\textwidth}
		\centering
		\includegraphics[width=\linewidth,height=\textheight,keepaspectratio]{CR_C_VLQM.pdf}
		\caption{}
		\label{fig:app:cr_c:VLQM}
	\end{subfigure}\hspace{0.6cm}
	\begin{subfigure}{.35\textwidth}
		\centering
		\includegraphics[width=\linewidth,height=\textheight,keepaspectratio]{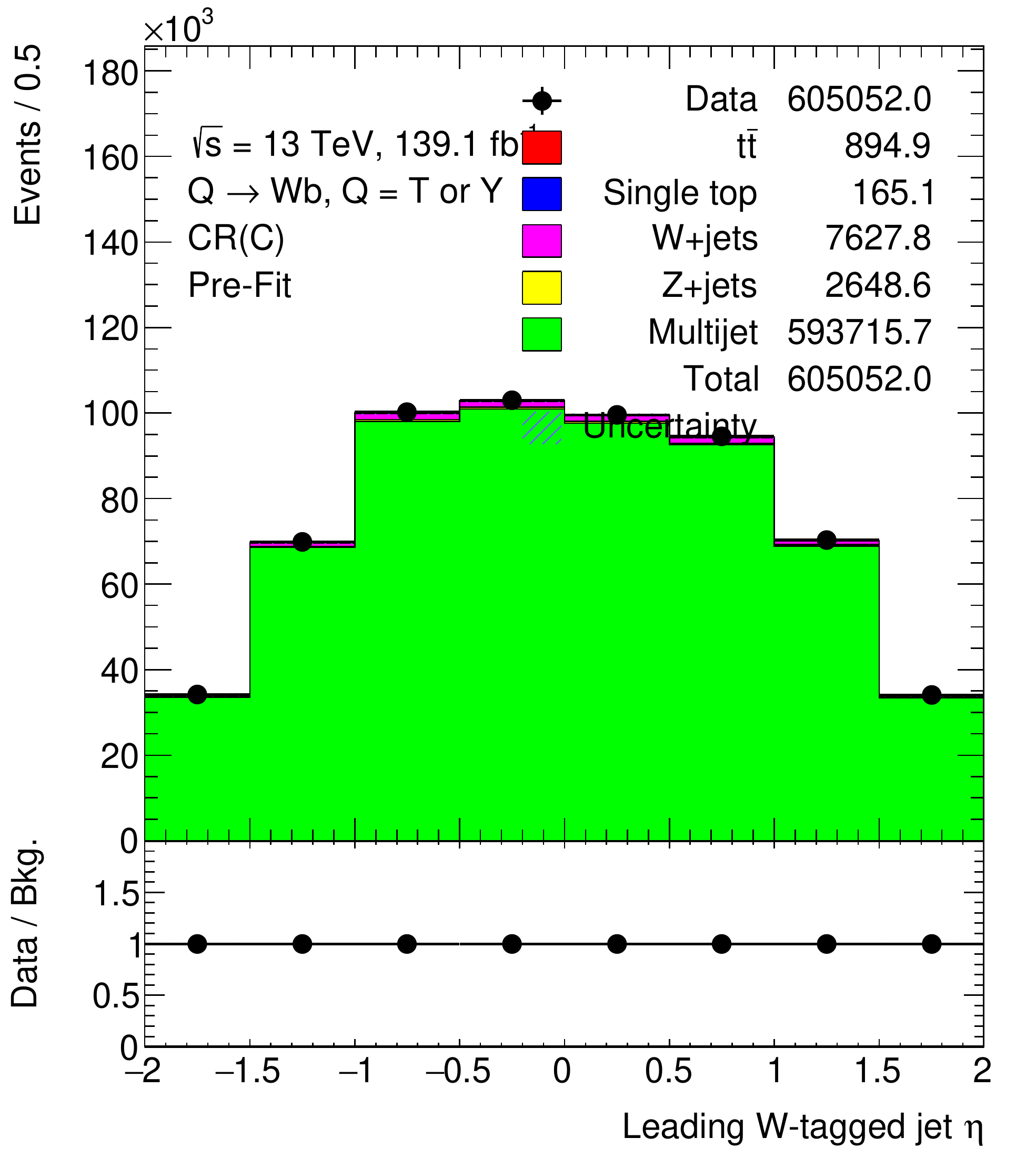}
		\caption{}
		\label{fig:app:cr_c:ljet_eta}
	\end{subfigure}
	\begin{subfigure}{.35\textwidth}
		\centering
		\includegraphics[width=\linewidth,height=\textheight,keepaspectratio]{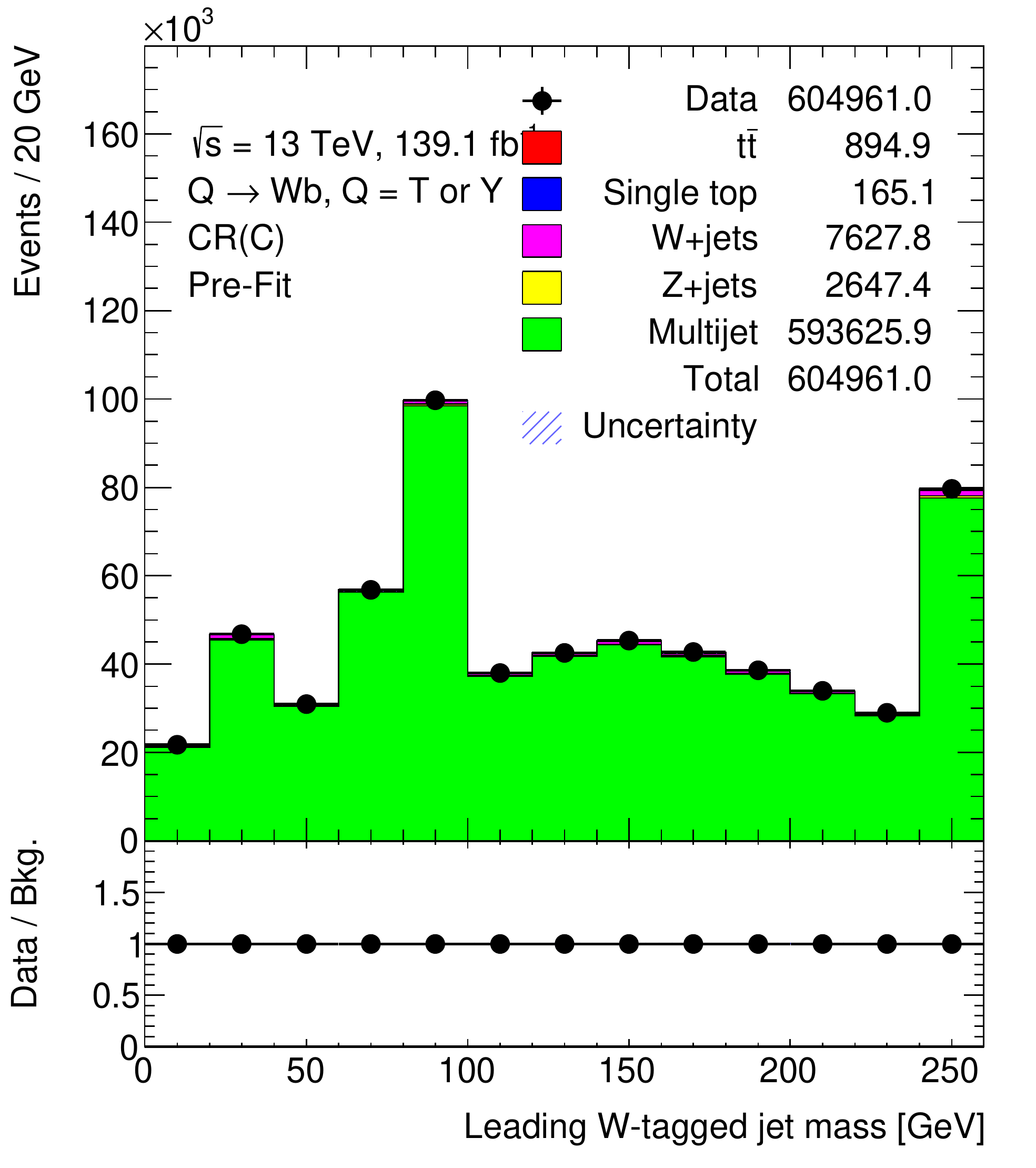}
		\caption{}
		\label{fig:app:cr_c:ljet_m}
	\end{subfigure}\hspace{0.6cm}
	\begin{subfigure}{.35\textwidth}
		\centering
		\includegraphics[width=\linewidth,height=\textheight,keepaspectratio]{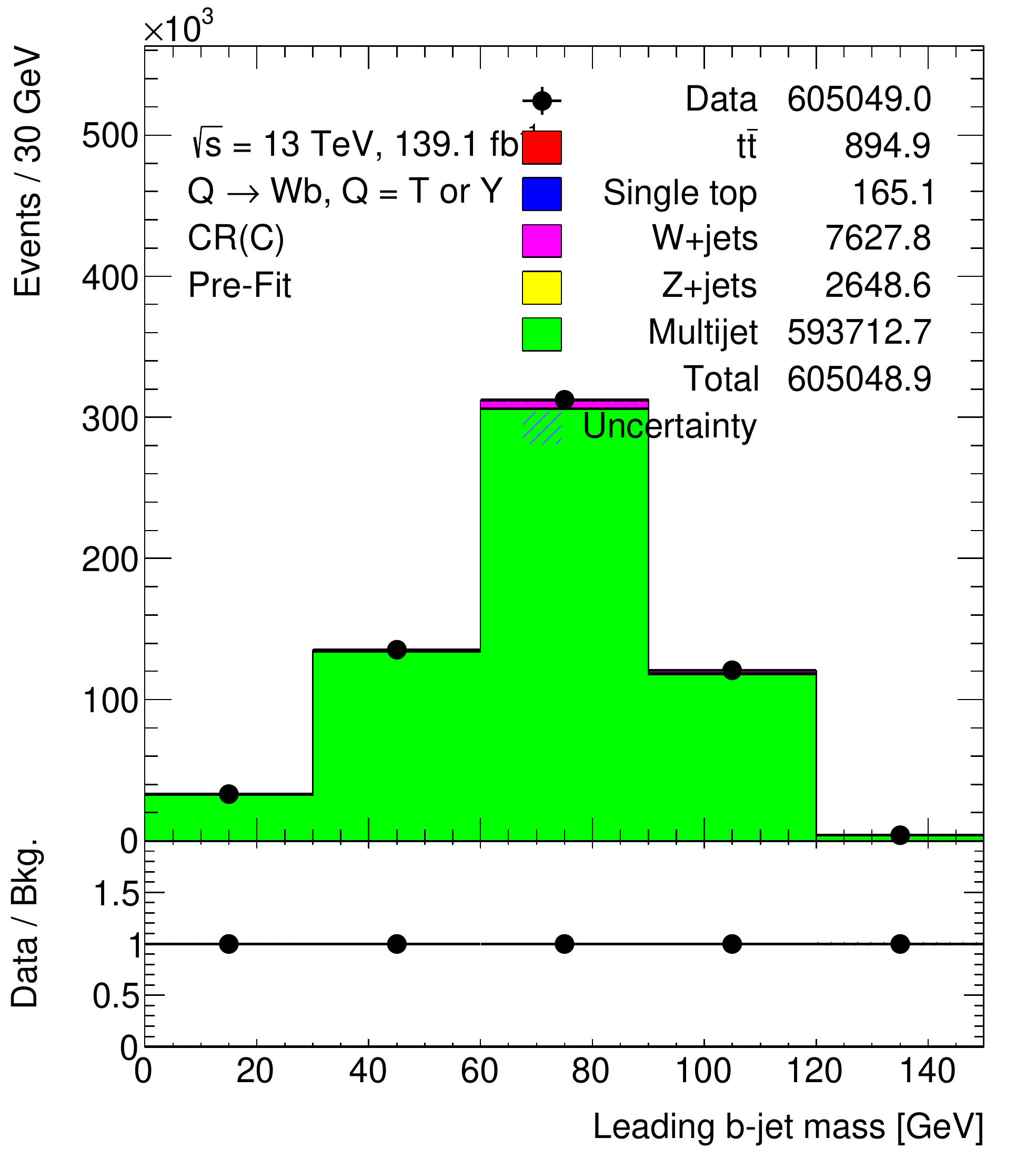}
		\caption{}
		\label{fig:app:cr_c:jet_m}
	\end{subfigure}
	\caption{A data/bkg.\ comparison of kinematic and reconstructed variables in CR C where the multijet background (in green) is calculated by Eqn.\ \ref{eqn:app} and the other backgrounds are from the MC simulation. The variables include (a) $p_{\text{T}}$ of $W$-tagged large-$R$ jet, (b) $p_{\text{T}}$ of leading $b$-tagged small-$R$ jet, (c) VLQ mass reconstructed from the kinematics of $W$-tagged large-$R$ jet and leading $b$-tagged small-$R$ jet, (d) $\eta$ distribution of $W$-tagged large-$R$ jet, (e) mass of $W$-tagged large-$R$ jet, and (f) mass of leading $b$-tagged small-$R$ jet.}
	\label{fig:app:cr_c}
\end{figure}

\begin{figure}[hbt!]
	\centering
	\graphicspath{{figs/appendix/CRD/}}
	\begin{subfigure}{.35\textwidth}
		\centering
		\includegraphics[width=\linewidth,height=\textheight,keepaspectratio]{CR_D_ljet_pt.pdf}
		\caption{}
		\label{fig:app:cr_d:ljet_pt}
	\end{subfigure}\hspace{0.6cm}
	\begin{subfigure}{.35\textwidth}
		\centering
		\includegraphics[width=\linewidth,height=\textheight,keepaspectratio]{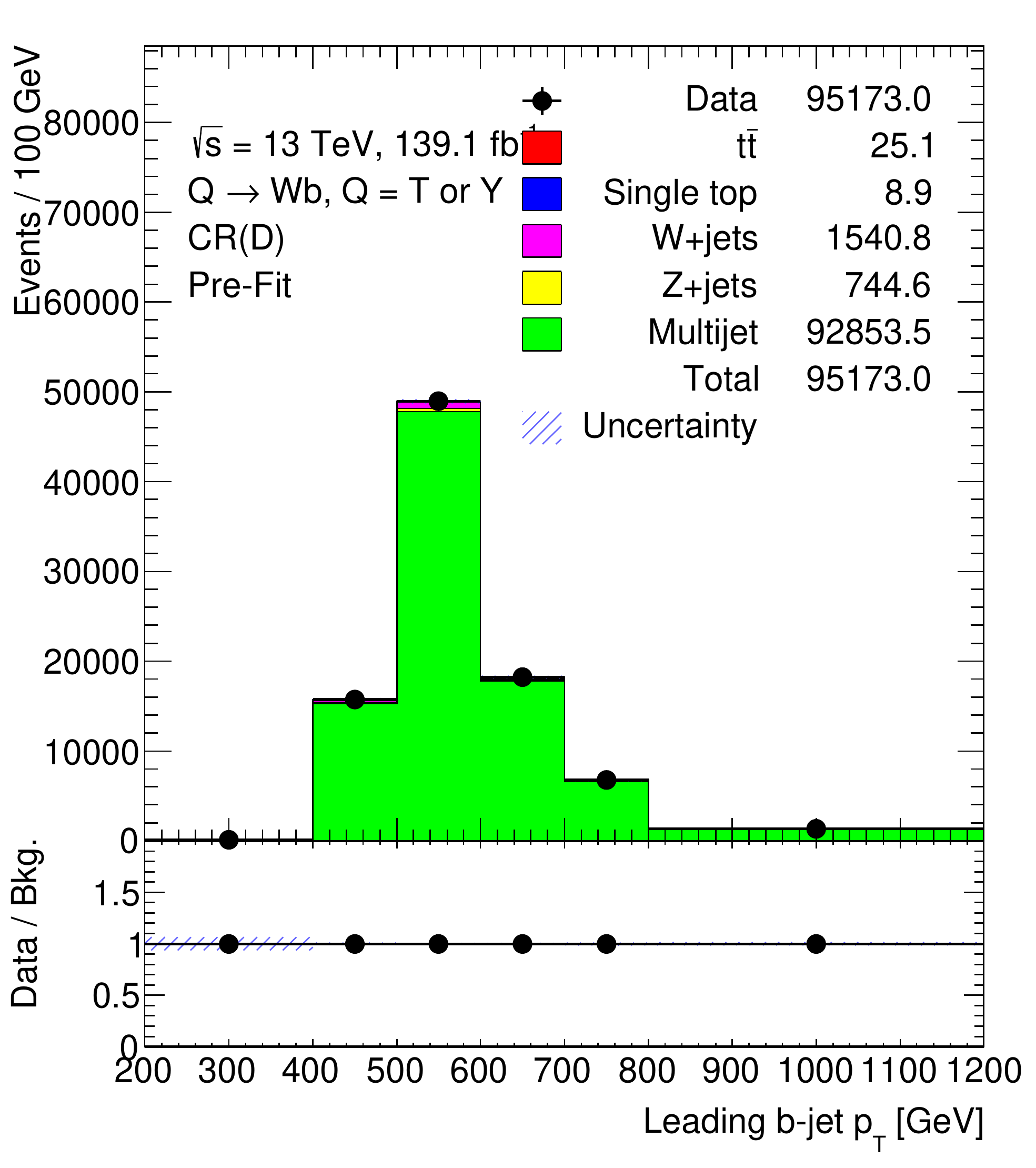}
		\caption{}
		\label{fig:app:cr_d:jet_pt}
	\end{subfigure}
	\begin{subfigure}{.35\textwidth}
		\centering
		\includegraphics[width=\linewidth,height=\textheight,keepaspectratio]{CR_D_VLQM.pdf}
		\caption{}
		\label{fig:app:cr_d:VLQM}
	\end{subfigure}\hspace{0.6cm}
	\begin{subfigure}{.35\textwidth}
		\centering
		\includegraphics[width=\linewidth,height=\textheight,keepaspectratio]{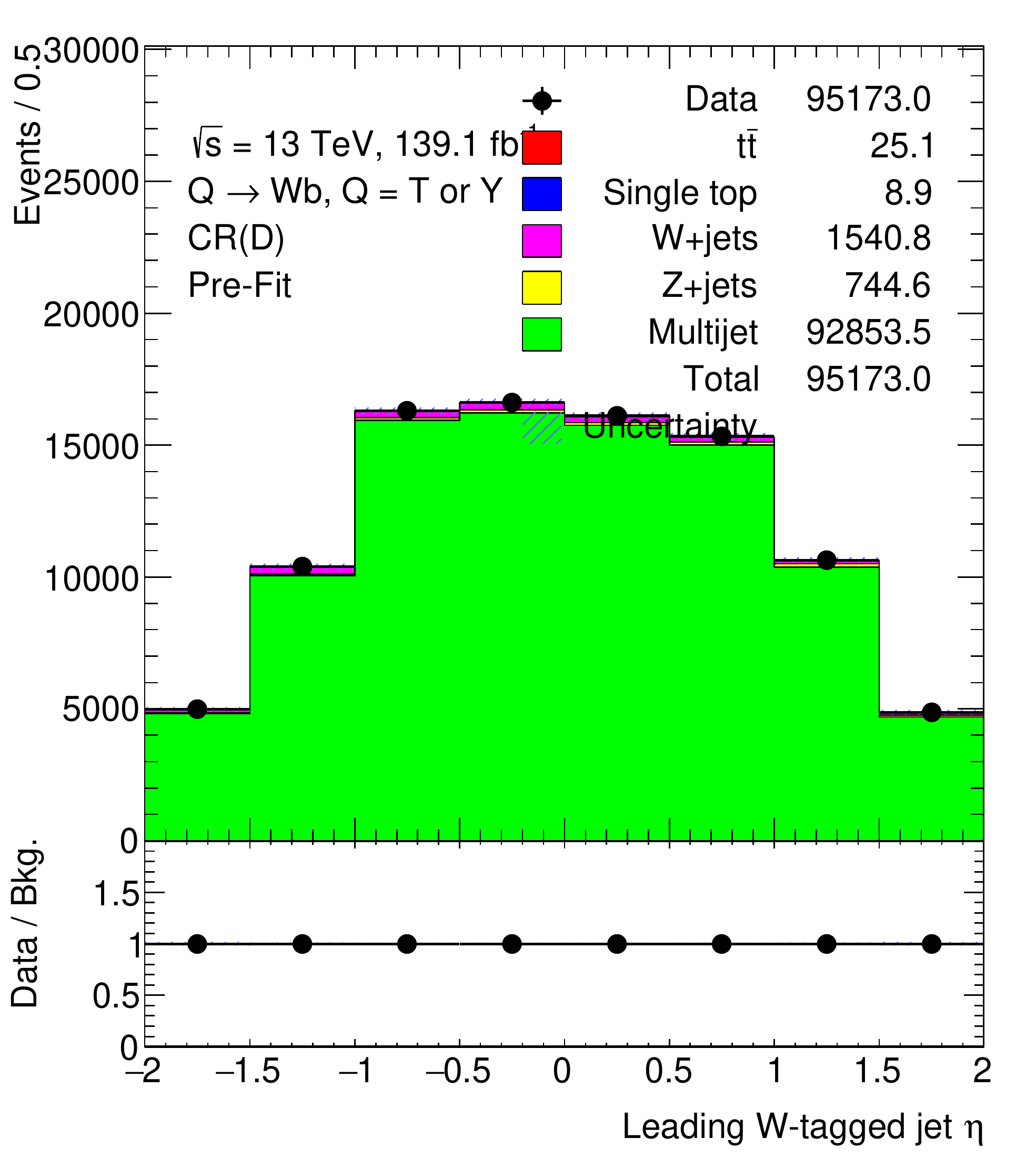}
		\caption{}
		\label{fig:app:cr_d:ljet_eta}
	\end{subfigure}
	\begin{subfigure}{.35\textwidth}
		\centering
		\includegraphics[width=\linewidth,height=\textheight,keepaspectratio]{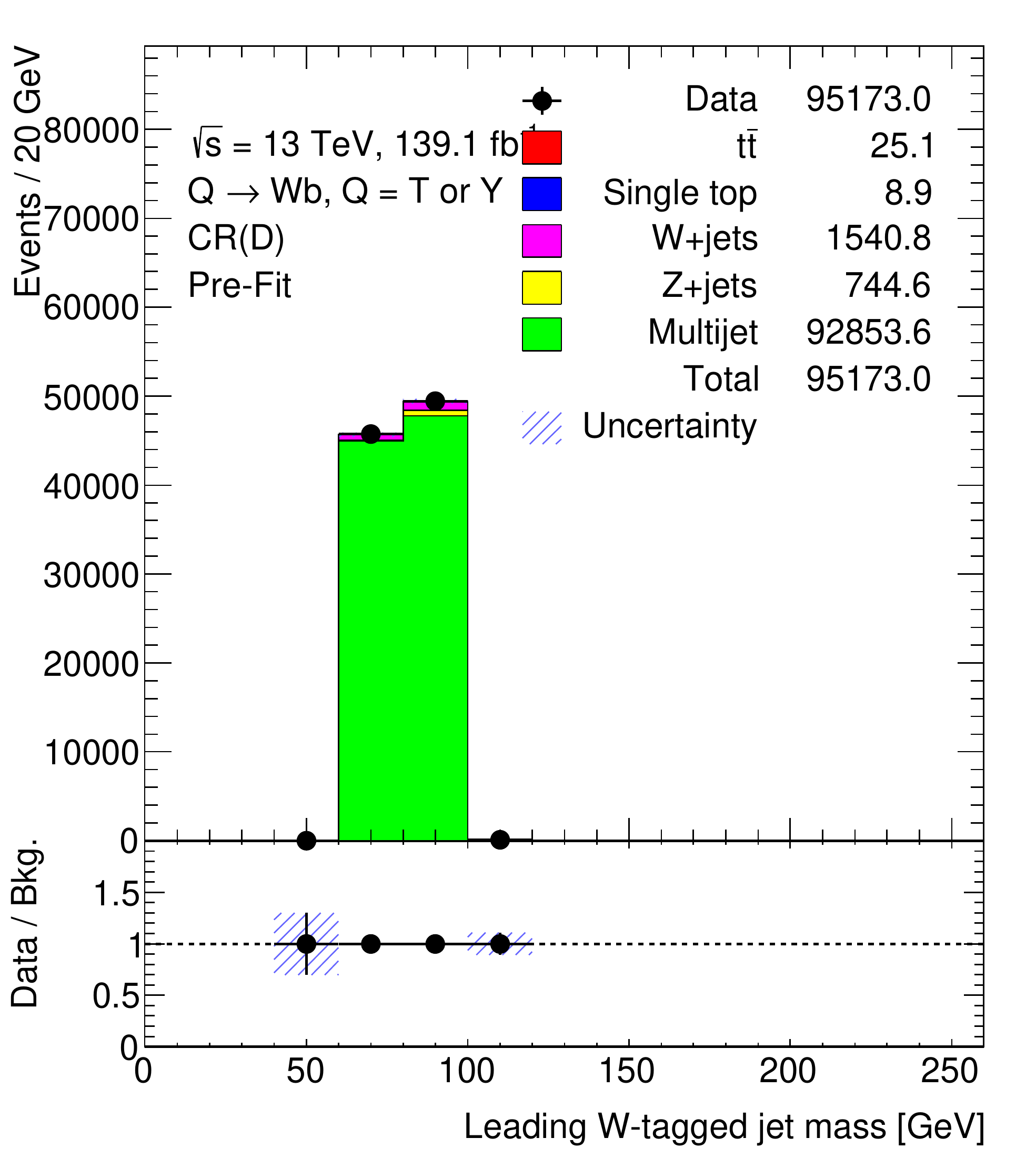}
		\caption{}
		\label{fig:app:cr_d:ljet_m}
	\end{subfigure}\hspace{0.6cm}
	\begin{subfigure}{.35\textwidth}
		\centering
		\includegraphics[width=\linewidth,height=\textheight,keepaspectratio]{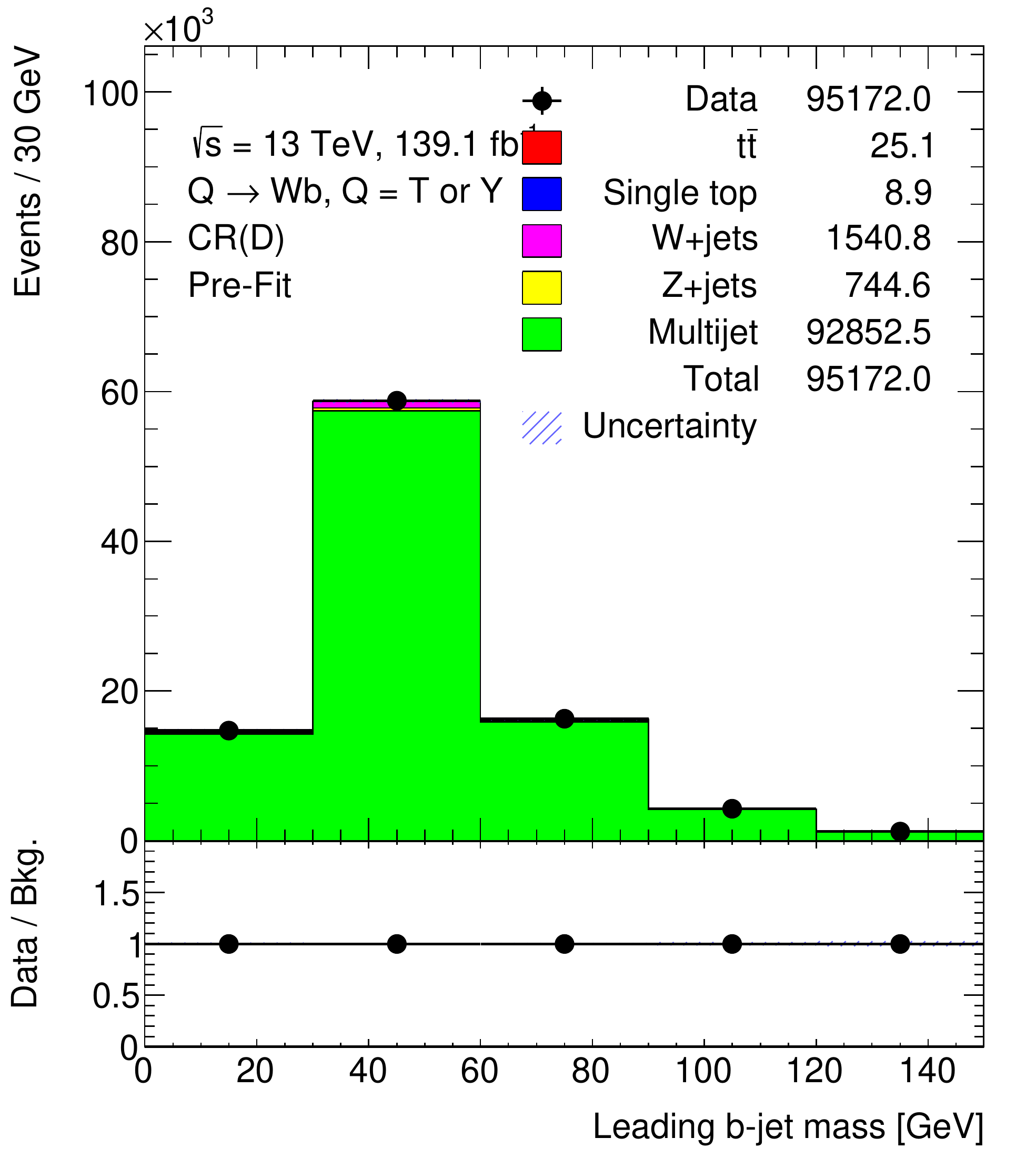}
		\caption{}
		\label{fig:app:cr_d:jet_m}
	\end{subfigure}
	\caption{A data/bkg.\ comparison of kinematic and reconstructed variables in CR D where the multijet background (in green) is calculated by Eqn.\ \ref{eqn:app} and the other backgrounds are from the MC simulation. The variables include (a) $p_{\text{T}}$ of $W$-tagged large-$R$ jet, (b) $p_{\text{T}}$ of leading $b$-tagged small-$R$ jet, (c) VLQ mass reconstructed from the kinematics of $W$-tagged large-$R$ jet and leading $b$-tagged small-$R$ jet, (d) $\eta$ distribution of $W$-tagged large-$R$ jet, (e) mass of $W$-tagged large-$R$ jet, and (f) mass of leading $b$-tagged small-$R$ jet.}
	\label{fig:app:cr_d}
\end{figure}

\begin{figure}[hbt!]
	\centering
	\graphicspath{{figs/appendix/CRD1/}}
	\begin{subfigure}{.35\textwidth}
		\centering
		\includegraphics[width=\linewidth,height=\textheight,keepaspectratio]{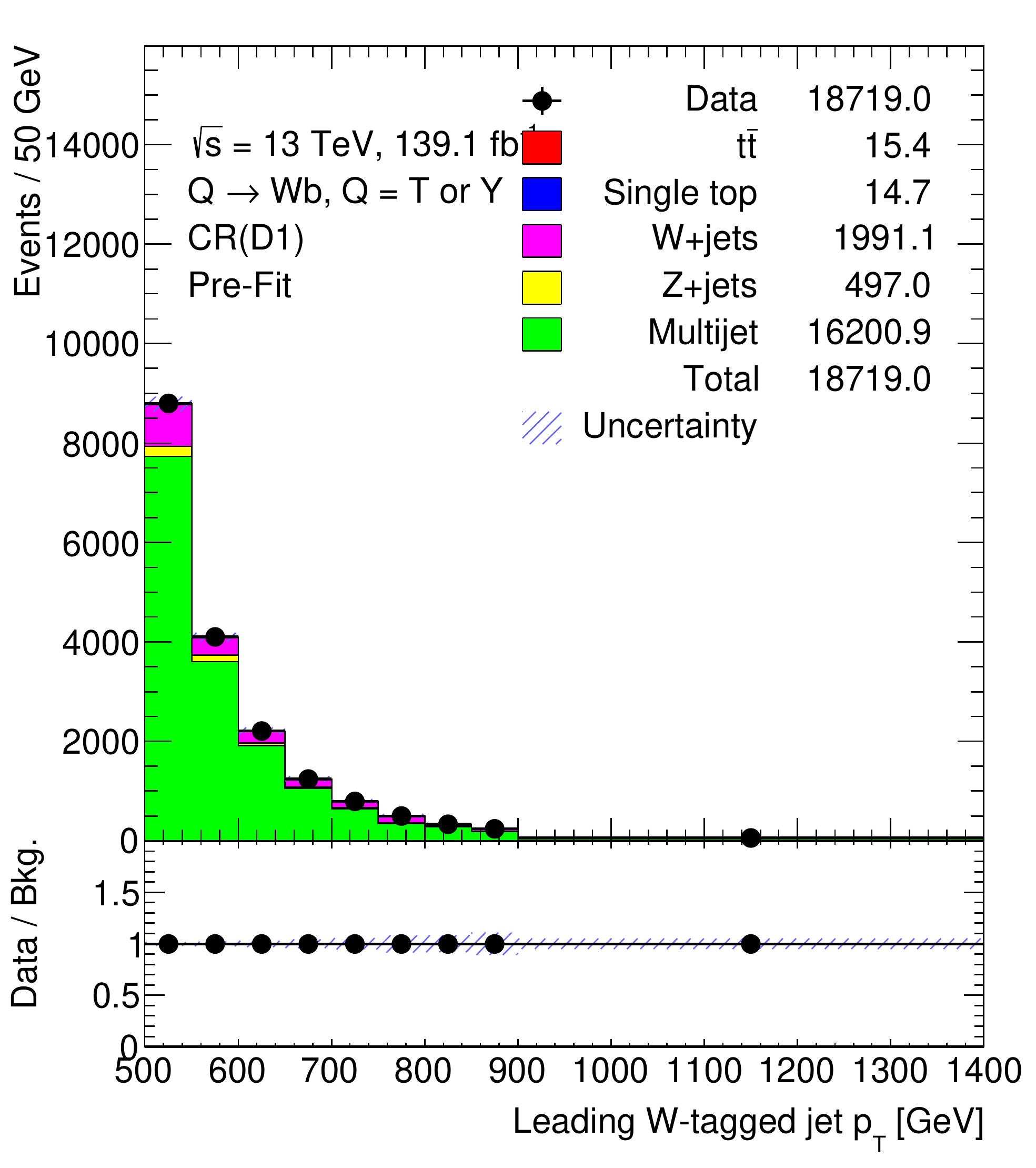}
		\caption{}
		\label{fig:app:cr_d1:ljet_pt}
	\end{subfigure}\hspace{0.6cm}
	\begin{subfigure}{.35\textwidth}
		\centering
		\includegraphics[width=\linewidth,height=\textheight,keepaspectratio]{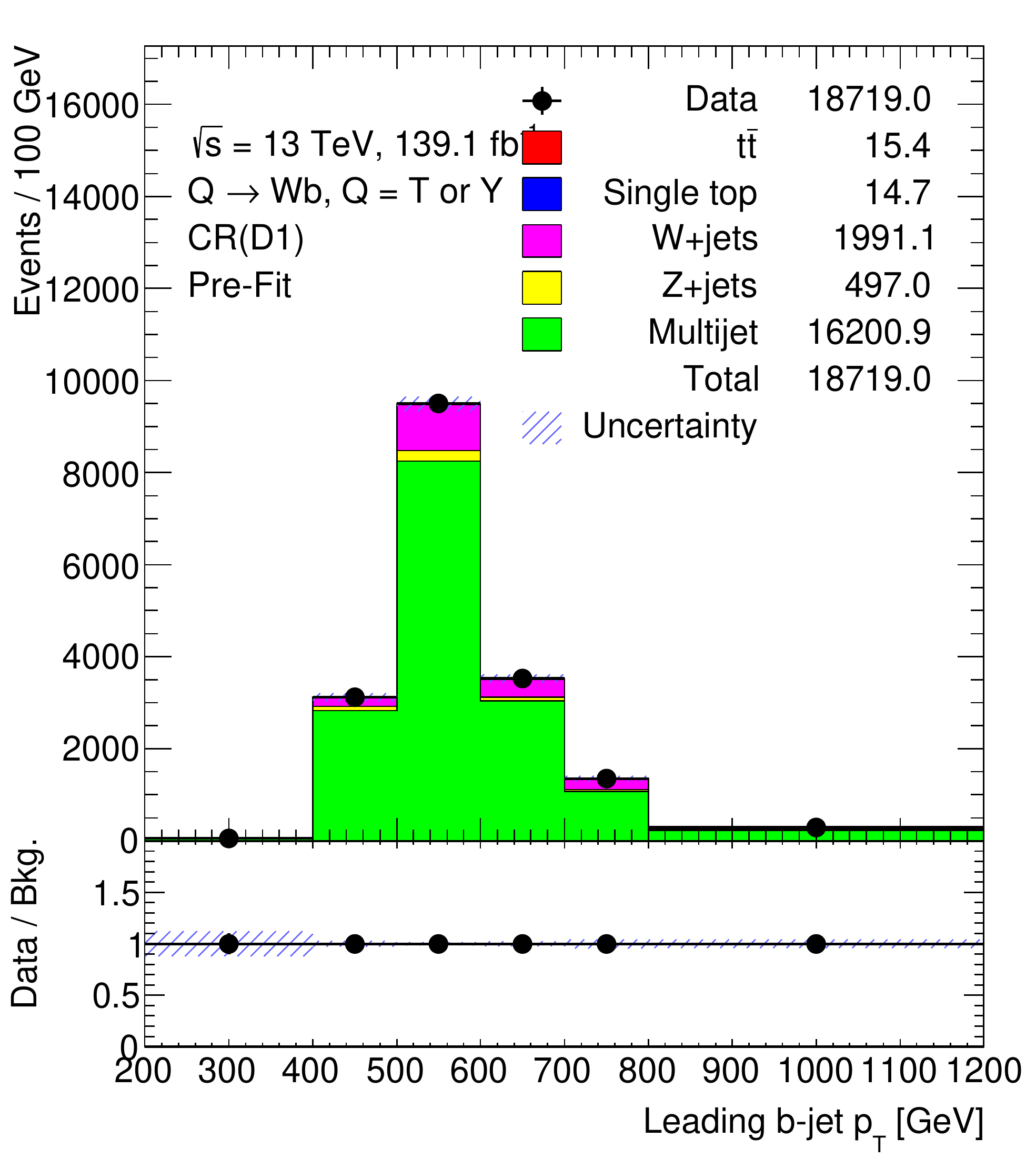}
		\caption{}
		\label{fig:app:cr_d1:jet_pt}
	\end{subfigure}
	\begin{subfigure}{.35\textwidth}
		\centering
		\includegraphics[width=\linewidth,height=\textheight,keepaspectratio]{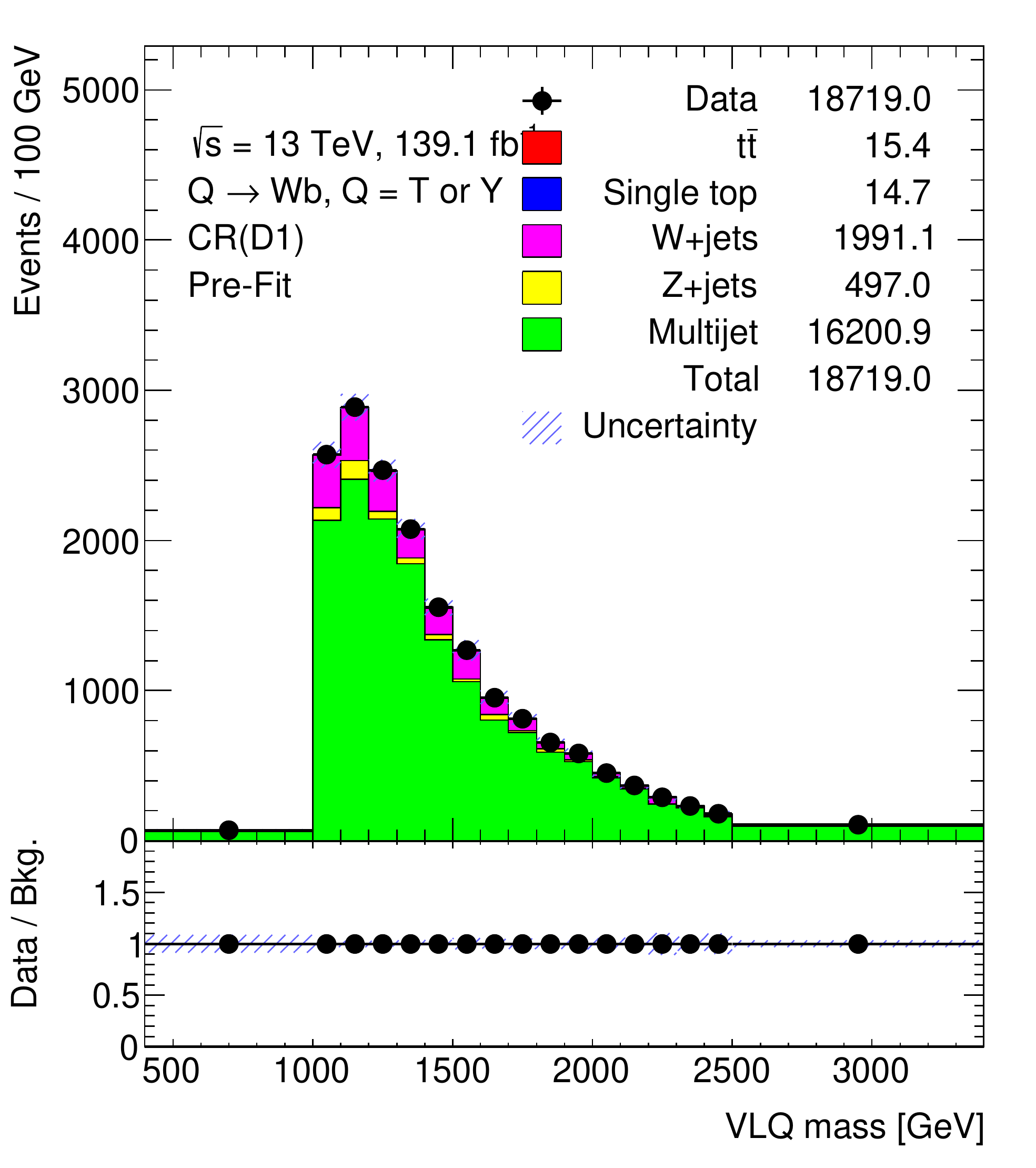}
		\caption{}
		\label{fig:app:cr_d1:VLQM}
	\end{subfigure}\hspace{0.6cm}
	\begin{subfigure}{.35\textwidth}
		\centering
		\includegraphics[width=\linewidth,height=\textheight,keepaspectratio]{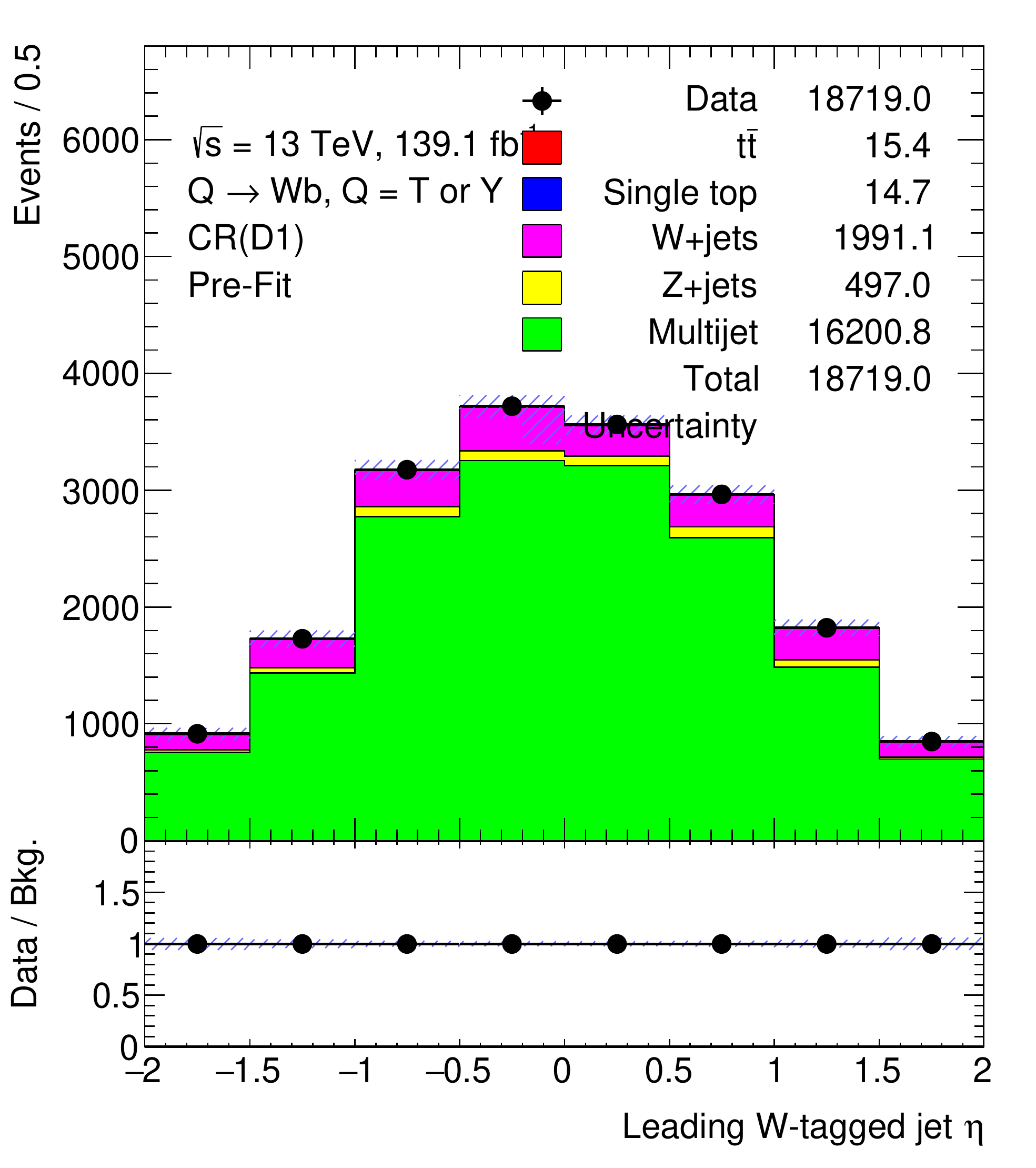}
		\caption{}
		\label{fig:app:cr_d1:ljet_eta}
	\end{subfigure}
	\begin{subfigure}{.35\textwidth}
		\centering
		\includegraphics[width=\linewidth,height=\textheight,keepaspectratio]{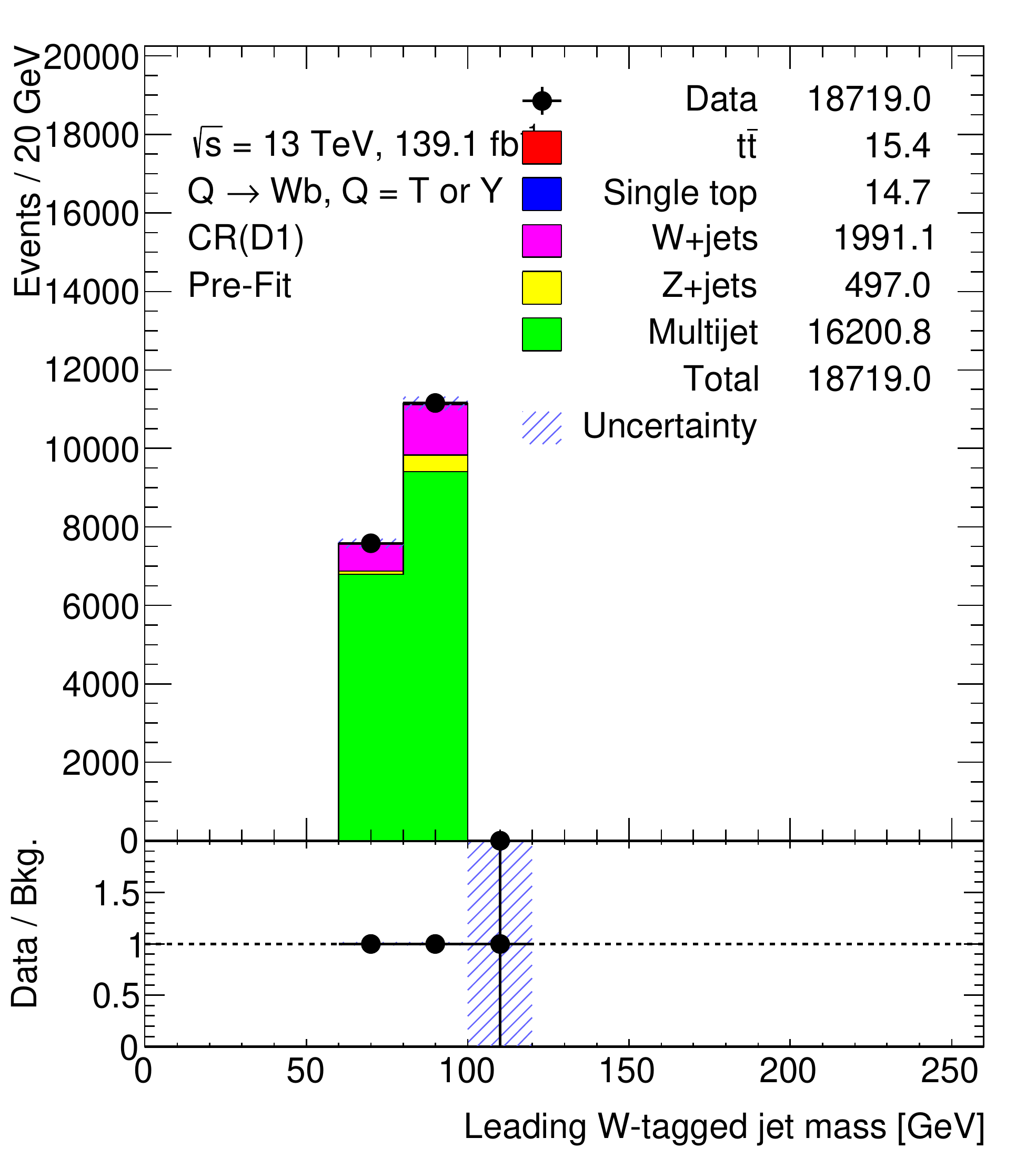}
		\caption{}
		\label{fig:app:cr_d1:ljet_m}
	\end{subfigure}\hspace{0.6cm}
	\begin{subfigure}{.35\textwidth}
		\centering
		\includegraphics[width=\linewidth,height=\textheight,keepaspectratio]{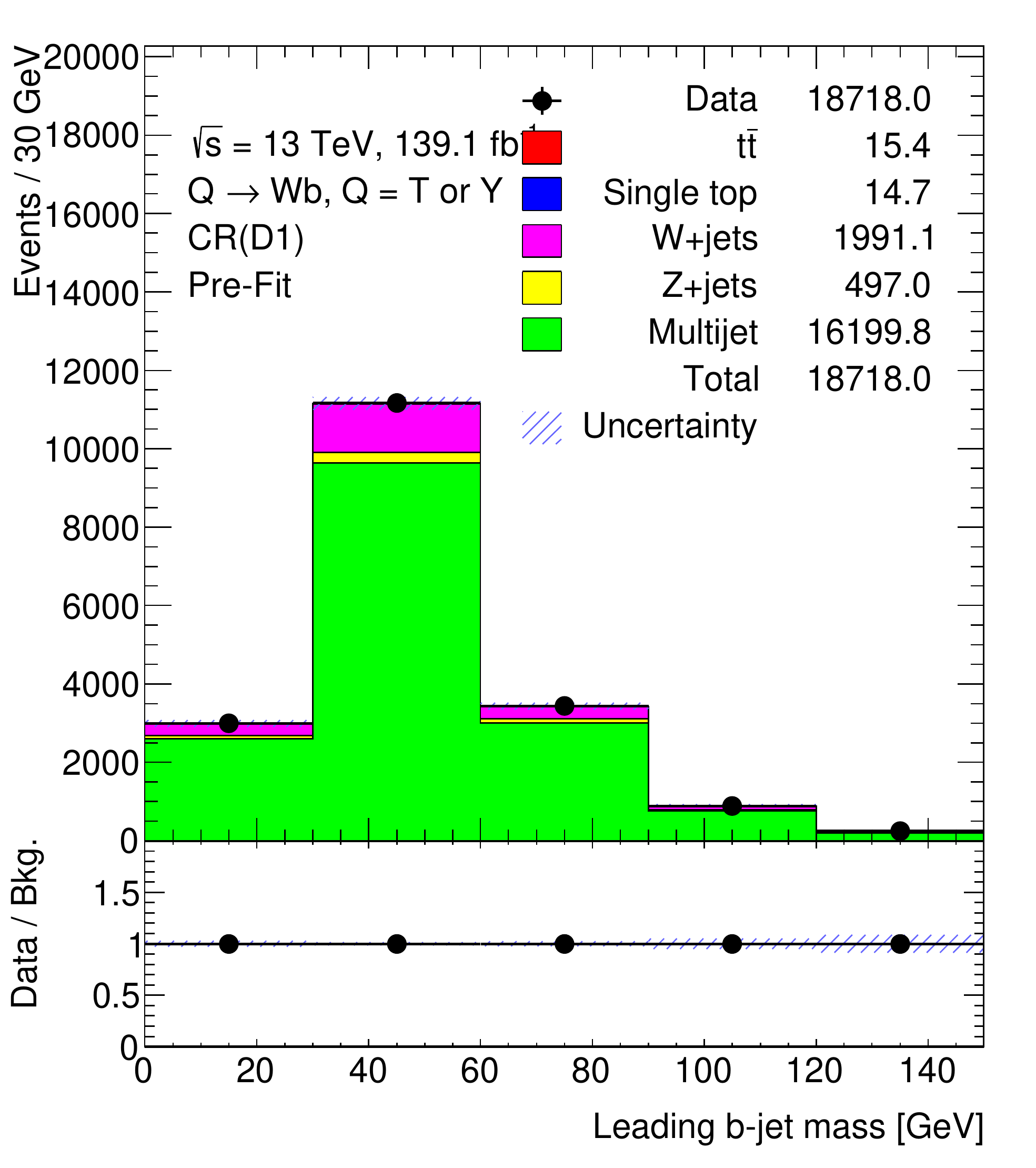}
		\caption{}
		\label{fig:app:cr_d1:jet_m}
	\end{subfigure}
	\caption{A data/bkg.\ comparison of kinematic and reconstructed variables in CR D1 where the multijet background (in green) is calculated by Eqn.\ \ref{eqn:app} and the other backgrounds are from the MC simulation. The variables include (a) $p_{\text{T}}$ of $W$-tagged large-$R$ jet, (b) $p_{\text{T}}$ of leading $b$-tagged small-$R$ jet, (c) VLQ mass reconstructed from the kinematics of $W$-tagged large-$R$ jet and leading $b$-tagged small-$R$ jet, (d) $\eta$ distribution of $W$-tagged large-$R$ jet, (e) mass of $W$-tagged large-$R$ jet, and (f) mass of leading $b$-tagged small-$R$ jet.}
	\label{fig:app:cr_d1}
\end{figure}

Fig.\ \ref{fig:app:cr_b} shows all the six distributions from CR B. Fig.\ \ref{fig:app:cr_c} shows all the six distributions from CR C. Fig.\ \ref{fig:app:cr_d} shows all the six distributions from CR D. Fig.\ \ref{fig:app:cr_d1} shows all the six distributions from CR D1.

\chapter{Scaled multijet MC distributions}
\label{sec:app2}
In this appendix, the data/bkg.\ comparison plots for two kinematic variables are shown, in which the multijet background is considered from both the multijet MC and the scaled multijet MC (as described in \ref{sec:abcd:furtherimprovement:scaledcorr}) and the other backgrounds are from the MC simulation.

\begin{figure}[hbt!]
	\centering
	\graphicspath{{figs/appendix/scaledmultijet/ABVLQM/}}
	\begin{subfigure}{.35\textwidth}
		\centering
		\includegraphics[width=\linewidth,height=\textheight,keepaspectratio]{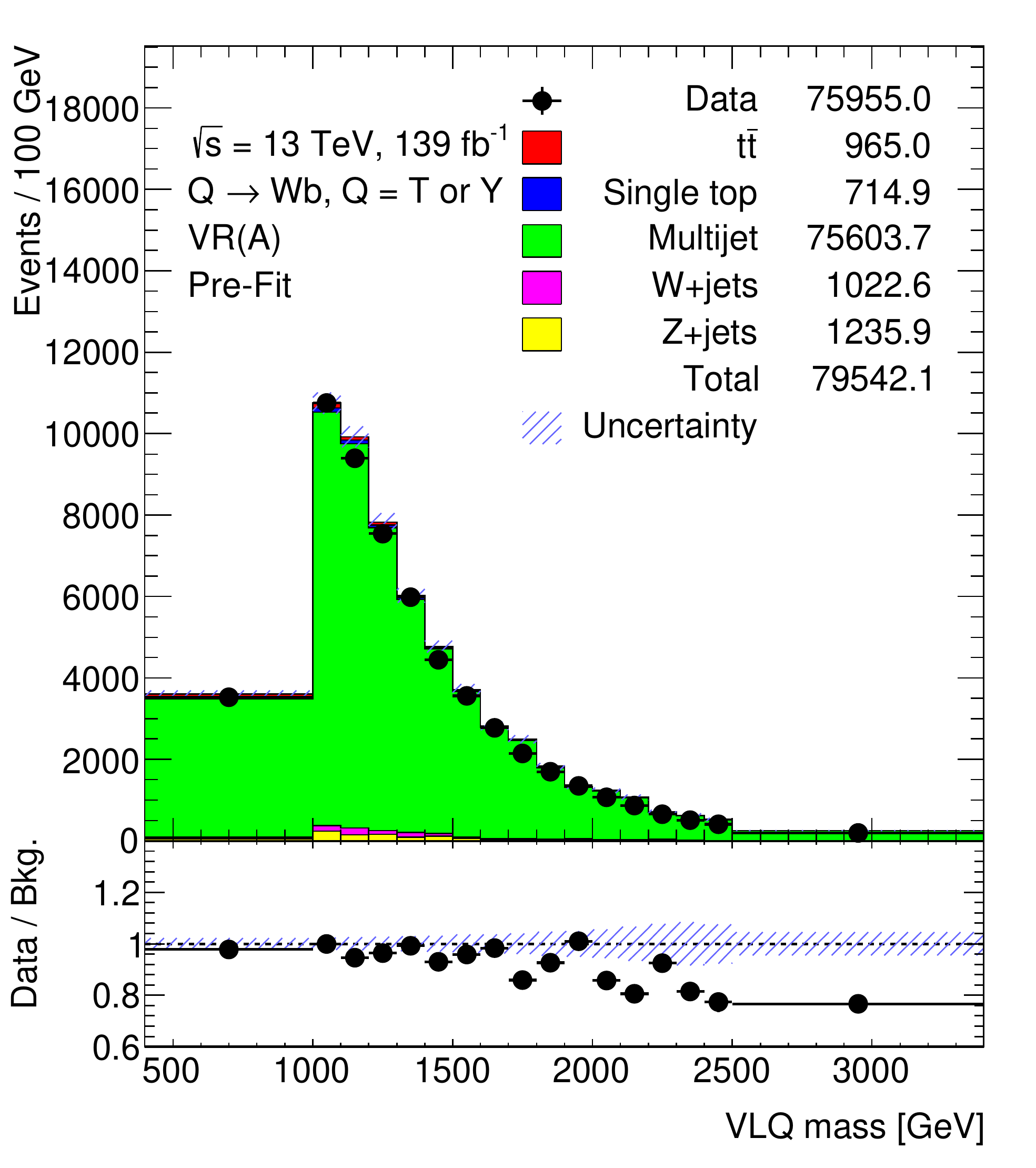}
		\caption{}
	\end{subfigure}\hspace{0.6cm}
	\begin{subfigure}{.35\textwidth}
		\centering
		\includegraphics[width=\linewidth,height=\textheight,keepaspectratio]{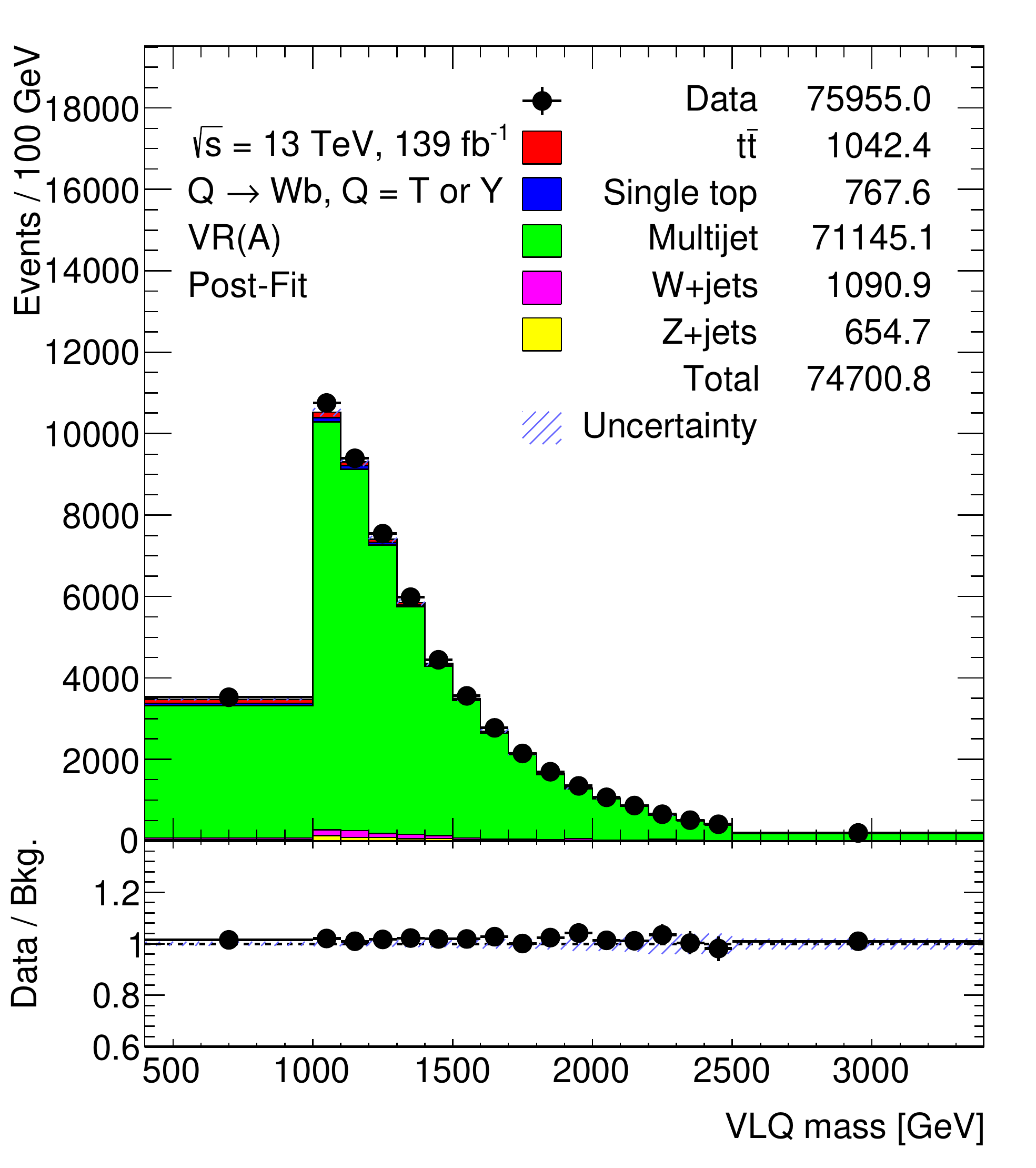}
		\caption{}
	\end{subfigure}
	\begin{subfigure}{.35\textwidth}
		\centering
		\includegraphics[width=\linewidth,height=\textheight,keepaspectratio]{CR_B_VLQM.pdf}
		\caption{}
	\end{subfigure}\hspace{0.6cm}
	\begin{subfigure}{.35\textwidth}
		\centering
		\includegraphics[width=\linewidth,height=\textheight,keepaspectratio]{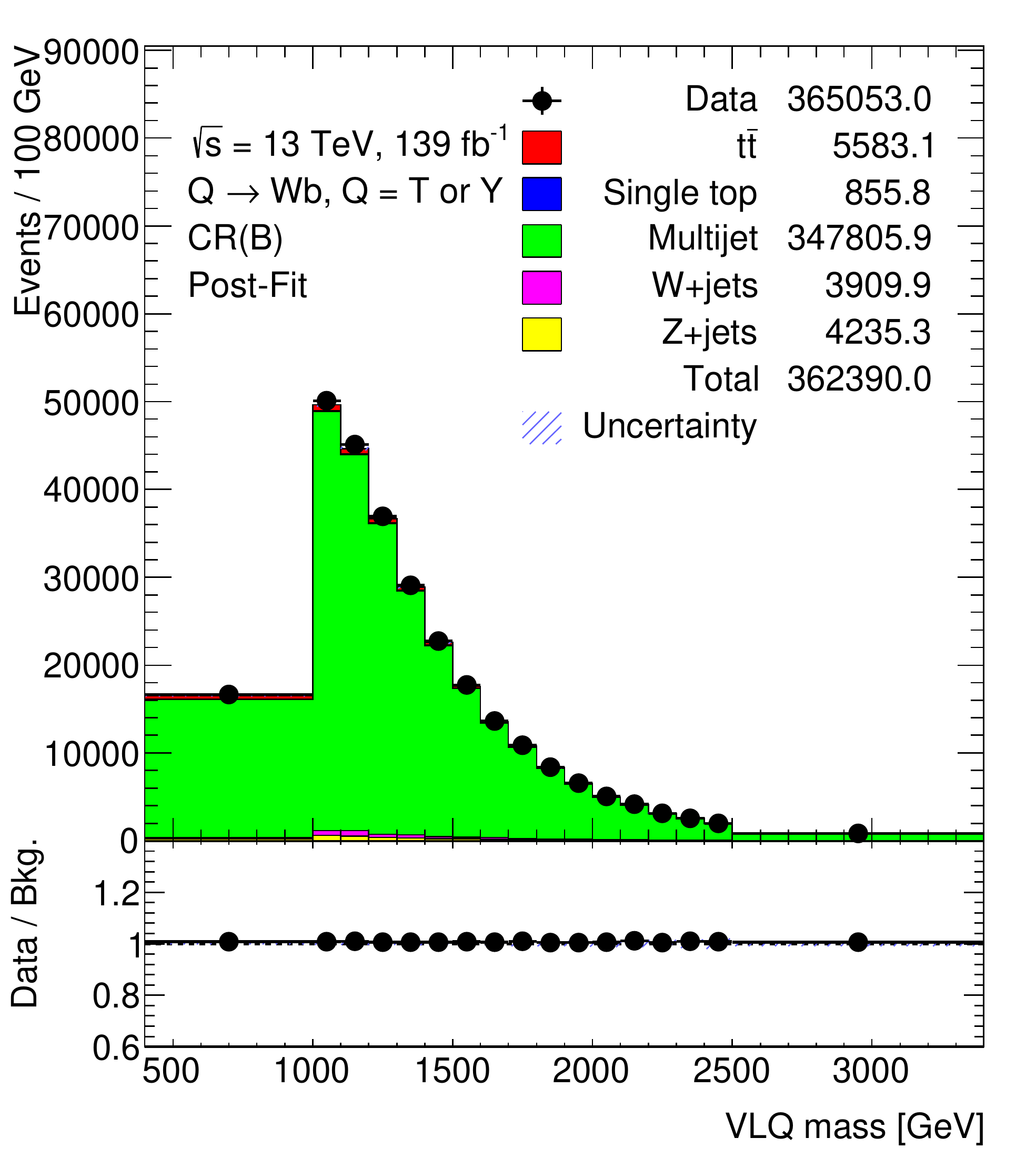}
		\caption{}
	\end{subfigure}
	\caption{VLQ mass distribution is shown when a likelihood fit is performed to fit regions A and B together in which the multijet MC is fitted to the data. It shows the VLQ mass distribution for (a) multijet MC and (b) scaled multijet MC in region A, and (c) multijet MC and (d) scaled multijet MC in region B.}
	\label{fig:app:ab:vlqm}
\end{figure}

\begin{figure}[hbt!]
	\centering
	\graphicspath{{figs/appendix/scaledmultijet/CDVLQM/}}
	\begin{subfigure}{.35\textwidth}
		\centering
		\includegraphics[width=\linewidth,height=\textheight,keepaspectratio]{CR_C_VLQM.pdf}
		\caption{}
	\end{subfigure}\hspace{0.6cm}
	\begin{subfigure}{.35\textwidth}
		\centering
		\includegraphics[width=\linewidth,height=\textheight,keepaspectratio]{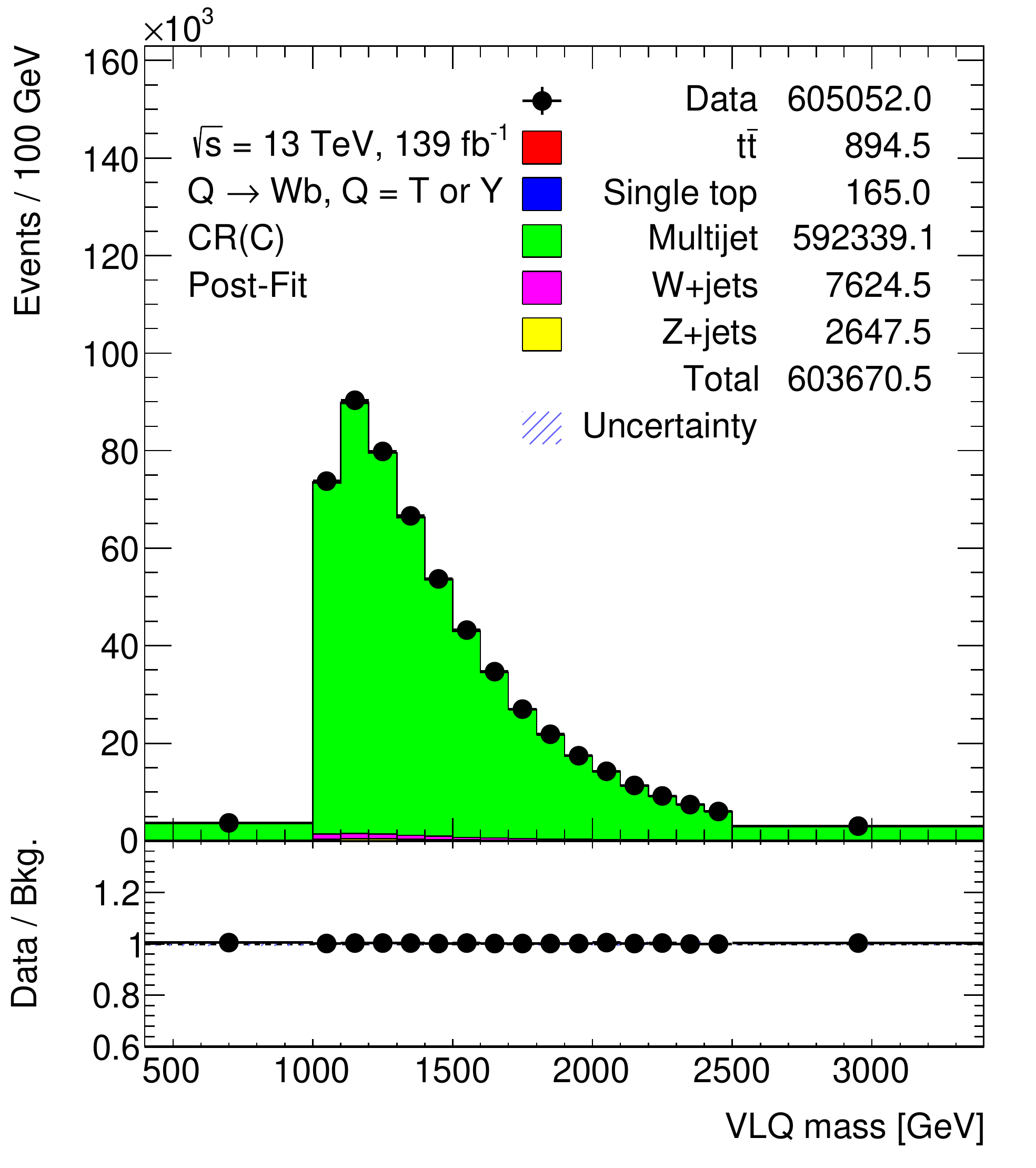}
		\caption{}
	\end{subfigure}
	\begin{subfigure}{.35\textwidth}
		\centering
		\includegraphics[width=\linewidth,height=\textheight,keepaspectratio]{CR_D_VLQM.pdf}
		\caption{}
	\end{subfigure}\hspace{0.6cm}
	\begin{subfigure}{.35\textwidth}
		\centering
		\includegraphics[width=\linewidth,height=\textheight,keepaspectratio]{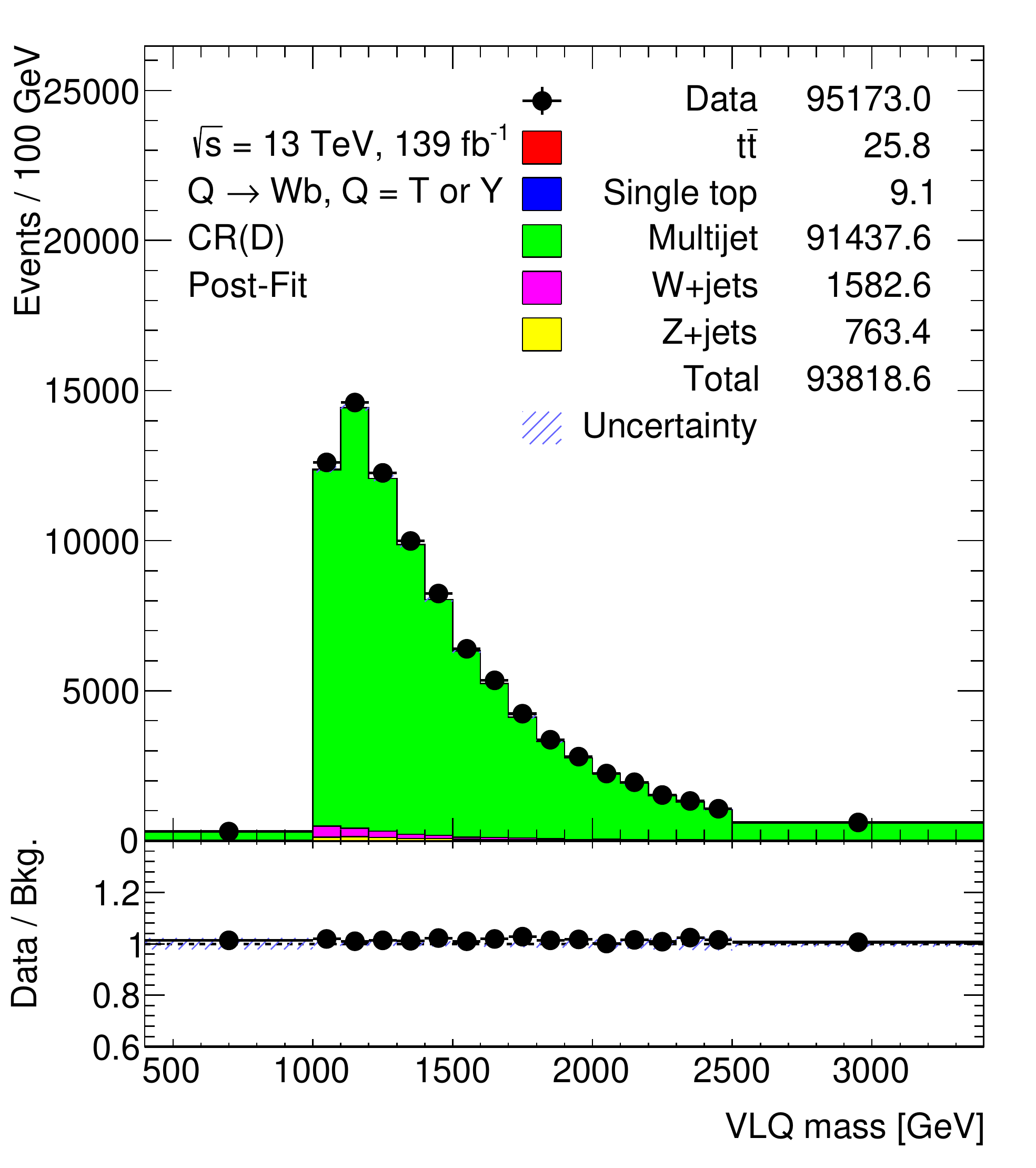}
		\caption{}
	\end{subfigure}
	\caption{VLQ mass distribution is shown when a likelihood fit is performed to fit regions C and D together in which the multijet MC is fitted to the data. It shows the VLQ mass distribution for (a) multijet MC and (b) scaled multijet MC in region C, and (c) multijet MC and (d) scaled multijet MC in region D.}
	\label{fig:app:cd:vlqm}
\end{figure}

Fig.\ \ref{fig:app:ab:vlqm} and Fig.\ \ref{fig:app:cd:vlqm} show the VLQ mass distribution in all the four regions A, B, C and D when the likelihood fit is performed on region A, B together and region C, D together. These distributions show both the multijet MC (which are mismodelled) and the scaled multijet MC.

\begin{figure}[hbt!]
	\centering
	\graphicspath{{figs/appendix/scaledmultijet/ABljetpt/}}
	\begin{subfigure}{.35\textwidth}
		\centering
		\includegraphics[width=\linewidth,height=\textheight,keepaspectratio]{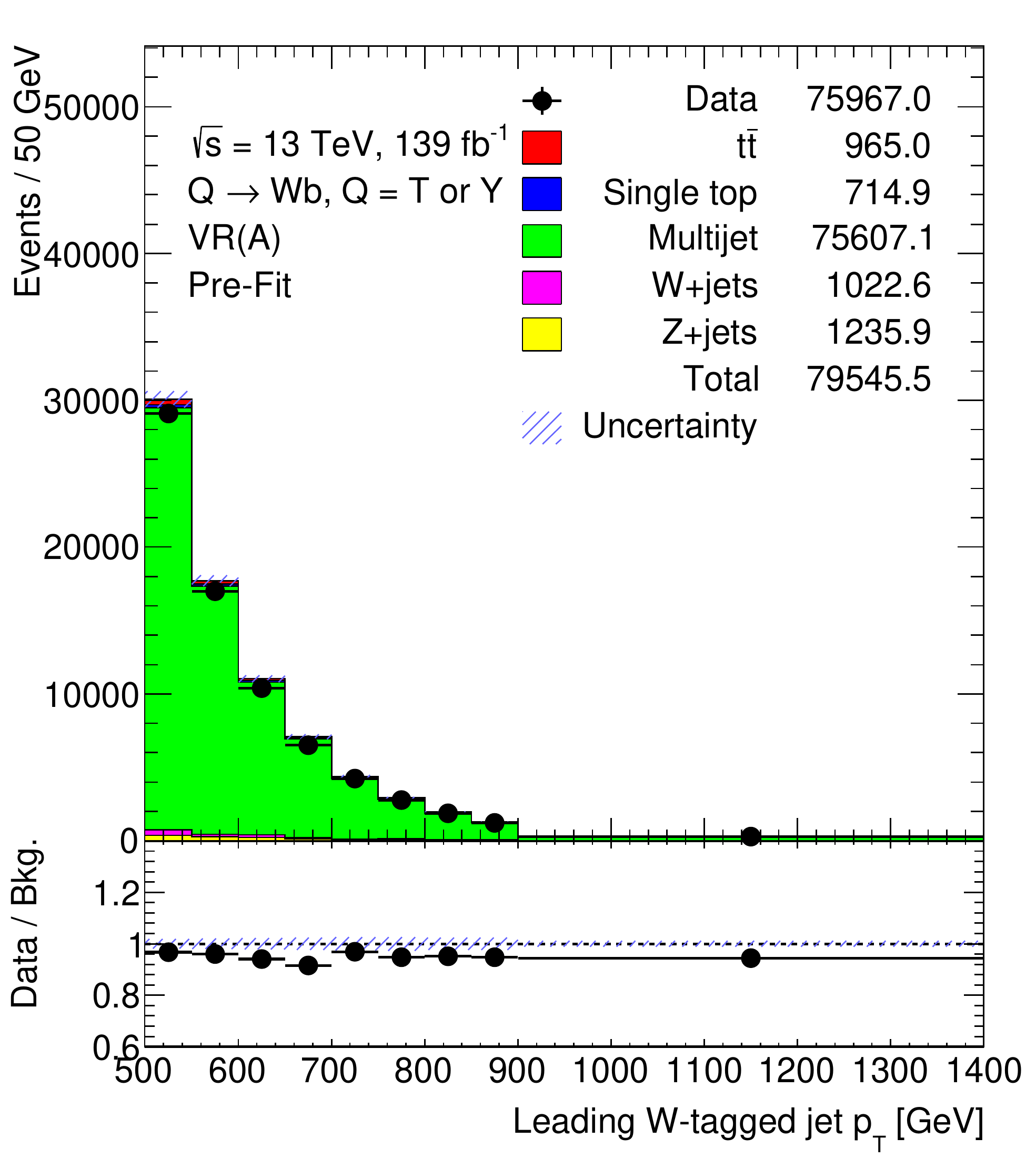}
		\caption{}
	\end{subfigure}\hspace{0.6cm}
	\begin{subfigure}{.35\textwidth}
		\centering
		\includegraphics[width=\linewidth,height=\textheight,keepaspectratio]{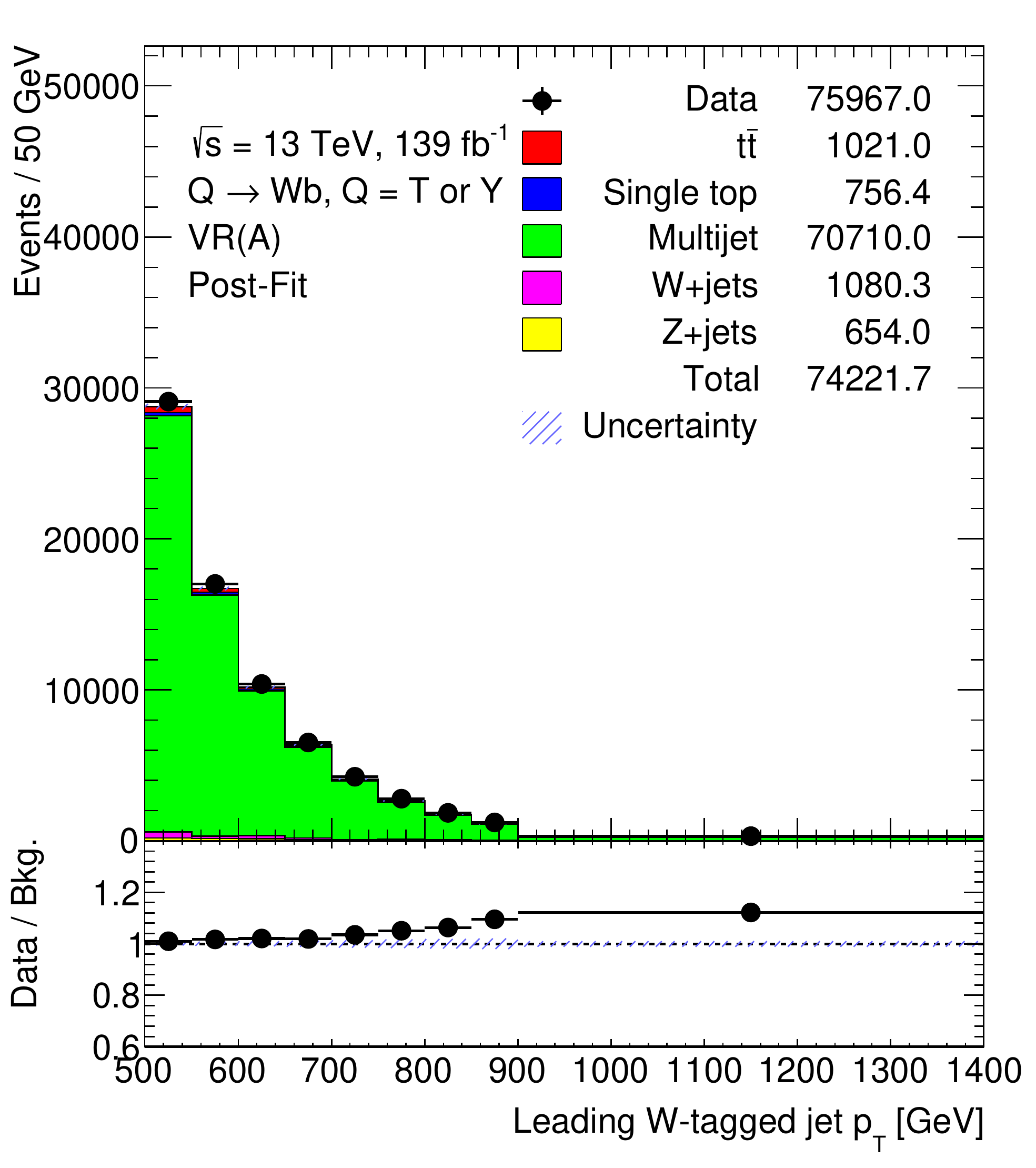}
		\caption{}
	\end{subfigure}
	\begin{subfigure}{.35\textwidth}
		\centering
		\includegraphics[width=\linewidth,height=\textheight,keepaspectratio]{CR_B_ljet_pt.pdf}
		\caption{}
	\end{subfigure}\hspace{0.6cm}
	\begin{subfigure}{.35\textwidth}
		\centering
		\includegraphics[width=\linewidth,height=\textheight,keepaspectratio]{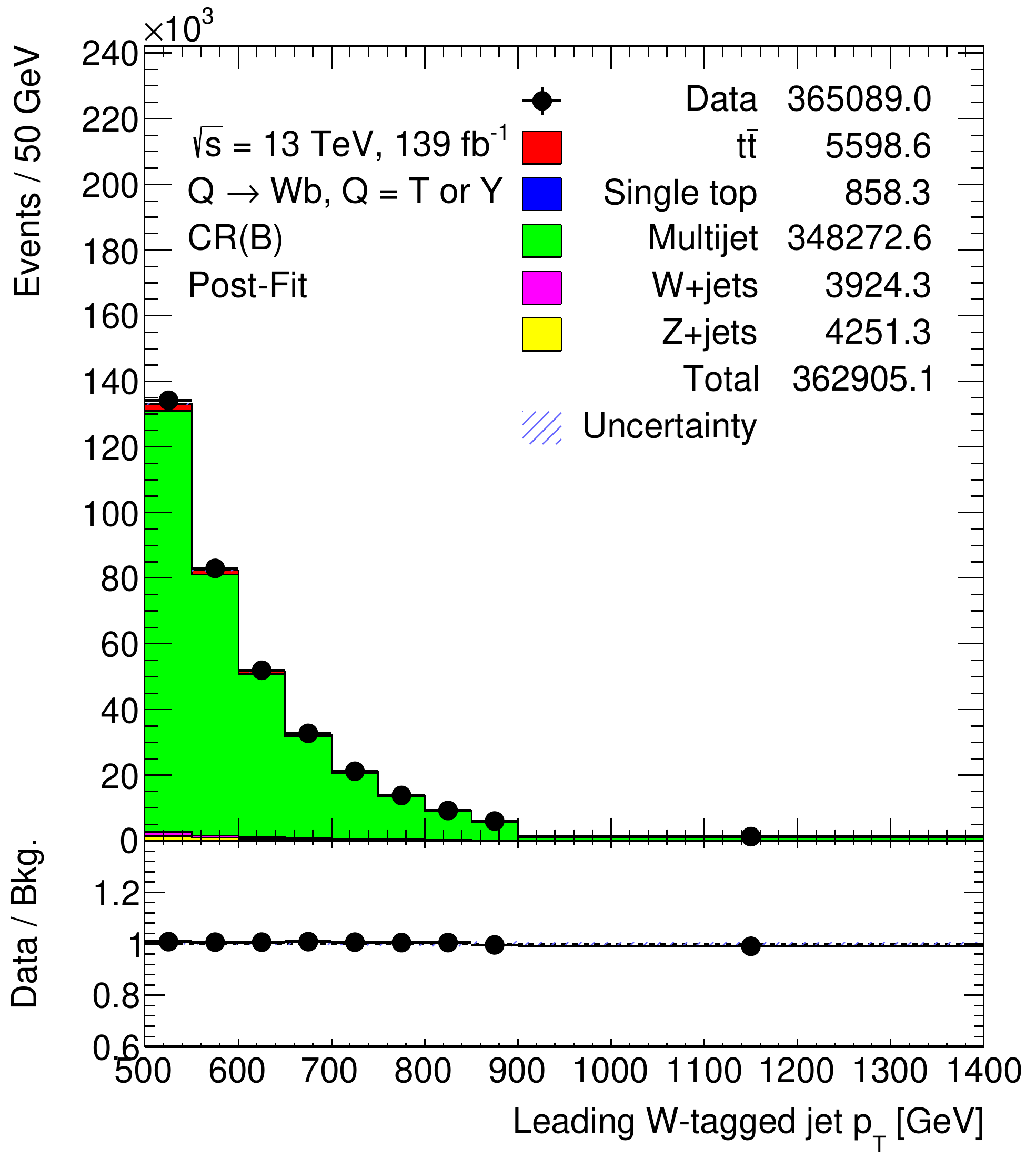}
		\caption{}
	\end{subfigure}
	\caption{\pt distribution of $W$-tagged jet is shown when a likelihood fit is performed to fit regions A and B together in which the multijet MC is fitted to the data. It shows \pt distribution of $W$-tagged jet for (a) multijet MC and (b) scaled multijet MC in region A, and (c) multijet MC and (d) scaled multijet MC in region B.}
	\label{fig:app:ab:ljetpt}
\end{figure}

\begin{figure}[hbt!]
	\centering
	\graphicspath{{figs/appendix/scaledmultijet/CDljetpt/}}
	\begin{subfigure}{.35\textwidth}
		\centering
		\includegraphics[width=\linewidth,height=\textheight,keepaspectratio]{CR_C_ljet_pt.pdf}
		\caption{}
	\end{subfigure}\hspace{0.6cm}
	\begin{subfigure}{.35\textwidth}
		\centering
		\includegraphics[width=\linewidth,height=\textheight,keepaspectratio]{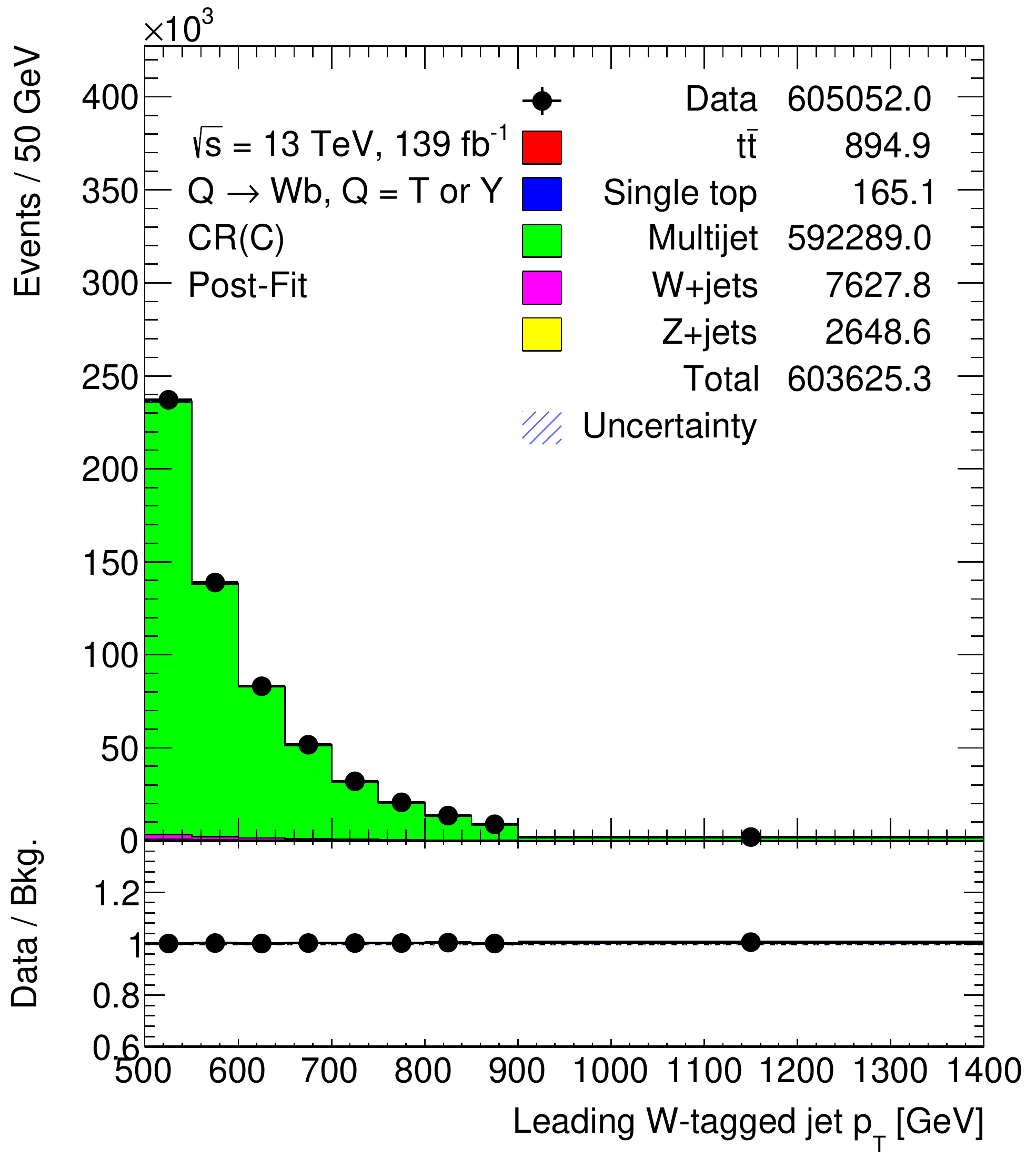}
		\caption{}
	\end{subfigure}
	\begin{subfigure}{.35\textwidth}
		\centering
		\includegraphics[width=\linewidth,height=\textheight,keepaspectratio]{CR_D_ljet_pt.pdf}
		\caption{}
	\end{subfigure}\hspace{0.6cm}
	\begin{subfigure}{.35\textwidth}
		\centering
		\includegraphics[width=\linewidth,height=\textheight,keepaspectratio]{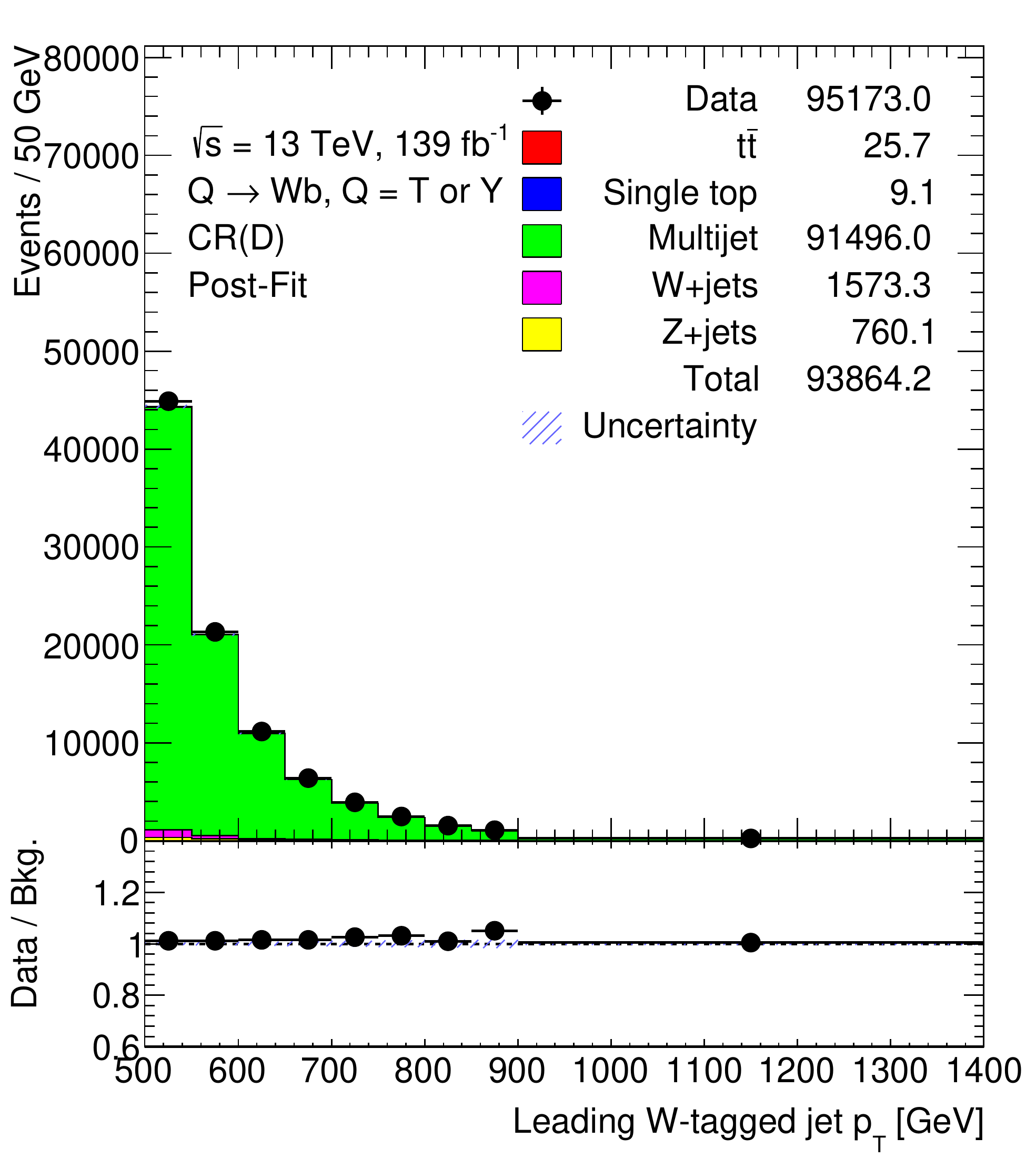}
		\caption{}
	\end{subfigure}
	\caption{\pt distribution of $W$-tagged jet is shown when a likelihood fit is performed to fit regions C and D together in which the multijet MC is fitted to the data. It shows \pt distribution of $W$-tagged jet for (a) multijet MC and (b) scaled multijet MC in region C, and (c) multijet MC and (d) scaled multijet MC in region D.}
	\label{fig:app:cd:ljetpt}
\end{figure}

Fig.\ \ref{fig:app:ab:ljetpt} and Fig.\ \ref{fig:app:cd:ljetpt} show the \pt distribution of $W$-tagged jet in all the four regions A, B, C and D when the likelihood fit is performed on region A, B together and region C, D together. These distributions show both the multijet MC (which are mismodelled) and the scaled multijet MC.

\backmatter
\listoffigures
\listoftables




\end{document}